\documentclass[10pt]{book}
\usepackage{cite}
\usepackage{epsfig}
\usepackage{amsfonts}
\usepackage{amssymb}

\def    \beq           	{\begin{equation}}
\def    \eeq           	{\end{equation}}
\def    \bea           	{\begin{eqnarray}}
\def    \eea           	{\end{eqnarray}}
\def    \nn            	{\nonumber}  
\def    \raw           	{\rightarrow}
\def    \Raw           	{\Rightarrow}

\def    \lraw          	{\leftrightarrow}
\def    \eps           	{\epsilon} 
\def    \ov            	{\overline}
\def    \wh           	{\widehat}

\def	\med		{\frac{1}{2}}
\def	\oh		{\frac{1}{2}}
\def	\sig		{\sigma}
\def	\Sig		{\Sigma}
\def 	\bi		{\be\begin{array}{rl}}
\def 	\ba		{\be\begin{array}}
\def 	\bis		{\be\begin{array}{c}}
\def	\bu		{\end{array}}
\def	\lam		{\lambda}
\def	\Lam		{\Lambda}
\def	\om		{\omega}
\def	\th		{\theta}
\def	\vt		{\vartheta}

\def	\a		{\alpha}
\def 	\O		{\Omega}
\def 	\Om		{\Omega}
\def 	\OR		{\Omega{\cal R}}
\def 	\R		{{\cal R}}

\def	\vp		{\varphi}
\def	\G		{\Gamma}
\def	\g		{\gamma}
\def	\F		{{\cal F}}
\def	\N		{{\cal N}}
\def	\M		{{\cal M}}
\def	\A		{{\cal A}}
\def	\S		{{\cal S}}

\def	\b		{\beta}
\def	\p		{\partial}

\def	\d		{\delta}
\def	\D		{\Delta}
\def 	\ti		{\times}
\def 	\inte		{{\bf Z}}
\def 	\nat		{{\bf N}}
\def 	\re		{{\bf R}}
\def 	\cpx		{{\bf C}}
\def 	\B		{{\bf B}}
\def 	\I		{{\bf I}}
\def	\real		{{\bf R}}
\def 	\q		{\vec{q}}
\def 	\tq		{\vec{\tilde q}}
\def	\arr		{\arrowvert}
\def	\ce		{{\cal E}}
\def	\cl		{{\cal L}}
\def	\cw		{{\cal W}}
\def	\preal		{{\rm Re\,}}
\def	\pim		{{\rm Im\,}}
\def	\k		{\kappa}

\def	\Tr		{{\rm Tr \,}}

\begin{document}

\begin{titlepage}
\pagestyle{empty}

\vspace*{1cm}
\begin{center}
{\Large {DEPARTAMENTO DE F\'ISICA TE\'ORICA}}\\
\vspace{0.2cm}
{\Large {UNIVERSIDAD AUT\'ONOMA DE MADRID}}\\
\end{center}

\vspace*{3cm}
\begin{center}
        \begin{tabular}{c}
       {\Huge {\bf Intersecting D-brane Models}}\\
	\end{tabular}
\end{center}
               
\vspace{11cm}

\begin{center}
{\large{Fernando G. Marchesano Buznego}}
\end{center}
\end{titlepage}

\newpage

\thispagestyle{empty}
\

\newpage

\begin{titlepage}
\pagestyle{empty}

\vspace*{1cm}
\begin{center}
{\Large {DEPARTAMENTO DE F\'ISICA TE\'ORICA}}\\
\vspace{0.2cm}
{\Large {UNIVERSIDAD AUT\'ONOMA DE MADRID}}\\
\end{center}

\begin{figure}[ht]
\centering
\epsfxsize=1in
\epsffile{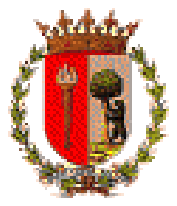}
\end{figure}

\begin{center}
        \begin{tabular}{c}
       {\Huge {\bf Intersecting D-brane Models}}\\
	\end{tabular}
\end{center}
               
\vspace{10cm}

\begin{center}
Memoria de Tesis Doctoral presentada ante la Facultad de Ciencias, \\
Secci\'on de Ciencias F\'{\i}sicas, de la Universidad Aut\'onoma de Madrid, \\
por {\bf Fernando G. Marchesano Buznego}\\
\vspace{0.2cm}
Trabajo dirigido por el {\bfseries{Dr. D. Luis E. Ib\'a\~nez Santiago}} \\
{{Catedr\'atico del Departamento de F\'{\i}sica Te\'orica}}\\
de la Universidad Aut\'onoma de Madrid, \\
\vspace{0.3cm}
{{Madrid, Mayo de 2003.}}\\
\end{center}
\end{titlepage}    

\newpage

\thispagestyle{empty}
\

\newpage

\newpage

\pagenumbering{roman}
\tableofcontents

\newpage

\ 
\newpage

\setcounter{page}{1}

\pagenumbering{arabic}


\chapter{Introduction}

The history of Theoretical Physics is full of interesting and exotic items, one of which is the birth of String Theory. Such birth took place in the late sixties, during the quest of a model describing hadron scattering amplitudes. Among the proposals that emerged on such research was the one from Veneziano \cite{Veneziano:yb}, who proposed an amplitude which resembled in several points the qualitative behaviour of hadrons at accelerator experiments. Such amplitude was generalized in the following years, and even it was concluded that it could naturally arise from a theory of relativistic strings \cite{Nambu:wf,Nielsen,Susskind:1970xm}. By the mid seventies such string theories had been studied and generalized, and lots of interesting theoretical results were already derived.

It was also around 1974 when the Parton Model and Quantum Chromodynamics were discovered. These theories offered a very accurate description of experimental data in the perturbative regime, then becoming a much more exiting possibility for describing strong interactions than the Veneziano proposal. By that time, however, a new possibility had emerged. Namely, it was conceived that a theory of strings could describe the world at a more fundamental level. The previous proposal was modified, and String Theory was presented as a model describing all the fundamental interactions of Nature. Nowadays, such proposal constitutes one of the few serious candidates on the unification of fundamental forces, or Theory of Everything (TOE).

Considering a theory of strings as a model describing Nature had, as a direct consequence, upgiving one of the fundamental assumptions of the quantum field theories considered so far. Indeed, in a theory of strings the fundamental object is not a point particle, but a one-dimensional object 
\footnote{Such a one-dimensional object has two possible topologies, corresponding to open and closed string theories.}.
The consequences of modifying such assumption are quite remarkable: 
althought more than thirty years have passed since such theories were seriously considered, every now and then unexpected theoretical discoveries appear, completely changing our understanding of the theory.

\begin{figure}[ht]
\centering
\epsfxsize=5.5in
\epsffile{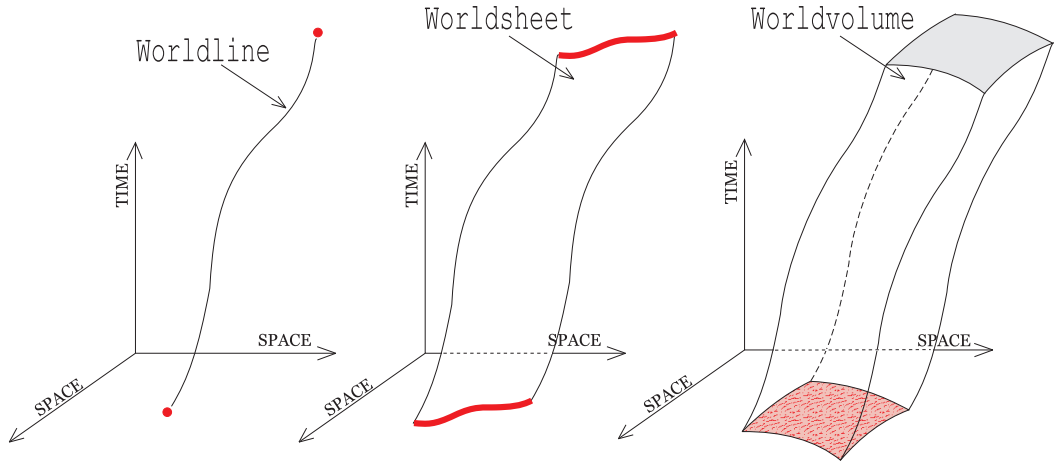}
\caption{Particle, open string and membrane propagation through spacetime, sweeping a one, two and three-dimensional trajectory on $M_{10}$, respectively. Figure taken from \cite{Duff:2001jp}.}
\label{psm}
\end{figure}
 
Is easy to understand that, if we expect String Theory to describe our universe, it should be possible to deduce from it the other theories that have been empirically tested, and that we know work rather well up to some rank of energies or distances. In particular, this implies that such theory of strings should recover in some definite limit the Standard Model of Particle Physics (SM) and Einstein's General Relativity. Moreover, up to energies reached at accelerators the fundamental matter components seem to have pointlike structure, so we should not `see' the stringy structure of Nature below some very high energy scale (stated in another terms, strings must have a typical length smaller than $10^{-18}$ m). In a string theory such condition is encoded in the only free parameter of the theory, which is the string scale. Such parameter is related to the intrinsic string tension of a classical string, so it tell us how much energy do we have to pay in order to stretch the string or make it vibrate.

We are thus led to the following picture. Fundamental physics is described by a quantum theory of one-dimensional objects (the strings) whose typical energy scale is above the ones reached experimentally. If we could reach this fundamental scale, we could then see the structure of such extended objects, such as the oscillating modes, etc. Much above these energies, physics would be described by an {\it effective field theory}, which is derived by considering the lightest oscillation modes of the string. 

One of the most important results on the seventies was discovering that the closed string sector of any string theory automatically includes gravitation. In particular, there exists an oscillation mode of a closed string propagating in the vacuum corresponding to a spin 2 particle which can be identified with the graviton \cite{Scherk:1974ca}. The Standard Model of Particle Physics should also be part of such effective theory and, ever since String Theory was postulated as a TOE, there have been a lot of work and effort devoted to find the SM `inside' a theory of strings. This is one of the most ambitious tasks involving String Theory, and essentially covers the branch known as String Phenomenology.

The reason of such quest for realistic physics to be so formidable is that, even if string theories are in some sense unique and do not admit free parameters, there is an infinite number of possible vacua where to define them. Around each of these vacua we should perform perturbative computations to be compared to low-energy observed physics. Not having a complete understanding of the non-perturbative dynamics of the theory, is impossible to know which of these vacua is the proper one.

Let us give an example of such problem. There exists five different theories of supersymmetric strings, also known as superstring theories, and each of them imposes a whole set of consistency conditions on the geometric background where a string probe should propagate. The so-called critical strings are those that, given these restrictions, do propagate in a ten dimensional Ricci-flat spacetime. Since our intention is to relate such constructions with the observed physics, we must then explain why we only see four of these ten dimensions in our everyday experience. A rather straightforward solution involves considering six of the spatial dimensions to be compact and very small, recovering the old ideas of Kaluza and Klein \cite{Kaluza:tu,Klein:tv}. However, the precise geometry and even the topology of these six compact dimensions is not fully determined by the above restrictions, and we find ourselves with a myriad of possible compactifications that, in principle, are all on equal footing.

On account of this situation, the simplest strategy seems to be studying the low energy physics of each of this infinity of vacua allowed by the theory. Once we have found a vacuum whose related perturbative physics is similar enough to the SM, we may ask ourselves why should it be preferred among all the possible vacua.

To sum up, if we expect String Theory to fulfill their mid seventies expectations and become Theories of Everything, it is necessary to find the SM at least in some specific vacuum configuration allowed by the theory itself.

The present work is based on one of the latest efforts towards this goal, the so-called Branes at Angles or Intersecting Brane World scenario. As we will see, intersecting branes contain several phenomenologically appealing general features, which suggest that they may be a promising scenario where to construct semirealistic superstring vacua. In particular, this class of configurations have provided for the first time the construction of explicit D-brane models which at low energies yield just the chiral spectrum and gauge group of the Standard Model.

The structure of the present work is as follows. In the remaining of this introduction we give a brief historical survey on the search of the Standard Model in the context of String Theory. We differentiate between the two periods which followed the two superstring revolutions. The aim of this survey is to motivate our work, as well as giving account of all the efforts made in the branch of Superstring Phenomenology. Finally, we give a short list of the features involving Intersecting Brane Worlds, and briefly argue why they may be good candidates for finding semi-realistic scenarios.

In the second chapter we introduce the basic class of objects upon the whole thesis is based: D-brane intersecting at angles. We first introduce them as non-compact higher dimensional objects in $M_{10}$, where we expose their most salient features. We then consider six different kinds of toroidal, orbifold and orientifold compactifications involving D-branes at angles, as well as the corresponding T-dual picture. We finally discuss the most general case of intersecting D-branes, for which many of the results of this thesis are valid.

The third chapter is dedicated to obtain the low-energy spectrum of the effective theory in these six different compactifications. We pay special attention to the massless open string sector of these theories, where unitary gauge groups and chiral fermions charged under them arise. For completeness, we also present (part of) the closed string sector, as well as some massive but light excitations which may put phenomenological constraints to a model.

The fourth chapter of the thesis is of more technical nature. There we study some consistency conditions that any D-brane model must satisfy in order to yield a consistent theory. In particular, we present the conditions Ramond-Ramond tadpole cancellation for the toroidal, orbifold and orientifold compactifications considered before. We also show how these conditions imply the cancellation of chiral anomalies in the effective theory, in some cases with the help of a generalized Green-Schwarz mechanism.

After developing this theoretical framework, we present some specific intersecting D-brane models in the fifth chapter. These are mainly D6 and D5-brane models yielding the chiral spectrum and gauge group of a three generation Standard Model, with no extra chiral matter nor $U(1)$ factors. In the case of D5-branes some Left-Right extensions of the Standard Model are also constructed. We discuss some phenomenological aspect of these models, paying special attention to model-independent features such as the structure of global $U(1)$ symmetries. Although we have achieved such spectrum in a very particular class of configurations involving D-branes at angles, we argue that the main results are also valid for more general D-brane constructions. In particular they hold for generic intersecting D-brane models, where the chiral spectrum and gauge groups can be encoded in terms of robust topological data.

We leave the study of supersymmetry for the sixth chapter. There we study the conditions for two non-compact D6-branes at angles to preserve a supersymmetry, which can be readily translated to toroidal and orientifold compactifications. We then briefly discuss the more general case of D6-branes wrapping 3-cycles on a Calabi-Yau threefold, with the help of calibrated and special Lagrangian geometry. Finally, we illustrate these concepts with a D6-brane example yielding a low-energy MSSM-like spectrum.

In the seventh chapter we address the computations of Yukawa couplings in Intersecting Brane Worlds. We first give a general view of this problem, and later compute the general form of such couplings for toroidal compactifications. There we show explicitly how Yukawas depend on the closed and open string moduli of the theory. We also comment on the deep relation of the computations of Yukawas in Intersecting Brane Worlds with Mirror Symmetry. 

We leave the last chapter for the conclusions of this work, whereas the appendixes contain technical details related to the main text.

Along this thesis we will use elements and notation already standard in string theory literature, and which are common to the basic texts on the field \cite{GSW1,GSW2,Bailin,Polchinski:1996na,Polchi1,Polchi2,Johnson:2000ch}. We refer the reader to such texts for greater detail and completeness on these basic aspects of the theory.

The present text mainly contains the research on Intersecting Brane Worlds where the author was involved in the last years, namely refs. \cite{Ibanez:2001nd,Cremades:2002dh,Cremades:2003qj}. It also contains some background taken from \cite{Blumenhagen:2000wh,Aldazabal:2000dg,Aldazabal:2000cn,Blumenhagen:2000ea}. The works \cite{Cremades:2002te,Cremades:2002cs} are also connected to this topic of research, and will form part of some other thesis \cite{Daniel}.

\section{The quest for the Standard Model}

Along the history of string theory, the more formal and the phenomenologically inclined branches of the field have been intimately related to each other. On the one hand, the motivation of having phenomenologically viable physics has driven purely theoretical research on one or another direction whereas, on the other hand, in several occasions a formal discovery has opened new possibilities to find realistic physics in the context of string theory. Two good examples of the latter are the celebrated string revolutions, each based in groundbreaking discoveries which influenced the whole understanding of the theory. On the following, we will briefly recall the attempts to find the SM in the context of string theory, following the historical line of these two string revolutions.

The First String Revolution took place around 1984, as Green and Schwarz discovered a new mechanism to formulate consistent superstring theories in ten dimensions \cite{Green:sg}. Until then, two such consistent theories have been constructed, namely type IIA and type IIB superstring theories. Both involved closed strings only, and the effective field theories derived from the low energy spectrum amounted to the two different $\N = 2$ Supergravity (SUGRA) theories in ten dimensions, named in the same manner. Both of these effective theories are free of inconsistencies such as chiral, mixed and gravitational anomalies. On the contrary, the superstring theory known as type I, which involved both open and closed strings, seemed to have such quantum anomalies. With the discovery of the Green-Schwarz mechanism, however, it was possible to show that if type I theory was endowed with a Yang-Mills theory with gauge group $SO(32)$, then the anomalies could factorise and be canceled, finally obtaining a consistent effective theory.

Soon after this the heterotic superstring theories were discovered, which are closed string theories whose construction is based on an hybrid of the superstring and the bosonic string \cite{Gross:1984dd,Gross:1985fr,Gross:1985rr}. These two theories involve a 10D SUGRA $\N=1$ effective theory and, as type I theory, are endowed with gauge groups which are, respectively, $SO(32)$ and $E_8 \ti E_8$, so that their construction also involves the Green-Schwarz mechanism.

The discovery of $E_8 \ti E_8$ heterotic string theory caused a big impact in the scientific community, for it was not only an example of a consistent quantum theory containing gravity, but also contained a gauge group big enough to nicely accommodate as a subgroup $SU(3) \ti SU(2) \ti U(1)$, i.e., the gauge group of the Standard Model. It was at this precise moment when string theory was sought as a candidate for a consistent description of all the fundamental physical interactions.

The first attempts to build realistic string models, then, were based in $E_8 \ti E_8$ heterotic string compactifications. A phenomenologically viable compactification requires obtaining and effective theory in four dimensions with a chiral spectrum and a gauge group containing $SU(3) \ti SU(2) \ti U(1)$. Since gravity was also to be a part of the low energy spectrum, the string scale was fixed at the order of the Planck scale, and the hierarchy problem was avoided by imposing local $\N = 1$ supersymmetry (SUSY). As it was shown in \cite{Candelas:en,Nemeschansky:1986yx}, such conditions required the six {\it extra} dimensions to fulfill some constraints, such as being a compact Riemannian manifold $\M$ with $SU(3)$ holonomy group. Such manifolds are known as Calabi-Yau threefolds, or ${\bf CY_3}$ \cite{Greene:1996cy,Huebsch:nu,Joyce:2001xt}.

Since, typically, the string scale $M_s \sim M_P$ and also the volume of $\M$ is of this order of magnitude, both the quantum massive oscillations of a fundamental string and the Kaluza-Klein model represent degrees of freedom that will decouple from the low energy physics. Our effective field theory will then be constructed from the massless string modes. By an analysis of the anomalies that such 4D theories could develop, it was seen that a simple way of obtaining an anomaly free effective theory was identifying the manifold $\M$ spin connection with the $E_8 \ti E_8$ gauge bundle connection. It is then necessary to identify the $SU(3)$ holonomy group with a subgroup of $E_8 \ti E_8$, in particular with a subgroup of one of the $E_8$'s. This naturally breaks the previous gauge symmetry down to the subgroup which commutes with the embedding of $SU(3)$, that is, down to $E_6 \ti E_8$. The group $E_6$ contains the SM gauge group being, in fact, the gauge group of one of the proposals known as Grand Unification Theories (GUT) \cite{Pati:1974yy,Georgi:sy}. In particular, a whole family of quarks and leptons can be embedded in the representation {\bf 27} of such group. The remaining $E_8$ is to be seen as a {\it hidden sector} of the theory, coupled to the visible sector only through gravitational interactions.

Over the infinity of possible Calabi-Yau manifolds, we would like to select those whose phenomenological features resemble the SM. For instance, it can be seen that the net number of representations {\bf 27} that appear under compactification is a topological quantity given by $\med |\chi (\M)|$, where $\chi$ is the Euler characteristic of the manifold. So, if we wish to construct a compactification yielding three families of quarks and leptons we may better stick to ${\bf CY_3}$ manifolds with Euler characteristic $\pm 6$. There are still other phenomenological restrictions that can be encoded into topological information of $\M$ such as the fundamental homotopy group $\pi(\M)$, which measures the ability of breaking $E_6$ to a more realistic gauge group by means of Wilson lines. For a more complete discussion on $E_8 \ti E_8$ heterotic compactifications on ${\bf CY_3}$ see \cite{GSW2,Polchi2}. For the realisation of an explicit model with three generations see \cite{Greene:bm,Greene:1986jb}.

Although $E_8 \ti E_8$ heterotic compactifications on Calabi-Yau manifolds have provided rather realistic models, it is difficult to perform computations in such manifolds where, in most cases, not even the metric is known. An interesting class of spaces where to compactify and where computations are much more tractable is given by toroidal orbifolds. A toroidal orbifold is defined by quotienting a toroidal manifold, in our case $T^6$, by a discrete group $P$ of isometries named point group. Such point group must act cristallographically on the lattice of translations $\Lam$ which defines the torus as $T^6 = \re^6/\Lam$, that is, its geometrical action must be well-defined on the torus. In general, such geometrical action will have some fixed points, where the orbifold quotient $T^6/P$ will develop a singularity. Even if this implies having a singular compact space, which would lead to some divergences in a point particle field theory, string propagation turns our to be well defined on orbifolds \cite{Dixon:jw,Dixon:1986jc}. Since the geometry is by far more simple than that of a {\bf CY} and the metric is flat outside the singularities, computations can be more fully and more easily carried out, and quantities of physical interest are thus more easily computable. As a result, orbifold compactification has become one of the main techniques in order to find phenomenologically appealing models \cite{Ibanez:1987sn,Font:1989aj}.

\begin{figure}[ht]
\centering
\epsfxsize=2.5in
\epsffile{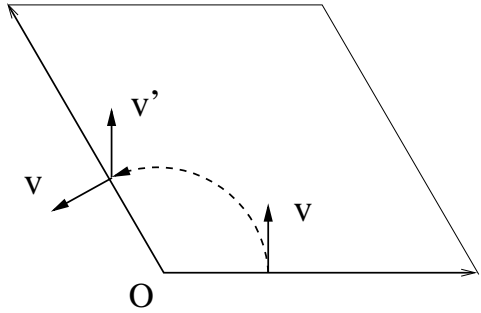}
\caption{$T^2/\inte_3$ orbifold. Both axis of the torus are identified under the action of the $\inte_3$ generator, which is a $2\pi/3$ rotation over the origin $O$. The vector $V$ is parallel transported to $V'$, which makes an angle of $2\pi/3$ with the image of $V$ under the orbifold action.}
\label{holon}
\end{figure}

The holonomy group of a toroidal orbifold $T^6/P$ is given by the embedding of the point group $P \subset SO(6)$ (see figure \ref{holon}). As in a ${\bf CY_3}$, if such discrete group is a subgroup of $SU(3)$ the four dimensional effective theory will yield $\N = 1$ SUSY. In fact, it turns out that toroidal orbifolds with this holonomy are nothing but a singular limit on the whole family of Calabi-Yau threefolds, where the curvature is concentrated on the fixed points.

Given that heterotic are theories of closed strings, the fields involved in the effective field theory will come from closed strings propagating on the compact manifold $\M$, as well as over the non-compact dimensions. In an orbifold compactification, a string can close its ends in several manners, providing several sectors to the theory. The {\it untwisted} sector consist of those strings which are closed both on the orbifold quotient and on the covering space $T^6$ from which it is constructed. On the contrary, the {\it twisted} sector of the theory is given by those strings which are open on $T^6$, but whose ends are related by the action of an element $g$ of the point group $P$ (see figure \ref{twisted}). Since these two points are identified in the quotient space $T^6/P$, this is a closed string on the orbifold. In general, each (conjugacy class of an) element $g \in P$ defines a different sector, the states corresponding to such sector propagating on the submanifold fixed under $g$. In particular, the four non-compact dimensions to be identified with $M_4$ are always fixed under each element. When the group element in question is the neutral element of $P$, which we will associate with the untwisted sector, the strings are allowed to propagate all over the ten dimensions or the full {\it target space}, in this case $M_4 \ti T^6/P$.

Generically, the untwisted sector will contain a $D=4$ $\N = 1$ supergravity multiplet and vector multiplets of the corresponding gauge group. In addition, chiral matter will usually be obtained from both the twisted and untwisted sectors. The toroidal orbifolds which have been studied in greater detail are those whose point group $P$ is abelian and of the form $\inte_N$ or $\inte_N \ti \inte_M$. One of the orbifolds with more promising phenomenology has turned out to be the simple $\inte_3$, which naturally yields matter triplication. For a detailed analysis of its spectrum see \cite{Ibanez:1987dw,Bailin}.

As we have seen, the first attempts to connect string theory with observed physics were based on a theory defined on ten dimensional, six of which were supposed to be compact and very small. Actually, these ten dimensions are needed only in order that their associated bosonic and fermionic degrees of freedom cancel the so-called {\it conformal anomaly}. Is however possible to to cancel such anomaly by means of degrees of freedom with no obvious geometrical interpretation, constructing the theory directly in four dimensions. This fact motivated a series of constructions, such as the Gepner models \cite{Gepner:1987qi} or the free-fermion models \cite{Kawai:1986ah,Antoniadis:1986rn}. Even if, in principle, these new constructions constitute a new family where to look at in the quest for realistic physics, many of the turned out to be equivalent to one of the previous ten-dimensional compactifications \cite{Font:1989gq}.

Any consistent heterotic compactification, either on a ${\bf CY_3}$ or on a toroidal orbifold, is faced to the problem of describing realistic physics where, necessarily, must present broken supersymmetry. Since we have explicitly defined a supersymmetric $D=4$ effective theory, it is impossible to generate a potential which breaks supersymmetry, at least perturbatively. Such supersymmetry breaking must then come from non-perturbative effects. However, string theory is currently formulated only at a perturbative level, and before the Second Superstring Revolution stringy non-perturbative phenomena were widely unknown. In this manner, SUSY breaking developed in terms of non-perturbative effective field theory physics.

Among these effects, the simplest and most studied is, probably, gaugino condensation. This proposal basically claims that the hidden sector of an heterotic compactification, whose gauge group is given by the unbroken $E_8$, becomes strongly interacting below some intermediate scale $\sim 10^{12}$ GeV. this provokes a gaugino condensation in this sector, which explicitly breaks SUSY. Since the hidden sector and the visible sector are coupled to each other only by means of gravitational interactions, communicating this effect will be suppressed, and the supersymmetry breaking will be felt in the visible sector at $\sim 10^{2}$ GeV, near the electroweak scale. See \cite{Quevedo:1996sv} for more details on gaugino condensation and other phenomenological aspects of these kind of compactifications.

\section{Dualities and D-branes}
The Second Superstring Revolution took place around 1995, and its main topic was non-perturbative aspects of string theory. Until then, string theory was understood as five different superstring theories, apparently independent, known as type I, type II (A and B) and the two heterotic theories. However, in the context of this second revolution it was learnt that they were all related to each other by a web of string dualities. Roughly speaking, a duality relates two a priori different theories, but whose derived physics is actually the same. In the special case of string theories, a duality establishes a one-to-one correspondence between parameters and fields defining one theory (compactification radii, coupling constants, etc.) and the same set of quantities defining its dual. A well-known example of these duality maps is T duality, which we will deal with in some detail below.

\begin{figure}[ht]
\centering
\epsfxsize=4.5in
\epsffile{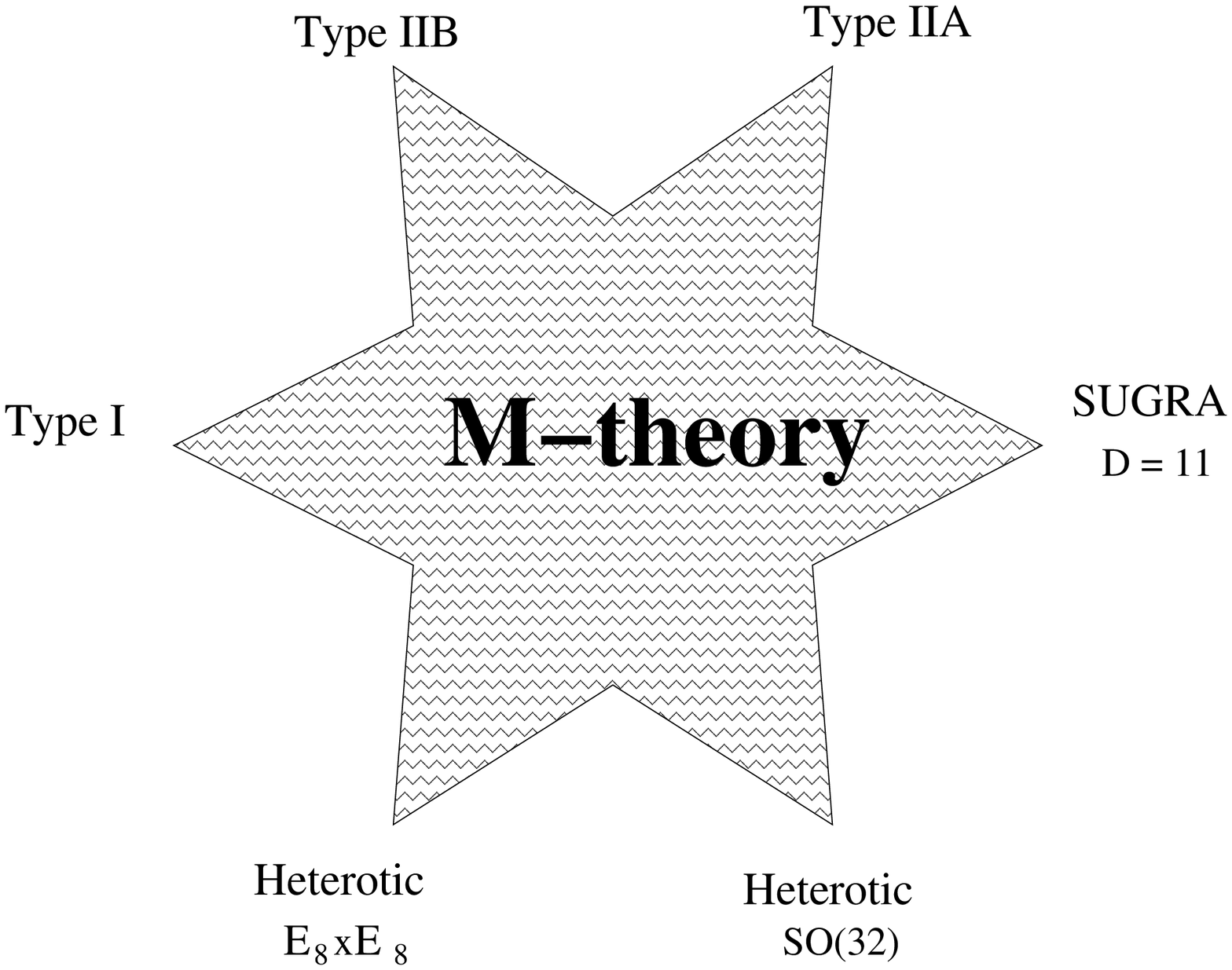}
\caption{Situation of string theory after the second superstring revolution. The previously disconnected five superstring theories are nothing but specific (limiting) points in the parameter space of a more fundamental theory: M-theory.}
\label{M2}
\end{figure}

The string duality web revealed that these five string theories were not isolated independent theories, but actually limiting cases of a deeper, more fundamental theory, named M-theory, whose precise nature has not yet been unraveled \cite{Witten:1995ex}. Such theory would be formulated in eleven spacetime dimensions, and its basic dynamical objects would be membranes rather than strings (see figure \ref{psm}). These membranes naturally appear as solitonic objects of $D=11$ SUGRA, which would be another limiting case of M-theory. The other five limiting cases, i.e., the five superstring theories, would then be obtained from compactifying the eleventh dimension in a very small length. Figure \ref{M2} shows an schematic representation of such situation.

Before this duality relations had been discovered, it had been possible to connect the infinity of supersymmetric vacua (i.e., Calabi-Yau compactifications) of the {\em same} superstring theory by simply varying the different compactification parameters. In this way, two different vacua of the same theory were connected by a continuous path in the space of compactifications. As a consequence of this new insight into stringy physics, it was possible to continuously connect every single string theory vacuum. This result is of undoubtedly theoretical interest. It was also was extremely appealing from the phenomenological point of view since, eventually, it could exist a dynamical mechanism that continuously selected one single vacuum from the whole set of superstring vacua.

Both in the formulation and in the evidence given of these new string dualities, non-perturbative objects of the theory such as the so-called D-branes played a central role. D-branes, D standing for Dirichlet, naturally emerge when considering a toroidal compactification of type I theory and performing a T-duality on one of the compact dimensions \cite{Polchinski:1996na}. Generically, a D$p$-brane is a BPS solitonic object of spatial dimension $p$ where the open strings localize their ends. Type IIA theory contains D$p$-branes with $p$ even, whereas on type IIB $p$ must be odd. One of the main features of these solitonic objects is that the supersymmetric effective theory arising from the worldvolume of a D-brane is endowed with a $U(1)$ gauge group. If we consider the possibility of having $N$ D-branes of the same kind on top of each other, then this gauge group will be $U(N)$. Moreover, the effective field theory is defined on the $p+1$ dimensions that the D-brane expands, and the fields on it are confined to propagate on such worldvolume.

The properties of these non-perturbative objects had a great relevance in many aspects of string theory and, in particular, for model building matters. Since a bunch of D-branes already contained Super Yang-Mills (SYM) gauge theories in their worldvolumes, it was not necessary to start from a string theory equipped with a gauge group, such as $E_8 \ti E_8$, in order to achieve a compactification containing the SM. This ended with the heterotic monopoly, and new semirealistic models started to be built based on type I and type II theories. Actually, the first consistent compactifications of type I theory on orbifold spaces were realised long time ago in \cite{Pradisi:1988xd,Bianchi:1990yu}, obtaining $D=6$ supersymmetric effective theories. In addition, such type I orbifolds were related with type II {\it orientifold} compactifications \cite{Sagnotti:1987tw,Horava:1989vt,Horava:1989ga,Dai:ua}. Roughly speaking, an orientifold is a generalization of the orbifold definition, where an element $\O$ changing string orientation is included. Such $D=6$ constructions were rediscovered in the modern language of D-branes in \cite{Gimon:1996rq,Dabholkar:1996zi,Gimon:1996ay}. Such class of compactifications was then generalized to orbifolds and orientifolds of type I and type II theories on six compact dimensions, yielding $\N=1$ chiral theories in four dimensions \cite{Berkooz:1996dw,Angelantonj:1996uy,Kakushadze:1997wx,Kakushadze:1997ku,Kakushadze:1997uj,Zwart:1997aj,Ibanez:1998xn,Aldazabal:1998mr}. At the same time, such compactifications were related with their heterotic duals. Some semi-realistic models were achieved in this particular context \cite{Kakushadze:1998ss,Kakushadze:1998dc,Lykken:1998ec}. A review of the phenomenology associated to these constructions can be found in \cite{Ibanez:1997rf}.

D-brane constructions not only allowed to rederive the previous achievements of heterotic compactification, but its properties as extended objects gave new possibilities into semi-realistic model-building, allowing to consider non-supersymmetric models. In heterotic models, both gauge and gravitational interactions arise from the untwisted sector of the theory, so they correspond to fields that propagate through the whole target space. If we wish to reproduce two energy scales which differ by several orders of magnitude, such as the Planck and the Electroweak scale, we need to have in our theory a mechanism such as supersymmetry, which prevents this hierarchy of scales to be destabilized by radiative corrections. If, alternatively, our gauge theory arises from a D$p$-brane with ($p < 9$), then gauge and gravitational interactions will propagate in different volumes. The gauge theory is confined to the $p + 1$ dimensions of the D-brane worldvolume, whereas gravitation, arising from the closed string sector, will propagate on the full ten-dimensional target space or {\it bulk} of the theory. As it was shown in \cite{Lykken:1996fj,Arkani-Hamed:1998rs,Antoniadis:1998ig,Shiu:1998pa}, from this simple observation we can obtain a difference of scales between gauge and gravitational interactions. In particular, we can obtain realistic compactifications where the string scale $M_s$ should not necessarily be of the order of the Planck scale, but as low at the TeV region or at some intermediate scale \cite{Ibanez:1998rf,Ibanez:1999bn}. In this way, we can consider non-supersymmetric models free from the scale hierarchy problem. Non-supersymmetric orientifold compactifications were first constructed in \cite{Antoniadis:1999xk,Aldazabal:1999jr}, whereas the semi-realistic models and the phenomenology associated to them were provided in \cite{Aldazabal:1999tw,Aldazabal:2000sk}.

The theoretical development in these new class of constructions, where D-branes played a central role, allowed to take one step further in semirealistic model building. So far, the quest for the SM had been based on considering a family of consistent compactifications in a certain superstring theory (as, e.g., ${\bf CY_3}$ heterotic compactifications) and exploring the parameter or moduli space of such family (Euler characteristic, Wilson lines, etc.) looking for a low energy theory which resembled as much as possible to the Standard Model. In \cite{Aldazabal:2000sa} a new strategy for finding the SM in a string-based model was proposed. Since the gauge group and chiral matter content of the Standard Model may arise as an effective theory from a set of D$p$-branes, and the physics of this (partial) effective theory is not very sensitive to the rest of the details of the compactification, we may conceive the construction of a realistic model in two steps. First, we will bother about a consistent D-brane configuration with the low-energy spectrum of the SM, either some field theory extension of it. Second, we will care about completing the construction, adding all the extra elements necessary to yield a fully-fledged compactification, including a four-dimensional gravity. This so-called {\it bottom-up} philosophy, enabled to find the simplest semi-realistic models up to that point. In such models, the SM was obtained from a bunch of D3-branes filling four-dimensional Minkowski spacetime and localized at an orbifold singularity in the compact space. Consistency conditions known as {\it tadpole} conditions imposed the presence of additional D-branes, namely D7-branes. The bottom-up philosophy can be applied in a wider class of models than D-branes at singularities. In particular, we will apply this idea in the construction of semi-realistic models of Chapter 5.

\section{Intersecting brane worlds}

The bottom-up philosophy has indeed produced a whole set of D-brane models whose semi-realistic effective theories contain either the Standard Model gauge group, either some extension of it. The nearest constructions to yield SM physics had achieved the following important features
\begin{itemize}

\item Chiral spectrum

\item Gauge group containing $SU(3) \ti SU(2) \ti U(1)$

\item Supersymmetry $\N = 1$ or $\N = 0$

\item Three quark-lepton generations

\end{itemize}
Although these general features get us quite close to obtain a realistic D-brane construction, in any of the previous models one always find extra chiral fermions and $U(1)$ gauge groups in the low energy spectrum. The usual procedure in the literature in order to get rid of these undesirable field content, is to abandon string theory techniques and use instead the low-energy effective Lagrangian. Then one tries to find some scalar field such that, giving it a v.e.v. in the appropriate direction, all extra gauge symmetries are broken and extra fermions become massive. Although this procedure is in principle valid, this requires a very complicated model-dependent analysis of the structure of the scalar potential and Yukawa couplings and, usually, the necessity of unjustified simplifying assumptions. It is clear that it would be nicer to find consistent string constructions with massless chiral spectrum identical to that of the SM and the gauge group just $SU(3) \ti SU(2) \ti U(1)$, everything already at the string theory level.

As we will see in the following chapters, configurations of D-branes at angles and, more generally, intersecting D-brane models are the first example of such constructions. Generically, these configurations yield a non-supersymmetric chiral low-energy spectrum. Each stack of $N$ D-branes will be endowed with a $U(N)$ gauge theory, so that the construction of the SM gauge group or some extension of it basically reduces to consider the appropriate set of D-brane stacks. Let us consider two different stacks of D-branes, each of them having, respectively, $N_a$ and $N_b$ D-branes on top of each other. This will give us the gauge group $U(N_a) \ti U(N_b)$ and, if these two stacks intersect at a single point in the compact six-dimensional space, then a chiral fermion will appear at their intersection, transforming in the bifundamental representation $(N_a, \bar{N_b})$ \cite{Berkooz:1996km}. From these two simple facts we will be able to construct semi-realistic models containing just the desired spectrum. 

These class of models, baptized as {\em Intersecting Brane Worlds} will present an interesting hierarchy on the different sectors of the effective theory. Namely,
\begin{itemize}

\item The gravity sector will propagate on the whole target space. That is, on the four non-compact dimensions and on the six compact dimensions.

\item The gauge sector will be confined to the D-brane worldvolumes, which will fill the four non-compact dimensions and a submanifold of the compact space.

\item Chiral matter will be localized at D-branes intersections so, generically, they will fill the non-compact dimensions and will be stuck at a point in the compact space.

\end{itemize}
This natural hierarchy will allow to implement the low string scale scenario discussed above, as well as consider non-supersymmetric models. Finally, intersecting brane worlds will provide an scenario where to address some well-known phenomenological problems and features of SM physics, translating them to a more geometrical language. Among others, we will see how one can address anomaly cancellation, family triplication, global $U(1)$ symmetries of the theory, proton stability, supersymmetry breaking by D-terms, computation of the holomorphic gauge function and computation of the Yukawa couplings.

\chapter{Branes intersecting at angles \label{atangles}}

In this chapter we introduce the basic class of objects upon the whole thesis is based: D-branes intersecting at angles. In order to study some of their salient features, which motivate their role as building bocks of semirealistic string-based constructions, we will first study the simplest case of two D-branes filling intersecting hyperplanes in euclidean space. We will incorporate this system into compact setups by means of considering simple toroidal compactifications and more complicated orbifold and orientifold twists of them. Finally, we will briefly discuss its generalization to intersecting branes in arbitrary compactifications.

The system of two type II flat D-branes intersecting at general angles was first studied in \cite{Berkooz:1996km,Arfaei:1996rg,Balasubramanian:1996uc,SheikhJabbari:1997cv}. In these works, the main interest was finding the conditions for this system to be supersymmetric, as well as the amount of supersymmetry that they could preserve. The full spectrum at the D-brane intersection was computed for this task, finding that it would generically contain a chiral fermion on it. Here we will focus on this last characteristic, which is essential for any semi-realistic effective theory and motivates the use of intersecting branes in model-building. We will leave the study of supersymmetry for chapter 6.

\section{Flat intersecting D-branes}

\begin{figure}[ht]
\centering
\epsfxsize=6in
\epsffile{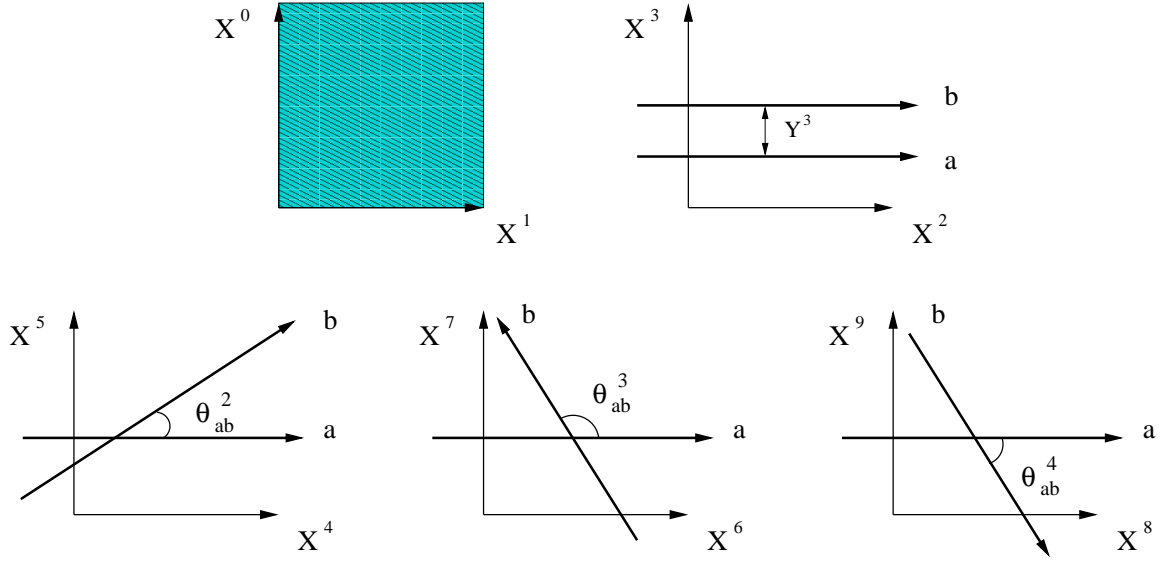}
\caption{Two D5-branes intersecting at angles in $M_{10}$. Both D-branes are parallel in the $2^{nd}$ complex plane, having a separation $Y^3$ in such plane and vanishing intersection. The angles are to be measured as indicated in the figure, going from brane $a$ to $b$ in the $ab$ sector, with the counterclockwise sense yielding a positive angle.}
\label{anglesth}
\end{figure}

Let us consider the system composed by two D$p$-branes $a$ and $b$ of type II theory in flat, non-compact, Minkowski space $M_{10}$. Let us further assume that each expands an oriented $p+1$-hyperplane, $p>1$, including the timelike  direction ($\mu = 0$) and a common spatial direction (say $\mu = 1$). Any such D$p$-branes is a $\med BPS$ soliton of the corresponding type II theory, so it is a stable object. In general, these two hyperplanes will be related by an isometry belonging to $(SO(8)/SO(p-1)) \ti T_{9-r}$, where $T_s$ is the group of translations in $s$ dimensions, and $r$ the dimension of the direct sum of both tangent spaces. Let us consider a vector basis where the rotation $R \in SO(8)/SO(p-1)$ can be diagonalized. In particular, let us choose complex coordinates $Z^\mu = X^{2\mu} + i X^{2\mu + 1}$ such that we can express such rotation as
\beq
R : Z^\mu \mapsto e^{\pi i \vt^\mu} Z^\mu.
\label{rotation}
\eeq
By our previous choices, $\vt^{0} = 0$ and $R$ can then be seen as an element of the subgroup $U(4) \subset SO(8)$, i.e., a rotation that preserves the complex structure $Z^\mu$. The rotation angles $\vt$, which we have chosen to measure in units of $\pi$, specify the system in this particular basis. Such situation is illustrated in figure \ref{anglesth} for the particular case $p=5$, $r = 8$.

Notice that we are dealing with oriented strings, so we can distinguish where the open strings begin and end. Let us call the $ab$ sector the one containing string oscillations with boundary conditions beginning ($\sig = 0$) at $a$ and ending ($\sig = \pi$) at $b$. The mass operator for such sector is \cite{Arfaei:1996rg}:
\beq
\alpha' M_{ab}^2 = {{Y^2}\over {4\pi^2\alpha^\prime}} + N_\nu + \nu \left(\sum_{\mu=1}^4 \vt_{ab}^{\ \mu} - 1 \right)\ ,
\label{mass}
\eeq
where $\alpha^\prime$ is the string tension, $Y^2$ is the squared length of the transverse separation of the two D-branes, and $N_\nu$ stands for the number operator of the oscillations, which on the Ramond ($\nu = 0$) and Neveu-Schwarz ($\nu = \med$) sectors is given by
\bea
N_0 &  =  & 
\sum_{\mu=1}^4 \left[\sum_{n>0}\left( \alpha^\mu_{-n_+} \alpha^\mu_{n_+} + \alpha^\mu_{-n_-} \alpha^\mu_{n_-} \right) + \alpha^\mu_{-\vt^i} \alpha^\mu_{\vt^\mu}\right] \nonumber \\
& + & \sum_{\mu=1}^4 \left[\sum_{r>0}\left( r_+\psi^\mu_{-r_+} \psi^\mu_{r_+} + r_-\psi^\mu_{-r_-} \psi^\mu_{r_-} \right) + \vt^\mu\psi^\mu_{-\vt^\mu} \psi^\mu_{\vt^\mu}\right] \label{numero1} \\
N_\med &  =  & 
\sum_{\mu=1}^3 \left[\sum_{n>0}\left( \alpha^\mu_{-n_+} \alpha^\mu_{n_+} + \alpha^\mu_{-n_-} \alpha^\mu_{n_-} \right) + \alpha^\mu_{-\vt^\mu} \alpha^\mu_{\vt^\mu}\right] \nonumber \\
& + & \sum_{\mu=1}^4 \left[\sum_{r>0}\left( r_+\psi^\mu_{-r_+} \psi^\mu_{r_+} + r_-\psi^\mu_{-r_-} \psi^\mu_{r_-} \right)\right] 
\label{numero2}
\eea
In the above formulae we are taking usual string theory notation. Operators $\a^\mu_{\pm n}$ y $\psi^\mu_{\pm r}$ come from the usual mode expansion of the bosonic $X^\mu$ and fermionic $\Psi^\mu$ worldsheet operators. Subindices stand for the corresponding oscillation mode: the bosonic index $n$ is always integer, whereas the fermionic index $r$ takes integer of half-integer values depending if we are on the Ramond or Neveu-Schwarz sector that is, $r \in \inte + \nu$. As before, $\mu$ indexes the target space complex dimensions. Notice that we have implicitly chosen the light-cone gauge, so that $\mu = 0$ does not appear in any expression above.  

The above number operators are quite similar to the ones arising from parallel branes, the only difference arising from non-trivial boundary conditions, which shift the oscillation modding as $n_\pm \equiv n \pm \vt^i$, $r_\pm \equiv r \pm \vt^i$ (we have simplified notation as $\vt^i \equiv \vt^i_{ab}$). As a consequence, creation and annihilation operators no longer have an integer or semi integer modding, just as in parallel or orthogonal D-brane systems, but they present a twist proportional to the corresponding angle. As expected, we recover the usual modding for parallel or orthogonal D-branes when choosing $\vt^\mu = 0, \med$. On the above expressions we have supposed $0 < \vt^\mu < 1$ so that oscillator modes are correctly normal ordered. For negative values one should replace $\vt^\mu \raw |\vt^\mu|$.

Just as in orbifold compactifications, this non-integer oscillator modding can be handled by means of bosonization language \cite{GSW1,Bailin}. This method give us an efficient and simple way of computing the open string spectrum, at least for the lightest part of it, which is the relevant part for our phenomenological purposes. In this language our rotation is encoded in the four-dimensional twist vector $v_\vt = (\vt^1,\vt^2,\vt^3,\vt^4)$ of, say, the $ab$ sector. Oscillation states are described by the sum $r + v_\vt$, where $r \in (\inte + \med +\nu)^4$. GSO projection \cite{Gliozzi:1976jf,Gliozzi:1976qd} is implemented by imposing $\sum_i r^i =$ odd, and the mass of any such state is given by
\beq
\a' M_{ab}^2 = {Y^2 \over 4\pi\a^\prime} + N_{bos}(\vt) + {(r + v_\vt)^2 \over 2} -\med + E_{ab},
\label{mass2}
\eeq
where $N_{bos}(\vt)$ stands for the bosonic oscillator contribution and $E_{ab}$ is the vacuum energy
\beq
E_{ab} = \sum_{\mu=1}^4 \med |\vt^\mu| (1 - |\vt^\mu|).
\label{vacuumenergy}
\eeq

Given these two formulae is easy to construct the associate low energy spectrum and see its explicit dependence on the four angles $\vt$. At this point, however, let us impose some phenomenologically motivated restriction which will set one of these four angles to zero. Indeed, as explained in the introduction, D-brane semirealistic models are usually realised by identifying some of the D-brane worldvolume dimensions with $M_4$. This implies that we have to deal with configurations of D$(3+n)$-branes, $n \in \nat$. In our specific context, we will take our four Minkowski dimensions to be identified with $(X^0, X^1, X^8, X^9)$ and every single D$p$-brane of a definite configuration will have to expand these four directions. This readily implies that a pair of D-branes will be related not by a general $SO(8)$ rotation but by an $SO(6)$ element (a $U(3)$ element in a suitably chosen complex basis). Put in a different way, we must set $\vt^4 \equiv 0$. In fact, the fourth entry of our vector $r + v_\vt$, which will be either an integer ($\nu = 0$) or half integer ($\nu = \med$) number, will provide the four-dimensional Lorentz quantum number of our state. Thus, a string oscillation on the Ramond (R) sector will yield four-dimensional fermions, whereas Neveu-Schwarz (NS) excitations will yield bosons from the four dimensional point of view.

Let us illustrate this by considering two intersecting D6-branes (i.e., $n = 3$) related by a $U(3)$ rotation with generic positive angles $(\vt^1, \vt^2, \vt^3, 0)$, $0 < \vt^i < 1$, $i = 1, 2, 3$. It is then easy to check that we will have a unique massless state coming from the Ramond sector, given by
\beq
r_R + v_\vt= ( -\med+\vt^1, -\med+\vt^2, -\med+\vt^3, +\med) ,
\label{fermion}
\eeq
which, in terms of four-dimensional physics, corresponds to a Weyl fermion of positive chirality. On the other hand, the lightest string excitations in the NS sector are given by
\beq
\begin{array}{c}
(-1+\vt^1, \vt^2, \vt^3, 0) \\
(\vt^1, -1+\vt^2, \vt^3, 0) \\
(\vt^1, \vt^2, -1+\vt^3, 0) \\
(-1+\vt^1, -1+\vt^2, -1+\vt^3, 0)
\end{array}
\label{tachyons}
\eeq
where we are still supposing $0 < \vt^i < 1$. These states will all correspond to $D=4$ scalars, as its associate helicity shows. Unlike the chiral fermion, their squared mass will depend on the specific value of the angles $\vt^i$, being positive, negative or null. The last case actually corresponds to a mass degeneracy with the fermionic sector of the theory, yielding a supersymmetric spectrum in the $ab$ sector. On the other hand, if the lightest scalar is not massless (which is the generic situation), then our low energy spectrum will be explicitly non-supersymmetric. In particular, a scalar with a negative squared mass corresponds to a tachyon in our $D=4$ theory, which physically signals an instability of the system. Actually, both the supersymmetry condition and the presence of a tachyon do have a clear geometrical meaning, which we will elucidate below (see chapters \ref{spectrum} and \ref{SUSY}).

Each of the $ab$ states (\ref{fermion}) and (\ref{tachyons}) are localized at the D-branes intersections \cite{Berkooz:1996km}, so both the chiral fermions and the scalars below are stuck at a single point in the dimensions additional to $M_4$. This accounts for the hierarchy of sectors discussed at the end of the introduction and which is characteristic of Intersecting Brane Worlds. Recall that, roughly speaking, this hierarchy arises because chiral matter, gauge interactions and gravitational interactions propagate in different dimensions of the target space.

\section{T-dual picture: D-branes with magnetic fluxes \label{Tdual}}

As mentioned in the introduction, in order to conceive a string-based model containing some semi-realistic physics in its low energy limit, we must compactify six of the nine spatial dimensions. Other features realistic physics demands involve a spectrum including chiral fermions, realistic gauge group and broken supersymmetry. In general, such characteristics are impossible to obtain from any of the five superstring theories under compactification on the simplest compact ${\bf CY_3}$, which is $T^6$. However, as was observed in \cite{Bachas:1995ik}, turning on a non-vanishing magnetic field in a plain toroidal compactification of type I string theory implies both chiral spectra and supersymmetry breaking.

These general features remind of the ones seen in the previous subsection, in the context of branes intersecting at angles. Such coincidence its not spurious since both systems are in fact related by T-duality. Roughly speaking, two T-dual theories yield the same physics: same interactions, same operators, same Hilbert space. In our case, this will imply a one-to-one correspondence between two sets of theories. On the one side of the T-dual picture we will have D-branes with magnetic fluxes, whereas on the other side we will have D-branes at angles.

Let us see this correspondence in detail. Consider the bosonic part of the two-dimensional open string worldsheet action, the so-called sigma model \cite{Polchi1,Johnson:2000ch}
\bea
S & = & {1 \over 4\pi\a^\prime} \int_\Sigma d^2\sigma \sqrt{-g}
\left[ \left( g^{ab}G_{\mu\nu}(X) + \eps^{ab}B_{\mu\nu}(X)\right) 
\partial_a X^\mu \partial_b X^\nu + \a^\prime R \Phi(X)
\right] \nonumber \\
& + & \int_{\partial\Sigma} d \tau A_i(X) \partial_\tau X^i,
\label{sigma}
\eea
where $g$, $R$ are the world-sheet $\Sig$ internal metric and the associate Ricci scalar, whereas $G_{\mu\nu}$, $B_{\mu\nu}$, $\Phi$ and $A_i$ are target-space background fields, namely the metric, $B$-field, dilaton and boundary gauge field. Let us suppose a D$p$-brane with worldvolume coordinates $X^i$, $i = 0,...,p$ and transverse coordinates $X^m$, $m = p+1,...,9$. We will also consider constant background fields: $G_{\mu\nu} = \eta_{\mu\nu}$, $\Phi =$ const., $B_{\mu\nu} =$ const. Varying the action (\ref{sigma}), we will find that the fields $X^\mu$ must satisfy the usual two-dimensional wave equation. Boundary conditions at $\sig = 0, \pi$ on the D-brane worldvolume coordinates $i$, however, will be slightly different from usual Neumann $(\p_\sig X^i  = 0)$ boundary conditions:
\beq
\partial_\sigma X^i + {\cal F}^{\ i}_j \partial_\tau X^j = 0, \quad  \quad X^m = x^m_0,
\label{bc}
\eeq
getting mixed Neumann and Dirichlet \footnote{Conditions $\p_\tau X^j = 0$ are not exactly Dirichlet conditions, in which positions should be fixed as in $X^j (0, \tau) = X^j (\pi, \tau) = x^j_0$. We interpret them as conditions on the conjugate momenta, which is associated to the fact that the D-brane `position' has not been specified. This is also applicable to boundary conditions (\ref{bc3})} conditions on the coordinates $X^i$, $i = 0,...,p$.

The quantity $\F$ producing this mixing is nothing but the gauge invariant flux on the brane, expressed in terms of an antisymmetric tensor as ${\F} = {B \over \sqrt G} + 2\pi \a'F$ \cite{Polchi1}, $F = d A$ being the gauge field strength. For simplicity, let us consider all the components of $\F$ to vanish with the exception of ${\F}_{12} = - {\F}_{21} \equiv {\F}$, which we take to be a constant. The boundary conditions on the plane $(1,2)$ will then be
\beq
\begin{array}{c}
\partial_\sigma X^1 + {\cal F} \partial_\tau X^2= 0, \\
{\cal F}\partial_\tau X^1 - \partial_\sig X^2 = 0.
\end{array}
\label{bc2}
\eeq
Let us focus on these two coordinates, which we will suppose to be compact and parametrizing a two-dimensional torus. Notice that our initial D-brane with flux $\F$ wraps on such $T^2$. In case $\F = 0$ we know that by performing a T-duality on any of such directions we will end up with a D$(p-1)$-brane in the T-dual picture. Let us now see the effect of T-duality in the more general case $\F \neq 0$.

A two-torus can be parametrized by two real coordinates $x_i \in (0, 2\pi R)$, $i = 1,2$ being defined by a metric $G$ and an antisymmetric $B$-field. Such closed string moduli can be expressed in terms of two complex scalars, namely the complex structure $\tau = \tau_1 + i \tau_2$
\beq
\tau_1 = {G_{12} \over G_{11}}, \  \ \tau_2 = {\sqrt{G} \over G_{11}},
\label{cpxstr}
\eeq 
and the complexified K\"ahler form $J = J_1 + i J_2$
\beq
J_1 = {B_{12} R^2\over \a^\prime}, \  \ J_2 = {A \over 4\pi^2 \a^\prime},
\label{kahler}
\eeq
where $A = 4 \pi^2 R^2 \sqrt{G}$ is the torus area. By standard T-duality prescriptions, when performing it on the direction $x_1$, we must interchange  $\tau \leftrightarrow J$ and $\p_\sigma X^1 \leftrightarrow \p_\tau X^1$. The former implies that the T-dual torus is described by a geometry of the form
\bea
\tau_1' = {B_{12} R^2 \over \a^\prime}, & & \tau_2' = {A \over 4 \pi^2 \a^\prime}, \nonumber \\
B_{12}' = {\a^\prime \over R^{\prime  2}} \tau_1, & & A^\prime = 4 \pi^2 \a^\prime \tau_2.
\label{dualtorus}
\eea
We thus see that if we initially choose a rectangular torus $(\tau_1 = G_{12} = 0)$, then the T-dual torus will have a vanishing $B$-field \footnote{As an interesting remark, notice that if we also choose a vanishing $B$-field $(\tau_1 = G_{12} = B_{12} = 0)$, then T-duality relations (\ref{dualtorus}) reduce to $R_1 \leftrightarrow {\a' \over R_1}$, with $R_1 \equiv R \sqrt {G_{11}}$. This simplified version is the usual T-duality rule found in the literature \cite{Polchinski:1996na}.}. In the following we will suppose D-branes wrapping $T^2$'s with non-vanishing fluxes $F$ and $B$ and with pure imaginary complex structure. In the T-dual picture (branes at angles' picture), this implies that we will be working with non-rectangular tori with vanishing $B$-field.

In order to see the effect of the latter T-duality interchange let us define the angle $\th$ such that ${\F} = {\rm cotg} \ \th$. Boundary conditions (\ref{bc2}) can then be expressed (after T-duality) as
\beq
\p_\sig \tilde X^1 = 0, \quad \quad \p_\tau \tilde X^2 = 0,
\label{bc3}
\eeq
\beq
\left(\begin{array}{c}
\tilde X^1 \\ \tilde X^2
\end{array}\right)
\quad = \quad
\left(\begin{array}{cc}
cos \ \th & -sin \ \th \\ sin \ \th & cos \ \th 
\end{array}\right)
\left(\begin{array}{c}
X^1 \\ X^2
\end{array}\right).
\label{rotation2}
\eeq
This shows that, in the T-dual picture to a D$p$-branes with magnetic flux, we have a slanted D$(p-1)$-brane yielding an angle $\th$ with the $x_1$ axis of the torus. Notice that in this picture we have no mixture of Neumann and Dirichlet conditions, so that there is no flux $\F$ on this slanted D$(p-1)$-brane.

\begin{figure}[ht]
\centering
\epsfxsize=5in
\epsffile{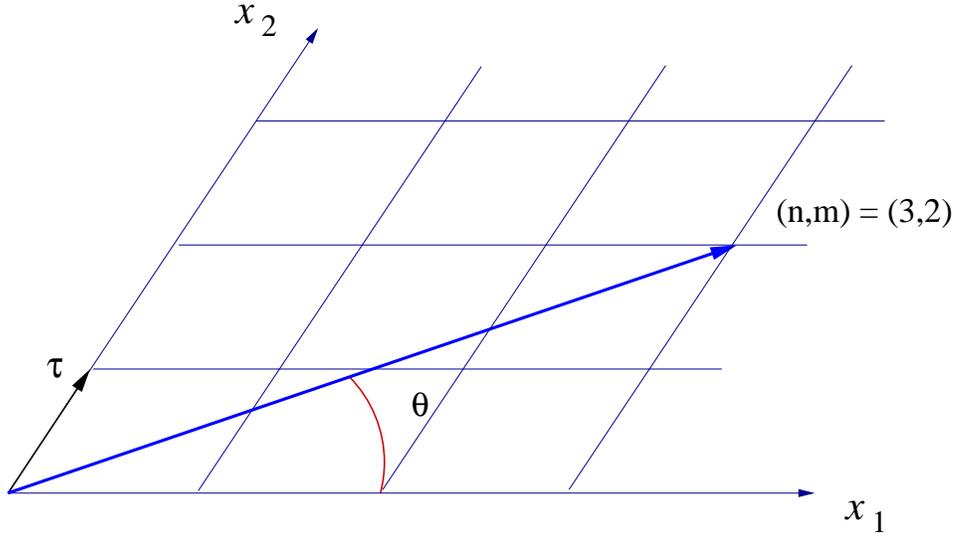}
\caption{D-brane wrapped on the minimum length 1-cycle on the homology class $[(n,m)] = [(3,2)] \in H_1(T^2,\inte)$. Fixed this homology class, the angle $\th$ between the D-brane and the $x_1$ axis will depend on the torus complex structure $\tau^\prime$.}
\label{toro2}
\end{figure}

In fact, the correspondence between both T-dual systems can be made more precise. Consider, in the branes at angles' picture, a D$(p-1)$-brane wrapped not densely on a two-torus, hence belonging to a definite homology class of 1-cycles of $T^2$ which is specified by two integer numbers $(n,m) \in H_1(T^2,{\bf Z})$. Let us consider a tension (i.e., length) minimizing representative in such class, which is a straight line. As can be seen in figure \ref{toro2}, the angle yielded by this line with the $x_1$ axis is
\beq
cotg \ \th = {n + m\tau_1' \over m\tau_2^\prime} = {n \over m\tau_2^\prime} + {\tau_1^\prime \over \tau_2^\prime}.
\label{cotg}
\eeq

%
Since this quantity has to be identified with the flux ${\F}_{12} = {B_{12} \over \sqrt G} + 2\pi \a'F_{12}$ on the T-dual image, using the identities (\ref{dualtorus}) we see that the factor $\tau_1'/\tau_2'$ has to be identified with the contribution of the $B$-field. On the other hand, the first factor of the r.h.s. of (\ref{cotg}) corresponds to the $U(m)$ gauge bundle field strength
\beq
F = 2\pi {n \over m A} I_m,
\label{flux}
\eeq
arising from a stack of $m$ D$p$-branes wrapping the whole torus, and with first Chern number $n$. Is possible to generalize this correspondence to more complicated systems, as D-branes wrapping half homology cycles of $T^{2n}$. The T-dual objects will then correspond to torons, arising from D-branes wrapping the whole $T^{2n}$ and with non-trivial flux on their worldvolume \cite{'tHooft:1979uj,'tHooft:1981sz,vanBaal:1982ag,vanBaal:1984ar,Guralnik:1997th,Troost:1999xn,Bogaerts:2000rx}. A dictionary relating both T-dual pictures, as well as other related questions regarding the preserved supersymmetry and stability of a certain configuration have been developed in \cite{Rabadan:2001mt}.

Finally, let us mention that, from the point of view of D-branes with fluxes, turning on a non-vanishing flux $\F$ on a D-brane wrapping a two-torus induces a internal non-commutative geometry \cite{Connes:1997cr,Douglas:1997fm,Ardalan:1998ce,Seiberg:1999vs}, whose non-commutativity parameters is given by
\beq
[X^1,X^2] = 2\pi i \a^\prime {{\F} \over 1 + {\F}^2}.
\label{nc}
\eeq
The non-commutative geometry formalism turns computations less intuitive in this T-dual picture with fluxes, even if it is as valid as the other picture and, in some aspects, simpler. Some theoretical aspects, as well as semi-realistic constructions have been analysed in this framework in \cite{Blumenhagen:2000fp,Blumenhagen:2000wh,Angelantonj:2000hi,Blumenhagen:2000vk,Angelantonj:2000rw,Blumenhagen:2000ea}. In the following, we will usually restrict ourselves to the T-dual picture of D-branes at angles, or intersecting D-branes, where no flux is induced on the worldvolume of the D-brane.

\section{Toroidal and orbifold compactifications}

Now that we have shown the basic theoretical features regarding flat intersecting D-branes, such as the spectrum at their intersection and their T-dual counterparts, let us try to implement this knowledge in the building of semi-realistic models. The usual procedure is to consider configurations involving D$(3+n)$-branes (with fixed $n$) which fill the four spacetime dimensions $(X^0, X^1, X^8, X^9)$, which are taken to be non-compact in order to identify them with $M_4$. The extra $n$ dimensions of each D-brane will be filling some subspace of the six remaining coordinates, which parametrize a compact six-dimensional Riemannian manifold. In general, we need to impose the compactness constraint in order to recover four-dimensional gravity at low energies. However, the bottom-up philosophy \cite{Aldazabal:2000sa} teach us that, in building up a realistic model from D-branes, we may first care about the gauge theory and chiral spectrum living at the D-branes worldvolume, and later bother about embedding such gauge theory in a fully-fledged string compactification. In our particular context, this implies that we must consider configurations of D$(3+n)$-branes whose extra $n$ dimensions are compact since, in that way, we will recover a four dimensional gauge group at low energies. If such extra worldvolume dimensions were not compact, then the corresponding gauge coupling constants would not vanish and would get global instead of local symmetries. Notice that this is a weaker constraint than imposing the full six dimensions to be compact, as we will presently see in explicit examples.

\begin{figure}[ht]
\centering
\epsfxsize=6in
\epsffile{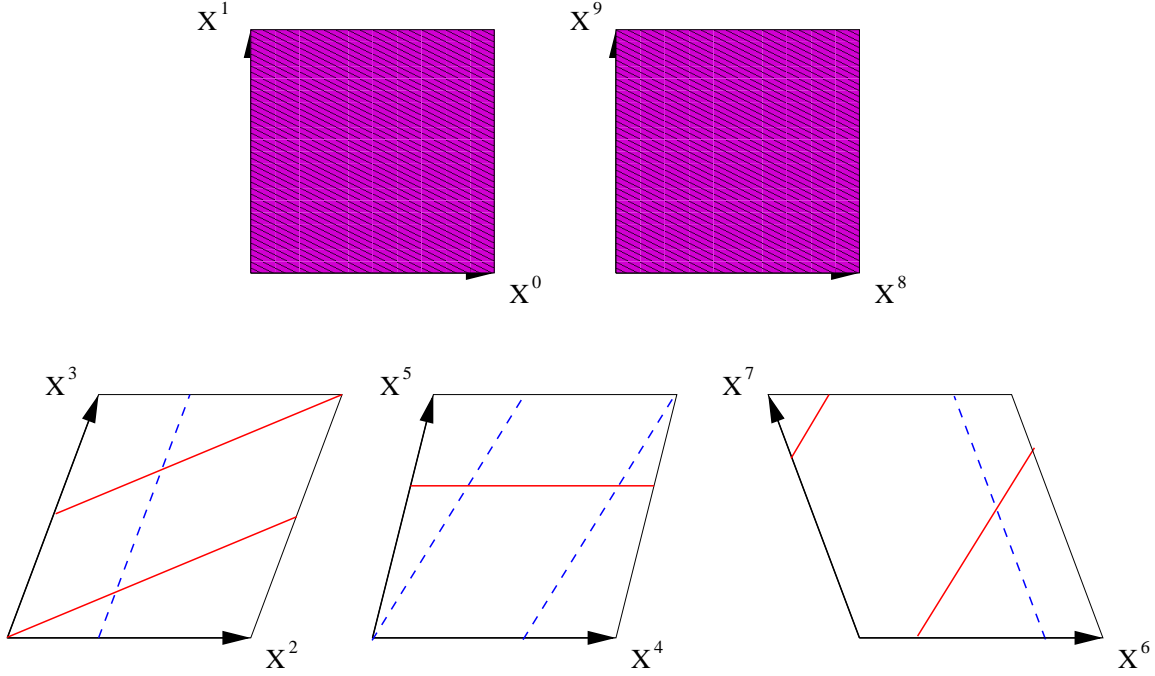}
\caption{Pair of D6-branes wrapped on factorisable 3-cycles on $T^6 = T^2 \times T^2 \times T^2$. Both branes fill the four non-compact dimensions $(X^0, X^1, X^8, X^9)$, so this will be the dimension of their intersections. Notice the analogies of this picture with figure \ref{anglesth}, the main difference being that now they intersect several times.}
\label{compact5}
\end{figure}

The simplest compactification meeting our requirements and yielding non-trivial physics is the plain toroidal one, i.e., considering the six extra dimensions to be $T^{2n} \ti \cpx^{3-n}$ ($n = 0,1,2,3$). In order to mimic the class of configurations of figure \ref{anglesth}, we will suppose that the $2n$-dimensional torus has a flat metric factorisable as a product of $n$ two-tori, i.e., $T^{2n} = T^2 \ti \dots \ti T^2$. We will also suppose the $n$ extra dimensions of each D$(3+n)$-brane to be wrapped on a $n$-cycle of such $T^{2n}$ and located at a single point in $\cpx^{3-n}$, thus being compact submanifolds. More specifically, we will suppose them to wrap $n$-cycles which can be written as a direct product of $n$ 1-cycles, each wrapping on a different $T^2$, that is
\beq
\Pi_a = \bigotimes_{r=1}^n \left(n_a^{(r)}, m_a^{(r)}\right),
\label{factorisable}
\eeq
where $(n^{(r)}, m^{(r)})$ are the wrapping numbers of the corresponding (length minimizing) 1-cycle on the $r^{th}$ two-torus, and $a$ labels the set of D$(3+n)$-branes. We will name this subclass of $n$-cycles as {\it factorisable} cycles, and a D-brane wrapped on it a factorisable brane. Notice that, by this choice, the worldvolume of each D-brane of the configuration will be given by $M_4 \ti T^n$, which is the nearest geometry to a flat D$(3+n)$-brane expanding a hyperplane while having $n$ compact dimensions. In particular, these factorisable branes are $\med BPS$ objects as well. An important difference with the non-compact setup, however, is that two D-branes may now intersect several times, yielding several replicas of the spectrum at the intersection. An example of such configuration involving two D6-branes (i.e., $n=3$) is presented in figure \ref{compact5}.

In the above discussion we have only considered configurations of D$(3+n)$-branes with $n \leq 3$. We could, in principle, also consider D$(3+n)$-branes wrapped on $T^6 = T^2 \ti T^2 \ti T^2$ and with $n > 3$. This would imply that a D-brane has to wrap, at least, a whole $T^2$. Configurations of this kind yielding interesting physics, however, are related by T-duality with the ones already considered, so in the following we will restrict to $n \leq 3$. 

Let us first consider the case $n=3$, that is, type IIA D6-branes wrapping factorisable 3-cycles on $T^2 \ti T^2 \ti T^2$. Generically, two such D-branes will intersect at several points, as shown in figure \ref{compact5}. Since the metric background as well as the induced metric on the D-branes worldvolumes is flat, the physics at each intersection is locally that of two intersecting D6-branes in flat ten-dimensional spacetime. As a consequence, we can readily apply our results from previous sections and compute the spectrum living at the intersection, obtaining a chiral fermion and several scalars. Notice, however, that the angles between two branes are not arbitrary, but depend both on the specific $n$-cycles they wrap and on the geometry of the compact space. 

Indeed, the geometry of a factorisable $2n$-torus is given by the complex structure $\tau^{(r)}$ and the K\"ahler form $J^{(r)}$ of each two-torus. Given some fixed homology classes the angle yielded between two D-branes will depend solely on the complex structures. In particular, a brane wrapped on the cycle $(n_a^{(r)}, m_a^{(r)})$ on the $r^{th}$ two-torus will yield with the $(1,0)^{(r)}$ cycle (i.e., the $r^{th}$ horizontal axis) the angle
\beq
\th_a^{r} = cotg^{-1} {n_a^{(r)} + m_a^{(r)} \ {\rm Re} \tau^{(r)}
\over m_a^{(r)} \ {\rm Im} \tau^{(r)}}.
\label{cotg2}
\eeq
The relevant angles in computing the spectrum at the intersection, however, are those between two branes $a$ and $b$. These are given by $\vt_{ab}^{r} \equiv \th_b^{r} - \th_a^{r}$, and in the case of D6-branes fill the twist vector $v_\vt^{ab} = (\vt_{ab}^1,\vt_{ab}^2,\vt_{ab}^3,0)$. In the same manner, it can be seen that a type IIA D6-brane $a$ described by the angles (\ref{cotg2}) is related by T-duality along the directions $(X^2, X^4, X^6)$ with a D9-brane of type IIB theory with magnetic fluxes ${\cal F}_{2r,2r+1} = {\rm cotg} \ \th_a^r$.\footnote{In case each of the angles (\ref{cotg2}) is given by $\pm \frac{\pi}{2}$, the T-dual D-brane is not a D9-brane but a lower dimensional D7, D5 or D3-brane of type IIB theory.}

Let us now turn into configurations of D$(3+n)$-branes with $n < 3$, compactified on $T^{2n} \ti \cpx^{3-n}$. A simple computation of the spectrum shows that such class of configurations will yield non-chiral effective theories. Let us, for instance, consider type IIA D4-branes wrapped on a 1-cycle of $T^2$ and pointlike in $\cpx^2$. The twist vector which is obtained in the bosonic language has a single non-vanishing component, and is given by $v_\vt^{ab} = (\vt_{ab}^1,0,0,0)$. Let us suppose, as before, that $0 < \vt_{ab}^1 < 1$. Then we obtain four massless fermions at each intersection, given by
\beq
\begin{array}{cc}
(- \med + \vt_{ab}^1, -\med, -\med, +\med),  
&  (-\med+\vt_{ab}^1, -\med, +\med, -\med), \\ \\
(- \med + \vt_{ab}^1, +\med, -\med, +\med),  
&  (-\med+\vt_{ab}^1, +\med, +\med, +\med). 
\end{array}
\label{nonchiral}
\eeq
By looking at the fourth entry of each vector, we see that we will get two pairs of fermions of opposite chirality, while having the same quantum numbers. As a consequence, there will be no net chirality on the low energy spectrum. The same argument can be repeated if we choose the opposite sign for $\vt_{ab}^1$, as well as for the cases where we have type IIB D3 and D5-branes ($n = 0$ and $n = 2$ cases, respectively).

This lack of chirality was to be expected from the existence of a Coulomb branch in the above configuration. Indeed, notice that, since both D4-branes $a$ and $b$ were located at a single point in the transverse space $\cpx^2$, it is the possible to continuously separate them in such space by a distance $Y$ (just as when two of them are (anti)parallel on one or several tori), which is a Coulomb branch on the effective field theory. Such distance will modify the mass operator (\ref{mass}), in such a way that the whole spectrum at the $ab$ sector will get a global shift in the mass of their states as we increase the D4-branes separation. This can only happen from the effective theory viewpoint if we have a non-chiral spectrum.

So, in a way, we may hope to get a chiral theory if we manage to `trap' the D-branes in the transverse directions where they are pointlike. A simple way to do this is considering D-branes sitting not on a smooth space homeomorphic to $\cpx^{3-n}$ but on an orbifold singularity of the kind $\cpx^{3-n}/\inte_N$, where they are stuck. From the effective theory point of view, chirality will arise from considering invariant states under the orbifold action, that is states that are well-defined on the orbifolded space. Indeed, we do not expect that all the states in (\ref{nonchiral}) survive the orbifold projection, most of them being projected out (see Chapter \ref{spectrum} for explicit computations).

To sum up, we find that we must consider configurations of type II theory  compactified on $T^{2n} \ti \cpx^{3-n}/\inte_N$, containing D$(3+n)$-branes wrapping factorisable $n$-cycles of $T^{2n}$ and sitting at the orbifold point (i.e., the origin) in the transverse dimensions of $\cpx^{3-n}/\inte_N$. Each of these class of compactifications, with $n = 0, 1, 2$ or $3$ may yield low energy spectra of a chiral $D=4$ theory. The case of $n=0$, i.e., D3-branes at orbifold singularities $\cpx^3/\inte_N$, was studied in \cite{Aldazabal:2000sa}, were the bottom-up philosophy that we follow in this work was also described. Notice that in this particular case no actual angles nor intersections appear, so we will not consider them in the following. It is important, however, to stress that semi-realistic constructions were already achieved in this setup. The rest of the cases $0 < n \leq 3$ were studied in \cite{Aldazabal:2000dg}, including the case $n = 3$ already discussed above, and which does not involve any orbifold projection. We will base part of the forthcoming discussion in the material presented there.

The simplest way to build an orbifold singularity $\cpx^n/\G$ is by identifying points of the space $\cpx^n$ under the geometrical action of $G: \cpx^n \raw \cpx^n$, which leaves fixed the origin of $\cpx^n$ and respects the multiplication table of the discrete group $\G$. In our case, we take $\G = \inte_n$, a cyclic group, so the whole action is completely fixed given that of a generator $\om$ of the group. The most general geometrical action on $\cpx^2$ is
\bea
G_\om: \left(
\begin{array}{c}
Z^2 \\ Z^3
\end{array}
\right) \mapsto
\left(
\begin{array}{cc}
e^{2\pi i b_1/N} & 0 \\
0 & e^{2\pi i b_2/N}
\end{array}
\right)
\cdot
\left(
\begin{array}{c}
Z^2 \\ Z^3
\end{array}
\right),
\label{rotaorbi}
\eea
where we have taken again $Z^\mu = X^{2\mu} + i X^{2\mu + 1}$ as complexified coordinates. Just as we associate a twist vector $v_\vt$ to the relative positions of two D-branes related by a general $U(n)$ rotation, we can represent the rotation (\ref{rotaorbi}) by a vector twist $v_\om$, which will correspond to the $\inte_N$ generator $\om$.
\footnote{To be accurate, this is a standard procedure in plain orbifold constructions, which were performed long before configurations of branes at angle were considered.} 
In the case of the action (\ref{rotaorbi}), the vector twist will be given by $v_\om = {1 \over N} (0,b_1,b_2,0)$, where we impose $b_1 \equiv b_2  {\rm \ mod \ } 2$ for the variety to admit spinors. The orbifold will be supersymmetric if $b_1 = \pm b_2 {\rm  \ mod \ } N$, so that the rotation (\ref{rotaorbi}) will be contained in a $SU(2)$ subgroup of the general rotation group. As a result, will the closed string spectrum (bulk spectrum) will preserve some amount of supersymmetry. Notice that, in this supersymmetric case, we may express the $\inte_N$ action by an alternative generator $\om^\prime$ such that the corresponding twist vector gets the simplified form $v_{\om^\prime} = {1 \over N} (0,1,\pm 1,0)$.

The orbifold group action, however, is not restricted to a mere identification of geometrical points. Since in general we will deal with D-branes sitting at an orbifold singularity, Chan-Paton degrees of freedom living on their worldvolume will also be transformed non-trivially under the action of $\G$. This broadens the possibilities from the model-building viewpoint, as we will see when computing the low-energy spectrum from the open string sector in Chapter \ref{spectrum}.

The last case under study is type IIB compactified on $T^4 \times \cpx/\inte_N$, with D5-branes wrapping factorisable 2-cycles on $T^4 = T^2 \times T^2$ (notice that the factorisability condition was trivial on the previous case $n=1$). By T-duality, we can relate this class of configurations to those of D7-branes wrapping the whole $T^4$ with non-trivial magnetic fluxes on their worldvolume, while again sitting at the orbifold singularity on $\cpx/\inte_N$. The orbifold twist vector, which represents the geometric action of the orbifold, is in this case given by $v_\om = {1 \over N} (0,0,-2,0)$, where the $2$ is necessary in order that spinors survive the orbifold projection. Such a twist is explicitly non-supersymmetric and, as a consequence, a tachyon will appear in the closed string NS-NS sector (see Chapter \ref{spectrum} for details).

\section{Orientifold compactifications \label{oricomp}}

Actually, the first consistent fully-fledged compactifications involving D-branes at angles were not plain toroidal nor orbifold compactifications, but {\it orientifold} compactifications \cite{Blumenhagen:1999md,Blumenhagen:1999ev,Blumenhagen:1999db} (see also \cite{Pradisi:1999ii} for related work). In general, an orbifold singularity can be obtained by quotienting a smooth space by a local geometrical action of a discrete group, such as $\cpx^n/\G$. An orientifold is a generalization of such quotient, including as an element of the group the parity reversal symmetry $\O$ which changes the orientation of the strings. A generic orientifold quotient can then be expressed as $\cpx^n/(\Gamma_1 + \O \Gamma_2)$, since $\O^2 = 1$. This more general class of quotients deserves a new name because the action of $\O$ is not of geometric nature (i.e., it does not act on the target space) but is a symmetry of the string theory whose action is taken on the worldsheet \footnote{For a detailed presentation of theoretical issues involving orientifolds, as well as an extensive collection of model-building techniques involving them, we recommend \cite{tesisraul}.}.

An important example of an orientifold is type I theory of open an unoriented strings. Such theory can be constructed from type IIB, as a theory of closed and oriented strings on ten non-compact dimensions. Indeed, type IIB has as an internal symmetry the orientation reversal $\O$, so we can quotient it by the orientifold group $\inte_2 \simeq \{1,\O\}$. As a result of this quotient, we will obtain a theory containing just unoriented strings, since only those are well invariant on the new theory, being invariant under the orientifold group action. On the other hand, $\O$ acts trivially on the target space, so the target space of the quotiented theory will again be $M_{10}$. In addition, by consistency conditions we will have to include 32 D9-branes to our theory, which implies that, at the end of the day, we will have a theory including open strings: type I theory (see, e.g., \cite{Dabholkar:1997zd}).

The compactifications built in \cite{Blumenhagen:1999md,Blumenhagen:1999ev,Blumenhagen:1999db} were based on type II permutational orientifolds of the kind $(\inte_N + \OR \inte_N)$ on four and six dimensional tori (here $\OR$ is a T-dual version of the operator $\O$, see below). Consistency conditions naturally demanded the presence of D-branes at angles. However, the massless open string spectrum involved was supersymmetric and non-chiral. The generalization to $\inte_N \ti \inte_M$ orientifolds followed in \cite{Forste:2000hx}, with similar results. 

We will not study these class of configurations, but the simpler ones considered in \cite{Blumenhagen:2000wh,Blumenhagen:2000ea} which exhibit phenomenologically appealing features of chirality and broken supersymmetry. Such configurations are nothing but a compactification of type I theory on a factorisable six-torus, with magnetic fluxes $\F$ on the D9-branes worldvolumes. As we have just described, type I can be seen as an orientifold of type IIB theory, so we are actually considering
\beq
\frac {{\rm Type \ IIB \ on} \ T^6}{\{1 + \O\}}.
\label{typeI}
\eeq
Since this construction involves nothing but an orientifold of type IIB D9-branes with magnetic fluxes, T-duality will relate it to an orientifold of type IIA D6-branes at angles. It is important, however, to know what is the T-dual analogue of $\O$.

\begin{figure}[ht]
\centering
\epsfxsize=4.6in
\epsffile{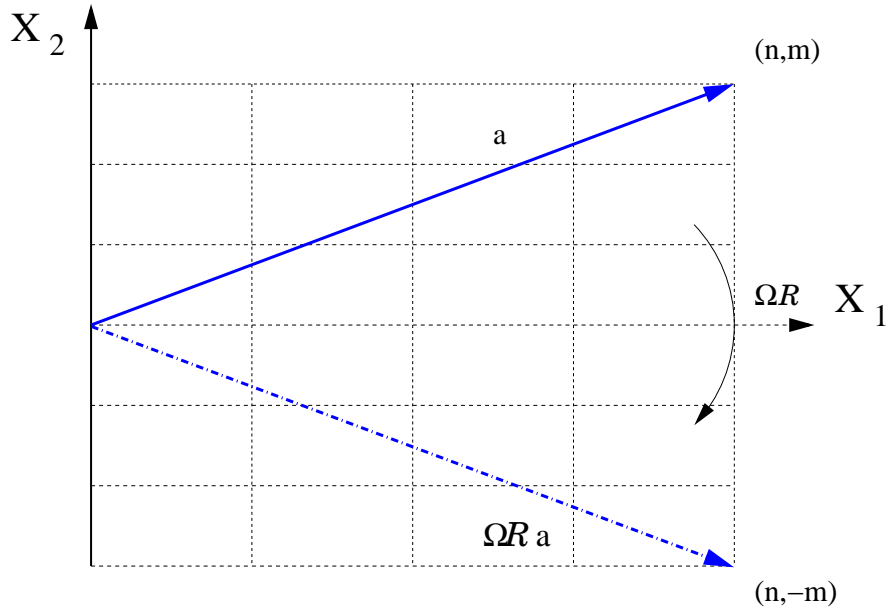}
\caption{Orientifold action on a D-brane wrapped on a straight 1-cycle with wrapping numbers $(n,m)$. Notice that, in general, the geometrical action of $\R$ will induce an action on homology spaces, which in this case of this two-torus is $\R : [(n, m)] \mapsto [(n, -m)]$.}
\label{mirror2}
\end{figure}

In general, if we consider type IIB theory compactified on a circle of radius $R$ on the $i^{th}$ coordinate, we can perform a T-duality on such direction and obtain type IIA compactified on a circle of radius $\a'/R$. Type IIB is invariant under $\O$, so parity reversal is a good symmetry of the theory. This is not the case of type IIA, and under T-duality $\O$ gets mapped to \cite{Dabholkar:1997zd}
\bea
\O \ {\it in \ type \ IIB} & 
\stackrel{T_{(i)}}\longrightarrow
& \OR_{(i)} \ {\it in \ type \ IIA},
\label{omegaR}
\eea
where ${\cal R}_{(i)}$ is the $i^{th }$ coordinate inversion $X^i \stackrel{{\cal R}_{(i)}}\longrightarrow -X^i$. This implies that the T-dual version of (\ref{typeI}) is
\beq
\frac {{\rm Type \ IIA \ on \ } \ T^6}{\{1 + \OR\}},
\label{dual}
\eeq
where ${\cal R}$ stands for the reflection of the three coordinates where the T-duality has been performed. Contrary to the toroidal case above, in orientifold compactifications we will choose to T-dualize coordinates $(3,5,7)$, i.e., $\R = \R_{(3)}  \R_{(5)} \R_{(7)}$. If we complexify our coordinates as $Z^i = X^{2i} + i X^{2i + 1}$, $\R_{(i)}$ can be expressed as a simple complex conjugation 
\beq
{\cal R}_{(i)}: Z^i \longleftrightarrow \bar Z^i, \ \ (i = 1,2,3).
\label{conjugation}
\eeq
The $\OR$ action will have two important effects. First, for each D-brane introduced on a configuration we will have to introduce its image under $\OR$, otherwise the whole construction will not be invariant under the orientifold group. Let us illustrate this by considering a rectangular $T^2$ (i.e., with a complex structure purely imaginary). If we introduce a D-brane $a$ lying on a straight 1-cycle with wrapping numbers $(n_a, m_a)$, then its image under $\OR$ can be obtained by mirroring it with respect to the horizontal axis and will have the wrapping numbers $(n_a, -m_a)$, as can be appreciated in figure \ref{mirror2}. This {\it mirror} D-brane will be denoted either by $\OR a$ or by $a$*.

\begin{figure}[ht]
\centering
\epsfxsize=6in
\epsffile{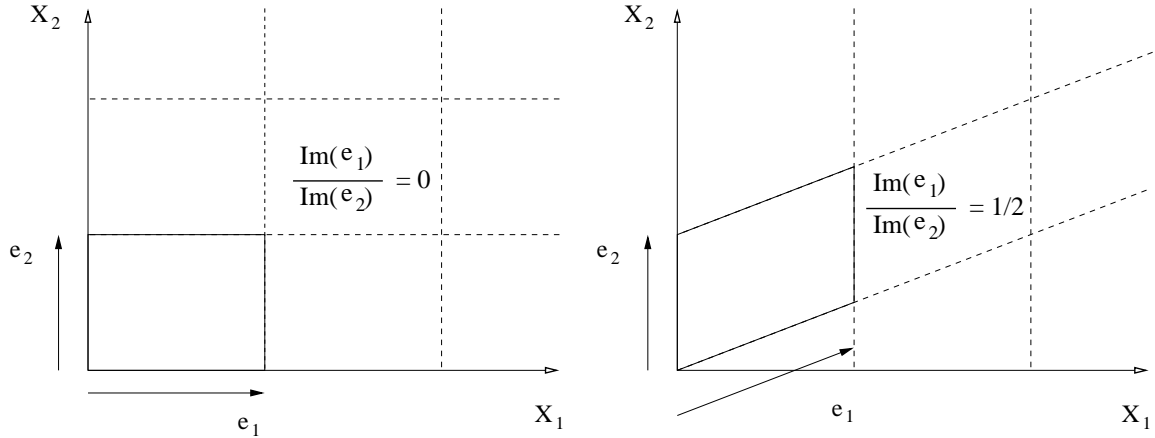}
\caption{A well-defined action of $\OR$ freezes one of the degrees of freedom of the complex structure of a $T^2$, leaving for it the two possible values Im$(e_1)$/Im$(e_2) = 0, 1/2$.}
\label{bflux}
\end{figure}

Second, not every complex structure is well-defined under the geometrical action of $\OR$. Usually we define a flat two-torus by considering a complex plane $\cpx$ and two complex numbers $\{e_1, e_2\}$, not related by real multiplication, so that they define a two-dimensional lattice $\Lam$. The torus is then defined as the quotient $T^2 \approx \cpx/\Lam$. In order for this lattice to be well-defined under the $\OR$ action, it has to be invariant under complex conjugation. As a result Im$(e_1)$/Im$(e_2)$ can only take values on $\inte/2$ and, actually, only two possibilities turn to be inequivalent under redefinitions of the torus lattice, say $0$ and $1/2$. Im$(e_1)$/Im$(e_2) = 0$ corresponds to the rectangular torus considered earlier, while Im$(e_1)$/Im$(e_2) = 1/2$ implies a tilted geometry. These two choices are depicted in figure \ref{bflux}. On the other hand, notice that Re$(e_1)$/Im$(e_2)$ is still completely unconstrained. 

Since Im$(e_1)$/Im$(e_2)$ is nothing but the imaginary part of the complex structure of the torus (after the redefinition $X_1 \lraw X_2$) the restriction on this geometrical degree of freedom translates under T-duality into a discretization of the $B$-field. That the $B$-field is a frozen NS-NS closed string moduli in type I theory is a well known fact \cite{Bianchi:1991eu,Bianchi:1997rf,Witten:1997bs,Angelantonj:1999jh,Angelantonj:1999xf,Kakushadze:2000hm,Blumenhagen:2000ea}. In particular, it is found that its integral over a torus surface may yield the two discrete values $b = 0, 1/2$ (in units of $4\pi \a'$). By abuse of language, we will denote by $b^{(i)} = 0, 1/2$ the two possibilities of having the $i^{th}$ two-torus of our configuration with a rectangular or tilted geometry.

It turns out that the action of $\OR$ on a 1-cycle in a tilted geometry is far more involved than in figure \ref{mirror2}, at least in terms of the wrapping numbers of such cycle. We can recover a simplified expression also in this case if we consider {\it fractional} 1-cycles, defined as
\beq
\left(n_a^{(i)},m_a^{(i)}\right)_{\rm frac} := \left(n_a^{(i)},m_a^{(i)}\right)_{\rm integ} + b^{(i)} \left(0,n_a^{(i)}\right)_{\rm integ}, 
\label{frac}
\eeq
where $i$ labels each of the two-tori on $T^6 = T^2 \ti T^2 \ti T^2$. This definition admits semi-integer $m$'s, and reduces to the usual one when $b$ vanishes. Fractional 1-cycles are nothing but a convenient basis for expressing our D-brane content, and we can produce fractional 3-cycles by tensoring them as in (\ref{factorisable}). In particular, we see that for fractional cycles the action of $\OR$ is simply given by $(n_a^{(i)},m_a^{(i)})_{\rm frac} \longrightarrow (n_a^{(i)},-m_a^{(i)})_{\rm frac}$. This makes fractional 1-cycles a useful convention when dealing with model-building matters, although their meaning is far less intuitive from $T^6$ homology viewpoint. From now on and unless otherwise stated, we will always use the fractional language (\ref{frac}), omitting the subindices.

Having deduced the natural orientifold generalization of type IIA D6-brane models, it is now a simple matter to extend this construction to configurations of intersecting D$(3+n)$-branes. The orientifold group will now involve a combined action of both the orbifold $\inte_N$ projection and the $\OR$ orientifold involution. More specifically, we will consider the class of theories given by
\beq
{\rm Type \ II \ on \ } \frac{\ T^{2n} \ti \cpx^{3-n}/\inte_N}{\{1 + \OR\}}.
\label{singuori}
\eeq
The case $n=3$ reduces to the type IIA compactifications already discussed, involving D6-branes at angles, and with $\R$ being the $\inte_2$ action composed by the three simultaneous reflections $\R_{(3)} \R_{(5)} \R_{(7)}$. On the other hand, $n=1$ features type IIA D4-branes at angles, and the $\inte_2$ action will now be given by $\R = \R_{(3)}\R_{(4)}\R_{(5)}\R_{(6)}\R_{(7)}$ (that is, we perform a complex conjugation on the torus and a reflection on the transverse dimensions to it). Finally, $n=2$ involves type IIB D5-branes, and the involution accompanying the worldsheet parity will be $\R = \R_{(3)}\R_{(5)}\R_{(6)}\R_{(7)}(-1)^{F_L}$, $F_L$ being the left fermion number\footnote{This new non-geometrical contribution of $(-1)^{F_L}$ to $\R$ is necessary in order to guarantee the invariance of the theory under the orientifold action \cite{Dabholkar:1997zd}.}. Again, we can simply express the geometrical part of these actions in terms of complexified coordinates as
\beq
\R :
\begin{array}{ll}
Z^i \longmapsto \bar Z^i & i = 1 {\rm \ to \ } n, \\
Z^i \longmapsto - Z^i & i = n + 1 {\rm \ to \ } 3.
\end{array}
\label{R}
\eeq
Again, we will not consider the class of models with $n = 0$, which involve type IIB D3-branes sitting at orientifold singularities. The search of phenomenologically appealing configurations in such context has been addressed in \cite{Alday:2002uc}.

In order to motivate the classes of configurations in (\ref{singuori}) with the specific parity action (\ref{R}), let us relate them to some previous constructions in the literature. In particular, let us consider the compactification
\beq
{\rm Type \ I \ on \ } T^{2} \ti T^4/\inte_3
= \frac{{\rm Type \ IIB \ on \ } T^{2} \ti T^4/\inte_3}{\{1 + \O\}},
\label{D8a}
\eeq
which is a well-defined theory, naturally including D9-branes filling the whole ten-dimensional spacetime. We will allow these D9-branes to have non-trivial fluxes $\F$ on their worldvolume, restricted to the unorbifolded $T^2$ complex dimension. By performing a T-duality on the vertical coordinate of such $T^2$ (which in our conventions is the $3^{rd}$ direction) we obtain
\beq
{\rm Type \ I \ on \ } T^{2} \ti T^4/\inte_3 \ \stackrel{T_{(3)}} \longleftrightarrow \ \frac{{\rm Type \ IIA \ on \ } T^{2} \ti T^4/\inte_3}{\{1 + \OR_{(3)}\}}
\label{D8b}
\eeq
where, by the usual arguments of section 2.2, instead of D9-branes with fluxes we have D8-branes at angles, i.e., wrapping 1-cycles on the unorbifolded $T^2$. Such type IIA compactifications involving D8-branes at angles have been considered in \cite{Honecker:2002hp}. In order to relate them to the constructions on (\ref{singuori}), we need to perform a T-duality on each of the directions of $T^4/\inte_3$, so that we get
\beq
\frac{{\rm Type \ IIA \ on \ } T^{2} \ti T^4/\inte_3}{\{1 + \OR_{(3)}\}} \
\stackrel{T_{(4,5,6,7)}}\longleftrightarrow \
\frac{{\rm Type \ IIA \ on \ } T^{2} \ti T^4/\inte_3}{\{1 + \OR\}},
\label{D8c}
\eeq
where now $\R = \R_{(3)}\R_{(4)}\R_{(5)}\R_{(6)}\R_{(7)}$. The former D8-branes at angles on the l.h.s. of (\ref{D8c}) will now be represented, in this T-dual version of the theory, by D4-branes localized at a point in $T^4/\inte_3$. In particular, they may be stuck at the origin of such orbifolded space, which is nothing but an orientifold singularity of the kind (\ref{singuori}) with $n = 1$ and $N = 3$, at least locally. Thus, we have recovered a particular construction in (\ref{singuori}) by simply considering type I theory on an space with orbifold singularities and with D9-branes with fluxes and performing a chain of T-dualities. Similar arguments can be made for the rest of the configurations.

Just as in the case $n = 3$ a D$(3+n)$-brane $a$ with $n = 1, 2$ will not, in general, be invariant under the $\OR$ action, but we will have to add a mirror brane $a$*. Considering again factorisable $n$-cycles, and expressing our brane content in terms of fractional cycles as in (\ref{frac}), the geometric action of the orientifold will be simply given by $(n_a^{(i)},m_a^{(i)}) \stackrel{\OR}\longrightarrow (n_a^{(i)},-m_a^{(i)})$. This will not be, though, the only relevant action of the orientifold parity, but also its action on the Chan-Paton degrees of freedom will have to be taken into account (see below).

\section{General intersecting D-branes \label{general}} 

In the previous sections of this chapter we have described the class of compactifications that we will study in the following. Roughly speaking, these involve type II superstring theory compactified on $T^{2n} \ti \cpx^{3-n}/\inte_N$, with D$(3+n)$-branes wrapping factorisable cycles of the flat $(T^2)^n$ and sitting at the orbifold singularity. We have also considered the possibility of an extra orientifold projection. This family of compactifications is a {\it minimal} modification of a system of flat type II D-branes intersecting in a non-compact flat space, which still allows us to consider D-branes with just $3+1$ non-compact dimensions. As we have explained, this is a must if we want to construct a configuration yielding semi-realistic phenomenology since, by means of Kaluza-Klein reduction, we must recover a $D=4$ field theory from the gauge theory living on the worldvolume of the D$(3+n)$-brane. They are minimal in the sense that we are still working with a flat background (modulo orbifold singularities, which are harmless for string propagation), and with D-branes of flat worldvolume. This implies that the associated Conformal Field Theory (CFT) can be exactly solved to all orders in the $\a'$ perturbative expansion. Moreover, we can take advantage of many computations performed in the non-compact setup and translate them to ours. In particular, we can exactly compute the spectrum at an intersection, by simply evaluating the mass operator (\ref{mass}), since the local geometry is exactly the same.

In addition, these simple configurations may help us to understand more involved ones. Let us, for instance, take the example of type IIA D6-branes. A general configuration would involve a compactification of the form $M_4 \ti \M_6$, with $\M_6$ a compact six-dimensional Riemannian manifold, and several stacks of D6-branes, each with worldvolume of the form $M_4 \ti \Pi_a$, with $\Pi_a \subset \M_6$ a closed 3-cycle \footnote{Supersymmetry and/or stability in the closed and open string sectors will impose several constraints on the geometry of $\M_6$ and on the 3-cycles. See Chapter \ref{SUSY} for details.}. Generically, a pair of such 3-cycles will intersect at a finite number of points $\Pi_a \cap \Pi_b$ in $\M_6$ and, locally, each of these intersections will look like two intersecting 3-planes in $\re^6$. In particular, a massless chiral fermion and several light scalars will appear from each such intersection. We thus find that hyperplanes intersecting at angles provide local models for general compactifications and not only for toroidal ones. However, it must be noted that it is extremely hard to compute the full low energy spectrum of such a compactification, since in order to compute the masses of the light scalars the local geometry of both branes at each intersection must be known, in particular the relative angles they yield and hence the local metric of the manifold. On the other hand, the spectrum of chiral fermions is much more easy to handle with. Indeed, although the number of intersections $\#(\Pi_a \cap \Pi_b)$ is not a topological quantity, and may depend on closed and open string moduli of the compactification, we may construct such invariant by subtracting the number of right-handed fermions to the left-handed ones. We then obtain the (signed) intersection number of the two 3-cycles $\Pi_a$ and $\Pi_b$, which only depends on their respective homology classes $[\Pi_a], [\Pi_b] \in H_3(\M_6, \inte)$. Thus, by knowing the half-homology of the compact manifold $\M_6$ and its associated intersection form $I_{ab} \equiv [\Pi_a] \cdot [\Pi_b]$, we are able to compute the {\em net} number of chiral fermions between two intersecting D6-branes. This net number of fermions are the ones which remain strictly massless (the rest of the fermions at intersections getting a mass by worldsheet instanton corrections, see Chapter \ref{yukint}), so are the relevant fermion content for building up an effective theory.

The net number of chiral fermions is nothing but an example among many features of toroidal intersecting branes that readily generalize to more complicated geometries. Along the study of theoretical and phenomenological aspects of the compactifications presented above, we will try to be as general as possible, emphasizing the results that also hold in a general setup. In fact, in order to deal with D6-branes wrapping complicated 3-cycles, we do not need to consider a very complicated six-manifold, but a simple factorisable $T^6$ will do. Indeed, notice that by restricting ourselves to configurations of factorisable 3-cycles we are not exploring the whole homology space $H_3(T^6, \inte)$, which is an integer lattice of dimension $b_3(T^3) = 20$, but only a small subset of the $(H_1(T^2,\inte))^3$ sublattice of dimension $8$. In addition, two factorisable cycles $[\Pi_a]$ and $[\Pi_b]$ will add up in homology space to a class of 3-cycles $[\Pi_\g] \in (H_1(T^2,\inte))^3$ which in general cannot be expressed as a product of three 1-cycles. Hence, factorisable 3-cycles do not form a sublattice of $H_3(T^6, \inte)$. As the addition of 3-cycles has a clear meaning as a physical process (brane recombination) we cannot simply neglect the existence of these non-factorisable classes of 3-cycles on $(H_1(T^2,\inte))^3$. Throughout our discussion, then, we will try to express the most part of our theoretical results in terms of a generalized formalism which includes factorisable and non-factorisable cycles on equal footing, and whose notation has been collected in the Appendix \ref{qbasis}. This formalism would also be useful for more general compactifications with $\M_6 \neq T^6$. Although most of the work regarding intersecting brane model building has been performed either on toroidal compactifications, either in orbifold and orientifold variants of them, some generalizations to Calabi-Yau threefold geometries have been performed in \cite{Blumenhagen:2002wn,Uranga:2002pg,Blumenhagen:2002vp}. Some of these geometries admit an orientifold projection as well, where the geometrical parity $\R = \R_{(3)}\R_{(5)}\R_{(7)}$ of $(T^2)^3$ is substituted by an anti-holomorphic involution $\bar\sig$, acting locally as (\ref{conjugation}). The main features of D6-branes orientifolded models of the last section are readily generalized in this way.

\begin{figure}
\centering
\epsfxsize=4.8in
\hspace*{0in}\vspace*{.2in}
\epsffile{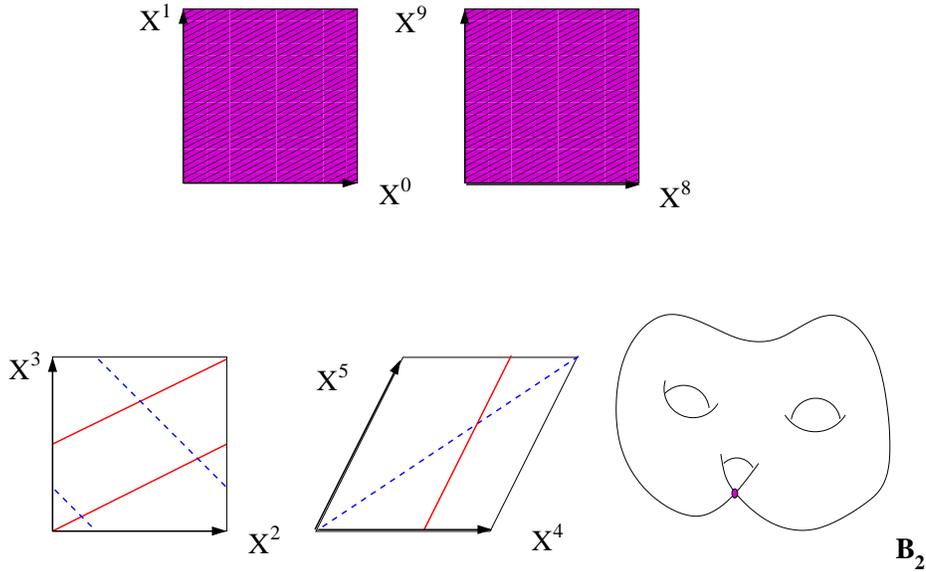}
\caption{\small{Intersecting brane world setup involving orbifold singularities. In the picture we consider configurations of D5-branes filling four-dimensional Minkowski spacetime, wrapping factorisable 2-cycles of ${\bf T^2 \ti T^2}$ and sitting at a singular point of some compact two-dimensional space
$\B_2$. }
\label{world2}}
\end{figure}

In some particular limits of the geometrical moduli such general threefolds may develop orbifold singularities. In particular, we can see the framework of intersecting D$(3+n)$-brane on $T^{2n} \ti \cpx^{3-n}/\inte_N$ for $n = 1, 2$ as a local description of this more general compactification. As an example we may consider the compactification on the r.h.s. of (\ref{D8c}) which, forgetting about the orientifold quotient, presents nine orbifold singularities with $n=1$ and $N=3$. The framework we foresee, however, is not constrained to local models for toroidal orbifolds. In general, we could consider a compactification such as $M_4 \ti T^{2n} \ti \B_{6-2n}$, where $\B_{6-2n}$ is a $(6-2n)$-dimensional manifold developing an orbifold singularity, where the D$(3+n)$-branes sit. An example of this setup for $n=2$ is depicted in figure \ref{world2}. As a general rule $\B_{6-2n}$ should be compact, so that when dimensionally reducing down to $M_4$ we recover four-dimensional gravity. An important point in this construction, is that we do not expect the physical features regarding the effective field theories arising from the D-branes worldvolumes to depend on the global geometry and topology of $\B_{6-2n}$. Indeed, they should only be sensitive to the local geometry of the singularity where they sit. This simple fact allows us to forget, at least in a first step of semi-realistic model-building, about the global properties of $\B_{6-2n}$ which would only affect to the gravity sector of the theory. This is a particular application of the bottom-up approach proposed in \cite{Aldazabal:2000sa}, and which we will follow in our constructions.

Configurations involving orbifold singularities do also admit further generalizations. Indeed, following the general discussion performed with intersecting D6-branes, is natural to generalize our constructions to $M_4 \ti {\bf A}_{2n} \ti \B_{6-2n}$. Here ${\bf A}_{2n}$ is not necessarily $T^{2n}$ but a general compact Riemannian $2n$-dimensional manifold where D$(3+n)$-branes wrap $n$-cycles, while sitting at a singular point of $\B_{6-2n}$. Even more generally, the complete compact space may not be a direct product ${\bf A}_{2n} \ti \B_{6-2n}$, but rather a fibration of ${\bf A}_{2n}$ over $\B_{6-2n}$, as long as the singular fibres are away from the D-brane location (i.e., the orbifold singularity in $\B_{6-2n}$). By the same token, we may also incorporate the orientifold projection in the geometries that admit an involution $\bar\sig$ as a symmetry.


\chapter{Effective theory spectrum \label{spectrum}}

Given the intersecting D-brane configurations described in the previous chapter, one may wonder which kind of $D=4$ effective theories may be obtained from them in the low energy limit. In order to shed some light into this problem, in the present chapter we will compute the low energy spectrum that we get in generic toroidal, orbifold and orientifold compactifications. We will only bother about the states whose tree level mass is under the string scale $M_s$, which we will suppose above the energy scales probed by accelerators. Both closed and open string sectors will be discussed, giving special emphasis to the latter, where the gauge and chiral content of the effective theory will arise. In particular, we will pay special attention to the open string excitations giving rise to the {\it massless} fermionic content of our theory, which we should identify with the SM chiral spectrum. In addition, we will care about {\it tachyonic} scalars. In general these are unwanted excitations, since they signal an instability of the configuration. However, in some cases they can be interpreted as symmetry breaking particles, the most famous case being the Standard Model Higgs particle. Finally, we will study the spectrum of {\it massive} particles below the string scale, whose presence may put phenomenological constraints to semi-realistic models. In particular, we will see how intersecting D-brane models present an specific sector of light particles, baptized as {\it gonions} in \cite{Aldazabal:2000cn}.

Usually, analysing the low energy spectrum of a string-based class of models is the first step towards building a semirealistic $D=4$ compactification from a superstring theory. The generic features of such spectrum will dictate its phenomenological viability. In particular, in order to build a semirealistic model important issues as chirality, family triplication and realistic gauge group must be possible to achieve. As we will see in this chapter, intersecting brane worlds naturally contain all these characteristics.

\section{Closed string spectrum}

The closed string spectrum of any superstring theory contains a massless bosonic sector which is universal, i.e., it does not depend on the specific theory under consideration. Such sector generates $D = 10$ Einstein-Hilbert action as a low energy effective theory \cite{GSW1,Polchi2,Johnson:2000ch}. In particular, there always exist a massless spin 2 excitation which at low energies can be identified with the ten-dimensional graviton. In addition, the degrees of freedom corresponding to a $D=4$ graviton will always survive under compactification, either in orbifold, orientifold, or more general six-dimensional manifolds. As a consequence, the closed string sector of our theory will contain gravity. Even if, for the purpose of finding the bosonic and fermionic content of the SM, the closed string sector plays a secondary role, we will briefly describe the spectrum we obtain from it under toroidal, orbifold and orientifold compactifications.

\subsection{Toroidal case}

In the case of a toroidal compactification, we can obtain the associated closed string spectrum from simple dimensional reduction on $T^6$ of the $D = 10$ type IIB supergravity action
\footnote{In the intersecting D-brane models discussed in the previous chapter, toroidal compactifications have been related with D6-branes at angles. One could hence expect that the closed string spectrum should be derived from type IIA supergravity instead of type IIB. However, we could always consider the T-dual picture involving type IIB D9-branes with fluxes. In fact, both type II theories yield the same spectrum by compactification on a circle, hence on a $T^6$.}. 
Such supergravity theory possess $\N = 2$ supersymmetry in $D=10$, with both gravitinos of the same chirality. The bosonic content is contained in the gravity multiplet, and is shown in table \ref{typeIIB}. In total they describe 128 bosonic degrees of freedom.
\begin{table}[htb]
\renewcommand{\arraystretch}{2.5}
\begin{center}
\begin{tabular}{cc|cc}
NSNS sector & degrees of freedom & RR sector & degrees of freedom \\
\hline
Dilaton $\phi$ & 1 & 0-form $A_0$ & 1 \\
2-form $B_{\mu\nu}$ & 28 & 2-form $A_2$ & 28 \\
Graviton $g_{\mu\nu}$ & 35 & 4-form $A_4$ & 35
\end{tabular}
\caption{Bosonic content of type IIB supergravity in $D = 10$.}
\label{typeIIB}
\end{center}
\end{table}

Each of these 128 bosonic degrees of freedom will be conserved under dimensional reduction down to $D=4$, becoming $70$ scalars, $28$ 1-forms or vectors and $1$ four-dimensional graviton. These particles and their supersymmetric partners precisely fit into a $D=4$ $\N = 8$ supergravity multiplet:
\beq
\left(-2, -\frac{3}{2}^8, -1^{28}, -\frac{1}{2}^{56}, 0^{70},
\frac{1}{2}^{56}, 1^{28}, \frac{3}{2}^8, 2\right).
\label{sugra8}
\eeq

\subsection{Orbifold case}

\begin{figure}[ht]
\centering
\epsfxsize=6in
\epsffile{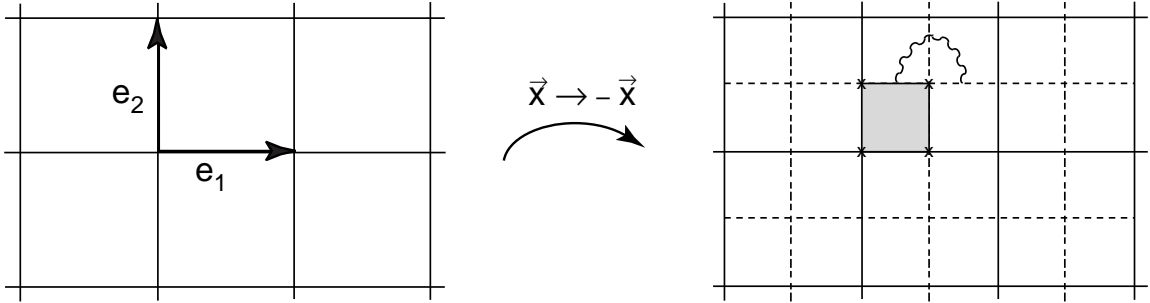}
\caption{Twisted sector in the $T^2/\inte_2$ orbifold. Figure taken from \cite{Quevedo:1996sv}.}
\label{twisted}
\end{figure}

The orbifold case is more involved than plain toroidal compactifications. In general, we can distinguish between two sectors: the {\it twisted} and the {\it untwisted} sector. The untwisted sector contains closed strings which are already closed without taking account of the orbifold identification of points. For instance, if we consider a toroidal orbifold $T^{6-2n}/\inte_N$, the untwisted strings correspond to closed loops on $T^{6-2n}$. On the contrary, the twisted sector contains strings which are closed up to an orbifold group element action, i.e., they are not closed before orbifold identification of points. See figure \ref{twisted} for an example in $T^2/\inte_2$. Closed string modes from the untwisted sector can propagate on the whole target space, including the non-compact dimensions $M_4$ as well as the six-dimensional manifold where we are compactifying our theory. On the contrary, closed strings from the twisted sector are confined to propagate on the orbifold singularity. In the class of orbifold compactifications introduced in the previous chapter, this orbifold singularity is of the form $M_ 4\ti T^{2n}$.

In the same way as described in the toroidal case, the untwisted sector will provide $D=10$ fields such as the graviton, dilaton and antisymmetric $B_{\mu\nu}$ tensor. Dimensional reduction of them down to $D=4$ will depend on the topology and geometry of the compact space $\M_6$ we are considering and, in general, some degrees of freedom will not be well-defined. In particular, the orbifold singularity imposes severe constraints on them, projecting out those fields which are not invariant under the $\G$ action. The $D=4$ d.o.f. of the graviton, $B_{\mu\nu}$ and dilaton will be always invariant under such action, so $D=4$ gravity is guaranteed in a generic compactification. On the other hand, reduction of NSNS massless fields in the compact dimensions will yield scalar fields with information on the geometry of $\M_6$, named as {\it moduli} fields. Finally, the number of supersymmetries preserved on such compactification will depend on the holonomy group of $\M_6$. In a generic ${\bf CY_3}$  compactification we will have all our closed string spectrum arranged in $D=4$, $\N =2$ representations. The global geometry of $\M_6$ is not, however, relevant at this stage of the construction, since it will not affect directly the local physics at the singularity, where our gauge theory will arise. The embedding of such singularity in a full compact manifold belongs to a second step in the bottom-up philosophy, which will not be discussed in the present thesis (see \cite{Aldazabal:2000sa} for some considerations on this second step). There is hence no point in considering the low-energy spectrum arising from the untwisted closed string sector. 

The closed string twisted sector can be easily computed by means of the bosonic language. Closed string states will be described by a tensor product of {\it left} and {\it right} oscillations, hence they will be of the form $r_L \otimes (-r_R)$, where the four dimensional vectors $r_{(L,R)} \in (\inte + \med + \nu_{(L,R)})^4$ describe left and right oscillators in bosonic language, respectively. Here $\nu_{L}$ ($\nu_{R}$) represents the Ramond or Neveu-Schwarz sectors for the left (right) part of the spectrum. We will then have four sectors, given by NSNS $(\nu_L = \med, \nu_R = \med)$, NSR $(\nu_L = \med, \nu_R = 0)$, RNS $(\nu_L = 0, \nu_R = \med)$ and RR $(\nu_L = 0, \nu_R = 0)$. GSO projection will be equal for both left and right states in type IIB theory and opposite in type IIA. Finally, in this bosonic language the geometric action of an orbifold group $\inte_N$ will be encoded in a twist vector $v_\om \in \re^4$, where $\om$ is a generator of the discrete group.

The spectrum on a generic orbifold construction is obtained by looking at invariant states under the orbifold action. Under the action of the generator $\om \in \inte_N$ a state $r_L$ will pick up a phase $exp (2\pi i \ r_L \cdot v_\om)$. If the right state $r_R$ picks up exactly the same phase, then the tensor product $r_L \otimes (-r_R)$ will be invariant under the geometrical action of $\om$, hence under the full action of $\G$. Moreover, we are interested in massless oscillations, so the state vectors $r$ will have to satisfy
\beq
\a^\prime M^2 = {(r_{L,R} + k v_\om)^2 \over 2} -\med + E_{0} = 0,
\label{mass3}
\eeq
where $k$ represents the $k^{th}$ twisted sector (i.e., the set of strings closed up to the action of $\om^k$) and the vacuum energy $E_0$ is given by 
\beq
E_{0} = \sum_{i=1}^3 \med |k v_\om^i| (1 - |k v_\om^i|).
\label{vacuum2}
\eeq

Let us consider orbifold compactifications of type IIA theory on $T^2 \ti \cpx^2/\inte_N$. In this case the most general twist vector will be given by $v_\om = {1 \over N} (0,b_1,b_2,0)$. If we suppose $b_1, b_2 > 0$, then the Ramond twisted sector $k = 1$ will contain the left states $r_{\eps,L} = \med(-\eps,-,-,\eps)$ and a right state $r_{\eps, R} = \med(\eps,-,-,\eps)$, with  $\eps = \pm$ (notice that GSO projection differs in both sectors). In general, the four massless states of $k = 1$ will pick up a phase $exp (-\pi i \ (|b_1| + |b_2|))$ under the action of $\om$. Tensoring any of the two left states with their right counterparts we will obtain an invariant state under the orbifold action. The combinations arising on the RR sector are shown in table \ref{twistspec}. We thus obtain two scalars and a vector.

\begin{table}[htb]
\renewcommand{\arraystretch}{2.5}
\begin{center}
\begin{tabular}{cc}
RR state & helicity \\
\hline
$r_{+,L} \otimes (-r_{-,R})$ & 1 \\
$r_{+,L} \otimes (-r_{+,R})$ & 0 \\
$r_{-,L} \otimes (-r_{-,R})$ & 0 \\
$r_{-,L} \otimes (-r_{+,R})$ & -1
\end{tabular}
\caption{Closed string twisted spectrum arising from the RR sector. The helicity of the tensor product of two states is given by the sum of individual helicities. Notice that the helicity of $-r$ is minus the helicity of $r$.}
\label{twistspec}
\end{center}
\end{table}

In case the orbifold preserves some bulk supersymmetry as, e.g., if $b_1 = b_2 = 1$, then there will be massless states on the NS sector as well, given by $r_{NS} = (0,-1,0,0)$ and $r_{NS} = (0,0,-1,0)$. Tensoring again left and right states we will obtain massless invariants arising on the NSNS sector. Repeating the procedure in the NSR and RNS sectors, we will finally obtain $D=4$ bosonic and fermionic states filling up an $\N = 4$ gauge supermultiplet
\beq
\left(-1, -\med^4, 0^6,
 \frac{1}{2}^{4}, 1\right),
\label{gauge4}
\eeq
corresponding to a $U(1)$ gauge group. We can repeat the above computations for arbitrary $k$, finding the same final result on each. Taking account of all the $N-1$ twisted sectors, then we obtain a gauge group $U(1)^{N-1}$.

If $|b_1| \neq |b_2|$, then the orbifold twist will not be supersymmetric, and one of the NS states will become tachyonic. This also happens when we consider type IIB compactifications on $T^4 \ti \cpx/\inte_N$, where the twist vector is $v_\om = {1 \over N} (0,0,-2,0)$. In this case, there are four massless states on the Ramond sector given by $r_{R} = \med(\pm,\pm,\pm,+)$ with the proper GSO projection implemented (notice that now this projection is the same for both left and right sectors). By tensoring states, in the RR sector we obtain a 0-form and a 2-form for each twisted sector $k$, both defined in the eight dimensions fixed by the action of the orbifold which, in the case under study, is $M_4 \ti T^4$. We will thus obtain four type IIB RR twisted $p$-forms of even $p$:
\beq
\begin{array}{cccc}
A_0^{(k)}, & A_2^{(k)}, & A_4^{(k)}, & A_6^{(k)},
\end{array}
\label{pforms}
\eeq
which are to be seen as $p$-forms defined in $M_4 \ti T^2 \ti T^2$. Here $k = 1, \dots, N-1$ denotes the $k^{th}$ twisted sector of the theory, whereas $A_p^{(k)}$ and $A_{6-p}^{(k)}$ field strengths are related by Hodge duality in $D = 8$.

On the other hand, in the NS sector we will obtain a scalar $r_{NS} = (0,0,0,1)$ whose squared mass is always negative. Indeed, on the $k^{th}$ twisted sector the mass formula reads $\a' M^2 = -\frac {k}{N}$ for this state, both in the left and right sector, and tensoring both we obtain a closed string NSNS scalar excitation with mass $\a' M^2 = -\frac {2k}{N}$. The antiparticle of such tachyonic state will come from the $(N-k)^{th}$ twisted sector, so we finally obtain $(N-1)/2$ complex tachyons (we are supposing $N$ odd, which will be the main case under study in this thesis).

A classification of compact non-supersymmetric orbifolds, both in type II and heterotic theories has been performed in \cite{Font:2002pq}, where detailed computations of the spectrum as the above presented have been done. In addition, condensation of closed type II tachyons has been analysed in the case of non-compact orbifolds in \cite{Adams:2001sv,Harvey:2001wm,Dabholkar:2001wn}. The fate of closed string tachyons in compact orbifolds has however not been studied in detail, and this is supposed to be the case involving a semirealistic construction. Since the dynamics of these tachyons is not totally understood, we will ignore their existence in the following.

\subsection{Orientifold case}

The main difference regarding orientifold compactifications as (\ref{dual}), compared to plain toroidal compactifications, is that the initial closed string massless spectrum in $D=10$ corresponds to type I supergravity, and hence preserves half of the supersymmetries present in the toroidal case. Again, since the whole theory is dual to the orientifold (\ref{typeI}) of type IIB theory, the bosonic content in the decompactification limit can be identified with the type IIB fields invariant under the action of $\O$. These fields are the dilaton, graviton and the RR 2-form, which add up to 64 bosonic degrees of freedom. Under dimensional reduction to $D=4$ we will obtain an $\N = 4 $ spectrum made of a supergravity multiplet
\beq
\left(-2, -\frac{3}{2}^4, -1^{6}, -\frac{1}{2}^{4}, 0\right) +
 \left(0, \frac{1}{2}^{4}, 1^{6}, \frac{3}{2}^4, 2\right)
\label{sugra4}
\eeq
and six gauge multiplets like (\ref{gauge4}).

Compactifications involving an orientifold singularity (\ref{singuori}) are related to plain orbifold constructions in the same manner as plain orientifold constructions are related to toroidal compactifications. In general, we should consider closed string states already present in type II orbifold compactification, and then impose a further identification under $\OR$. In a $\inte_N$ orbifold, this translates into the identification of $k$ and $N - k$ sectors, obtaining half of the previous spectrum. For instance, in case we are dealing with an orbifold $\cpx^2/\inte_N$ we will obtain a $U(1)^{\frac{N-1}{2}}$ gauge group for odd $N$ and $U(1)^{\frac{N}{2}}$ for $N$ even. By the same token, the $N-1$ twisted tachyonic scalars will not be complex but real fields. In general, orbifold and orientifold singularities with arbitrary group $\G$ can be studied by means of quiver diagrams \cite{tesisraul,Douglas:1996sw,Johnson:1996py,Douglas:1997de,Uranga:2000ck}.

\section{Open string spectrum}

We will now discuss the open string spectrum of the compactifications described above. This sector of the theory turns out to be of utmost interest from the model-building perspective since, as already indicated, the lightest excitations of open strings localized between two D-branes intersections generically yield chiral fermions charged under unitary gauge groups. Moreover, the full spectrum at this intersection is generically non-supersymmetric. This provides us with quite an appealing framework for finding the SM particle content. As in the closed string part, we will first describe the spectrum arising in a plain toroidal compactification, and then we will study the variations we have to perform in order to deal with the orbifold and orientifold cases.

\subsection{Toroidal case}

A specific toroidal compactification will contain several stacks of type IIA D6-branes, say $K$ of them, each stack $a$ containing $N_a$ coincident D6-brane wrapping a 3-cycle $\Pi_a$ of homology class $[\Pi_a] \in H_3(T^6, \inte)$ ($a = 1, \dots, K$). Usually we will consider factorisable configurations, which are those that allow us to express these 3-cycles as $\Pi_a = (n_a^{(1)}, m_a^{(1)}) \otimes (n_a^{(2)}, m_a^{(2)}) \otimes (n_a^{(3)}, m_a^{(3)})$, i.e., as product of three 1-cycles, each wrapped on a different $T^2$ factor in $T^6 = T^2 \ti T^2 \ti T^2$.

In general, each stack $a$ will have associated some Chan-Paton degrees of freedom, represented by an $N_a \ti N_a$ unitary matrix. These will give rise to a $U(N_a)$ gauge theory localized at the D6$_a$-brane worldvolume, i.e., at $\Pi_a$ \cite{Polchinski:1996na,Johnson:2000ch}. These massless states of the theory arise from strings with both endpoints on the same D6-brane, which will be denoted by D6$_a$D6$_a$ sector. Being $\med BPS$ states, a D6-brane on a flat factorisable 3-cycle breaks half of the $D=4$ $\N=8$ supersymmetries of a toroidal compactification of type IIA, i.e., it yields a $D=4$ $\N=4$ spectrum upon dimensional reduction. Consequently, a factorisable D6-brane will contain a Super Yang-Mills $\N=4$ spectrum, which in terms of $D=4$ fields of the gauge supermultiplet (\ref{gauge4}) can be expressed as the $U(N_a)$ gauge bosons plus six real scalars and four Majorana fermions transforming in the adjoint of the unitary group. The six scalars have a definite geometrical interpretation in terms of the D6-brane configuration. Namely, three of their v.e.v.'s parametrize the D6-brane position on $T^2 \ti T^2 \ti T^2$, whereas the other three v.e.v.'s represent the value of the flat connection background around the each 1-cycle of the D6-brane worldvolume, i.e., they parametrize Wilson lines on the compactified gauge theory.

Chiral fermions will arise from the D6$_a$D6$_b$ sector, which corresponds to (oriented) strings starting on the stack of D6-branes $a$ and ending on the stack $b$. The lowest string excitations on this sector will always be localized at the intersections of both D6-branes, i.e., at $\Pi_a \cap \Pi_b$. As mentioned on the previous chapter, generically the lightest state on the R sector will be given by a massless fermion. Indeed, let us consider some generic twist vector $v_\vt^{ab} = (\vt_{ab}^1,\vt_{ab}^2,\vt_{ab}^3,0)$ with non-vanishing angles, which specifies the boundary conditions on the D6$_a$D6$_b$ sector. By simple inspection of the mass formula, is easy to see that such fermion will be unique and given by
\beq
r_R = \med \left(-s(\vt_{ab}^1), -s(\vt_{ab}^2), -s(\vt_{ab}^3), \prod_{i=1}^3 s(\vt_{ab}^i) \right),
\label{fermion2}
\eeq
where $s(\vt_{ab}^i) \equiv {\rm sign \ }(\vt_{ab}^i)$, and the fourth component is fixed by applying GSO projection. If, as before, we consider $0 < \vt_{ab}^i < 1$, this expression reduces to (\ref{fermion}), which in the effective theory represents a fermion with four-dimensional helicity $\med$. We now see that we may also get a fermion of opposite chirality by choosing different angles. An efficient way to compute the chirality of such fermions is by means of the intersection number $I_{ab}$ between the two 3-cycles $\Pi_a$ and $\Pi_b$. In general, the intersection number of a pair of 3-cycles in a six-dimensional manifold $\M_6$ is a topological invariant which only depends on the homology classes $[\Pi_a]$, $[\Pi_b] \in H_3(\M_6, \inte)$ \cite{candelas}. In the particular case of factorisable 3-cycles on $T^2 \ti T^2 \ti T^2$, such number can be expressed as a product of three intersection numbers, each corresponding to a pair of 1-cycles on a different $T^2$ factor
\beq
I_{ab} \equiv [\Pi_a]\cdot[\Pi_b] = \prod_{i=1}^3 I_{ab}^{(i)} = \prod_{i=1}^3 \left(n_a^{(i)} m_b^{(i)} - m_a^{(i)} n_b^{(i)} \right).
\label{inter}
\eeq
Since $s(\vt_{ab}^i) = s(I_{ab}^{(i)})$, the four-dimensional chirality of (\ref{fermion2}) will be given by $s(I_{ab})/2$. The convention we will use in the following will identify left-handed fermions with the quantum number $\med$, i.e., with a positive intersection number, whereas right-handed fermions will be identified with a negative intersection number. Notice that the D6$_b$D6$_a$ sector, which has a vector twist $v_\vt^{ba} = - v_\vt^{ab}$, contains the antiparticles of the D6$_a$D6$_b$ sector, consistently with the fact that $I_{ba} = - I_{ab}$. Hence, when identifying SM matter with fermions with the same chirality, say left-handed, we should always look at the sectors with positive intersection number.

The intersection number $I_{ab}$ not only gives information about the chirality of the massless fermions at the intersection, but also provide us with their multiplicity. Indeed, in case of factorisable (straight) $n$-cycles on a factorisable $T^{2n}$, we find that $|I_{ab}| = \# (\Pi_a \cap \Pi_b)$, which means that stacks $a$ and $b$ will intersect exactly $|I_{ab}|$ times. Moreover, the same boundary conditions will hold on each intersection, hence each will provide the same open string spectrum, in particular the same massless fermions in the $R$ sector. As a result, we find that the massless spectrum on the D6$_a$D6$_b$ sector will contain $I_{ab}$ left-handed fermions transforming in the bifundamental representation $(N_a, \bar N_b)$ of the corresponding gauge groups 
\footnote{Is easy to find the representation under which every state transforms, by simply performing an arbitrary gauge transformation on the Chan-Paton degrees of freedom and then checking how does this open string state transform under it.}. 
In a general situation $\Pi_a, \Pi_b \subset \M_6$, however, the intersection number will not be the (signed) number of intersections between two stacks, that is, $|I_{ab}| \leq \# (\Pi_a \cap \Pi_b)$. It easy to see that $\# (\Pi_a \cap \Pi_b)$ is not a topological invariant, and we can change it by continuously deforming $\Pi_a$ and  $\Pi_b$. A topological invariant can be constructed by considering the number of intersections in $(\Pi_a \cap \Pi_b)$ yielding left-handed fermions and subtracting from them the number of intersections yielding right-handed fermions. The final quantity $\#_L (\Pi_a \cap \Pi_b) - \#_R (\Pi_a \cap \Pi_b)$ is nothing but the intersection number $I_{ab}$ of the two cycles. In general, then, the intersection number $I_{ab}$ provides us with the {\em net} chiral content of the D6$_a$D6$_b$ sector. This is, however, the relevant part of the spectrum in building up the effective theory, since we expect the rest of the fermions to couple among each other and get a vector-like mass by means of world-sheet instantons (see Chapter \ref{yukint} for details on this).

Besides chiral fermions there will exist several light scalars at each intersection, arising from the NS sector of the theory. Their gauge quantum numbers will be the same as the chiral fermions, i.e., they will transform in bifundamental representations. Their (tree level) mass will now depend on the particular angles yielded between the pair of D6-branes. If we consider two factorisable stacks $a$ and $b$, the four states which can become the lightest scalar can be expressed in bosonic language as
{\bea
\begin{array}{cc}
\vspace{0.2cm}
{\rm \bf State} \ (r_{NS}) \quad & \quad {\bf Mass^2} \\
\vspace{0.1cm}
(-s(\vt^1),0,0,0) & \alpha' M^2 =
\frac 12(-|\vartheta^1|+|\vartheta^2|+|\vartheta^3|) \\
\vspace{0.1cm}
(0,-s(\vt^2),0,0) & \alpha' M^2 =
\frac 12(|\vartheta^1|-|\vartheta^2|+|\vartheta^3|) \\
\vspace{0.1cm}
(0,0,-s(\vt^3),0) & \alpha' M^2 =
\frac 12(|\vartheta^1|+|\vartheta^2|-|\vartheta^3|) \\
(-s(\vt^1),-s(\vt^2),-s(\vt^3),0) & \alpha' M^2
= 1-\frac 12(|\vartheta^1|+|\vartheta^2|+|\vartheta^3|)
\label{scalars}
\end{array}
\eea}
where for simplicity we have only consider $r_{NS}$ when defining our state (in bosonic language such state would be actually defined by $r_{NS} + v_{\vt}^{ab}$), and we have suppressed the subscripts $ab$.

In the limit $\vt^i \raw 0$ three of the scalars in (\ref{scalars}) and some other three more massive scalars at the intersection become massless, signaling that $\N = 4$ supersymmetry has been recovered. In this sense, we can consider the above scalars as superpartners of the chiral fermion at the same D6$_a$D6$_b$ intersection, at least in some limits of the configuration. In general, if any of the scalars (\ref{scalars}) becomes massless then the whole sector D6$_a$D6$_b$ will present in its spectrum a mass degeneration between bosons and fermions, the states arranging themselves in $D=4$ $\N = 1$ supermultiplets. It may well also happen that not only one but two of the above scalars become massless, which implies that the whole spectrum fills up $D=4$ $\N = 2$ representations. These degeneracies of the open string spectrum are not a coincidence, but signal the fact that two stacks of D-branes may preserve one or several supersymmetries in common depending on their relative position. We postpone a more detailed study of the supersymmetry structure on D-branes at angles to Chapter \ref{SUSY}.

Supersymmetry will be broken if none of the `supersymmetric partners' of the chiral fermion at the intersection is massless. It may well happen that the four of the scalars in (\ref{scalars}) are massive, which will yield a non-supersymmetric spectrum. The pair of D6-brane stack will then be a non-supersymmetric configuration, in principle stable. On the contrary, one of the scalars may also become tachyonic. Although the spectrum is again non-supersymmetric, the presence of this tachyon signals a (tree level) instability of the system (D6$_a$, D6$_b$). Just as in the case of the (D-brane, anti-D-brane) system \cite{Sen:1999mg,Lerda:1999um,Schwarz:1999vu,Gaberdiel:2000jr}, we should expect (D6$_a$, D6$_b$) to decay to some other configuration of lower energy, while preserving some definite quantum numbers (RR charges). Otherwise stated, we expect the system of two branes $a$ and $b$ to lower its tension (which is the sum of their individual tensions) by decaying to a bound state $\g$ with less tension and same RR charges. We call such a process D-brane recombination. Since the configurations we are considering have no fluxes on the D-branes worldvolumes, the tension of a D-brane is proportional to its volume, and the whole problem reduces to a geometrical problem: Does it exist a 3-cycle $\Pi_\g$ such that $[\Pi_g] = [\Pi_a] + [\Pi_b]$ in $H_3(T^6,\inte)$ and with Vol ($\Pi_\g$) $<$ Vol ($\Pi_a$) + Vol ($\Pi_b$)? This problem, known in the mathematical literature as the  {\it angle criterion}, was addressed by Nance and Lawlor on the eighties \cite{Nance,Lawlor,Harvey}, and basically states that the existence of $\Pi_\g$ corresponds to the appearance of a tachyon at some intersection.

To sum up, we can express the gauge group and chiral fermion content arising from a general configuration of factorisable D6-branes on $T^6$ as
\bea
& \prod_a U(N_a) \nn \\ & \sum_{a<b} I_{ab} (N_a, \bar N_b)
\label{spec}
\eea
Notice that this formula is actually valid for a general configuration of intersecting D6-branes on an arbitrary compact manifold $\M_6$.

\subsection{Orbifold case}

Let us now consider the new features that the orbifold case introduces with respect to a toroidal compactification. Just as in the closed string sector, when computing the open string states that are present in an orbifolded theory we must only consider those that are invariant under the orbifold group action. In the open string sector we must consider both the geometrical action of $\G$ and its action on the internal degrees of freedom of the D-branes, i.e., the Chan-Paton factors. The former action will induce an action on the oscillation modes, which will pick up a definite phase depending on their spacetime labels. The latter action, on the other hand, will affect the Chan-Paton states, again contributing with a phase to the string wavefunction. If both phases compensate each other, then the corresponding open string state will be well defined on the orbifolded theory.

Let us be more specific and denote an open string state by $|\psi, ij \rangle$, where $\psi$ represents the string oscillation mode and $i, j$ represent the Chan-Paton state. That is, $i$ stands for the D-brane where the oriented open string begins and $j$ the one where it ends. When considering the D$p_a$D$p_b$ sector, then these labels can take the values $i \in \{1, \dots, N_a\}$ and $j \in \{1, \dots, N_b\}$. The action of the orbifold group $\G$ on this open string state is
\bea
g: & |\psi, ij \rangle \  \longmapsto \ \left(\g_g\right)_{i i^\prime} |g \cdot\psi, i' j' \rangle \left(\g_g^{-1}\right)_{j^\prime j},
\label{orbiaction}
\eea
where $g \in \G$ and $\g_g$ is a unitary matrix associated to this group element \footnote{To be more precise, $\g_g$ is a projective representation of $\G$. That is, the image under a projective homomorphism $\g : g \mapsto \g_g$ from the group $\G$ to the set of unitary matrices. In particular, this implies that the matrices $\g_g$ must satisfy the group law up to phases \cite{Gimon:1996rq}.}.

In case $\G = \inte_N$, then the geometric action $(g\cdot\psi)$ acting on the oscillation modes can be represented by means of a twist vector $v_\om$ \cite{Aldazabal:1998mr}. Namely, if the state $\psi$ is represented in the bosonic language by a vector $r$, then under the action of $\inte_N$'s generator $\om$ it will pick up a phase exp$(2\pi i r \cdot v_\om)$. The action on the Chan-Paton degrees of freedom, in turn, can also be determined by the action of such generator $\om$. Given a stack $a$ of D-branes, the most general form for a $U(N_a)$ element representing it is given by
\beq
\gamma_{\om,a} = {\rm diag} \left( {\bf 1}_{N_a^0}, e^{2 \pi i \frac{1}{N}}{\bf 1}_{N_a^1}, \ldots, e^{2 \pi i \frac{N-1}{N}}{\bf 1}_{N_a^{N-1}} \right),
\label{gamma}
\eeq
where we have to impose $\sum_{i=0}^{N-1} N_a^i = N_a$. A way of interpret (\ref{gamma}) is to consider that, from our stack of $N_a$ coincident D-branes, $N_a^0$ of them are invariant under the action of $\om$, $N_a^1$ of them pick up a phase $e^{2 \pi i \frac{1}{N}}$ under it, etc \dots

From the above considerations, is possible to compute in general the open string spectrum. Let us first consider the case of D4-branes sitting at an orbifold singularity of the kind $\cpx^2/\inte_N$. Recall that the most general twist vector is given by $v_\om = {1 \over N} (0,b_1,b_2,0)$, encoding the geometric action (\ref{rotaorbi}) on $\cpx^2$.

\begin{itemize}

\item{D4$_a$D4$_a$ sector}

The massless states surviving the GSO projection, along with their behaviour under the $\inte_N$ twist $\om$ are
{\bea
\begin{array}{cccc}
\vspace{0.1cm}
{\rm \bf NS \ state} \quad & \quad {\bf Z_N \ phase}  \quad & \quad {\rm \bf R \ state} \quad & \quad {\bf Z_N \ phase} \\
(\pm1,0,0,0) & 1 & \pm\med(-,+,+,+) & e^{\pm\pi i \frac{b_1 + b_2}{N}}\\
(0,\pm1,0,0) & e^{\pm2\pi i \frac{b_1}{N}} & \pm\med(+,-,+,+) & e^{\mp\pi i \frac{b_1 - b_2}{N}}\\
(0,0,\pm1,0) & e^{\pm2\pi i \frac{b_2}{N}}  & \pm\med(+,+,-,+) & e^{\pm\pi i \frac{b_1 - b_2}{N}}\\
(0,0,0,\pm1) & 1  & \pm\med(+,+,+,-) & e^{\pm\pi i \frac{b_1 + b_2}{N}}
\label{sector4aa}
\end{array}
\eea}

The open string spectrum is obtained by keeping states $|\psi, ij \rangle$ invariant under the combined geometric plus Chan-Paton $\inte_N$ action. Those states which are not invariant are not well-defined on the orbifolded theory, and are hence projected out from the spectrum. On the D4$_a$D4$_a$ sector, then, the resulting $D=4$ gauge group and matter fields after the $\inte_N$ projection is performed are given by
{\beq
\begin{array}{rl}
\vspace{0.1cm}
{\rm\bf Gauge\; Bosons} & \quad \prod_a \prod_{i=1}^N U(N_a^i) \\
\vspace{0.1cm}
{\rm\bf Complex\; Scalars} & \quad \sum_a \sum_{i=1}^N \;
[\, (N_a^i,{\ov N}_a^{i+b_1}) + (N_a^i,{\ov N}_a^{i+b_2}) \, 
+ {\bf Adj}_a^i ] \\
\vspace{0.1cm}
{\rm\bf Left\; Fermions} & \quad \sum_a \sum_{i=1}^N \; [\;
(N_a^i,{\ov N}_a^{i-(b_1-b_2)/2}) + (N_a^i,{\ov N}_a^{i+(b_1-b_2)/2}) ] \\
{\rm\bf Right\; Fermions} & \quad \sum_a \sum_{i=1}^N \; [\;
(N_a^i,{\ov N}_a^{i+(b_1+b_2)/2}) + (N_a^i,{\ov N}_a^{i-(b_1+b_2)/2})]
\label{spectrum4aa}
\end{array}
\eeq}
where the index $i$ is defined mod $N$. This spectrum is non-chiral and, generically, non-supersymmetric. In case we have a supersymmetric twist,  $v_\om = {1 \over N} (0,1,-1,0)$, this sector will preserve $\N = 2$ supersymmetry in terms of the four-dimensional effective theory. The above fields will arrange in $\N= 2$ multiplets as
{\beq
\begin{array}{rl}
\vspace{0.1cm}
{\rm\bf Vector \ Multiplet} &\quad \prod_a \prod_{i=1}^N U(N_a^i) \\
{\rm\bf Hypermultiplet} & \quad \sum_a \sum_{i=1}^N (N_a^i,{\ov
N}_a^{i+1})
\label{multiplets4aa}
\end{array}
\eeq}

\item{D4$_a$D4$_b$ sector}

In this sector open strings are twisted by the D4-branes on a single $T^2$, hence the corresponding twist vector will have just one nonvanishing entry and will be of the form $(\vt_{ab},0,0,0)$. For a generic angle $\vt$ we obtain one tachyonic scalar and four massless fermions. Let us suppose that $0 < \vt_{ab} < 1$. Then these states are
{\beq
\begin{array}{ccc}
\vspace{0.1cm}
{\rm\bf Sector} & {\rm\bf State} & {\bf Z_N \ phase}  \\
\vspace{0.1cm}
{\rm NS} & (-1+\vartheta,0,0,0) & 1 \\
{\rm R}  & (-\med+\vartheta,-\med,-\med,+\med) &
e^{-\pi i\frac{(b_1+b_2)}{N}} \\
         & (-\med+\vartheta,-\med,+\med,-\med) &
e^{-\pi i\frac{(b_1-b_2)}{N}}\\
         & (-\med+\vartheta,+\med,-\med,-\med) & 
e^{\pi i\frac{(b_1-b_2)}{N}}\\
         & (-\med+\vartheta,+\med,+\med,+\med) & 
e^{\pi i\frac{(b_1+b_2)}{N}}
\end{array}
\label{sector4ab}
\eeq}

This piece of the spectrum is explicitly non-supersymmetric, even for supersymmetric $\inte_N$ twists. The NS state is always tachyonic, with $\a' M^2 = -\med |\vt_{ab}|$, which again signals an instability against recombining intersecting D4-branes with same Chan-Paton eigenvalue. More precisely, this means that two D4-branes $a$ and $b$ wrapping two different 1-cycles $\Pi_a$ and $\Pi_b$ on $T^2$, cover a length which is greater than that of a third D4-brane wrapped on $\Pi_c$, and such that $[\Pi_c] = [\Pi_a] + [\Pi_b]$. This fact is nothing but a restatement of the well-known triangle inequality. On the other hand, the fermions arising from the R sector will be massless, and will usually provide us with a chiral spectrum. Notice that the antiparticles of these states appear in the D4$_b$D4$_a$ sector, which is twisted by $(-\vartheta_{ab},0,0,0)$. Finally, we must consider the replication of such spectrum by the intersection number $I_{ab}$ of the two 1-cycles $[\Pi_a] = [(n_a,m_a)]$ and $[\Pi_b] = [(n_b,m_b)]$, defined as
\beq
I_{ab} \equiv [\Pi_a]\cdot[\Pi_b] = n_a m_b - m_a n_b
\label{interD4}
\eeq

After considering the Chan-Paton contribution to the orbifold action and projecting out the non-invariant states, we obtain the final spectrum
{\beq
\begin{array}{rl}
\vspace{0.1cm}
{\rm\bf Tachyons} & \sum_{a<b} \sum_{i=1}^N \; I_{ab}\cdot
(N_a^i,{\ov N}_b^i) \\
\vspace{0.1cm}
{\rm\bf Left\; Fermions} & \sum_{a<b} \sum_{i=1}^N \;  I_{ab}\cdot \;
[\; (N_a^i,{\ov  N}_b^{i-(b_1+b_2)/2}) + (N_a^i,{\ov N}_b^{i+(b_1+b_2)/2})
\; ] \\
{\rm\bf Right\; Fermions} & \sum_{a<b} \sum_{i=1}^N \; I_{ab}\cdot
[\; (N_a^i,{\ov  N}_b^{i-(b_1-b_2)/2}) + (N_a^i,{\ov  N}_b^{i+(b_1-b_2)/2})
\; ]
\end{array}
\label{spectrum4ab}
\eeq}

\end{itemize}

Let us now consider the case of D5-branes wrapping factorisable two-cycles on $T^2 \times T^2$ and sitting at the orbifold singularity $\cpx/\inte_N$. The orbifold twist will now be fixed to be $v_\om = {1 \over N} (0,0,-2,0)$. Being explicitly non-supersymmetric, this twist will involve a more economic spectrum.

\begin{itemize}

\item{D5$_a$D5$_a$ sector}

The massless states surviving the GSO projection in both R and NS sectors are
{\beq
\begin{array}{cccc}
\vspace{0.1cm}
{\rm \bf NS\ State} \quad & \quad {\bf \inte_N \ phase}  \quad 
& \quad {\rm \bf R \ State} \quad & \quad {\bf \inte_N \ phase} \\
(\pm1,0,0,0) & 1 & \pm\oh(-,+,+,+) & e^{\mp2\pi i \frac{1}{N}}\\
(0,\pm1,0,0) & 1 & \pm\oh(+,-,+,+) & e^{\mp2\pi i \frac{1}{N}}\\
(0,0,\pm1,0) & e^{\mp4\pi i \frac{1}{N}} & \pm\oh(+,+,-,+) 
& e^{\pm2\pi i \frac{1}{N}}\\
(0,0,0,\pm1) & 1 & \pm\oh(+,+,+,-) &  e^{\mp2\pi i \frac{1}{N}}
\label{sector5aa}
\end{array}
\eeq}

Already at this level we can see that the resulting spectrum is explicitly non-supersymmetric. Indeed, the fourth NS state has the Lorentz quantum numbers of the gauge bosons, while the first and the second will correspond to adjoint scalars, also invariant under the $\inte_N$ action. On the R sector, however, there are no adjoint fermions invariant under $\inte_N$, so that the corresponding gauginos will be projected out. The final spectrum of this sector will be given by
{\beq
\begin{array}{rl}
\vspace{0.1cm}
{\rm\bf Gauge\; Bosons} & \prod_a \prod_{i=1}^N U(N_a^i) \\
\vspace{0.1cm}
{\rm\bf Complex \; Scalars} & \sum_a \sum_{i=1}^N
[\; (N_a^{i},{\ov N}_a^{i-2}) + 2\times  {\bf Adj}_a^i\; ] \\
\vspace{0.1cm}
{\rm\bf Left\; Fermions} 
& \sum_a \sum_{i=1}^N 2 \ti (N_a^{i},{\ov N}_a^{i-1}) \\
{\rm\bf Right\; Fermions} 
& \sum_a \sum_{i=1}^N 2 \ti (N_a^{i},{\ov N}_a^{i-1})
\label{spectrum5aa}
\end{array}
\eeq}

\item{D5$_a$D5$_b$ sector}

Again the chiral and tachyonic spectrum will arise from the mixed sectors. In this case the twist vector will be given by $(\vt_{ab}^1,\vt_{ab}^2,0,0)$, and the lightest states will now be
{\beq
\begin{array}{cccc}
\vspace{0.1cm}
{\rm\bf Sector} & {\rm\bf State} & {\rm\bf \inte_N \ phase} 
& {\rm\bf \a' Mass^2} \\
{\rm NS} & (-1+\vartheta^1,\vt^2,0,0) & 1 & -\oh(\vt^1 - \vt^2) \\
         & (\vt^1,-1+\vartheta^2,0,0) & 1 & \oh(\vt^1 - \vt^2) 
\vspace{0.1cm}\\
{\rm R}  & (-\oh+\vartheta^1,-\oh+\vartheta^2,-\oh,+\oh) &
e^{2\pi i\frac {1}{N}} & 0\\
         &  (-\oh+\vartheta^1,-\oh+\vartheta^2,+\oh,-\oh) &
e^{-2\pi i\frac{1}{N}} & 0
\end{array}
\label{sector5ab}
\eeq}
where $\vt^i \equiv \vt^i_{ab}$ and we have supposed  $0 < \vt^i < 1$, $i = 1,2$. In any case, one of the NS states will be necessarily tachyonic, unless $|\vt^1| = |\vt^2|$ and both are massless. We must also consider the intersection number of both branes
\beq
I_{ab} \equiv [\Pi_a]\cdot[\Pi_b] = I_{ab}^{(1)} I_{ab}^{(2)} 
= (n_a^{(1)} m_b^{(1)} - m_a^{(1)} n_b^{(1)}) (n_a^{(2)} m_b^{(2)}- m_a^{(2)} n_b^{(2)}).
\label{interfive}
\eeq

The final spectrum arising from this sector is thus
{\beq
\begin{array}{rl}
\vspace{0.1cm}
{\rm\bf Tachyons} & \quad \sum_{a<b} \sum_{i=1}^N \; I_{ab} \cdot
(N_a^i,{\ov N}_b^i) \\
\vspace{0.1cm}
{\rm\bf Left\; Fermions} & \quad \sum_{a<b} \sum_{i=1}^N \;
I_{ab} \cdot (N_a^i,{\ov  N}_b^{i+1}) \\
{\rm\bf Right\; Fermions} & \quad \sum_{a<b} \sum_{i=1}^N \;  I_{ab} \cdot
(N_a^i,{\ov  N}_b^{i-1})
\end{array}
\label{spectrum5ab}
\eeq}

\end{itemize}

\subsection{Orientifold case}

The low energy spectrum arising from an orientifold compactification will be of particular interest for our model-building purposes. Since the $\OR$ action maps a D-brane stack $a$ into its mirror image $\OR a$, generically with some other wrapping numbers, we must consider new sectors in our theory. These new sectors, in particular the new representations associated with them, will play an essential role when looking for the minimal chiral content of the Standard Model, as we will see in Chapter \ref{models}. It is also important to keep in mind that in orientifold constructions the final spectrum of our theory will arise from combination of states under $\OR$, and not only under the orbifold group. 

Let us first compute the spectrum in the case of type IIA D6-branes in $T^6/(1 + \OR)$. The relevant open string sectors are

\begin{itemize}

\item{D6$_a$ D6$_a$ sector ($aa$)}

The $\OR$ action takes this sector to the $\OR$D6$_a$ $\OR$D6$_a$ ($a$*$a$*) sector, so we should only count the spectrum of one of these two sectors. Just as in the plain toroidal case, they will contain the particle content of a vector multiplet in $D=4$ $\N=4$ $U(N_a)$ Super Yang-Mills theory. In the particular case that a stack $a$ is its own image under $\OR$, that is, if $\OR$D6$_a=$ D6$_a$, then an orientifold projection must be performed on the Chan-Paton degrees of freedom, and either an $SO(2N_a)$ or $USp(2N_a)$ gauge group will arise.\footnote{In general, $SO(N)$ gauge groups will arise from a stack of $N$ coincident D$p$-branes on top of an orientifold plane or O$p$-plane. In the particular case at hand, and when dealing with square tori ($b^{(i)} = 0$ $i = 1, 2, 3$) this means that $\Pi_a = (1,0)(1,0)(1,0)$, and that the stack must be located at a fixed point under the $\R$ action. Similarly, $USp(N)$ gauge groups will arise from 3-cycles as $(1,0)(0,1)(0,\pm 1)$ and permutations on the three two-tori. For more details on this see \cite{Blumenhagen:2000wh}.}

\item{D6$_a$ D6$_b$ sector ($ab$)}

$\OR$ takes this sector to $\OR$D6$_b$ $\OR$D6$_a$ ($b$*$a$*), so no orientifold projection is needed and things will work just as in the toroidal case. We will again obtain massless chiral fermions in the bifundamental representation $(N_a, \bar N_b)$, plus some light scalars in the same representation. This result must be multiplied by the intersection number $[\Pi_a]\cdot[\Pi_b]$ given by (\ref{inter}).

\item{D6$_a$ $\OR$D6$_b$ sector ($ab$*)}

This sector, not present in the toroidal case, is taken to the D6$_b$ $\OR$D6$_a$ ($ba$*) sector. Chiral fermions and scalars will now transform in the representation $(N_a, N_b)$, \footnote{One can understand why this representation is $(N_a, N_b)$ instead of $(N_a, \bar N_b)$ by noticing that $\O$ identifies conjugate representations of the gauge groups arising from D6$_b$ and $\OR$D6$_b$ stacks. Indeed, if $T_b$ is a generator of the gauge group $U(N_b)$ arising from the sector D6$_b$D6$_b$, then $\OR$ maps it to the generator $-T_{b^*}^*$ on the sector $\OR$D6$_b$$\OR$D6$_b$ (see \cite{Polchi2}).} and their multiplicity will be given by
\beq
I_{a\b^*} := [\Pi_a]\cdot[{\cal R} \Pi_b] = 
-(n_a^{(1)} m_b^{(1)} + m_a^{(1)} n_b^{(1)}) (n_a^{(2)} m_b^{(2)} 
+ m_a^{(2)} n_b^{(2)}) (n_a^{(3)} m_b^{(3)} + m_a^{(3)} n_b^{(3)}).
\label{inter2}
\eeq

\item{D6$_a$ $\OR$D6$_a$ sector ($aa$*)}

Contrary to the previous ones, this sector is always invariant under the $\OR$ action. We must then perform the appropriate orientifold projection in order to get the correct spectrum. Dealing with factorisable cycles and in the fractional formalism (\ref{frac}), there will be $I_{aa*} = - 8 n_a^{(1)} n_a^{(2)} n_a^{(3)} m_a^{(1)} m_a^{(2)} m_a^{(3)}$ intersections on this sector, $8 m_a^{(1)} m_a^{(2)} m_a^{(3)}$ of which are fixed by the action of $\R$. We must then perform the $\O$ projection on the latter, after what we will obtain fermions transforming in the antisymmetric representation of the group $U(N_a)$ if $\prod_{i=1}^3 n_a^{(i)} \geq 1$ and in the symmetric representation if $\prod_{i=1}^3 n_a^{(i)} \leq -1$.
\footnote{Notice that when $\prod_{i=1}^{3} n_a^{(i)} = 0$ and $\prod_{i=1}^{3} m_a^{(i)} \neq 0$ we still have a $U(N_a)$ gauge group with chiral fermions living on it. In general there will be $4 \prod_{i=1}^{3} m_a^{(i)}$ fermions in the antisymmetric and the same number of fermions in the symmetric, but with opposite chirality. This will give us the same contribution to chiral  $SU(N_a)$ anomalies as the general formula (\ref{specori}).} 
No $\O$ projection should be performed on the rest of the intersections of this sector, but only half of them should be counted, since the other half is related by the action of $\R$. They will then contribute with $- 4 m_a^{(1)} m_a^{(2)} m_a^{(3)} (n_a^{(1)} n_a^{(2)} n_a^{(3)} - 1)$ chiral fermions in symmetric plus antisymmetric representations. With each of the above fermions there will also be present four light scalars, also transforming on the symmetric and antisymmetric representations. For a more detailed deduction of this spectrum see \cite{Blumenhagen:2000wh}.
\end{itemize}

After considering all these different sectors, we can summarize the general spectrum arising from D6-brane orientifold configurations to be
\bea
{\rm\bf Gauge\; Group} & &\prod_a  U(N_a) \nonumber \\
{\rm\bf Fermions\; } & &\sum_{a<b} \left[I_{ab} (N_a, \bar N_b)  
+ I_{ab^*} (N_a, N_b)\right] \label{specori}\\
& & + \sum_a \left[ 
- 4 \prod_{i=1}^3 m_a^{(i)} \left(\prod_{i=1}^3 n_a^{(i)} + 1\right)
 ({\bf A}_a)
- 4 \prod_{i=1}^3 m_a^{(i)} \left(\prod_{i=1}^3 n_a^{(i)} - 1\right) 
({\bf S}_a)  \right] \nonumber
\eea
where each gauge group contains the full $\N=4$ vector multiplet, and to each chiral fermion present in (\ref{specori}) we must add a set of light scalars whose masses depend on the particular intersection angles, just as in (\ref{scalars}).

Let us now present the spectrum regarding D4 and D5-branes sitting on orientifold singularities. The main novelty from the previous D6-brane construction comes from the fact that now $\OR$ also acts on Chan-Paton matrices. The geometrical action of $\R$ is trivial in these internal degrees of freedom, while $\O$ complex conjugates the orbifold phases. More specifically, if a stack $a$ is described by
\beq
\begin{array}{c}
\vspace{0.2cm}
\bigotimes_{i=1}^n \left(n_a^{(i)}, m_a^{(i)}\right), \\ \vspace{0.25cm}
\g_{\om,a} = {\rm diag} \left( {\bf 1}_{N_a^0}, \a {\bf 1}_{N_a^1}, \ldots, \a^{N-1} {\bf 1}_{N_a^{N-1}} \right),
\end{array}
\label{brane_a}
\eeq
then the stack of mirror D-branes $a$* will be described by
\beq
\begin{array}{c}
\vspace{0.2cm}
\bigotimes_{i=1}^n \left(n_a^{(i)}, -m_a^{(i)}\right), \\
\vspace{0.25cm}
\g_{\om,a} = {\rm diag} \left( {\bf 1}_{N_a^0}, \a^{N-1} {\bf 1}_{N_a^1}, \ldots, \a {\bf 1}_{N_a^{N-1}} \right),
\end{array}
\label{brane_a*}
\eeq
where we have defined $\a \equiv e^{2\pi i \frac{1}{N}}$ and we are again expressing our D-brane content in term of fractional wrapping numbers.

From these general considerations we can readily compute the spectrum of an arbitrary configuration of D4-branes, by just adding the contributions from the new sectors on the orientifolded theory. The sector D4$_a$D4$_a$ will give the same contribution (\ref{spectrum4aa}) as in the orbifold case, whereas the rest of the sectors will give the contribution
\footnote{In case $n_a = 0$, the tachyonic spectrum from the sector $aa$* will be reduced to an antisymmetric representation $({\bf A}_a^0)$ of the $U(N_a^0)$ gauge group. This is just a T-dual orbifolded version of
the non-BPS D-brane systems constructed in \cite{Witten:1998cd,Rabadan:2000ma,Loaiza-Brito:2001ux}.}
{\beq
\begin{array}{l}
\vspace{1.5mm}
{\rm\bf Tachyons} \\ \sum_{a<b} \sum_{i=1}^N \; 
[\; I_{ab}(N_a^i,{\ov N}_b^i) + 
I_{ab^*}(N_a^i,N_b^{-i})\;]\\
\sum_a [\; |m_a| (|n_a|+1)({\bf A}_a^0) + |m_a| (|n_a| - 1) 
({\bf S}_a^0)\;]
\vspace{3.5mm}\\ \vspace{1.5mm}
{\rm\bf Left\; Fermions}  \\ \sum_{a < b} \sum_{i=1}^N \; 
I_{ab} [\; (N_a^i,{\ov  N}_b^{i+\med{(b_1+b_2)}}) 
+ (N_a^i,{\ov N}_b^{i-\med{(b_1+b_2)}}) \;]\\
\sum_{a \leq b} \sum_{i=1}^N \; 
I_{ab^*} [\; (N_a^i,N_b^{-i-\med{(b_1+b_2)}}) 
+ (N_a^i,N_b^{-i+\med{(b_1+b_2)}}) \;] \\
\sum_a \sum_{i=1}^N \; (\d_{i,-i-\med(b_1+b_2)}+\d_{i,-i+\med(b_1+b_2)})
[\; - m_a (n_a + 1) ({\bf A}_a^i) - m_a (n_a - 1) ({\bf S}_a^i)\;]
\vspace{3.5mm}\\ \vspace{1.5mm}
{\rm\bf Right\; Fermions} \\ \sum_{a<b} \sum_{i=1}^N \; 
I_{ab} [\; (N_a^i,{\ov  N}_b^{i+\med{(b_1-b_2)}}) 
+ (N_a^i,{\ov N}_b^{i-\med{(b_1-b_2)}}) \;]\\
\sum_{a \leq b} \sum_{i=1}^N \; 
I_{ab^*} [\; (N_a^i,N_b^{-i-\med{(b_1-b_2)}}) 
+ (N_a^i,N_b^{-i+\med{(b_1-b_2)}}) \;]\\
\sum_a \sum_{i=1}^N \; (\d_{i,-i-\med{(b_1-b_2)}}+\d_{i,-i+\med{(b_1-b_2)}})
[\; - m_a (n_a - 1) ({\bf A}_a^i) - m_a (n_a - 1) ({\bf S}_a^i)\;]
\end{array}
\label{spectrum4ab*}
\eeq}
where the index $i$ is defined mod $N$. In obtaining this spectrum we have supposed that the D4-branes are sitting in an orbifold singularity $\cpx^2/\inte_N$ with odd $N$. In principle, we could also consider the case of even $N$. The only new contribution arising in this case, though, would be present in the tachyonic sector of the spectrum.

The case involving D5-branes is entirely analogous. From the D5$_a$D5$_a$ sector we will again obtain the spectrum (\ref{spectrum5aa}), whereas from the rest we obtain
{\beq
\begin{array}{l}
\vspace{1.5mm}
{\rm\bf Complex\; Scalars} \\ 
2 \times \sum_{a<b} \sum_{i=1}^N \; 
[\; \arr I_{ab}\arr (N_a^i,{\ov N}_b^i) + 
\arr I_{ab^*}\arr (N_a^i,N_b^{-i})\;]\\
2 \times \sum_a [\;2 \arr m_a^1 m_a^2\arr (\arr n_a^1 n_a^2\arr + 1) 
({\bf A}_a^0)
+ 2 \arr m_a^1 m_a^2\arr (\arr n_a^1 n_a^2\arr - 1) ({\bf S}_a^0)\;] 
\vspace{3.5mm}\\ \vspace{1.5mm}
{\rm\bf Left\; Fermions} \\ \sum_{a \leq b} \sum_{i=1}^N \;
[\;I_{ab}(N_a^i,{\ov  N}_b^{i+1}) + I_{ab*}(N_a^i,N_b^{-i-1})\;]  \\
\sum_a \sum_{i=1}^N \; \d_{i,-i-1}
[\;2 m_a^{(1)} m_a^{(2)} (n_a^{(1)} n_a^{(2)} + 1) ({\bf A}_a^i)
+ 2 m_a^{(1)} m_a^{(2)} (n_a^{(1)} n_a^{(2)} - 1) ({\bf S}_a^i)\;] 
\vspace{3.5mm}\\ \vspace{1.5mm}
{\rm\bf Right\; Fermions} \\ \sum_{a \leq b} \sum_{i=1}^N \;
[\;I_{ab}(N_a^i,{\ov  N}_b^{i-1}) + I_{ab*}(N_a^i,N_b^{-i+1})\;]  \\
\sum_a \sum_{i=1}^N \; \d_{i,-i+1}
[\;2 m_a^{(1)} m_a^{(2)} (n_a^{(1)} n_a^{(2)} + 1) ({\bf A}_a^i)
+ 2 m_a^{(1)} m_a^{(2)} (n_a^{(1)} n_a^{(2)} - 1) ({\bf S}_a^i)\;] 
\end{array}
\label{spectrum5ab*}
\eeq}

Finally, let us point out some subtlety regarding the computation of the {\em scalar} spectrum arising from the $aa$* sector. This subtlety arises when the intersection number on some two-torus $I_{aa*}^{(i)}$  vanishes. In this case we must compute the $aa$* scalar spectrum forgetting about the $i^{th}$ two-torus and applying the multiplicity formulas for the D-branes of one lower dimension. Notice that if $m_a^i = 0$ there is an extra contribution to the mass$^2$ of the whole spectrum arising from $aa^*$, coming from the separation $Y$ that both mirror branes may have in the $i^{th}$ two-torus.

\section{Intermediate spectrum}

Usually, when computing the low energy effective field theory spectrum of a string-based compactification, we only take care of massless oscillations of the string, as well as tachyonic modes which may signal some instability. This is because all the rest of the string excitations are supposed to have a tree level mass of the order of the string scale $M_s$, which is taken beyond experimental bounds, and hence should be considered as massive particles to be integrated out in order to get the effective theory. In string constructions based on D-branes at arbitrary angles, however, it is possible to obtain open string states which are light enough to be relevant at low energies. A clear example can be found in models of intersecting D6-branes, where each chiral fermion stuck at an intersection has four scalar partners (\ref{scalars}) with same quantum numbers. Since these four (squared) masses vary continuously with the intersection angles, they can get any value between $-\med M_s^2$ and $M_s^2$.

In this section we will briefly study the generic spectrum of states appearing in intersecting brane models whose tree level mass is at an intermediate scale, i.e., below the string scale. This spectrum of light particles may be of some interest when imposing phenomenological restrictions to semi-realistic models, as well as for finding signals of new physics beyond the SM at next generation particle accelerators. This fact is particularly true in the low string scale scenarios proposed in \cite{Arkani-Hamed:1998rs,Antoniadis:1998ig}, where the string scale is of order 1-10 TeV, and that can be naturally realised by means of intersecting brane worlds. Here we will present the different sectors of the theory contributing to this spectrum of light particles, paying special attention to configurations of D6-branes which hold the most complex structure. A more detailed analysis will be given in Chapter \ref{models}, once explicit semi-realistic examples are constructed.
\begin{figure}[ht]
\centering
\epsfxsize=6in
\epsffile{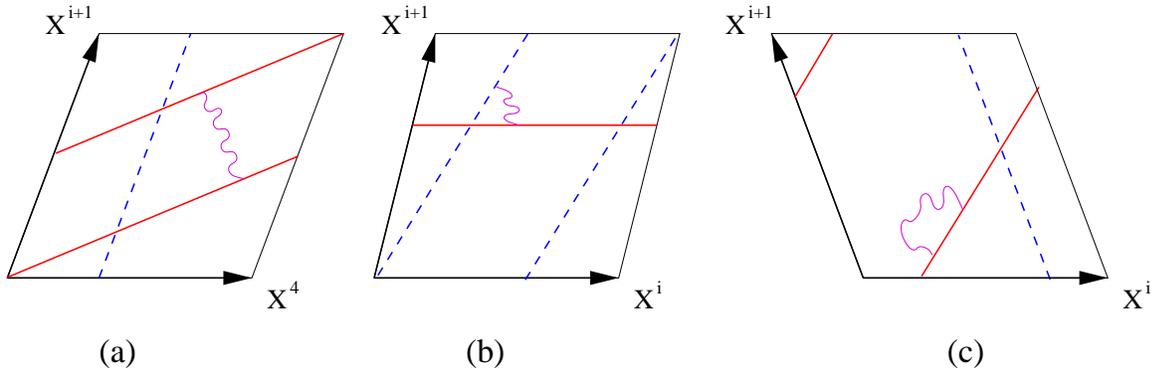}
\caption{Intermediate spectrum arising from branes at angles. In the figure, we show the geometry related to (a) winding, (b) gonionic and (c) Kaluza-Klein excitations.\label{stringies}}
\end{figure}

\subsection{D$_a$D$_a$ sector}

\begin{itemize}

\item{Kaluza-Klein states}

In general, gauge interactions are sensitive to the presence of {\em some} of the extra dimensions in the theory, which are those that the corresponding D-brane wraps on the compact space. In the models of D($3+n$)-branes at angles under study, each brane will wrap $n$ compact extra dimensions from $M_4$, and hence Kaluza-Klein replicas of the D$_a$D$_a$ sector will appear, including massive copies of the gauge bosons. These masses will be of the form
\beq
M_{KK} = \sum_{i=1}^n \frac{k_i}{l_i},
\label{KK}
\eeq
where $k_i \in \inte$ is the Kaluza-Klein mode on the $i^{th}$ torus and $l_i$ is the length of the D-brane on such torus. Since the masses of such KK excitations will depend on the volume that the D-branes wrap on the compact dimensions, a compactification involving a very large $T^{2n}$ will necessarily imply very light KK modes.

\item{Winding states}

The geometry defining our compact space may as well imply winding states under the string scale. Winding modes arise from strings with both ends on the same stack of D-branes, but whose lenght cannot be continously contracted to a point (see figure \ref{stringies}.(a)). Just as with KK states, these excitations will consist of massive copies of the massless D$_a$D$_a$ spectrum, its mass given by
\beq
M_{W} = \sum_{i=1}^n \om_i \frac{A_i}{l_i},
\label{winding}
\eeq
where $\om_i \in \inte$ and $A^{(i)}$ is the area of the $i^{th}$ two-torus. Contrary to KK excitations, winding modes are lighter as smaller is the volume of $T^{2n}$.

\end{itemize}

We then conclude that Vol ($T^{2n}$), i.e., the compactification scale of the D-brane sector, should have an intermediate value so that any of the previous states is not too light to be observed in accelerators.

\subsection{D$_a$D$_b$ sector}

We have seen that one of the salient features involving D-branes at angles is the existence of chiral fermions living at intersections. Accompanying this fermion there exist a series of scalar states whose mass depends on the specific angles at the intersection, as well a tower of massive fermionic states whose mass also depends on the angles \cite{Berkooz:1996km}. These massive replicas whose mass depend on the intersection angles were baptized as {\it gonions} in \cite{Aldazabal:2000cn}, and are exclusive excitations from intersecting brane world models. Since they have the same quantum number as chiral fermions, in a realistic model they would constitute massive replicas of the Standard Model fermionic content.

Since the richest gonion structure arises from D6-brane configurations, let us consider this class of compactifications. In (\ref{gonionsfer}) we list, in bosonic language, the lightest gonion states arising on the Ramond sector for a generic twist $v_\vt = (\vt^1,\vt^2,\vt^3,0)$, $\vt^i > 0$.
{\beq
\begin{array}{cc}
{\rm \bf State}  \ (r_{R}) \quad & \quad {\bf \a^\prime(Mass)^2} 
\vspace{0.2cm} \\ \vspace{0.05cm}
\med \left(+,-,-,-\right) & 0 \\ \vspace{0.05cm}
\med \left(-,+,-,-\right) & \vt^1 \\ \vspace{0.05cm}
\med \left(-,-,+,-\right) & \vt^2 \\ 
\med \left(-,-,-,+\right) & \vt^3 
\vspace{0.1cm} \\ \vspace{0.05cm}
\med \left(-,-3,-,-\right) & 1-\vt^1 \\  \vspace{0.05cm}
\med \left(-,-,-3,-\right) & 1-\vt^2 \\
\med \left(-,-,-,-3\right) & 1-\vt^3 
\vspace{0.1cm} \\ \vspace{0.05cm} 
\a_{-\vt^i}\med \left(+,-,-,-\right) & \vt^i \\
\a_{\vt^i-1}\med \left(+,-,-,-\right) & 1-\vt^i
\label{gonionsfer}
\end{array}
\eeq}

The last two lines correspond to excitations involving twisted bosonic oscillator operators, acting on the massless chiral state. Combined with the states of same mass and opposite chirality, these excitations form massive vector states on the effective field theory. 

The lightest gonions on the Neveu-Schwarz sector involve scalars as well as $D=4$ vector fields
{\bea
\vspace{0.2cm}
\begin{array}{cc}
{\rm \bf State}  \ (r_{NS}) \quad & \quad {\bf \a^\prime(Mass)^2} 
\vspace{0.2cm} \\ \vspace{0.05cm} 
\left(0,-1,0,0\right) & r - \vt^1 \\  \vspace{0.05cm}
\left(0,0,-1,0\right) & r - \vt^2 \\  \vspace{0.05cm}
\left(0,0,0,-1\right) & r - \vt^3 \\
\left(0,-1,-1,-1\right) & 1 - r
\vspace{0.1cm} \\ \vspace{0.05cm}
\a_{-\vt^i}\left(0,-1,-1,-1\right) & (1 - r) + \vt^i \\
\a_{\vt^i-1}\left(0,-1,-1,-1\right) & (1 - r) + (1 - \vt^i) 
\vspace{0.1cm} \\ \vspace{0.05cm}
\left(\pm 1,0,-1,-1\right) & (1 - r) + \vt^1 \\ \vspace{0.05cm}
\left(\pm 1,-1,0,-1\right) & (1 - r) + \vt^2 \\
\left(\pm 1,-1,-1,0\right) & (1 - r) + \vt^3 \vspace{0.2cm}
\label{gonionsbos}
\end{array}
\eea}
where we have defined $r \equiv \med  \left(\vt^1 +\vt^2 + \vt^3\right)$. It is interesting to see how all these massive states will arrange themselves in supermultiplets whenever any of the supersymmetry conditions is satisfied, i.e, whenever any of the scalars in (\ref{scalars}) become massless. Indeed, let us impose $\vt^3 = \vt^1 + \vt^2$, $\vt^i > 0$. Then the above spectrum can be arranged in $D=4$ $\N = 1$ supermultiplets, as

{\small \bea
\begin{array}{cccc}
{\rm \bf States} \ r_{R} \quad & \quad {\rm \bf States} \ r_{NS} \quad 
& \quad {\bf \a^\prime(Mass)^2} \quad & \quad {\bf Multiplets} 
\vspace{0.2cm} \\
\med \left(+,-,-,-\right) \quad & \quad \left(0,0,0,-1\right) & 0
& {\rm  Chiral \ +}
\vspace{0.2cm} \\ \vspace{0.05cm}
\med \left(-,+,-,-\right) \quad & \quad \left(0,0,-1,0\right) & \vt^1 &
 {\rm  Chiral \ -} \\
\a_{-\vt^1}\med \left(+,-,-,-\right) \quad 
& \quad  \a_{-\vt^1} \left(0,0,0,-1\right) & \vt^1 &  {\rm  Chiral\ +}
\vspace{0.2cm} \\  \vspace{0.05cm}
\med \left(-,-,+,-\right) \quad & \quad \left(0,-1,0,0\right) & \vt^2 &
 {\rm  Chiral\ -} \\
\a_{-\vt^2}\med \left(+,-,-,-\right) \quad 
& \quad \a_{-\vt^2} \left(0,0,0,-1\right) & \vt^2 &  {\rm  Chiral\ +}
\vspace{0.2cm} \\ \vspace{0.05cm}
\med \left(-,-,-,-3\right) \quad & \quad \left(0,-1,-1,-1\right) 
& 1-\vt^1-\vt^2 & {\rm  Chiral\ -} \\
\a_{\vt^3-1}\med \left(+,-,-,-\right) \quad & \quad \a_{\vt^3-1} 
\left(0,0,0,-1\right) & 1-\vt^1-\vt^2 & {\rm  Chiral\ +} 
\vspace{0.2cm} \\  \vspace{0.05cm}
\med\left(-,-3,-,-\right) \quad & \quad  \left(-1,-1,0,-1\right) 
& 1-\vt^1 & {\rm Vector\ -}\\
\med\left(+,-,+,-3\right) \quad & \quad \left(+1,-1,0,-1\right)
& 1-\vt^1 & {\rm Vector\ +}
\label{susyspec}
\end{array}
\eea}


\subsection{Extra massless states}

In connecting string-based models with phenomenology, we sould pay special attention to massless states generically present in the theory and that we do not observe in everyday physics. On the closed string sector of the theory, we expect to obtain such states both from the twisted and untwisted part of the spectrum. In particular, in case we compactify on a $T^6$ (which is the case we are considering for D6-branes configurations) the untwisted sector will yield a massless spectrum containing a full $D = 4$ $\N = 8$ supergravity multiplet, which is far beyond experiment. However, if the D6-brane content breaks all these bulk supersymmetries (and generically it will) we expect this supersymmetry breaking to be transmitted to the gravity sector at some level in perturbation theory, giving masses to all the supersymmetric partners of the graviton. Notice that some of the scalar states present on the supergravity multiplet (\ref{sugra8}) have a precise geometrical meaning. Namely, their v.e.v.'s determine the background geometry of the internal manifold $T^6$, being closed string moduli. For simplicity, we will consider such moduli as free parameters of the theory, although if the scalar superpartners of the graviton had a mass term, these moduli would be stabilized to a fixed value. A similar story would apply to configurations involving orbifold and orientifold singularities. However, since we have not specified the geometry of the compact dimensions tranverse to the D-branes, which largely determines the untwisted sector of the closed string spectrum, it is pointless to address this issue here.

Some extra massless spectrum will also arise from the open string sector. Indeed, we have seen that flat D6-branes yield a $D=4$ $\N=4$ gauge theory on their worldvolume. Hence the gauge bosons of a SM interaction, say $SU(3)$ gluons, would appear accompanied by the full vector multiplet. These unwanted particles should also get massive by radiative corrections, by a similar mechanism giving mass to the bulk extra particles. See \cite{Ibanez:2001nd}, Appendix I, for a computation of these corrections in terms of the effective field theory. Again, we can draw similar arguments regarding the case of D-branes at angles in orbifold singularities. D4-branes at angles with a supersymmetric orbifold twist will yield $\N=2$ vector multiplets in the low energy spectrum instead of $\N=4$, whereas D5-branes will present a simpler spectrum, with only some extra scalars in the adjoint representation. These adjoint scalars, as well as those present in $\N=2$ and $\N=4$ supermultiplets, are open string moduli of the theory, and determine the D-brane positions on each two-torus through their v.e.v.'s. Producing a mass term for them would then imply fixing the position of the D-branes. Since this moduli stabilization mechanism is widely unknown and difficult to guess at this level of the analysis, we will consider the D-brane positions as free parameters.

\chapter{RR tadpoles and anomalies \label{t&a}}

In the present chapter we will study the consistency conditions that any intersecting D-brane configuration must satisfy. Such restrictions are known as tadpole cancellation conditions, and play a crucial role when constructing a consistent effective field theory. In particular, RR tadpole conditions imply the cancellation of potential anomalies arising from the low energy chiral spectrum, as we will explicitly see in the particular class of models under study. We will study the cancellation of cubic non-Abelian, Abelian and mixed chiral anomalies, the last two involving a generalized Green-Schwarz mechanism. As a consequence of such mechanism, some Abelian gauge bosons will get massive, eliminating the corresponding $U(1)$ gauge symmetry from the effective theory. This fact will be of great importance when constructing semi-realistic models.

\section{Tadpoles}

Type II theory compactifications have a series of restrictions known as {\it tadpole} cancellation conditions. Physically, these are nothing but consistency relations of the equations of motion of certain massless fields present in the theory. Such fields arise from RR and NSNS closed string sectors and, as was shown in \cite{Polchinski:1995mt}, D-branes are sources of them. In general, a D$p$-brane is source of a ($p+1$)-form Ramond-Ramond $A_{p+1}$, \footnote{Notice that a D$p$-brane expands a ($p+1$)-dimensional submanifold $W_{p+1}$ on the target space. A ($p+1$)-form $A_{p+1}$ naturally couples to such object via the disk coupling $\int_{W_{p+1}} A_{p+1}$. Here we are supposing that the D$p$-brane has no magnetic fluxes turned on its worldvolume, which would induce some couplings involving RR $s$-forms with $s < p+1$.} as well as a source of gravitons and dilatons (the antisymmetric NSNS tensor $B_{\mu\nu}$, however, only couples to the fundamental string). In general, NSNS tadpoles generate a potential for NSNS fields, and can be reabsorbed by means of background redefinitions \cite{Fischler:ci,Fischler:tb,Dudas:2000ff,Blumenhagen:2000dc,Rabadan:2002wy}, on the contrary RR tadpoles are more dangerous and signal an inconsistency of the theory. Hence, in order to perform a consistent string-based compactification we should require their cancellation.

\begin{figure}[ht]
\centering
\epsfxsize=6in
\hspace*{0in}\vspace*{.2in}
\epsffile{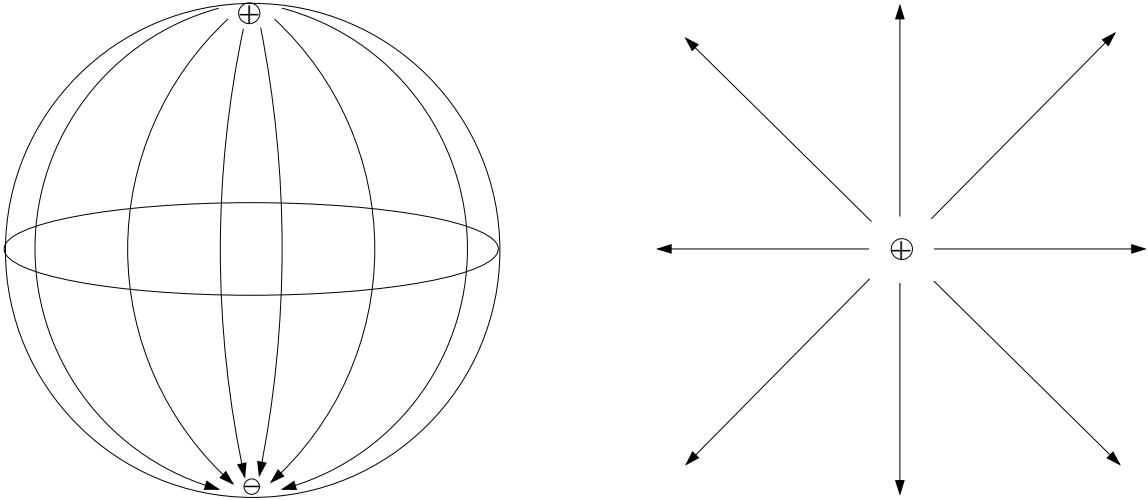}
\caption{The sum of all the charges on a compact space must vanish, since otherwise there is no place for the fluxlines to go. On non-compact spaces this condition is not present because the flux can always escape to infinity.}
\label{gauss}
\end{figure}

Since a D$p$-brane is charged with respect to a RR ($p+1$)-form $A_{p+1}$, it will act as a source in the equations of motions of such field \cite{Polchinski:1995mt}
\beq
dH_{p+2} = * J^e_{p-7} \quad \quad d*H_{p+2} = * J^m_{p+1},
\label{source}
\eeq
where $H_{p+2} = dA_{p+1}$ is the ($p+2$)-form field strength and $*$ is usual Hodge duality which, in a $D$-dimensional manifold relates $p$-forms and ($D-p$)-forms \cite{candelas}. $J^e$ represents the electric charge and $J^m$ the magnetic charge for this RR field. The equations of motion then impose the following constraints
\beq
\int_{\Sigma_k} * J_{10-k} = 0
\label{source2}
\eeq
where $\Sig_k$ is an arbitrary compact submanifold of dimension $k$. Physically, (\ref{source2}) amounts to requiring no net electric nor magnetic charge on this compact submanifold. As a consequence, if the field $A_{p+1}$ is confined to propagate on a compact submanifold $\Sig$, then the objects charged under such field must add their charges up to a zero total charge. The electrostatic analogue of such constraint is given by the Gauss law, which states that an electrically charged particle as a positron cannot exist on its own in a compact space as $S^2$, since it will be the source of fluxlines that must end somewhere, as for instance an electron. In a non-compact space, however, those lines can always scape to infinity (see figure \ref{gauss}).

\begin{figure}[ht]
\centering
\epsfxsize=4in
\hspace*{0in}\vspace*{.2in}
\epsffile{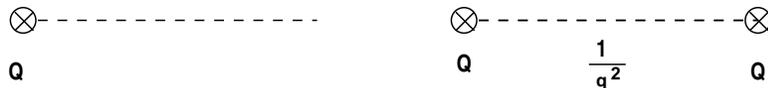}
\caption{Photon emission versus photon exchange. Figure taken from \cite{Dabholkar:1997zd}.}
\label{dabol}
\end{figure}

The charge of a D$p$-brane with respect to the RR field $A_{p+1}$ can be computed by looking how two D$p$-branes exchange RR closed string fields at large distances \cite{Polchinski:1996na,Dabholkar:1997zd}. The same idea can be applied to Quantum Electrodynamics, i.e., we can compute the coupling of a charged object to the electromagnetic field by simply measuring the photon exchange with another object whose charge is known. Indeed, let us consider a very massive particle with charge $Q$. One can compute the value of $Q$ by calculating the amplitude for vacuum going into a single photon in the background of this charge (i.e., the amplitude for photon emission or tadpole diagram). Alternatively, one can calculate the interaction between two particles each of charge $Q$ (see figure \ref{dabol}). From the Feynman diagram one would then obtain $Q^2/p^2$, $p$ being the momentum of the exchanged photon. This can be rewritten as
\beq
\frac{Q^2}{p^2} = Q^2 \int_0^\infty e^{-p^2 l} dl,
\label{exchange}
\eeq
where we see that the $p \raw 0$ infrared divergence corresponds to the divergence of this integral on large propagation distances $l \raw \infty$.

\vspace{0.3cm}

One can describe the interaction of two D-branes in a similar fashion, just by considering closed string exchanges at large distances. Notice that a D-brane will usually have both RR and NSNS charge, the latter given by the worldvolume tension. If a D-brane preserves some of the bulk supersymmetry on its worldvolume, then both RR and NSNS charges will be equal, thus being a non-perturbative BPS state of the theory. A system of two BPS D-branes may or may not preserve a bulk supersymmetry in common. If they do, then the exchange of RR fields exactly compensates the NSNS interaction, and no net force arises between them. As a result, we are allowed to simply add the NSNS charges (i.e., the tensions) of the two D-branes in order to compute the NSNS charge of the composite system \footnote{Notice that this fact is always true for RR charges.}, which is BPS as a whole. Hence, when dealing with a supersymmetric configuration of D-branes (and possibly some other charged objects) computation of the total NSNS charge of the system parallels that of the RR charge, and the former vanishes if the latter does. On the other hand, this will not happen in non-supersymmetric configurations, where we will generically have some NSNS tadpoles after canceling RR tadpoles.

Computation of RR charges by means of closed string exchange reveals the existence of a new class of non-perturbative objects, named orientifold planes or O$q$-planes. Let us consider an orientifold theory with gauge group $\G_1 + \O\G_2$. For each element $g \in \G_2$ such that $g^2 =$ Id we will have an O-plane on the theory, localized on the fixed point set of the geometrical action of $g$. An illustrative example of this is given by type I theory seen as the orientifold (\ref{typeI}). Here $\G_2=$ Id, so the fixed point set of $g \in \G_2$ is the whole target space $M_{10}$. This implies the existence of an O9-plane filling the whole target space, charged under the RR form $A_{10}$. Since the source of these fields occupies the whole target space, there is no place these fields could escape to, so it is necessary to cancel the corresponding RR tadpoles in order to get a consistent theory \footnote{Strictly speaking, RR tadpoles should be canceled because the field strength of $A_{10}$ would be a 11-form, which is a forbidden object in $M_{10}$. This RR field has hence no kinetic term and cannot propagate, so its equation of motion implies RR charge cancellation.}. By exchange of closed string modes is possible to compute the relative charge between an O9-plane and a D9-brane, finally finding that its charge can be canceled by adding 32 D9-branes to the theory. After performing the orientifold projection on the degrees of freedom of these D9-branes, we obtain an open string theory with gauge group $SO(32)$.

In the next subsections we will study which restrictions do tadpole cancellation conditions impose in the intersecting D-brane configurations described in Chapter \ref{atangles}. After this, we will show how these conditions imply anomaly cancellation in the $D=4$ chiral effective theories derived from these configurations. Actually, RR tadpole cancellation conditions are slightly stronger than the results we will derive in this chapter, involving some extra constraints of K-theoretical origin. These extra conditions are quite mild and will not play any role in the following chapters, so we postpone their discussion to Appendix \ref{Ktheory}.

\subsection{Toroidal case}

A plain toroidal compactification will not contain any O-plane, so the only RR charged objects in the model will be given by the D6-branes, charged under the type IIA RR 7-form $A_7$. The transverse space to any such D6-brane is contained in $T^6$, which is compact. Hence, it will be necessary to impose that the total RR charge on $T^6$ vanishes in order to insure RR tadpole conditions. 

Recall that a D6-brane configuration is specified by $K$ stacks of $N_a$ D6-branes, $a = 1, \dots, K$, wrapping 3-cycles $\Pi_a$ on $\M_6$. The terms of the spacetime action depending on the RR 7-form $A_7$ are
\bea
S_{A_7} & = & \int_{M_4 \ti \M_6} H_8 \wedge *H_8 + \sum_a N_a \int_{M_4 \ti \Pi_a} A_7 \nonumber \\
& = & \int_{M_4 \ti \M_6} A_7 \wedge d H_2 + \sum_a N_a \int_{M_4 \ti \M_6} A_7 \wedge \d_{\Pi_a},
\label{a7}
\eea
where $H_8$ is the 8-from field strength of $A_7$, $H_2$ its Hodge dual, and $\d_{\Pi_a}$ is the the Poincar\'e dual of $\Pi_a$ on $\M_6$ \cite{candelas}. The equations of motion are simply
\beq
d H_2 = \sum_a N_a \delta_{\Pi_a},
\label{cohomology}
\eeq
which are nothing but a special case of the second eq. in (\ref{source}). The integrability condition (\ref{source2}) can be translated to an equation in homology, yielding
\beq
\sum_a N_a [\Pi_a] = 0,
\label{tadpoleD6}
\eeq
where $[\Pi_a] \in H_3(\M_6, \inte)$ is the corresponding homology class of $\Pi_a$. In some sense, the homology class $[\Pi_a]$ encodes the RR charges of a single D6-brane wrapping $\Pi_a$. A stack of $N_a$ D6-branes will have $N_a$ such RR charge. Condition (\ref{tadpoleD6}) then implies that D6-branes RR charges must cancel each other, in a more sophisticated version of Gauss' law.

Notice that the above discussion is valid for an arbitrary compact manifold $\M_6$ and 3-cycles $\Pi_a$. Let us now consider the particular case where $\M_6 = T^2_{(1)} \ti T^2_{(2)} \ti T^2_{(3)}$, and the D6-branes wrap factorisable 3-cycles
\beq
[\Pi_a] = \left(n_a^{(1)} [a_1] + m_a^{(1)} [b_1]\right) \otimes
\left(n_a^{(2)} [a_2] + m_a^{(2)} [b_2]\right) \otimes
\left(n_a^{(3)} [a_3] + m_a^{(3)} [b_3]\right),
\label{factorisable2}
\eeq
where $[a_i]$, $[b_i]$ are a basis of the homology space $H_1(T^2_{(i)}, \inte)$. Imposing this particular decomposition we obtain as tadpole cancellation conditions
{\beq
\begin{array}{ll} \vspace{0.1cm}
\sum_a N_a \ n_a^{(1)} n_a^{(2)} n_a^{(3)} = 0 \quad &  \quad 
\sum_a N_a \ m_a^{(1)} n_a^{(2)} n_a^{(3)} = 0 \\ \vspace{0.1cm}
\sum_a N_a \ n_a^{(1)} m_a^{(2)} m_a^{(3)} = 0 \quad & \quad 
\sum_a N_a \ n_a^{(1)} m_a^{(2)} n_a^{(3)} = 0 \\ \vspace{0.1cm}
\sum_a N_a \ m_a^{(1)} n_a^{(2)} m_a^{(3)} = 0 \quad & \quad 
\sum_a N_a \ n_a^{(1)} n_a^{(2)} m_a^{(3)} = 0 \\
\sum_a N_a \ m_a^{(1)} m_a^{(2)} n_a^{(3)} = 0 \quad & \quad 
\sum_a N_a \ m_a^{(1)} m_a^{(2)} m_a^{(3)} = 0
\label{tadpoleD6b}
\end{array}
\eeq}

\subsection{Orbifold case}

In case we are dealing with a $\inte_N$ orbifold singularity, the number of $p$-forms arising from the RR sector will be effectively multiplied by $N$. Namely, for each of the RR $p$-forms of type II theory, we must consider one copy of them coming from the untwisted sector and $N-1$ copies from the twisted sector. Here we will not bother about untwisted tadpoles, since they correspond to an RR field that propagates on the whole target space and, in particular, in the bulk of the $\inte_N$ singularity. For our model building purposes, we are supposing such space to be $\cpx^{3-n}/\inte_N$, which is non-compact, and hence the fluxlines of the untwisted fields can escape to infinity. As a consequence, the RR untwisted tadpoles will not have any effect on the local physics of the orbifold singularity, and in particular on the gauge sector of the effective field theory. A more detailed discussion and several examples on this general fact can be found in \cite{Aldazabal:1999nu}. However, when performing a fully-fledged compactification, in which the global properties of the compact manifold ${\bf B}_{6-2n}$ that contains the orbifold singularity are specified, such untwisted tadpole conditions should be taken into account.

Let us first consider the case of D4-branes sitting in a $\cpx^2/\inte_N$ orbifold singularity. We will suppose a general orbifold twist vector $v_\om = \frac 1N (0,b_1,b_2,0)$ and Chan-Paton action (\ref{gamma}) for each stack of $N_a$ D4-branes wrapping the 1-cycle $(n_a, m_a)$. The charge of such D-brane stack with respect of the $k^{th}$ twisted sector 5-form $A_5^{(k)}$ is proportional to ${\rm Tr \ } \g_{w^k,a} = {\rm Tr \ } (\g_{w,a})^k$, so tadpole conditions read
{\beq
\begin{array}{c}
c_k^2 \ \sum_a n_a \ {\rm Tr \ }\g_{\om^k,a} = 0, \\
c_k^2 \ \sum_a m_a \ {\rm Tr \ }\g_{\om^k,a} = 0,
\label{tadpoleD4}
\end{array}
\eeq}
for $k = 0, \dots, N-1$, and where we have defined $c_k^2 = \prod_{r=1,2} {\rm sin } (\pi k b_r/N)$. This prefactor usually appears in twisted tadpole conditions. Notice that $c_{k=0} = 0$, which implies that the untwisted tadpole condition is trivially satisfied. Such condition would involve the trace  Tr $\g_{w^N,a}$ = Tr Id$_{N_a} = N_a$, so it would impose a constraint on the total number of D4-branes in the singularity. Is easy to convince oneself that such a condition cannot exist, since a stack of $N$ D-branes on the regular representation of $\inte_N$ (i.e., a stack with one D-brane on each phase $\a^k = e^{2\pi i \frac kN}$ of the orbifold) constitutes a non-fractional D4-brane that can escape the orbifold singularity. Such continuous deformation would be seen as a Coulomb branch on the effective theory \cite{Douglas:1996sw}. 

Conditions (\ref{tadpoleD4}) allow for a geometric interpretation, at least for supersymmetric twists $v_\om$. In that case, a fractional \cite{Douglas:1996xg} D4-brane of Chan-Paton phase $e^{2\pi i\frac{s}{N}}$ can be seen as a D6-brane wrapped on the 1-cycle  $[\Pi_a]=n_a[a] + m_a[b]$ in $T^2$ times the $s^{th}$ collapsed two-cycle $[\Sigma_s]$ in the singularity. A D4-brane with Chan-Paton matrix (\ref{gamma}) can be then understood as a superposition of such D6-branes wrapped on $[\Pi_a] \otimes [\Sigma_s]$, and the conditions above amount to the vanishing of the total homology class
\bea
\sum_{a=1}^K \sum_{s=0}^{N-1} N_a^s\, [\Pi_a]\otimes [\Sigma_s] = 0.
\label{geointe}
\eea
Since $\sum_{s=0}^{N-1} [\Sigma_s]=0$, one can increase the $N_a^s$ by an
$s$-independent (but possibly $a$-dependent) amount and still satisfy the
homological condition. Hence, the Chan-Paton matrices for $k=0$ are
unconstrained.

Tadpole conditions are quite analogous for D5-branes sitting at $\cpx/\inte_N$ orbifold singularities. Again only twisted tadpole conditions for the 6-forms $A_6^{(k)}$ will be relevant, and in general they will read
\beq
d_k^2 \ \sum_a [\Pi_a] \ {\rm Tr \ }\g_{\om^k,a} = 0, 
\label{tadpoleD5}
\eeq
where again $k = 0, \dots, N-1$ and the prefactor is given by $d_k^2 ={\rm sin \ } \frac{2\pi k}{N}$. Here  $[\Pi_a] \in H_2(\M_4,\inte)$ represents the homology class of a general 2-cycle where a D5-brane stack $a$ wraps, in a compact four dimensional manifold $\M_4$. If we stick to factorisable toroidal configurations, then tadpoles conditions reduce to 
{\beq
\begin{array}{cc} \vspace{0.1cm}
d_k^2 \ \sum_a n_a^{(1)} n_a^{(2)} \ {\rm Tr \ }\g_{\om^k,a} = 0 \quad 
& \quad d_k^2 \ \sum_a n_a^{(1)} m_a^{(2)} \ {\rm Tr \ }\g_{\om^k,a} = 0 \\
d_k^2 \ \sum_a m_a^{(1)} m_a^{(2)} \ {\rm Tr \ }\g_{\om^k,a} = 0 \quad 
& \quad d_k^2 \ \sum_a m_a^{(1)} n_a^{(2)} \ {\rm Tr \ }\g_{\om^k,a} = 0.
\label{tadpoleD5b}
\end{array}
\eeq}

Actually, it is possible to express all the above tadpole conditions in one single closed formula, which is inspired on the general D6-brane formula (\ref{tadpoleD6}) and the geometric interpretation of the supersymmetric orbifold case (\ref{geointe}). We will refer to this general language as the $q$-basis formalism and, roughly speaking, consist of associating to each stack $a$ of D$(n+3)$-branes on a $\inte_N$ singularity a $2^n \ti N$ vector $\q_a$, so that tadpole conditions reduce to
\beq
\sum_{a,i} N_a^i \q_{a,i} = 0,
\label{tadpolesqorbi}
\eeq
where the internal index $i = 0,\dots,N-1$ is only present in orbifold compactifications. The $\q$ vector expression will depend on the family of intersecting D-branes we are dealing with. When dealing with factorisable configurations, it will reduce to
{\beq
\begin{array}{ccc}
\q_a(D6) = 
\left( 
\begin{array}{c}
n_a^{(1)} n_a^{(2)} n_a^{(3)} \\
m_a^{(1)} m_a^{(2)} m_a^{(3)} \\
m_a^{(1)} m_a^{(2)} n_a^{(3)} \\
n_a^{(1)} n_a^{(2)} m_a^{(3)} \\
m_a^{(1)} n_a^{(2)} m_a^{(3)} \\
n_a^{(1)} m_a^{(2)} n_a^{(3)} \\
n_a^{(1)} m_a^{(2)} m_a^{(3)} \\
m_a^{(1)} n_a^{(2)} n_a^{(3)} 
\end{array} 
\right), &
\q_{a,i} (D5) = 
\left( 
\begin{array}{c}
n_a^{(1)} n_a^{(2)} \\
m_a^{(1)} m_a^{(2)} \\
n_a^{(1)} m_a^{(2)} \\
m_a^{(1)} n_a^{(2)}
\end{array} 
\right) \a^i, &
\q_{a,i} (D4) = 
\left( 
\begin{array}{c}
n_a \\
m_a 
\end{array} 
\right) \a^i.
\nonumber
\end{array}
\label{vectoresq}
\nonumber
\eeq}

The $q$-basis does not introduce any new feature into the constructions under study, but allows us to express tadpole conditions in a simpler and more unified way. Moreover, it will turn to be quite useful when studying chiral anomaly cancellation on the next section. A more detailed exposition of this formalism is given in Appendix \ref{qbasis}.

\subsection{Orientifold case}

Orientifold compactifications posses many attractive features from the model building point of view, as we will discover on the next chapter. Regarding tadpoles, the main item involves the appearance of a new RR charged object: the O$p$-plane. In the orientifold compactifications under study, such O$p$-planes will have RR and NSNS charges opposite to those of D$p$-branes
\footnote{Strictly speaking, an O$p$-plane has RR charges of opposite sign to a D$p$-brane $a$ wrapped on the same homology cycle, that is, such that $[\Pi_a] = [\Pi_{ori}]$. If, on the contrary, we consider a D$p$-brane $b$ such that $[\Pi_b] = - [\Pi_{ori}]$, then both RR charges will have the same sign. Here D$p$-branes $a$ and $b$ are objects of same NSNS and opposite RR charge, since $[\Pi_a] + [\Pi_{b}] = 0$, and so are antibranes of each other \cite{Sen:1999mg}. The NSNS charge of the O$p$-planes considered in this thesis will always be negative to any D$p$-brane NSNS charge, being a objects of negative tension.}, 
so it will be necessary to add some amount of D$p$-branes in order to satisfy tadpoles. This is an important point. Indeed, in either toroidal or orientifold models RR D-brane charges have to cancel among each other, and the simplest (and with lowest energy) such configuration is the trivial case where no D-brane is present.

In general, tadpole cancellation conditions for an orientifold compactification will be identical to its orbifold counterpart, except that now we must include the O-plane RR charge contribution. In the case of D6-branes wrapping square tori (i.e., when $b^{(i)}=0$, $i = 1, 2, 3$.) the condition (\ref{tadpoleO6}) gets modified to
\beq
\sum_a N_a \left([\Pi_a] + [\Pi_{a^*}]\right) = 32 [\Pi_{O6}],
\label{tadpoleO6}
\eeq
where we have included the contribution of the mirror D6-branes $[\Pi_{a^*}] = [\R\Pi_a]$ and the orientifold plane, which wraps the homology cycle $[\Pi_{O6}]$. As mentioned above, the O-plane is located at the fixed point set under the involution $\R$ which accompanies $\O$. This action can be described by complex conjugation as in (\ref{conjugation}), and in a square torus the locus of fixed will be given by the 7-dimensional submanifold $M_4 \ti \{ {\rm Im \ } Z_i = 0, \med | \ i = 1,2,3\}$. Such object can be identified to 8 O6-planes, each wrapping the same homology class of 3-cycles $[\Pi_{O6}] = \bigotimes_{i=1}^3 [a_i]$ on $T^6$. As the relative charge between an O6-plane and a D6-brane is $-4$, we finally obtain a total negative RR charge which can be canceled by 32 D6-branes wrapped on $[\Pi_{O6}]$. This is exactly what (\ref{tadpoleO6}) states and, actually, it is nothing but type I tadpole conditions on the T-dual version involving type IIA D6-branes. In the more general case where we allow for tilted tori, the homology cycle each O6-plane wraps will be given by
\beq
[\Pi_{O6}] = \bigotimes_{i=1}^3 \left({1 \over 1 - b^{(i)}}[a_i] - 2 b^{(i)} [b_i]\right).
\label{cycleori}
\eeq
The number of independent O6-planes is given by the number of disconnected pieces in the fixed point set of $\R$. This also depends on the geometry of each $T^2$ being rectangular ($b^{(i)} = 0$) or tilted ($b^{(i)} = \med$), as in figure \ref{bflux}. In general the number of O6-planes will be given by $8\b^1\b^2\b^3$, where we have defined $\b^i \equiv 1 - b^{(i)}$. Tadpole conditions in this more general case are
\beq
\sum_a N_a \left([\Pi_a] + [\Pi_{a^*}]\right) = 32\b^1\b^2\b^3 [\Pi_{O6}].
\label{tadpoleO6c}
\eeq
Recall that $b^{(i)}$ in the T-dual picture of D9-branes with fluxes represents a discrete $b$-field turned on the $i^{th}$ two-torus. Hence, this formula reproduces the well-known fact that the relative charge between an O-plane and a D-brane gets reduced in the presence of a $b$-field \cite{Bianchi:1997rf,Witten:1997bs,Kakushadze:1998bw,Blumenhagen:1998tj}.

Since the expression (\ref{cycleori}) is quite complicated, it turns to be more convenient to reexpress everything in terms of fractional 1-cycles (\ref{frac}), where the O6-plane homology class is given by 
\beq
\left[\left({1}/{\b^1},0\right)\right] \otimes 
\left[\left({1}/{\b^2},0\right)\right] \otimes 
\left[\left({1}/{\b^3},0\right)\right].
\label{cicloorifrac}
\eeq
In this formalism is evident that the O6-plane is invariant under the $\OR$ action. In case that we restrict to D6-branes wrapping factorisable 3-cycles, the fractional language turns out to be particularly convenient in order to express tadpole conditions, which now read \cite{Blumenhagen:2000wh,Blumenhagen:2000ea}
{\beq
\begin{array}{c} \vspace{0.1cm}
\sum_a N_a \ n_a^{(1)}n_a^{(2)}n_a^{(3)} = 16 \\ \vspace{0.1cm}
\sum_a N_a \ n_a^{(1)}m_a^{(2)}m_a^{(3)} = 0 \\ \vspace{0.1cm}
\sum_a N_a \ m_a^{(1)}n_a^{(2)}m_a^{(3)} = 0 \\ 
\sum_a N_a \ m_a^{(1)}m_a^{(2)}n_a^{(3)} = 0
\label{tadpoleO6b}
\end{array}
\eeq}
where now the sum on $a$ does not run over mirror branes $a$*.

Tadpole cancellation conditions for the orientifold on the r.h.s. of (\ref{D8b}) have been explicitly derived in \cite{Honecker:2002hp}. On the untwisted sector one obtains
\bea \vspace{0.4cm}
\left[ \sum_a n_a N_a - 16\right]^2 & = & 0,
\label{untwisted1} \\
{\rm Tr} \left(\gamma^{-1}_{\OR_{(5)},a^*}\gamma^{T}_{\OR_{(5)},a}\right) 
& = & N_a,
\label{untwisted2}
\eea
and on the twisted sector
\beq \vspace{-0.4cm}
\sum_{k=1}^{2}\Big|\sum_{a}(m_a+bn_a)
\left({\rm Tr} \gamma_{k,a} - {\rm Tr} \gamma_{k,a^*}\right)\Big|^2 = 0,
\label{twisted1}
\eeq
\beq
\sum_{k=1}^{2}\Bigl(
8^2+\Big|\sum_{a}n_a\left({\rm Tr} \gamma_{k,a} 
+ {\rm Tr} \gamma_{k,a^*}\right)\Big|^2
- 16
\sum_{a} n_a \left(g_{2k} {\rm Tr} \gamma_{2k,a}
+\tilde{g}_{2k}{\rm Tr} \gamma_{2k,a^*}\right)
\Bigr) = 0,
\label{twisted2}
\eeq
where we have simplified notation as ${\rm Tr }\g_{k,a} \equiv {\rm Tr }\g_{\om^k,a}$. This construction is related by the T-duality (\ref{D8c}) to a compactification involving D4-branes on a $\inte_3$ singularity, so we would expect similar tadpole conditions. Recall that untwisted tadpoles should however not be considered at this level, since they are not relevant for the local physics at the singularity. Hence, we must not pay attention to condition (\ref{untwisted1}). Condition (\ref{untwisted2}), however, is related to an algebraic consistency condition between open and closed string sectors and must be taken into account. Such condition amounts to
\beq
\gamma_{\OR_{(5)},a^*} = \gamma^{T}_{\OR_{(5)},a},
\label{algebra}
\eeq
and deals with how the Chan-Paton actions $\g_{k,a}$ and $\g_{k,a^*}$ involving a D-brane and its mirror should be related, just as (\ref{brane_a}), (\ref{brane_a*}) show. Notice that from these expressions we deduce that ${\rm Tr} \gamma_{k,a^*} = {\rm Tr} \gamma_{-k,a} = \ov{{\rm Tr} \gamma_{k,a}}$. 

Since in the orientifold compactification studied in \cite{Honecker:2002hp} contains a $\inte_3$ singularity, $2k \equiv -k \ {\rm mod} \ 3$ and condition (\ref{twisted2}) can be expressed as a perfect square under the choice of phases $g_{2k} = \tilde{g}_{2k} = 1$. We can then extract tadpole conditions to be 
\bea
\sum_a n_a \left({\rm Tr} \gamma_{k,a} + {\rm Tr} \gamma_{k,a^*} \right) 
& = & 2 \sum_a n_a \ {\rm Re} \left({\rm Tr} \gamma_{k,a}\right) = 8, 
\label{twisted1b} \\
\sum_a m_a \left({\rm Tr} \gamma_{k,a} - {\rm Tr} \gamma_{k,a^*} \right) 
& = & 2 \sum_a m_a \ {\rm Im} \left({\rm Tr} \gamma_{k,a}\right) = 0,
\label{twisted2b}
\eea
where we have made the substitution $m_a + b n_a \raw m_a$ (i.e., we have expressed everything in terms of fractional cycles) and $k = 1,2$.

Given these results, let us now consider the (T-dual) case of D4-branes at angles in a $\cpx^2/\inte_N$ orientifold singularity, but now with $N$ an arbitrary odd integer and an arbitrary twist vector $v_\om = \frac 1N (0,b_1,b_2,0)$. Tadpole conditions are quite similar to (\ref{tadpoleD4}), now including the charge associated to an O4-plane siting on the singularity. Again in terms of fractional cycles they read
\bea
& & c_k^2 \ \sum_a n_a \ \left({\rm Tr} \gamma_{k,a} 
+ {\rm Tr} \gamma_{k,a^*} \right) = 8  
\prod_{r = 1}^2 {\rm sin } \left(\frac{\pi k b_r}{2N}\right),
\label{tadpoleO4n}\\
& & c_k^2 \ \sum_a m_a \left({\rm Tr} \gamma_{k,a} 
- {\rm Tr} \gamma_{k,a^*} \right) = 0,
\label{tadpoleO4m}
\eea
where again $c_k^2 = \prod_{r = 1}^2 {\rm sin }(\pi k b_r/N)$. We can rewrite them more elegantly as
\beq
c_k^2 \ \sum_a \left([\Pi_a] \ {\rm Tr} \gamma_{k,a} 
+ [\Pi_{a^*}] \ {\rm Tr} \gamma_{k,a^*} \right)
= [\Pi_{O4}] \ 8 \b 
\prod_{r = 1}^2 {\rm sin} \left(\frac{\pi k b_r}{2N}\right),
\label{tadpoleO4}
\eeq
where $\b \equiv 1 - b$ discriminates between rectangular and tilted tori. 

For model building purposes, it is useful to convert tadpole conditions into more tractable expressions. First notice that expressions (\ref{tadpoleO4n}), (\ref{tadpoleO4m}) can be rewritten as
\bea
& & \sum_a n_a \ \left({\rm Tr} \gamma_{2k,a} 
+ {\rm Tr} \gamma_{2k,a^*} \right) = 
{2 \over \prod_{r = 1}^2 {\rm cos } \left( \frac{\pi k b_r}{N}\right)} 
\label{tadpoleO4bn} \\
& & \sum_a m_a \left({\rm Tr} \gamma_{2k,a} 
- {\rm Tr} \gamma_{2k,a^*} \right) = 0,
\label{tadpoleO4bm}
\eea
which correctly reproduce (\ref{twisted1b}), (\ref{twisted2b}) in case we take $N=3$ and $|b_1| = |b_2| = 1$. Actually, condition (\ref{tadpoleO4bn}) can be expressed in terms of orbifold phases. Let us show it for the supersymmetric twist $|b_1| = |b_2| = 1$. Taking $2k \equiv 1  \ {\rm mod} \ N$ in (\ref{tadpoleO4bn}), we can easily read the condition that has to be imposed to the Chan-Paton matrix $\gamma_{\om,a}$
\beq
\sum_a{n_a \left({\rm Tr} \gamma_{\om,a} 
+ {\rm Tr} \gamma_{\om,a^*}\right)} =
{8 \over \left({\alpha^{N+1 \over 4}  + \bar\alpha^{N+1 \over 4}}\right)^2},
\label{generatortwist2}
\eeq
where $\alpha = e^{2\pi i/N}$. We can finally reexpress (\ref{generatortwist2}) as a sum of orbifold phases by using
\beq
{1 \over {\alpha^{N+1 \over 4}  + \bar\alpha^{N+1 \over 4}}} =
\iota \left(1 + \sum_{l=1}^r{(\alpha^{l}+ \bar\alpha^{l})}\right),
\label{decomp2}
\eeq
\beq
\iota = \left\{\begin{array}{l}
+1 \ {\rm if} \ N = 4r+1 \\
-1 \ {\rm if} \ N = 4r+3
\end{array}\right.
\label{iota}
\eeq

Let us finally consider twisted tadpole conditions for D5-branes sitting on $\cpx/\inte_N$. In terms of D5-branes wrapping 2-cycles on $T^4$, such tadpoles read
\beq
d_k^2 \ \sum_a \left([\Pi_a] \ {\rm Tr} \gamma_{k,a} 
+ [\Pi_{a^*}] \ {\rm Tr} \gamma_{k,a^*} \right)
= [\Pi_{O5}] \ 16 \b^1\b^2 \
{\rm sin} \left(\frac{\pi k}{N} \right),
\label{tadpoleO5}
\eeq
where $d_k^2 = {\rm sin \ } \frac{2\pi k}{N}$, and where again we are supposing odd $N$. This formula is easily generalizable to D5-branes wrapping arbitrary 2-cycles $\Pi$ on a compact four-dimensional manifold $\M_4$ at the tip of the orientifold singularity. Indeed, we only have to notice that $4 \b^1\b^2$ is the number of O5-planes which sit on the homology class $[\Pi_{O5}]$, and all the other quantities in (\ref{tadpoleO5}) have an straightforward generalization. If, on the other hand, we are interested on D5-branes wrapping factorisable 2-cycles of $T^2 \ti T^2$, then (\ref{tadpoleO5}) reduces to a slight modification of (\ref{tadpoleD5b})
{\beq
\begin{array}{rcl} \vspace{0.1cm}
d_k^2 \ \sum_a n_a^{(1)} n_a^{(2)} \ 
\left({\rm Tr} \gamma_{k,a} + {\rm Tr} \gamma_{k,a^*} \right)
& = & 16 \ {\rm sin} \left(\frac{\pi k}{N} \right) \\  \vspace{0.1cm}
d_k^2 \ \sum_a m_a^{(1)} m_a^{(2)} \ 
\left({\rm Tr} \gamma_{k,a} + {\rm Tr} \gamma_{k,a^*} \right) & = & 0\\  
\vspace{0.1cm}
d_k^2 \ \sum_a n_a^{(1)} m_a^{(2)} \ 
\left({\rm Tr} \gamma_{k,a} - {\rm Tr} \gamma_{k,a^*} \right) & = & 0\\
d_k^2 \ \sum_a m_a^{(1)} n_a^{(2)} \ 
\left({\rm Tr} \gamma_{k,a} - {\rm Tr} \gamma_{k,a^*} \right) & = & 0
\label{tadpoleO5b}
\end{array}
\eeq}

Just as we have done with D4-brane tadpoles, let us rewrite the upper set of equations in (\ref{tadpoleO5b}) as
\beq
\sum_a{n_a^{(1)} n_a^{(2)} \left({\rm Tr} \gamma_{2k,a} 
+ {\rm Tr} \gamma_{2k,a^*}\right)} =
{16 \over {\alpha^k + \alpha^{-k}}},
\label{decomp}
\eeq
Taking $2k \equiv 1  \ {\rm mod} \ N$, the condition to be imposed on $\gamma_{\om,a}$ is
\beq
\sum_a{n_a^{(1)} n_a^{(2)} \left({\rm Tr} \gamma_{\om,a} 
+ {\rm Tr} \gamma_{\om,a^*}\right)} =
{16 \over {\alpha^{N+1 \over 2}  + \alpha^{N-1 \over 2}}} =
16 \ \eta \sum_{l=1}^r{(\alpha^{2l-1}+ \bar\alpha^{2l-1})},
\label{generatortwist}
\eeq
\beq
\eta = \left\{\begin{array}{l}
+1 \ {\rm if} \ N = 4r-1 \\
-1 \ {\rm if} \ N = 4r+1
\end{array}\right.
\label{eta}
\eeq

Decompositions (\ref{generatortwist2}), (\ref{generatortwist}) give us a hint on the $\inte_N$ orbifold phases that O4 and O5-planes posses for different $N$. In the $q$-basis formalism is natural to define a vector $\q_{Op,i}$ containing all the O-plane RR charges with phase $\a^i$. In such a way that tadpole conditions could be written as 
\beq
\sum_{a,i} N_a^i\left(\q_{a,i}+\q_{a^*,-i}\right) = Q_{Op} N_{Op} 
\sum_i\q_{Op,i}.
\label{tadpolesqori}
\eeq
where $- Q_{Op}  = - 4$ is the relative RR (twisted) charge between an O$p$-plane and a D$p$-brane, and $N_{Op}$ is the number of O$p$-planes in the compact manifold. In the case of D$(3+n)$-branes wrapping $n$-cycles of $T^{2n}$ this number is given by $N_{O(3+n)} = 2^n \prod_{i=1}^n \b^i$.

Moreover, it is also possible to define a linear operator $\OR$, that in the $q$-space relates the vectors $\q_a$ and $\q_{a*}$ of two mirror branes as $\OR \q_a = \q_{a*}$. We could then reexpress our tadpole conditions as
\beq
2 \sum_{a,i} N_a^i \ P_+ \q_{a,i} = 
\sum_{a,i} N_a^i \ \left(1 + \OR\right)\q_{a,i} = Q_{Op} N_{Op} 
\sum_i\q_{Op,i},
\label{tadpolesqorib}
\eeq
where we have defined $P_+ = \med\left(1 + \OR\right)$. As is proven in Appendix \ref{qbasis}, $P_+$ is a projector operator. See also this appendix for explicit expressions for $\OR$ and $\q_{Op,i}$. Here we just present their expression for the particular example of D4-branes wrapped on square tori and sitting on a $\inte_5$ orientifold singularity
{\beq
\begin{array}{c}
\vspace{0.2cm}
\q_{a,i} = 
\left( 
\begin{array}{c}
n_a \\
m_a
\end{array} 
\right) \a^i, \ \
\OR = \left(
\begin{array}{cccc}
1 & 0 \\
0 & 1 \\
& & -1 & 0 \\
& & 0 & -1 
\end{array} 
\right) \ \otimes \ (\a \mapsto \bar\a), \nonumber \\
\q_{O5,0} =
\left( 
\begin{array}{c}
2 \\
0 
\end{array} 
\right), \ \
\q_{O5,1} =
\left( 
\begin{array}{c}
1 \\
0 
\end{array} 
\right) \a,  \ \ 
\q_{O5,-1} =
\left( 
\begin{array}{c}
1 \\
0 
\end{array} 
\right) \bar \a,  \ \ 
\q_{O5,2} = \q_{O5,-2} =
\left( 
\begin{array}{c}
0 \\
0 
\end{array}
\right),
\end{array}
\eeq}

\section{Anomalies}

In a quantum theory, an anomaly corresponds to the violation of a classical symmetry at the quantum level. A symmetry that we want to be preserved should, then, be free of anomalies. This is the case of gauge or local symmetries in a quantum field theory, where if anomalies do not cancel then non-physical degrees of freedom will not decouple from the theory. 

When we build-up an effective field theory from the low-energy spectrum of a string based model, is natural to ask oneself if such field theory will develop some kind of anomalies, which would signal an inconsistency of the construction. In general, in a type I or type II compactification, consistency associated to RR tadpole cancellation imply cancellation of anomalies in the effective theory. In fact, tadpole conditions are usually stronger than simply requiring cancellation of non-Abelian chiral anomalies, which are those more directly related to them \cite{Aldazabal:1999nu,Uranga:2000xp}. Indeed, the cancellation of these non-Abelian anomalies is a direct consequence of the tadpole conditions, as we will see in some examples below. On the other hand, cancellation of other class of anomalies such as mixed and gravitational is a less trivial matter, and in general will involve the presence of counterterms contributing to them. A consistent string theory, however, will always provide such counterterms. A paradigmatic example is given by the cancellation of gravitational anomalies in $D = 10$ supergravity effective theories arising from type I and heterotic strings. The mechanism involved in such cancellation is known as Green-Schwarz mechanism \cite{Green:sg}.

In the present section we will address the computation of quantum chiral anomalies which arise in the effective field theories derived from intersecting brane worlds \footnote{In the class of models that we will present on the next chapter there will not be any contribution to the gravitational anomaly, so we will not address its cancellation. A detailed computation involving such anomaly in intersecting D6-brane worlds can be found in the Appendix of \cite{Cvetic:2001nr}.}. In particular, we will see how the mixed and cubic $U(1)$ chiral anomalies get canceled by a generalized version of the Green-Schwarz mechanism. As we will see later on, the exact way in which such mechanism works will have important consequences in the search of the Standard Model though intersecting brane worlds.

\begin{figure}[ht]
\centering
\epsfxsize=6in
\epsffile{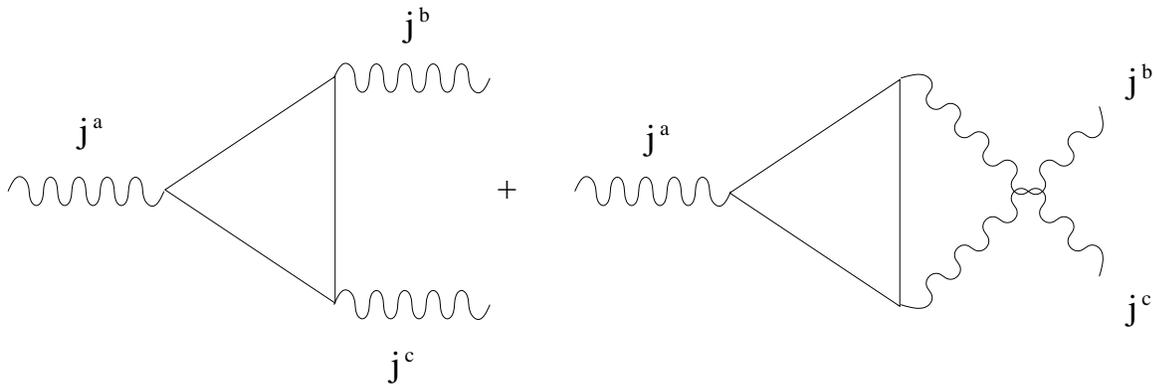}
\caption{Field theory diagrams that contribute to the anomaly of a gauge theory in a $D=4$ chiral theory. In the triangular internal loop runs a chiral fermion charged under each gauge current involved in the external legs of the diagram.}
\label{anomaly}
\end{figure}

\subsection{Cubic non-Abelian anomalies}

The cubic non-Abelian anomaly appears as a result of the coupling of chiral fermions to non-Abelian gauge groups such as $SU(N)$. In a chiral $D=4$ gauge theory there exist one-loop triangular diagrams which contribute to the anomaly of a gauge group $G$, spoiling the conservation of the associated current $j$. Such diagrams generate, for instance, divergent gauge boson mass terms and in general spoil Ward identities, so that cancellation of non-physical states and $S$-matrix unitarity are no longer guaranteed. 

In the particular case of a non-Abelian cubic anomaly, the three currents $j$ involved in figure \ref{anomaly} correspond to the same non-Abelian gauge group $G$, and the contribution of such diagram is proportional to the trace over group matrix generators in the representation $r$, i.e., to the quantity \cite{Peskin}
\beq
\A^{abc}(r) \equiv {\rm Tr} \left(t_r^a\{t_r^b,t_r^c\} \right).
\label{cubicnonab1}
\eeq
In general, $\A^{abc}$ is a group invariant, totally symmetric in $(a,b,c)$ and with each index in the adjoint representation. In the particular case of $SU(2)$ such invariant does not exist, so its cubic anomaly trivially vanishes. For $SU(N)$ groups with $N \geq 3$ there exist a unique invariant $d^{abc}$ with the required properties, so we can define an anomaly coefficient $A(r)$ such that
\beq
\A^{abc}(r) = \med A(r) d^{abc}.
\eeq
In the table \ref{casimir} we have listed the value of this Cubic Casimir $A(r)$, as well as the Quadratic Casimir $C(r)$, for the chiral representations appearing in the low-energy spectrum of the previous chapter. Notice that a sector of the theory with non-chiral spectrum will never contribute to such anomaly, since $\A^{abc}(r) = - \A^{abc}(\bar r)$ if $r$ and $\bar r$ are conjugate representations.

\begin{table}[htb]
\renewcommand{\arraystretch}{2.5}
\begin{center}

\begin{tabular}{c||c|c|c|c}
$r$  & $D(r)$ & $Q(r)$  & $C(r)$  
& $A(r)$ \\
\hline
{\bf F} & $N$ & 1 & $\med$ & 1 \\
${\bf \bar F}$ & $N$ & -1 & $\med$ & -1 \\
{\bf S} & ${N(N+1) \over 2}$ & 2 & ${N+2 \over 2}$ & $N+4$ \\
${\bf \bar S}$ & ${N(N+1) \over 2}$ & -2 & ${N+2 \over 2}$ & $-N-4$ \\
{\bf A} & ${N(N-1) \over 2}$ & 2 & ${N+2 \over 2}$ & $N-4$ \\
${\bf \bar A}$ & ${N(N-1) \over 2}$ & -2 & ${N+2 \over 2}$ & $-N+4$ \\
\end{tabular}
\caption{Casimirs of several $SU(N)$ representations. The representations of interest are the fundamental {\bf F}, symmetric {\bf S}, and antisymmetric {\bf A}. Here $D(r)$ is the dimension of the representation $r$, $Q(r)$ is the $U(1)$ charge of $r$ under the decomposition $U(N) = SU(N) \ti U(1)$ and $C(r)$, $A(r)$ are the quadratic and cubic casimir coefficient, respectively.}
\label{casimir}
\end{center}
\end{table}

Just as has been analysed in Chapter 3, the chiral sector of the constructions under study will be localized at D-branes intersections, so we will only have to bother about this part of the spectrum in order to compute the chiral anomalies. Let us first address the case of D6-branes wrapping 3-cycles on $T^6$. Given the spectrum (\ref{spec}), $SU(N_a)$ cubic anomaly will simply reduce to having the same number of fermions transforming on the fundamental as on the antifundamental of $SU(N_a)$, that is
\beq
\A_{SU(N_a)^3} = \sum_{b} I_{ab} N_b = \sum_{b} [\Pi_a]\cdot[\Pi_b] N_b
= [\Pi_a]\cdot \left(\sum_{b} N_b [\Pi_b]\right) = 0,
\label{chiralD6}
\eeq
where in the last equality we have made use of (\ref{tadpoleD6}). Thus we see that RR tadpole cancellation conditions readily imply cancellation of cubic non-Abelian anomalies, being even more restrictive that the latter. Notice also that in the previous computation we have not made use of any geometrical or topological feature related to $T^6$. In fact, we can generalize our result to any configuration of D6-branes wrapping 3-cycles on a generic compact manifold $\M_6$.

The computation of $SU(N)$ cubic anomalies in the orientifolded version of $T^6$ is quite similar, but in this case we will make use of the $q$-basis formalism in order to show its utility. Let us first write the spectrum (\ref{specori}) in a more general way:
\beq
\begin{array}{rl}
{\rm\bf Gauge\; Group} & \prod_a  U(N_a)
\vspace{2mm}\\
{\rm\bf Fermions\; } & \sum_{a<b} \left[I_{ab} (N_a, \bar N_b)  
+ I_{ab^*} (N_a, N_b)\right] \vspace{0.1cm}\\
& \sum_a \left[ 
\med \left(I_{aa^*} + N_{O6} I_{a,O6} \right)\ ({\bf A}_a)\ +
\med \left(I_{aa^*} - N_{O6} I_{a,O6} \right) ({\bf S}_a)  
\right],
\label{specori2}
\end{array}
\eeq
where $N_{O6} = 8\b^1\b^2\b^3$ is the number of O6-planes in the $T^2 \ti T^2 \ti T^2$ orientifold. From this spectrum we obtain the following contribution to the $SU(N_a)$ cubic anomaly
\bea
\A_{SU(N_a)^3} & = & 
\sum_{b \neq a} N_b \left(I_{ab} + I_{ab^*} \right) +
\med \left(I_{aa^*} + N_{O6} I_{a,O6} \right)\ (N_a - 4)\ +
\med \left(I_{aa^*} - N_{O6} I_{a,O6} \right) (N_a + 4)  
\nonumber \\
& = & \sum_{b} N_b \left(I_{ab} + I_{ab^*} \right) 
- 4 \ N_{O6} I_{a,O6} 
\  = \ \sum_{b} N_b \left(\q_a^{\ t \ }\I \ \q_b 
+ \q_a^{\ t\ }\I\OR \ \q_b \right)
- Q_{O6} N_{O6} \ \q_a^{\ t\ }\I \ \q_{O6}\nonumber \\
& = & \q_a^{\ t\ }\I \left(\sum_{b} N_b (1 + \OR)\q_b 
- Q_{O6} N_{O6} \ \q_{O6}\right) = 0.
\label{chiralO6}
\eea
In the r.h.s. of the second line we have introduced the (homology) intersection matrix $\I$, and we have expressed intersection numbers as $I_{ab} = \q_a^{\ t\ }\I\ \q_b$. We have also made use of the fact that the relative RR charge of and O6-plane and a D6-brane is $Q_{O6} = 4$. Finally, in the last equality we have imposed the tadpole condition (\ref{tadpoleO6c}). See Appendix \ref{qbasis} for a more detailed description of this notation.

Cancellation of cubic non-Abelian anomalies in orbifold D4-brane models has been shown in \cite{Aldazabal:2000dg}. Let us briefly repeat their results here. The $SU(N_a^j)$ cubic anomaly given the chiral matter content of (\ref{spectrum4ab}) is given by
\bea
\A_{SU(N_a^j)^3} & = &\sum_{b,k} N_b^k I_{ab} \ \d(j,k), \label{chiralD4} \\
\d(j,k) & \equiv & \d_{j,k+\frac{b_1+b_2}{2}} 
+ \d_{j,k-\frac{b_1+b_2}{2}} -
\d_{j,k+\frac{b_1-b_2}{2}} -  \d_{j,k-\frac{b_1-b_2}{2}},
\label{delta}
\eea
where $j, k \in \inte_N$, i.e.,  $j, k$ are defined mod $N$. Notice that $\d(j,k) = \d(k,j)$. Just as done in \cite{Leigh:1998hj,Aldazabal:2000dg}, we can make use of the discrete Fourier transform $\d_{ij} = \frac 1N \sum_{k=1}^N e^{\frac{2\pi i k}{N} (j-i)}$ to rewrite (\ref{chiralD4}) as 
\beq
\A_{SU(N_a^j)^3} = {-4 \over N} \sum_{k=1}^N e^{2\pi i\frac{k\cdot j}{N}} 
\left( c_k^2 \ \sum_a I_{ab} \ {\rm Tr \ }\g_{k,a}\right),
\label{chiralD4b}
\eeq
where the term in parentheses will vanish after imposing tadpole conditions (\ref{tadpoleD4}). There is, however, a more intuitive way of relating chiral anomalies with tadpoles. Notice that if we define an intersection matrix $\I$ acting on the $q$-basis vectors (\ref{vectoresq}) in such a way that $\q_{a,j}^{\ \dagger\ } \I \q_{b,k} \equiv I_{ab} \d(j,k)$, then we could express (\ref{chiralD4}) as
\beq
\A_{SU(N_a^j)^3} = \q_{a, j}^{\ \dagger\ } \I \ \sum_{b,k} N_b^k\q_{b,k}.
\label{chiralD4c}
\eeq
This anomaly automatically vanishes if $\sum_{b,k} N_b^k \ \q_{b,k} = 0$, which is nothing but the tadpole condition advanced in (\ref{tadpolesqorbi}), expressed in the $q$-basis formalism. One can see that (\ref{chiralD4c}) and (\ref{tadpoleD4}) are equivalent by use of the discrete Fourier transform.

This $q$-basis formalism can also be applied to D4-branes in orientifold singularities. In this case, and given the spectrum (\ref{spectrum4ab*}), the  $SU(N_a^j)$ cubic anomaly is given by
\beq
\A_{SU(N_a^j)^3} = \sum_{b,k} N_b^k \left(I_{ab} \ \d(j,k) 
+ I_{ab^*} \ \d(j,-k)\right) - 8\b \ I_{a,O4}\ \d(j,-j), 
\label{chiralO4}
\eeq
where $\d(j,k)$ is defined in (\ref{delta}). By use of the discrete Fourier transform, we can again rewrite this expression in an analogous way to (\ref{chiralD4b}) and see that it vanishes after imposing tadpole conditions (\ref{tadpoleO4}). Let us, however, make use of the $q$-basis formalism and of the definition of the intersection matrix $\I$, and convert this expression to
\beq
\A_{SU(N_a^j)^3} = \q_{a, j}^{\ \dagger\ } \I \left(\sum_{b,k} N_b^k 
(1 + \OR)\q_{b,k} - Q_{O4} N_{O4} \ \q_{O4,k}\right) = 0,
\label{chiralO4b}
\eeq
which vanishes because of the tadpole condition (\ref{tadpolesqori}). The precise value of $\q_{O4,k}$ can be easily deduced from orbifold phase decompositions as (\ref{generatortwist2}).

Cancellation of cubic anomalies in D5-branes models, either involving orbifold or orientifold singularities, follow the same pattern as the D4-brane case. Indeed, the expressions (\ref{chiralD4}) to (\ref{chiralO4b}) are almost identical, with the only difference that now $\d(j,k) = \d_{j,k-1} - \d_{j,k+1}$ and that $Q_{O4} = 4\b^1\b^2$. The explicit computations, by means of discrete Fourier transformations, have been performed in \cite{Aldazabal:2000dg} and \cite{Cremades:2002dh} for the orbifold and orientifold case, respectively.

\subsection{Mixed and cubic $U(1)$ anomalies}

Mixed $U(1)$ anomalies also emerge from a triangular diagram like those in figure \ref{anomaly}, but now one gauge boson belongs to a gauge group  $U(1)_a$ and the other two to a non-Abelian gauge group $SU(N_b)$. We denote such anomaly as $U(1)_a-SU(N_b)^2$. The cubic $U(1)$ anomaly, on the other hand, involves three  $U(1)$ gauge groups: $U(1)_a-U(1)_b-U(1)_c$. Since chiral fermions in intersecting brane constructions are, at most, charged under two independent $U(1)$'s\footnote{This seems to be the case in any (D-brane) string-based construction, and it comes from the fact that an open string has only two ends.} we should only care about cubic anomalies of the form $U(1)_a-U(1)_b^2$. The general expressions for mixed and cubic $U(1)$ anomalies are, in terms of table \ref{casimir}, given by
\bea
\A_{U(1)_a-SU(N_b)^2} = \sum_r Q_a(r)\cdot C_b(r), & &
\A_{U(1)_a-U(1)_b^2} = \sum_r Q_a(r)\cdot Q_b(r)^2,
\label{mixta}
\eea
where $r$ stands for the representation in which a fermion charged under both groups transforms. In general, we will see that tadpole conditions imply the {\em partial} cancellation of these anomalies. The remaining contribution has to be canceled by a series of counterterms which come from closed string fields propagating between two different branes. This effect is known in the literature as generalized Green-Schwarz (GS) mechanism. A formal discussion on this mechanism and several examples involving orbifold and orientifold compactifications see \cite{Sagnotti:1992qw,Ibanez:1998qp,Antoniadis:1999ge,Scrucca:1999zh}.

We can easily compute such anomalies in models of D6-branes wrapping 3-cycles of $T^6$ (or some more general compact manifold $\M_6$) by simple inspection of the spectrum (\ref{spectrum4ab}), which gives the contribution
\bea
\A_{U(1)_a-SU(N_b)^2} &=& \med \d_{ab} \sum_c N_c I_{ac} + \med N_a I_{ab},
\\ \A_{U(1)_a-U(1)_b^2} &=& \frac 13 \d_{ab} N_a 
\sum_c N_c I_{ac} + N_a N_b I_{ab},
\label{mixedD6}
\eea
where the factor of $1/3$ comes from the extra symmetry of the triangular diagram when $U(1)_a = U(1)_b$. 

The first term of both expressions vanishes after imposing tadpole conditions, whereas the second term will generically not vanish. The remaining anomaly should then be canceled by means of a generalized Green-Schwarz mechanism. In the particular case of D6-branes this mechanism is mediated by the exchange of RR untwisted fields between D6-branes, as was shown in \cite{Aldazabal:2000dg} for the case of the $U(1)$ mixed anomaly. Let us now see it for the $U(1)$ cubic anomaly, closely following \cite{Aldazabal:2000dg}.

Let us consider a D6-brane $a$ wrapping a 3-cycle $[\Pi_a]$. Due to the Chern-Simons action, this D-brane will couple to the RR 3-form $A_3$ and its ten-dimensional Hodge dual 5-form $A_5$ as \cite{Li:1995pq,Douglas:1995bn}
\bea
\med \mu_6\int_{D6_a} A_3 \wedge {\rm Tr \ }\left(\F_a \wedge \F_a\right), & & 
\mu_6\int_{D6_a} A_5 \wedge {\rm Tr \ }\F_a,
\label{couplingsD6}
\eea
where $\mu_6 = (2\pi)^{-6} (\a^\prime)^{-\frac 72}$ is the D6-brane RR charge and $\F_a = {B \over \sqrt{G}} + 2\pi\a^\prime F_a$ is the gauge invariant flux considered previously. In order make contact with low-energy effective physics is necessary to perform a dimensional reduction down to $D=4$. Let us then introduce two sets of 3-cycle homology classes, $\{[\Sig_i]\}_{i=1}^{b_3}$ and $\{[\Lam_i]\}_{i=1}^{b_3}$, both being a basis of the homology vector space $H_3(\M, \inte)$ and dual to each other, i.e., they satisfy $[\Lam_i]\cdot [\Sig_j] = \d_{ij}$. Given a general 3-cycle, we can expand its homology class in both bases
\bea
[\Pi_a] = \sum_i q_a^i [\Lam_i], 
& & [\Pi_a] = \sum_i \tilde q_a^i [\Sig_i].
\label{expansion}
\eea
Notice that these expressions are naturally related to the $q$-basis language. Indeed, the coefficients $q_a^i$ are nothing but the entries of the vector $\q_a$, whereas the dual coefficients $\tilde q_a^i$ are the entries of a vector $\tq$ related to the previous one by $\tq_a = - \I \q_a$.

\begin{figure}[ht]
\centering
\epsfxsize=3.5in
\epsffile{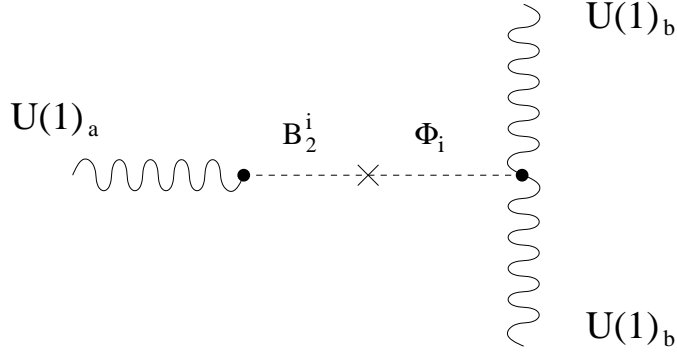}
\caption{Green-Schwarz diagram providing the counterterms necessary for cancellation of mixed and cubic $U(1)$ anomalies.}
\label{mechanism}
\end{figure}

After dimensional reduction, couplings (\ref{couplingsD6}) will translate into couplings between $D=4$ gauge bosons of the group $SU(N_a)$ and several RR untwisted fields. This set of RR fields can be expressed in $D=4$ physics by simply integrating $A_3$ and $A_5$ on 3-cycles of the compact six-dimensional manifold. Namely, we can define the $D=4$ fields
\bea
\Phi_i \equiv (4\pi^2 \a^\prime)^{-\frac 32} \int_{[\Lam_i]} A_3,
& & B_2^i \equiv (4\pi^2 \a^\prime)^{-\frac 32} \int_{[\Sig_i]} A_5,
\label{fieldsD6}
\eea
in terms of which the couplings (\ref{couplingsD6}) translate into $D=4$ effective field theory couplings
\bea
N_a \sum_i q_a^i \int_{M_4} \Phi_i {\rm Tr \ }\left(\F_a \wedge \F_a\right), 
& &
N_a \sum_i \tilde q_a^i \int_{M_4} B_2^i \wedge {\rm Tr \ }\F_a,
\label{couplingsD64D}
\eea
where the prefactor $N_a$ arises from normalization of the $U(1)_a$ generator, with respect to the $U(N_a)$ that contains it, as in \cite{Ibanez:1998qp}. Notice that fields (\ref{fieldsD6}) are Hodge duals in $D=4$, so these couplings can be combined in a Green-Schwarz diagram where $U(1)_a$ couples to the $i^{th}$ untwisted field $B_2^i$, which by means of duality with  $\Phi_i$ then couples to two $U(1)_b$ gauge bosons (see figure \ref{mechanism}). The coefficient of this amplitude is (modulo an $a$, $b$ independent numerical factor)
\beq
\A_{ab} = \sum_i N_a N_b \ \tilde q_a^i q_b^i =
N_a N_b \ \tq_a^{\ t} \cdot \q_b = 
N_a N_b \ \left( -\I \q_a \right)^t \cdot \q_b =
N_a N_b \ \q_a^{\ t\ }\I \q_b,
\label{gsD6}
\eeq
where in the last equality we have used that $\I$ is an antisymmetric matrix. This amplitude is precisely of the form required to cancel the residual $U(1)_a$-$U(1)_b^2$ anomaly in (\ref{mixedD6}).

The generalized Green-Schwarz mechanism will work in a similar fashion when canceling $U(1)$ anomalies in the other intersecting brane constructions presented in Chapter 2. Since this mechanism will have important phenomenological consequences in, for instance, giving mass to anomalous $U(1)$'s, let us briefly study the rest of the cases. 

In the case of the orientifolds of $T^6$, the spectrum (\ref{specori2}) provides the mixed and cubic $U(1)$ anomaly with the following contributions
\bea
\A_{U(1)_a-SU(N_b)^2} & = & \med \d_{ab} \left(\sum_c N_c 
\left(I_{ac} + I_{ac^*}\right) - Q_{O6} N_{O6} \ I_{a,O6} \right)
+ \med N_a \left(I_{ab} + I_{ab^*}\right) \nonumber \\
& = & \d_{ab} \ \q_a^{\ t\ }\I \left(\sum_c N_c P_+ \q_c 
- Q_{O6} N_{O6} \ \q_{O6} \right) + N_b \ \q_a^{\ t\ } 
\I P_+ \q_b, \\
\A_{U(1)_a-U(1)_b^2} & = & \frac 13 \d_{ab} \ N_a \left(\sum_c N_c 
\left(I_{ac} + I_{ac^*}\right) - Q_{O6} N_{O6} \ I_{a,O6} \right)
+ N_a N_b \left(I_{ab} + I_{ab^*}\right) \nonumber \\
& = & \frac 23 \d_{ab} \ N_a \q_a^{\ t\ }\I \left(\sum_c N_c P_+ \q_c 
- Q_{O6} N_{O6} \ \q_{O6} \right) + 2 N_a N_b \ \q_a^{\ t\ }\I P_+ \q_b.
\label{mixedO6}
\eea

Again the first term on each expression vanishes after imposing tadpoles. The residual contribution is identical to the one in the toroidal case, but for the insertion of the $P_+$ projector. The theory, then, will accommodate itself in order to produce the right counterterms canceling such residual anomalies. Indeed, in the orientifolded theory the couplings (\ref{couplingsD6}) will be completed by adding those involving the brane $a$*, where $U(N_a)$ gauge bosons live as well. However, is important to notice that the map relating $U(N_a)$ to $U(N_{a*})$ is not trivial, but the generators of both groups are related by the $\O$ action by conjugation. That is, their field strengths will be related by $\F_{a^*} = - \F_a$. As a result, the 5-form couples with different sign to the D6-brane $a$ and to $a$*. By a straightforward modification of the toroidal case, we conclude that the $D=4$ $U(1)_a$ gauge boson couplings to $\Phi_i$ are given by the components of $N_a(1 + \OR)\q_a$, while the couplings to $\B_2^i$ are given by $- N_a \I (1 - \OR) \q_a$. We then arrive at the Green-Schwarz contribution
\bea
\A_{ab} & = & 
\med N_a N_b \left( -\I (1 - \OR) \q_a \right)^t \cdot (1 +\OR) \q_b
\nonumber \\
& = & 2 N_a N_b \q_a^{\ t\ }P_-^t \I P_+ \q_b 
= 2 N_a N_b \q_a^{\ t\ }\I P_+ \q_b,
\label{gsO6}
\eea
where we have made use of the projectors' properties $P_\mp^t \cdot I = I \cdot P_\pm$ y $P_\pm^2 = P_\pm$ deduced in Appendix \ref{qbasis}, and the $\med$ factor comes from identifying $ab$ and $\O b \O a$ sectors. As we see, the GS contribution has again the appropriate form to cancel the residual cubic $U(1)$ anomaly in (\ref{mixedO6}).

What is the meaning of this $P_+$ projector? In order to shed some light on this, let us turn to the T-dual version of the theory, that is, type I D9-branes with magnetic fluxes. In type I theory, the only RR fields are the 2-form $A_2$ and its Hodge dual $A_6$. The couplings of these fields to D9 branes will be
\bea
\mu_9 \frac{1}{2!}\int_{D9_a} A_6 \wedge {\rm Tr \ }\F_a^2, & & 
\mu_9 \frac{1}{4!}\int_{D9_a} A_2 \wedge {\rm Tr \ }\F_a^4, 
\label{couplingsO6}
\eea
where $\mu_9 = (2\pi)^{-9} (\a^\prime)^{-5}$, and the wedge product is implicit in $\F^n_a$. We now define the four-dimensional RR fields
\bea
C^0 \equiv (4\pi^2 \a^\prime)^{-3} \int_{T^6} A_6, & &
C^I \equiv (4\pi^2 \a^\prime)^{-1}\int_{T^2_I} A_2, \\
B_2^0 \equiv A_2, & &
B_2^I \equiv (4\pi^2\a^\prime)^{-2} \int_{(T^2_J)\ti(T^2_K)} A_6,
\label{fieldsO6}
\eea
which are Hodge duals in four dimensions. Indeed, the duality relations are
\bea
dC^0 = - * dB_2^0, & &
dC^I = - * dB_2^I.
\label{HodgeO6}
\eea

Just as in the T-dual picture of D6-branes at angles, to compute the GS contribution we must perform a dimensional reduction down to $D=4$. In doing this, one should take into account that integration of Tr $\F_a$ along the $i^{th}$ two-torus yields a factor $4 \pi \a' m_a^{(i)}$ (fractional $m$), proportional to the flux in such two-torus. Similarly, integrating the pullback of the $D=10$ RR forms on such two-torus yields a factor $n^i_a$, which is he number of times that the D9$_a$-brane wraps it. We finally obtain the couplings
\bea
\begin{array}{ccc}
N_a\, m^{(1)}_a\, m^{(2)}_a\, m^{(3)}_a \int_{M_4} B_2^0 
\wedge {\rm Tr} \F_a, & \quad  \quad
& N_b\ n^{(1)}_b\, n^{(2)}_b\, n^{(3)}_b \int_{M_4} C^0 
\wedge {\rm Tr}\left(\F_b \wedge \F_b\right), \vspace{0.2cm} \\
N_a\, n^{(J)}_a\, n^{(K)}_a\, m^{(I)}_a \int_{M_4} B_2^I 
\wedge {\rm Tr}\F_a, & \quad  \quad
& N_b\ n^{(I)}_b\, m^{(J)}_b\, m^{(K)}_b \int_{M_4} C^I 
\wedge {\rm Tr} \left(\F_b\wedge \F_b\right),
\end{array}
\label{Tdualcouplings}
\eea
where again we have taken into account the correct $U(1)$ normalization. The amplitude of the GS diagram in figure \ref{mechanism} then yields
\beq
-N_a N_b m^{(1)}_a m^{(2)}_a m^{(3)}_a n^{(1)}_b n^{(2)}_b n^{(3)}_b \ -\
N_a N_b \sum_{I\neq J,K}  n^{(I)}_a n^{(J)}_a m^{(K)}_a 
n^{(K)}_b m^{(I)}_b m^{(J)}_b, 
\label{gsO6dual}
\eeq
which again provide us with the right contribution to cancel the cubic $U(1)$ anomaly. This same computation has been performed in \cite{Aldazabal:2000dg} for type IIB D9-branes with magnetic fluxes, obtaining a result similar to the T-dual toroidal case (\ref{gsD6}). The main difference of both computations comes from the fact the number of RR fields in type I theory is half of those in type IIB. In particular, the type IIB (toroidal) case has 8 $D=4$ RR fields (plus duals) mediating the GS mechanism, whereas type I (orientifold) has only 4 (plus duals) such fields. The significance of projectors $P_\pm$ is then the existence of half number of independent RR fields in the orientifold than in the toroidal compactification. This is not only present in the GS mechanism, but also on the tadpole conditions (\ref{tadpoleO6b}), which are half the amount of conditions of the toroidal case. In the picture of magnetic fluxes this comes from the fact that the only invariant RR form under $\O$ is $A_2$. In the T-dual picture of branes at angles, on the contrary, comes from the identification of homology and cohomology classes under $\R$.

Orbifold compactifications have also some new features with respect to compactification on smooth manifolds. In this case GS mediators are not untwisted but twisted RR fields, which are confined to the $\cpx^n/\inte_N$ orbifold singularity and propagate on $M_4 \ti T^{2n}$. This means that the local physics at the singularity manages to solve its potential quantum anomalies, without help of the fields propagating on the whole target space.

From the chiral spectrum in (\ref{spectrum4ab}), one can compute both mixed and cubic $U(1)$ anomalies involving D4-branes at an $\cpx^2/\inte_N$ singularity, obtaining
\bea
\A_{U(1)_{a,j}-SU(N_b^l)^2} & = & 
\med \d_{ab} \d_{jl} \sum_{c,r} N_c^r I_{ab} \ \d(j,r) 
+ \med N_a^j I_{ab} \ \d(j,l), \\
\A_{U(1)_{a,j}-U(1)_{b,l}^2} & = & 
\frac 13 \d_{ab} \d_{jl} \ N_a^j \sum_{c,r} N_c^r I_{ab} \ \d(j,r) 
+ N_a^j N_b^l I_{ab} \ \d(j,l),
\label{mixedD4}
\eea
where $\d(j,l)$ is defined in (\ref{delta}). The first contribution on each case vanishes because of tadpoles, while the second will be canceled by a GS mechanism. This necessary counterterm can be expressed in the $q$-basis language as
\beq
N_a^j N_b^l\q_{a,j}^{\ \dagger\ }\I \ \q_{b,l}.
\label{counterD4}
\eeq
Paralleling the discussion on the toroidal case, we would expect this term to appear from a decomposition of D4-branes RR charges in two 'dual' bases, from where $D=4$ RR twisted fields would be defined by integration. These RR fields would then couple to $U(1)_{a,j}$ $D=4$ gauge bosons with coupling proportional to the entries of the vector $\q_{a,j}$. This analogy can be made precise in case we have a supersymmetric twist $v_\om$.

The extension to orientifold singularities is straightforward. The only difference comes from the presence of the projectors $P_\pm$ inserted in (\ref{counterD4}). This reflects the fact that gauge bosons couple to the $D=4$ RR twisted 2-forms as $P_-\q_{a,j}$ and to the 0-forms as $P_+\q_{a,j}$. Finally, the argument is identical when dealing with D5-branes.

Although $q$-basis formalism provides us with a simple and elegant way of showing anomaly cancellation, for completeness let us show how this mechanism works in a slightly different, equivalent way. This formalism was used in \cite{Aldazabal:2000dg} to show explicitly the mixed anomaly cancellation in orbifold models of D$(3+n)$-branes. We will again work in the T-dual picture of type IIB D-branes with fluxes, where GS mediators are $2n(N-1)$ RR twisted fields (plus Hodge duals), defined upon dimensional reduction on $T^{2n}$. In case of type IIB theory on $T^2 \ti \cpx^2/\inte_Z$ these fields are
\beq
\begin{array}{lcl}
\vspace{0.1cm}
B_0^{(k)} = A_0^{(k)}, 
& & B_2^{(k)} = \int_{\bf T^2}  A_4^{(k)}, \\
C_0^{(k)} = \int_{\bf T^2} A_2^{(k)}, 
& & C_2^{(k)} =  A_2^{(k)}.
\end{array}
\label{pformsD4}
\eeq
The couplings in the case of D4-branes are then
\beq 
\begin{array}{cc}
c_k N_a^i\, n_a \int_{M_4} {\rm Tr} \left( \g_{k,a}\lam_i\right) \
B_2^{(k)}\wedge {\rm Tr} \F_{a,i},  
& c_k N_b^j\, m_b \int_{M_4} {\rm Tr} \left(\g_{k,b}^{-1}\lam_j^2\right) \
B_0^{(k)} \wedge {\rm Tr} \left(\F_{b,j}\wedge \F_{b,j}\right), \\
c_k N_a^i\, m_a \int_{M_4} {\rm Tr} \left( \g_{k,a}\lam_i\right) \
C_2^{(k)}\wedge {\rm Tr} \F_{a,i},
& c_k N_b^j\, n_b \int_{M_4} {\rm Tr} \left(\g_{k,b}^{-1}\lam_j^2\right) \
C_0^{(k)} \wedge {\rm Tr} \left(\F_{b,j}\wedge \F_{b,j}\right),
\end{array}
\label{dualcouplingsD4}
\eeq
where $k$ labels the $k^{th}$ twisted sector and $\lam_i$ denotes the Chan-Paton wavefunction for the gauge boson state. In general we will have ${\rm Tr} \left( \g_{k,a}\lam_i\right) = \a^{k i}$ \cite{Ibanez:1998qp}. Using the Hodge duality relations that relate $C_2^{(k)}$ with $C_0^{(k)}$ and $B_2^{(k)}$ with $B_0^{(k)}$ is easy to check that the GS contribution will cancel the $U(1)_{a,i}-U(1)^2_{b,j}$ anomaly. 

We can extend theses couplings to the orientifold case by simply adding the contributions of the mirror branes:
\bea & &
\begin{array}{c}
c_k N_a^i\, n_a \int_{M_4} {\rm Tr} \left(\g_{k,a}-\g_{k,a^*}\right)\lam_i
\ B_2^{(k)}\wedge {\rm Tr} \F_{a,i}, \\
c_k N_a^i\, m_a \int_{M_4} {\rm Tr} \left(\g_{k,a}+\g_{k,a^*}\right)\lam_i \
C_2^{(k)}\wedge {\rm Tr} \F_{a,i},
\end{array} 
\label{dualcouplingsO4}\\ & &
\begin{array}{c}
c_k N_b^j\, m_b \int_{M_4} \left(\g_{k,b}^{-1}-\g_{k,b^*}^{-1}\right)\lam_j^2 \
B_0^{(k)} \wedge {\rm Tr} \left(\F_{b,j}\wedge \F_{b,j}\right), \\
c_k N_b^j\, n_b \int_{M_4} \left(\g_{k,b}^{-1}+\g_{k,b^*}^{-1}\right)\lam_j^2 \
C_0^{(k)} \wedge {\rm Tr} \left(\F_{b,j}\wedge \F_{b,j}\right),
\end{array}
\label{dualcouplings2O4}
\eea
where now ${\rm Tr} \left( \g_{k,a^*}\lam_i\right) = \a^{-ki}$. Notice that couplings to RR fields in the sector $k$ are similar to those in the sector $-k$, which was to be expected since the $\O$ action relates conjugate phases. As a consequence we will have half the amount of twisted RR fields coupling to gauge bosons as in the orbifold case. This is easy to understand from the presence of the projectors $P_\pm$ in (\ref{counterD4}).

Finally, let us list the same couplings in the case of the ${\bf (T^2)_1} \ti{\bf  (T^2)_2} \ti \cpx/\inte_N$ orbifold models, corresponding to D5-branes in the intersecting picture. We first define the fields
\beq
\begin{array}{lcl}
B_0^{(k)} = A_0^{(k)}, 
& & B_2^{(k)} = \int_{\bf (T^2)_1 \times (T^2)_2} A_6^{(k)}, \\
C_0^{(k)} = \int_{\bf (T^2)_1 \times (T^2)_2} A_4^{(k)}, 
& & C_2^{(k)} =  A_2^{(k)}, \\
D_0^{(k)} = \int_{\bf (T^2)_2} A_2^{(k)},
& & D_2^{(k)} = \int_{\bf (T^2)_1} A_4^{(k)}, \\
E_0^{(k)} = \int_{\bf (T^2)_1} A_2^{(k)}, 
& & E_2^{(k)} = \int_{\bf (T^2)_2} A_4^{(k)},
\end{array}
\label{pformsD5}
\eeq
in terms of which the four dimensional couplings are, in the orbifold case
\bea & &
\begin{array}{c}
c_k N_a^i\, n^{(1)}_a n^{(2)}_a \int_{M_4} {\rm Tr} 
\left(\g_{k,a}\lam_i\right) \ B_2^{(k)}\wedge {\rm Tr} \F_{a,i}, \\
c_k N_a^i\, m^{(1)}_a m^{(2)}_a \int_{M_4} {\rm Tr} 
\left(\g_{k,a}\lam_i\right) \ C_2^{(k)}\wedge {\rm Tr} \F_{a,i}, \\
c_k N_a^i\, n^{(1)}_a m^{(2)}_a \int_{M_4} {\rm Tr} 
\left(\g_{k,a}\lam_i\right) \ D_2^{(k)}\wedge {\rm Tr} \F_{a,i}, \\
c_k N_a^i\, m^{(1)}_a n^{(2)}_a \int_{M_4} {\rm Tr} 
\left(\g_{k,a}\lam_i\right) \ E_2^{(k)}\wedge {\rm Tr} \F_{a,i},
\end{array} 
\label{dualcouplingsD5}
\\ & &
\begin{array}{c}
c_k N_b^j\, m^{(1)}_b m^{(2)}_b \int_{M_4} \left(\g_{k,b}^{-1}\lam_j^2\right) \
B_0^{(k)} \wedge {\rm Tr} \left(\F_{b,j}\wedge \F_{b,j}\right), \\
c_k N_b^j\, n^{(1)}_b n^{(2)}_b \int_{M_4} \left(\g_{k,b}^{-1}\lam_j^2\right) \
C_0^{(k)} \wedge {\rm Tr} \left(\F_{b,j}\wedge \F_{b,j}\right), \\
c_k N_b^j\, m^{(1)}_b n^{(2)}_b \int_{M_4} \left(\g_{k,b}^{-1}\lam_j^2\right) \
D_0^{(k)} \wedge {\rm Tr} \left(\F_{b,j}\wedge \F_{b,j}\right), \\
c_k N_b^j\, n^{(1)}_b m^{(2)}_b \int_{M_4} \left(\g_{k,b}^{-1}\lam_j^2\right) \
E_0^{(k)} \wedge {\rm Tr} \left(\F_{b,j}\wedge \F_{b,j}\right), 
\end{array}
\label{dualcouplings2D5}
\eea
and in the orientifold case,
\bea & &
\begin{array}{c}
c_k N_a^i\, n^{(1)}_a n^{(2)}_a \int_{M_4} 
{\rm Tr} \left(\g_{k,a}-\g_{k,a^*}\right)\lam_i \
B_2^{(k)}\wedge {\rm Tr} \F_{a,i}, \\
c_k N_a^i\, m^{(1)}_a m^{(2)}_a \int_{M_4} 
{\rm Tr} \left(\g_{k,a}-\g_{k,a^*}\right)\lam_i \
C_2^{(k)}\wedge {\rm Tr} \F_{a,i}, \\
c_k N_a^i\, n^{(1)}_a m^{(2)}_a \int_{M_4} 
{\rm Tr} \left(\g_{k,a}+\g_{k,a^*}\right)\lam_i \
D_2^{(k)}\wedge {\rm Tr} \F_{a,i}, \\
c_k N_a^i\, m^{(1)}_a n^{(2)}_a \int_{M_4} 
{\rm Tr} \left(\g_{k,a}+\g_{k,a^*}\right)\lam_i \
E_2^{(k)}\wedge {\rm Tr} \F_{a,i},
\end{array} 
\label{dualcouplingsO5} \\ & &
\begin{array}{c}
c_k m^{(1)}_b m^{(2)}_b \int_{M_4} 
{\rm Tr} \left(\g_{k,b}^{-1}+\g_{k,b^*}^{-1}\right)\lam_j^2 \
B_0^{(k)} \wedge {\rm Tr} \left(\F_{b,j}\wedge \F_{b,j}\right), \\
c_k n^{(1)}_b n^{(2)}_b \int_{M_4} 
{\rm Tr} \left(\g_{k,b}^{-1}+\g_{k,b^*}^{-1}\right)\lam_j^2 \
C_0^{(k)} \wedge {\rm Tr} \left(\F_{b,j}\wedge \F_{b,j}\right), \\
c_k m^{(1)}_b n^{(2)}_b \int_{M_4} 
{\rm Tr} \left(\g_{k,b}^{-1}-\g_{k,b^*}^{-1}\right)\lam_j^2 \
D_0^{(k)} \wedge {\rm Tr} \left(\F_{b,j}\wedge \F_{b,j}\right), \\
c_k n^{(1)}_b m^{(2)}_b \int_{M_4} 
{\rm Tr} \left(\g_{k,b}^{-1}-\g_{k,b^*}^{-1}\right)\lam_j^2 \
E_0^{(k)} \wedge {\rm Tr} \left(\F_{b,j}\wedge \F_{b,j}\right).
\end{array}
\label{dualcouplings2O5}
\eea

\subsection{Massive $U(1)$'s \label{masauunos}}

Cancellation of mixed and cubic $U(1)$ anomalies by means of a GS mechanism has some direct consequences in low energy physics, most of them quite relevant when discussing the phenomenology of semi-realistic models. The most notorious of these is the fact that some Abelian gauge fields get a mass term and, as a result, the corresponding $U(1)$ factors must be removed from the gauge group. Since the mechanism by which the gauge boson acquires a mass does not involve a non-vanishing v.e.v. for a scalar field, these massive $U(1)$ symmetries will remain as global symmetries of the effective Lagrangian.

In general, the Green-Schwarz mechanism involves some set of RR $D=4$ 2-forms $B_2^i$ which couple to the field strength of the $U(1)$ gauge groups as
\beq
\sum_{i} c^{\alpha}_i  \ B_2^i \wedge  F_\a,
\label{gsuno}
\eeq
where $F_\a$ is the the field strenght of the Abelian factor $U(1)_\a \subset U(N_\a)$. On the other hand, the Hodge duals RR 0-forms $\Phi_i$ couple to either Abelian or non-Abelian groups as
\beq
\sum_{i} d^{\b}_i \ \Phi_i \ {\rm Tr} (F_\b \wedge F_\b).
\label{gsdos}
\eeq
Finally, the proper combination of both couplings enters into the amplitude of the GS conterterm in figure \ref{mechanism}, cancelling the corresponding $U(1)_\a$ anomaly.
\beq
\A_{\a\b}\ +\ \sum_{i} c^{\alpha}_i d^{\b}_i \ =\ 0 .
\label{gstres}
\eeq

The important point is that the coupling (\ref{gsuno}) gives a mass to the gauge boson $A_\mu^\a$. In particular, every anomalous $U(1)$ will necessarily couple to a two-form $B_2$, so in the effective theory a mass term will be induced for the corresponding gauge boson. To understand the basis of the mechanism giving masses to the $U(1)$'s, let us consider the following Lagrangian coupling an Abelian gauge field $A_\mu$ to an antisymmetric tensor $B_{\mu\nu}$:
\begin{equation}
\label{dualuno} 
{\cal L}\ =\ -\frac{1}{12} H^{\mu\nu\rho}
H_{\mu\nu\rho}-\frac{1}{4g^2} F^{\mu\nu} F_{\mu\nu} 
+ \frac{c}{4}\ \epsilon^{\mu\nu\rho\sigma} B_{\mu\nu}\ F_{\rho\sigma},
\end{equation}
where 
\beq
\label{definicion}
H_{\mu\nu\rho}=\partial_\mu B_{\nu\rho}+\partial_\rho B_{\mu\nu}
+\partial_\nu B_{\rho\mu}, \qquad  F_{\mu\nu}=\partial_\mu
A_\nu-\partial_\nu A_\mu,
\eeq
and $g, c$ are arbitrary  constants. This corresponds to the kinetic term for the fields $B_{\mu\nu}$ and $A_{\mu}$ together with the $B\wedge F$ term. We will now proceed to dualize this Lagrangian and rewrite it in terms of an (arbitrary) field $H_{\mu\nu\rho}$, imposing the constraint $H=dB$ by the standard introduction of a Lagrange multiplier field $\eta$:
\begin{equation}
\label{dualdos}
{\cal L}_0=\ -\frac{1}{12} H^{\mu\nu\rho}\
H_{\mu\nu\rho}-\frac{1}{4g^2} F^{\mu\nu}\ F_{\mu\nu} 
- \frac{c}{6}\ \epsilon^{\mu\nu\rho\sigma} H_{\mu\nu\rho}\ A_{\sigma} 
-\frac{c}{6}\eta \epsilon^{\mu\nu\rho\sigma} \partial_\mu H_{\nu\rho\sigma}.
\end{equation}
Notice that integrating out $\eta$ implies d*$H=0$ which in turn implies that (locally) $H=dB$ and then we recover (\ref{dualuno}). Alternatively, integrating by parts the last term in (\ref{dualdos}) we are left with a quadratic action for $H$ which we can solve immediately to find
\begin{equation}
H^{\mu\nu\rho}= - {c}\ \epsilon^{\mu\nu\rho\sigma}
\left(A_\sigma+\partial_\sigma \eta\right).
\end{equation}
Inserting  this back into (\ref{dualdos}) we find:
\begin{equation}
{\cal L}_{A}\ =\ -\frac{1}{4g^2}\ F^{\mu\nu}\ F_{\mu\nu} -
\frac{c^2}{2} \left(A_\sigma+\partial_\sigma \eta\right)^2
\end{equation}
which is just a mass term for the gauge field $A_\mu$ after ``eating''  the scalar $\eta$ to acquire a mass $m^2 = g^2 c^2$. Notice that this is similar to the St\"uckelberg mechanism where we do not need a scalar field with a  vacuum expectation value to give a mass to the gauge boson, nor do we have a massive Higgs-like field at the end.

Notice that this mechanism requires the presence of the Green-Schwarz term $B\wedge F$  but not necessarily the anomaly cancellation term ($\Phi F\wedge F$). Therefore as long as a $U(1)$ field has a Green-Schwarz coupling $B\wedge F$, it does not have to be anomalous in order to acquire a mass. In general, the condition for a generic $U(1)$ generator
\beq
Q = \sum_\a \xi_\a Q_\a
\label{generator}
\eeq
to remain massless in the low energy theory is that it does not couple to closed string RR fields through a $B \wedge F$ coupling. Looking at the general expression (\ref{gsuno}) this translates into
\bea
\sum_\a c_i^\a \xi_\a = 0 \ \ \forall i & \sim & {\bf C} \cdot \vec{\xi} = 0,
\label{condunos}
\eea
where we have defined the matrix $({\bf C})_{i\a} = c_i^\a$, whose rows label a basis of RR forms and its colums a basis of $U(1)$'s in the particular configuration under study. Hence, the vector space of massless $U(1)$'s is given by Ker {\bf C}. Notice that the number of massive $U(1)$'s is nothing but Rank \cpx, so it could never be greater than the number of RR 2-forms coupling non-trivially to the open string sector.

Since the precise content of $B \wedge F$ couplings will determine the final gauge group of each D-brane based model, it seems worth to study the general form of these coupling in the intersecting D-brane models we have been considering. Let us first address the case of D6-branes wrapping 3-cycles on $\M_6$. Expressions (\ref{fieldsD6}) tell us that in this case $c_i^a = N_a \tq_a^i$, with $\tq_a^i \in \inte$ being the expansion coefficients of the D6-brane homology class $[\Pi_a] \in H_3(\M_6, \inte)$ in the basis $\{[\Sig_i]\}_{i=1}^{b_3}$. Let us then consider a full configuration containing $K$ stacks of $N_a$ D6-branes wrapped on the 3-cycles $\Pi_a$ of $\M_6$, ($a = 1, \dots, K$). We can define the sublattice $\tilde \Lam \subset H_3(\M_6, \inte)$ generated by $\{N_a [\Pi_a]\}_{a=1}^K$. Is then easy to see that \# (Massive $U(1)$'s) = Rank \cpx \ = dim $\tilde \Lam$, so we can already know the number of massive $U(1)$'s by just looking at the topological features of the configuration. Actually, the matrix \cpx \  can also be constructed using the topological information encoded in $\{N_a [\Pi_a]\}_{a=1}^K$ and choosing an appropiate basis $\{[\Sig_i]\}_{i=1}^{b_3}$. Of course, the final answer for the massless $U(1)$'s, which is Ker \cpx, does not depend on this choice. Finally, let us notice that tadpole conditions imply that
\bea
\sum N_a [\Pi_a] = 0 & \Raw & \sum_a N_a \tq_a^i = 0 \ \ \forall i,
\label{diagonal}
\eea
so if we choose the diagonal combination $\xi_a = 1$ $\forall a$, the corresponding $U(1)$ will not couple to any RR 2-form and will hence be massless. Tadpole conditions then imply that Ker \cpx \ is nontrivial.

The orientifold analogue of this construction will be slightly more complicated. As discussed in the previous section, the $B \wedge F$ couplings of each stack $a$ are given by the entries of the vector $N_a P_- \q_a$ instead of $N_a \q_a$, so only half of the RR 2-forms will be relevant for the GS mechanism. Moreover, tadpole conditions do not generically imply the existence of any massles $U(1)$. 

Let us be more specific and describe the $B \wedge F$ couplings in the particular case of D6-branes wrapping factorisable 3-cycles of $T^2 \times T^2 \times T^2$. Recall that although $b_3(T^6) = 20$, these factorisable cycles can only expand an 8-dimensional sublattice of $H_3(T^6, \inte)$ so there will be, at most eight relevant RR fields mediating the GS mechanism. Each stack $a$ of $N_a$ D6-branes wraps a 3-cycle (\ref{factorisable2}) and contains a $U(1)_a$ factor. The $B \wedge F$ couplings of such $U(1)_a$ are given by
\beq
N_a \q_a = 
N_a \left( 
\begin{array}{c}
n_a^{(1)} n_a^{(2)} n_a^{(3)} \\
m_a^{(1)} m_a^{(2)} m_a^{(3)} \\
m_a^{(1)} m_a^{(2)} n_a^{(3)} \\
n_a^{(1)} n_a^{(2)} m_a^{(3)} \\
m_a^{(1)} n_a^{(2)} m_a^{(3)} \\
n_a^{(1)} m_a^{(2)} n_a^{(3)} \\
n_a^{(1)} m_a^{(2)} m_a^{(3)} \\
m_a^{(1)} n_a^{(2)} n_a^{(3)} 
\end{array} 
\right). 
\label{couplingsBD6}
\eeq
The orientifold case only involves four such couplings for each D6-brane, which are given by
\beq
N_a P_-^{\ \prime}  \q_a = 
N_a \left( 
\begin{array}{c}
0 \\
m_a^{(1)} m_a^{(2)} m_a^{(3)} \\
0 \\
n_a^{(1)} n_a^{(2)} m_a^{(3)} \\
0 \\
n_a^{(1)} m_a^{(2)} n_a^{(3)} \\
0 \\
m_a^{(1)} n_a^{(2)} n_a^{(3)} 
\end{array} 
\right),
\label{couplingsBO6}
\eeq
where now we are taking fractional wrapping numbers. 

The case involving either orbifold or orientifold singularities parallels the above discussion, with the only difference that now the space of $U(1)$'s is given by $U(1)_\a = U(1)_{a,i}$, with  $a$ labeling the D-brane stacks and $i$ the orbifold phases. In the orbifold case, the matrix of $B^{(k)} \wedge F$ couplings of $U(1)_{a,i}$ with the $k^{th}$ twisted sector is given by the rows of $c_k N_a^i \q_{a,ki}$, the vector $\q_{a.ki}$ depending if we are dealing with D5 or D4-branes models. The orientifold case will be just the same, with the inclusion of the projector operator $P_-^{\ \prime} $, as defined in Appendix \ref{qbasis}. Alternatively, one can extract these coupling from (\ref{dualcouplingsD4}), (\ref{dualcouplingsO4}), (\ref{dualcouplingsD5}) and (\ref{dualcouplingsO5}).

\chapter{Looking for the Standard Model \label{models}}

In this chapter we will apply the framework described in the previous chapters, as well as the theoretical results derived in them, in order to construct intersecting D-brane models whose low energy spectrum is as realistic as possible. In particular, we will center on the construction of semi-realistic orientifold configurations of D6 and D5-branes at angles. By exploring these constructions, we will be able to achieve D-brane configurations whose low-energy limit yields the chiral content and gauge group of just a three generation Standard Model, with no extra massless fermions nor $U(1)$'s. This minimal content of particles appears already at the level of the string construction, and no effective field theory assumption is needed in order to get rid of unwanted extra fermion nor $U(1)$ factors. This fact implies some progress with respect to previous D-brane model-building in the literature, where extra chiral fermions or extended gauge groups always appeared at the string level. Indeed, the field theory techniques used in models where extra matter appears requires a very complicated model dependent analysis of the structure of the scalar potential and Yukawa couplings and, usually, the necessity of unjustified simplifying assumptions. Hence, the construction of models with minimal chiral content is much more satisfying and highly simplifies the study of the phenomenology related to such models.

After constructing these semi-realistic models, we will address some phenomenological features derived from them. In fact, one of the most appealing features involving Intersecting Brane Worlds is the fact that it is relatively easy to compute the main physical quantities defining the effective field theory. Moreover, these quantities have a clear geometrical meaning in the string-based construction, which allow us to get a new intuition of their significance.

\section{The Standard Model intersection numbers}

In our search for a string-theory description of the Standard Model we are going to consider configurations of D-branes wrapping on cycles on the six extra dimensions, which we will assume to be compact. Our aim is to find configurations with just the SM group $SU(3)\times SU(2)\times U(1)_Y$ as a gauge symmetry and with three generations of fermions transforming like the five representations:
\beq
\begin{array}{c}
Q_L^i=(3,2)_{\frac 16} \\  
U_R^i =({\bar 3},1)_{-\frac 23} \\
D_R^i=({\bar 3},1)_{\frac 13} \\
L^i=(1,2)_{-\frac 12} \\ 
E_R^i=(1,1)_1 .
\end{array}
\label{gensm}
\eeq
Now, in general, D-branes will give rise to $U(N)$ gauge factors in their world-volume, rather than $SU(N)$. Thus, if we have $r$  different stacks with $N_i$ parallel branes  we will expect gauge groups in general of the form $U(N_1)\times U(N_2)\times .... \times U(N_r)$. At points where the D-brane cycles intersect one will have in general massless fermion fields transforming like bifundamental representations, i.e., like $(N_i,{\bar N}_j)$ or $(N_i,N_j)$. Thus the  idea will be to identify these fields with the SM fermion fields.

A first obvious idea is to consider three types of branes giving rise in their world-volume to a gauge group $U(3)\times U(2)\times U(1)$. This in general turns out not to be sufficient. Indeed, as we said,  chiral fermions like those in the SM appear from open strings stretched between D-branes with intersecting cycles. Thus e.g., the left-handed quarks $Q_L^i$ can only appear from open strings stretched between the $U(3)$ branes and the $U(2)$ branes. In order to get the right-handed leptons $E_R^i$ we would need a fourth set with one brane giving rise to an additional $U(1)'$: the right-handed leptons would come from open strings stretched between the two $U(1)$ branes. Thus we will be forced to have a minimum of four stacks of branes with $N_a = 3, N_b = 2, N_c = 1$ and $N_d = 1$ yielding a $U(3)\times U(2)\times U(1)\times U(1)$ gauge group. Although apparently such a structure would yield four gauged  $U(1)$'s, we will show below that we expect three of these $U(1)$'s to become massive and decouple from the low-energy spectrum.

In the class of models we are considering the fermions come in bifundamental representations:
\beq
\sum_{a,b}
n_{ab}(N_a,{\overline N}_b)+m_{ab}  (N_a,N_b) + n_{ab}^* ({\overline N}_a,N_b)
+m_{ab}^* ({\overline N}_a , {\overline N}_b),
\label{bifundamentals}
\eeq
where here $n_{ab},n_{ab}^*,m_{ab},m_{ab}^*$ are integer non-negative coefficients which are model dependent, and $n_{ab}^*, m_{ab}^* \neq 0$ only if we consider orientifold constructions\footnote{As we have seen, in orientifold models fermions transforming like antisymmetric or symmetric representations may also appear. For the case of the SM group those states would give rise to exotic chiral fermions which have not been observed. Thus we will not consider these more general possibilities any further.}. As usual, cancellation of RR tadpoles will impose strong constraints for constructing a consistent D-brane model, and will also guarantee the cancellation of gauge anomalies. Notice that in the case of the D-brane models here discussed anomaly cancellation just requires that there should be as many fundamental ($N_a$) as antifundamental (${\overline N}_a$) representations for any $U(N_a)$ group. Any consistent model should also satisfy this weaker constraint, since otherwise tadpoles could not possibly be canceled.

An important fact for our discussion is that tadpole cancellation conditions impose this constraint even if the gauge group is $U(1)$ or $U(2)$. The constraint in this case turns out to be required for the cancellation of $U(1)$ anomalies. Let us now apply this to a possible D-brane model yielding $U(3)\times U(2)\times U(1)\times U(1)$ gauge group. Since  $U(2)$ anomalies have to cancel we will make a distinction between $U(2)$ doublets and antidoublets. Now, in the SM only left-handed quarks and leptons are $SU(2)$ doublets. Let us assume to begin with that the three left-handed quarks $Q_L^i$ were antidoublets $(3,{\bar 2})$. Then there are altogether 9 anti-doublets and $U(2)$ anomalies would never cancel with just three generations of left-handed leptons. Thus all models in which all left-handed quarks are $U(2)$ antidoublets (or doublets) will necessarily require the presence in the spectrum of 9 $U(2)$ lepton doublets (anti-doublets) \footnote{Indeed this can be checked for example in the semi-realistic models constructed in ref.\cite{Aldazabal:1999tw,Aldazabal:2000sk,Aldazabal:2000sa,Cvetic:2000st,Aldazabal:2000dg,Aldazabal:2000cn,Berenstein:2001nk}.}. There is, however, a simple way to cancel $U(2)$ anomalies sticking to the fermion content of the SM. They cancel if two of the left-handed quarks are antidoublets and the third one is a doublet. Then there is a total of six doublets and antidoublets and $U(2)$ anomalies will cancel. In order to accommodate this fermion content in an intersecting D-brane scheme we need both class of bifundamental fermion representations $(N_a,{\overline N}_b)$ and $(N_a,N_b)$ to appear in the chiral spectrum of our theory. This possibility will only be possible in orientifold models. 

Notice that in this case it is crucial that the {\it  number of generations equals the number of colours}. There is no way to build a D-brane configuration with the gauge group of the SM and e.g., just one complete quark/lepton generation. Anomalies (RR tadpoles) cannot possibly cancel.  We find this connection between the number of generations and colours quite attractive. 

It is now clear what are we looking for. We are searching for D-brane configurations with four stacks of branes yielding an initial  $U(3) \times U(2) \times U(1) \times U(1)$ gauge group. The multiplicity of the stacks and the associated gauge group of such D-brane configuration is shown, in table \ref{SMbranes}, along with a name that, as we will see, is related to their associated SM quantum numbers.

\begin{table}[htb]
\renewcommand{\arraystretch}{2}
\begin{center}

\begin{tabular}{|c|c|c|c|} 
\hline 
Label & Multiplicity & Gauge Group & Name \\ 
\hline 
\hline 
stack $a$ & $N_a = 3$ & $SU(3) \times U(1)_a$ & Baryonic brane\\ 
\hline 
stack $b$ & $N_b = 2$ & $SU(2) \times U(1)_b$ & Left brane\\ 
\hline 
stack $c$ & $N_c = 1$ & $U(1)_c$ & Right brane\\ 
\hline 
stack $d$ & $N_d = 1$ & $U(1)_d$ & Leptonic brane \\ 
\hline 
\end{tabular} 
\caption{Brane content yielding the SM spectrum. \label{SMbranes}}

\end{center}
\end{table}

Each of these stacks will be wrapping some cycles $\Pi_\a$, $\a = a, b, c, d$ in the internal compact space and will intersect each other a number of times. The chiral fermionic spectrum will be encoded in the intersection number of such cycles $I_{\a\b} = [\Pi_\a] \cdot [\Pi_\b]$, and in order to obtain a minimal three generation SM spectrum they should be\footnote{In case we are dealing with compactifications involving orbifold singularities, the Chan-Paton phase of each D-brane should also be taken into account, see below.}
\beq 
\begin{array}{lcl} 
I_{ab}\ =   \ 1, & & I_{ab*}\ =\ 2, \\ 
I_{ac}\ =   \ -3, & & I_{ac*}\ =\ -3, \\ 
I_{bd}\ =   \ -3,  & & I_{bd*}\ =\ 0, \\ 
I_{cd}\ =   \ 3, & & I_{cd*}\ =\ -3, 
\end{array} 
\label{intersec2} 
\eeq 
either some equivalent set which involves changing $b$ with $b$*, etc. All other intersection numbers should vanish, avoiding the presence of chiral matter in the $ad$, $ad$* sectors (leptoquarks), $aa$* sector (exotic matter), etc. which are ruled out experimentally. Here a negative number denotes that the corresponding fermions should have opposite chirality to those with positive intersection number. The chiral spectrum arising from these intersection numbers is presented in table \ref{SMcontent}, along with their associated SM quantum numbers.
\begin{table}[htb]
\renewcommand{\arraystretch}{1.5}
\begin{center}

\begin{tabular}{|c|c|c|c|c|c|c|c|} 
\hline Intersection & 
 Matter fields  &   &  $Q_a$  & $Q_b $ & $Q_c $ & $Q_d$  & Y \\ 
\hline\hline (ab) & $Q_L$ &  $(3,2)$ & 1  & -1 & 0 & 0 & 1/6 \\ 
\hline (ab*) & $q_L$   &  $2( 3,2)$ &  1  & 1  & 0  & 0  & 1/6 \\ 
\hline (ac) & $U_R$   &  $3( {\bar 3},1)$ &  -1  & 0  & 1  & 0 & -2/3 \\  
\hline (ac*) & $D_R$   &  $3( {\bar 3},1)$ &  -1  & 0  & -1  & 0 & 1/3 \\ 
\hline (bd) & $L$    &  $3(1,2)$ &  0   & -1   & 0  & 1 & -1/2  \\ 
\hline (cd) & $N_R$   &  $3(1,1)$ &  0  & 0  & 1  & -1  & 0  \\ 
\hline (cd*) & $E_R$   &  $3(1,1)$ &  0 & 0 & -1 & -1  & 1 \\ 
\hline 
\end{tabular} 
\end{center}
\caption{Standard model spectrum and $U(1)$ charges. The hypercharge generator is defined as $Q_Y = \frac 16 Q_a - \frac 12 Q_c - \frac 12 Q_d$.\label{SMcontent} }
\end{table}

As we discussed, cancellation of $U(N_\a)$ anomalies requires:
\bea
\sum_\b\, N_\b\, (I_{\a\b} + I_{\a\b*})\, = 0
\label{anomdsix}
\eea
which is indeed obeyed by the spectrum of table \ref{SMcontent}, although to achieve this cancellation we have to add three fermion singlets $N_R$. As shown below these have quantum numbers of right-handed neutrinos (singlets under hypercharge). Thus this is a first prediction of the present approach: {\em right-handed neutrinos must exist}.

The structure of the $U(1)$ gauge fields is quite relevant in what follows. The following important points are in order:

\begin {itemize}

\item
The four $U(1)$ symmetries $Q_a$, $Q_b$, $Q_c$ and $Q_d$ have clear interpretations in terms of known global symmetries of the Standard Model. Indeed, $Q_a$ is $3B$, $B$ being the baryon number and $Q_d$ is nothing but the lepton number. Concerning $Q_c$, it is twice $I_R$, the third component of right-handed weak isospin familiar from left-right symmetric models. Finally $Q_b$ has the properties of a Peccei-Quinn symmetry (it has mixed $SU(3)$ anomalies). We thus learn that {\it all these known global symmetries of the SM are in fact gauge symmetries} in this class of theories.

\item
The mixed $U(1)$ anomalies of these four $U(1)_\a$'s with the non-Abelian groups $SU(N_\b)$ are given by $\A_{\a\b} = \med  N_\a (I_{\a\b}+I_{\a\b*})$. It is easy to check that $(Q_a - 3Q_d)$ (which is $3(B - L)$) and $Q_c$ are free of triangle anomalies. In fact the hypercharge is given by the linear combination\footnote{Actually, is possible to choose the hypercharge as $Q_Y  = {1\over 6} Q_a \pm {1\over 2} Q_c  + {1\over 2} Q_d$, although both choices give equivalent physics.\label{alter}}
\beq
Q_Y \ =\ {1\over 6} Q_a\ -\ {1\over 2} Q_c \ -\ {1\over 2} Q_d,
\label{hyper}
\eeq
and is, of course, anomaly free. On the other hand the other orthogonal combinations $(3Q_a+Q_d)$ and $Q_b$ have triangle anomalies. As we have seen in the previous chapter, these anomalies will be canceled by a generalized GS mechanism that will render the associate gauge bosons massive. That is, imposing the intersection numbers (\ref{intersec2}) implies that the $(3Q_a + Q_d)$ and $Q_b$ gauge bosons will become massive. On the other hand the other two anomaly free combinations (including hypercharge) may be massive or not, depending on the couplings $c^{\alpha}_i$ in (\ref{gsuno}). Thus, in order to really obtain a Standard Model gauge group with the right standard hypercharge we will have to insure that it does not couple to any closed string mode which would render it massive, i.e., one should have
\beq
{1\over 6} c_{i}^a \ -\
{1\over 2}c_{i}^c \ -\  {1\over 2} c_{i}^d \ \  =\ 0, \quad \forall i
\label{ortohiper}
\eeq
This turns out to be an important constraint in the specific models constructed in the following sections. But an important conclusion is that in those models generically only three of the four $U(1)$'s can become massive and that in a large subclass of models it is the SM hypercharge which remains massless. Thus even though we started with four $U(1)$'s we are left at the end of the day with just the SM gauge group.

\item
Let us also remark that the symmetries whose gauge boson become massive {\em will persist in the low-energy spectrum as global symmetries}. This has important obvious consequences, as we will discuss below.

\end{itemize}

Up to now we have been relatively general and perhaps a structure like this may be obtained in a variety of string constructions. We believe that the above discussion identifies in a clear way what we should be looking for in order to get a string construction with a massless sector identical to the SM. In the following sections we will be more concrete and show how this philosophy may be followed in the simple settings discussed in previous chapters. Specifically, in the next sections we will be considering type IIA D6-branes wrapping at angles on a six torus $T^6$, and then type IIB D5-branes wrapping at angles on $T^4$ and sitting at an orbifold singularity $\cpx/\inte_N$. We will see how even in such a simple settings the desired structure can be implemented, and later discuss the specific properties and phenomenological features of each construction. 

\section{D6-brane models}

Let us try to construct a specific model of D6-branes wrapping factorisable 3-cycles of $T^6 = T^2 \ti T^2\ti T^2$ with low-energy spectrum given by that in table \ref{SMcontent}. This task will correspond to finding a configuration with the D-brane content of table \ref{SMbranes} reproducing the intersection numbers (\ref{intersec2}). We find that getting the spectrum of the SM is quite a strong constraint. Finding a solution to such intersection numbers involves making the right choices of wrapping numbers $n_\a^{(r)}$, $m_\a^{(r)}$, $\a = a, b, c, d$, $r = 1, 2, 3$ as well as considering two-tori with rectangular or tilted complex structure. To motivate the form of these solutions let us enumerate some of the constraints we have to impose.

\begin{itemize}

\item
We will require that for any brane $\a$ one has $\Pi_{r=1}^3\ m_\a^{(r)}\ = 0$ because of two reasons. First, to avoid the appearance of matter at the intersections of one brane to its mirror. This matter (transforming like symmetric or antisymmetric representations of the gauge group) has exotic quantum numbers beyond the particle content of the SM which we are trying to reproduce. In addition, there are tachyonic scalars at those intersections which would destabilize
the brane configuration.

\item
$\Pi_{r=1}^3\ m_\a^{(r)}$ is one of the coefficients of the $B \wedge F$ couplings on the $D=4$ effective theory, as (\ref{couplingsBO6}) shows. Hence, if $\Pi_{r=1}^3\ m_\a^{(r)} = 0$ is verified for every $\a$, then in these toroidal models there will be at most three RR fields $B_2^i$, $i = 1, 2, 3$ that couple to the Abelian groups. Thus at most three $U(1)$'s may become massive by the mechanism described in \ref{masauunos}. This implies that we should consider only models with four sets of branes at most, since otherwise there would be additional massless $U(1)$ gauge bosons beyond hypercharge.

\item
We further impose that we should reproduce the spectrum in table \ref{SMcontent}. This is the most constraining condition. It implies that the branes $a$ should be parallel to the $d$ brane in at least one of the three complex planes and that the $b$ branes are parallel to the $c$ brane. Getting $I_{ab} = 1$ and $I_{ab^*}=2$ requires that at least one of the three tori, say the third, should be tilted (see figure \ref{bflux}). Getting the other intersection numbers correct gives us also further constraints.

\end{itemize}

Imposing these conditions, we find a general class of solutions, given by the wrapping numbers shown in table \ref{solution}.

\begin{table}[htb]
\renewcommand{\arraystretch}{2.5}
\begin{center}
\begin{tabular}{|c||c|c|c|} 
\hline 
 $N_\a$  &  $(n_\a^{(1)},m_\a^{(1)})$  &  $(n_\a^{(2)},m_\a^{(2)})$   
&  $(n_\a^{(3)},m_\a^{(3)})$ \\ 
\hline
\hline $N_a = 3$ & $(1/\beta^1, 0)$  &  $(n_a^{(2)}, \epsilon \beta^2)$ & 
 $(1/\rho, \tilde \eps/2)$  \\ 
\hline $N_b=2$ &   $(n_b^{(1)}, -\tilde \eps\epsilon \beta^1)$     
&  $ (1/ \beta^2,0)$  & $(1,3\rho \tilde\eps /2)$   \\ 
\hline $N_c=1$ & $(n_c^{(1)},3\rho \epsilon \beta^1)$  &    
 $(1/\beta^2,0)$  & $(0,1)$  \\ 
\hline $N_d=1$ &   $(1/\beta^1,0)$  &  $(n_d^{(2)},\epsilon\beta^2/\rho)$  & 
$(1, -3\rho \tilde\eps /2)$ \\
\hline 
\end{tabular} 
\end{center}

\caption{D6-brane wrapping numbers giving rise to a SM spectrum. The general solutions are parametrized by two phases $\epsilon, \tilde \eps =\pm1$, the NS background on the first two tori $\beta^i = 1-b^{(i)} = 1,1/2$, four integers $n_a^{(2)}, n_b^{(1)}, n_c^{(1)}, n_d^{(2)}$ and a parameter $\rho = 1,1/3$. In order to obtain the correct hypercharge massless $U(1)$ those parameters have to verify the extra constraint (\ref{condhiper}).\label{solution}}  
\end{table}

In this table we have several discrete parameters. First we consider $\beta^i =1 - b^{(i)}$, with $b^{(i)} = 0, 1/2$ being the T-dual NS B-background field discussed in Chapter 2. As shown in \cite{Blumenhagen:2000ea}, the addition of this background is required in order to get and odd number of quark-lepton generations. From the point of view of branes at angles $\beta^i = 1$ stands for a rectangular lattice for the $i^{th}$ torus, whereas $\beta^i = 1/2$ describes a tilted lattice allowed by the $\OR$ symmetry. Notice that we have chosen the third torus to have $\beta^3 = 1/2$, following the points described above. We also have two phases $\epsilon, \tilde \eps = \pm 1$ and the parameter $\rho$ which can only take the values $\rho = 1, 1/3$. Furthermore, each of these families of D6-brane configurations depend on four integers ($n_a^{(2)}, n_b^{(1)}, n_c^{(1)}$ and $n_d^{(2)}$). Any of these choices lead exactly to the same massless fermion spectrum of table \ref{SMcontent}. 

In order to have a consistent construction, it is not enough an anomaly free effective theory, but one has now to ensure that these choices are consistent with the tadpole cancellation conditions (\ref{tadpoleO6b}). It turns out that all but the first of those conditions are automatically satisfied by the above families of configurations. The first tadpole condition reads in the present case:
\beq
\frac{3 n_a^{(2)}}{\rho \beta^1} \ +\ \frac{2n_b^{(1)}}{\beta^2} \ +\
\frac{n_d^{(2)}}{\beta^1} \ = \ 16.
\label{tadsm}
\eeq
Note  however that one can always relax this constraint by adding extra D6-branes with no intersection with the SM ones and not contributing to the rest of the tadpole conditions. For example, a simple possibility would be the addition of $N_h$ D6-branes parallel to the orientifold plane in the three two-tori, i.e., with wrapping numbers of the form 
\beq 
N_h\, (1/\beta_1,0)( 1/\beta _2,0)(2,0)  
\label{hidden} 
\eeq 
which does not intersect the branes from table \ref{SMbranes} whereas contributes to the first tadpole condition without affecting the other three. We will call such D6-brane a {\it hidden brane}. In this case the above condition would be replaced by the more general one
\beq
\frac{3n_a^2}{\rho \beta^1} \ +\ \frac{2n_b^1}{\beta^2} \ +\
\frac{n_d^2}{\beta^1} \ +\ 2N_hn_h^1n_h^2\  = \
16 \ .
\label{tadsm2}
\eeq
Thus the families of standard model configurations we have found are very weakly constrained by tadpole cancellation conditions. This is not so surprising. Tadpole cancellation conditions are closely connected to cancellation of anomalies. Since the SM is anomaly-free, it is not surprising that the solutions we find are almost automatically tadpole-free.

\subsection{Massive $U(1)$'s}

Let us now analyse the general structure of $U(1)$ anomaly cancellation in this class of models. As remarked, there are two anomalous $U(1)$'s given by the generators $(3Q_a+Q_d)$ and $Q_b$ and two anomaly-free ones which are $(Q_a - 3Q_d)$ and $Q_c$. From (\ref{Tdualcouplings}) one can see that $B_2^0$ does not couple to any D6-brane, due to the fact that $\Pi_{r=1}^3\ m_\a^{(r)} = 0$ for each D6-brane $\a$. The three remaining RR fields $B_2^I$, $I = 1, 2, 3$ couple to the $U(1)$'s in the models as follows:
\bea
B_2^1 &\wedge & \ {{2\tilde\eps \epsilon \beta^1}\over{\beta^2}}F_{b} 
\nonumber \\
B_2^2 &\wedge  & \ \frac{\epsilon  \beta^2 }
{\rho \beta^1}(3F_a\ +\  F_{d})
\\ \nonumber
 B_2^3 & \wedge  &  \ {\tilde\eps\over {2 \beta^1\beta^2 } }
\left(3\beta^2 n_a^{(2)} F_a\ +\ 6\rho \beta^1 n_b^{(1)} F_{b} \ -\ 
2 \b^1 \tilde\eps n_c^{(1)} F_c \ +\ 3\rho \beta^2 n_d^{(2)} F_d\right) \
\label{bfs}
\eea
On the other hand, the dual scalars $C^I$ and $C^0$ have couplings:
\bea
 C^1\ & & 
\left(\frac{\epsilon \tilde\eps \beta^2}{2\beta^1}\right)  
\left(F_a\wedge F_a\ -\ 3F_d\wedge F_d\right)\nonumber \\
 C^2\ &  & 
\left(\frac{3\rho\epsilon\beta^1}{2\beta^2}\right) 
\left(-F_b\wedge F_b\ +\ 2 F_c\wedge F_c\right)\\ \nonumber
C^0 & &
\left(\frac{n_a^{(2)}}{\rho \beta^1} F_a\wedge F_a
\ +\ \frac{n_b^{(1)}}{\beta^2} F_b\wedge F_b \ +\ \frac{n_d^{(2)}}{\beta^1}
F_d\wedge F_d \right) \
\label{cff}
\eea
and the RR scalar $C^3$ does not couple to any $F\wedge F$ term. It is easy to check that these terms cancel all residual $U(1)$ anomalies in the way described in the previous chapter. Notice in particular how only the exchange of the $B_2^1,B_2^2$ fields (and their duals $C^1,C^2$) can contribute to the cancellation  of anomalies since the $C^3$ field does not couple to $F\wedge F$ and $B_2^0$ does not couple to any $F_i$. The exchange of those RR fields proceeds in a universal manner (i.e., independent of the particular choice of  $n$'s) and hence the mechanism for the $U(1)$ anomalies to cancel is also universal. On the other hand the $B_2^3$ field does couple to a linear combination of the four $U(1)$'s and hence will render that combination massive. The $U(1)$ which remains light is given by the linear combination 
\footnote{In the particular case with $n_c^{(1)}=0$ one can have both anomaly-free $U(1)$'s remaining in the massless spectrum as long as one also has $n_a^{(2)} = n_d^{(2)} = 0$.}
\beq 
Q_0\ =\ {1\over 6} Q_a  +  {r\over 2} Q_c - {1\over 2} Q_d   
\,  , \quad  
r = {-\tilde\eps\beta^2 \over 2n_c^{(1)}\beta^1}(n_a^{(2)}+3\rho n_d^{(2)}) 
\label{unolig} 
\eeq 
If we want to have just the standard hypercharge at low energies this should be proportional to the hypercharge generator. This is the case for $r = -1$, i.e., as long as:
\beq
n_c^{(1)} \ =\ {{\tilde\eps\beta^2}\over {2\beta^1} } 
\left(n_a^{(2)} + 3\rho n_d^{(2)}\right)
\label{condhiper}
\eeq
which is an extra condition the four integers should fulfill in order to really obtain a SM at low energies. Thus we have found families of toroidal models with D6-branes wrapping at angles in which the residual gauge group is just the standard model $SU(3)\times SU(2)\times U(1)_Y$ and with three standard generations of quarks and leptons and no extra chiral fermions (except for three right-handed neutrinos which are singlets under hypercharge).

These models are specific examples of the general approach in the previous section. Notice in particular that in these models Baryon number ($Q_a)$, and lepton number ($Q_d$) are unbroken gauged  $U(1)$ symmetries. The same is the case of the symmetry $Q_b$ which is a (generation dependent) Peccei-Quinn symmetry. Once the RR-fields give masses to three of the $U(1)$'s of the models, the corresponding $U(1)$'s remain as effective global symmetries in the theory. This has the following important physical consequences:

1) Baryon number is an exact perturbative symmetry of the effective Lagrangian. Thus the proton should be stable. This is a very interesting property which is quite a general consequence of the structure of the theory in terms of D-branes intersecting at angles and which was already advanced in \cite{Aldazabal:2000cn}. Notice that this property is particularly welcomed in brane scenarios with a low energy string scale \cite{Arkani-Hamed:1998rs,Antoniadis:1998ig,Dienes:1998vh,Kakushadze:1998ss,Shiu:1998pa,Sundrum:1998sj,Kakushadze:1998dc,Sundrum:1998ns,Bachas:1998kr,Kakushadze:1998wp,Benakli:1998pw,Burgess:1998px,Ibanez:1998rf,Delgado:1998qr,Ibanez:1999it,Accomando:1999sj,Ghilencea:2000dg,Antoniadis:2000en,Abel:2000bj} in which stability of the proton is an outstanding difficulty. But it is also a problem in standard scenarios like the MSSM  in which one has to impose by hand discrete symmetries like R-parity or generalizations in order to have a sufficiently stable proton.

2) Lepton number is an exact symmetry in perturbation theory. This has as a consequence that Majorana masses for the neutrinos should be absent. Any neutrino mass should be of standard Dirac type. They can however be naturally small as we discuss below.

3) There is a gauged $U(1)$ symmetry of the Peccei-Quinn type ($Q_b$) which is exact at this level. Thus, at this level the $\theta_{QCD}$ parameter can be rotated to zero. 

These properties seem to be quite model independent, and also seem to be a generic property of any D-brane model which gives rise to {\it just} a SM spectrum at the intersections.

In general, the structure of anomalous, massive and massless $U(1)$'s is one of the most important issues involved in the construction of semi-realistic D-brane models. Moreover, their generic existence creates a whole set of model-independent phenomenological features which may be of some interest when checking the validity of intersecting D-brane models with particle physics experimental bounds, as well as when looking for physics beyond the Standard Model at particle physics next generation accelerators. The phenomenological implications associated with the existence of such massive $U(1)$'s have been addressed in \cite{Ghilencea:2002da}.

As a final comment note that the pseudoscalar $C^0$ remains massless at this level and has axionic couplings (eq.(\ref{cff})) to the gauge fields of the SM (and also to the fields coming from the extra branes added to cancel tadpoles, if present). It would be interesting to study the possible relevance of this axion-like field concerning the strong CP problem.

\subsection{Absence of tachyons}

We have been concerned up to now with the massless chiral fermions at the D6-brane intersections. As we know, in addition to these fermionic states there are several scalars at each intersection which in some sense may be considered as `SUSY-partners', squarks and sleptons, of the massless chiral fermions, since they have the same multiplicity $|I_{\a\b}|$ and carry the same gauge quantum numbers. The four lightest of these scalar states have masses
{\bea
\begin{array}{cc}
\vspace{0.2cm}
{\rm \bf State} & {\bf Mass^2} \\
\vspace{0.1cm}
t_1 = (-s(\vt^1)+\vt^1,\vt^2,\vt^3,0) & \alpha' M^2 =
\frac 12(-|\vartheta^1|+|\vartheta^2|+|\vartheta^3|) \\
\vspace{0.1cm}
t_2 = (\vt^1,-s(\vt^2)+\vt^2,\vt^3,0) & \alpha' M^2 =
\frac 12(|\vartheta^1|-|\vartheta^2|+|\vartheta^3|) \\
\vspace{0.1cm}
t_3 = (\vt^1,\vt^2,-s(\vt^3)+\vt^3,0) & \alpha' M^2 =
\frac 12(|\vartheta^1|+|\vartheta^2|-|\vartheta^3|) \\
t_4 = (-s(\vt^1)+\vt^1,-s(\vt^2)+\vt^2,-s(\vt^3)+\vt^3,0) & \alpha' M^2
= 1-\frac 12(|\vartheta^1|+|\vartheta^2|+|\vartheta^3|)
\label{tachdsix}
\end{array}
\eea}
in the notation of Chapter 3. Here $\vartheta^i$ are the intersection angles (in units of $\pi$) at each of the three subtori, and $s(\vt^i) \equiv {\rm sign \ }(\vt^i)$. As is obvious from these formulae, the masses depend on the angles at each intersection and hence on the complex structure of each two-torus. Thus, in  principle some of the scalars may be tachyonic, which would signal an instability of the system against D-brane recombination. 

For a D-brane configuration to be stable there should be no tachyons at {\it none} of the intersections. Notice that, in principle, we may only bother about avoiding tachyons in sectors with non-vanishing intersection number $I_{\a\b} \neq 0$. Indeed, if the $\a\b$ sectors has intersection number $I_{\a\b} = 0$ we can always separate branes $\a$ and $\b$ by a distance $Y$ in the subtorus where they are parallel, and then the whole spectrum in the $\a\b$ sector will be shifted in mass$^2$ by $Y^2/4\pi\a^\prime$, rendering every potential tachyon in this sector massive.

\begin{figure}
\centering
\epsfxsize=6.2in
\hspace*{0in}\vspace*{.2in}
\epsffile{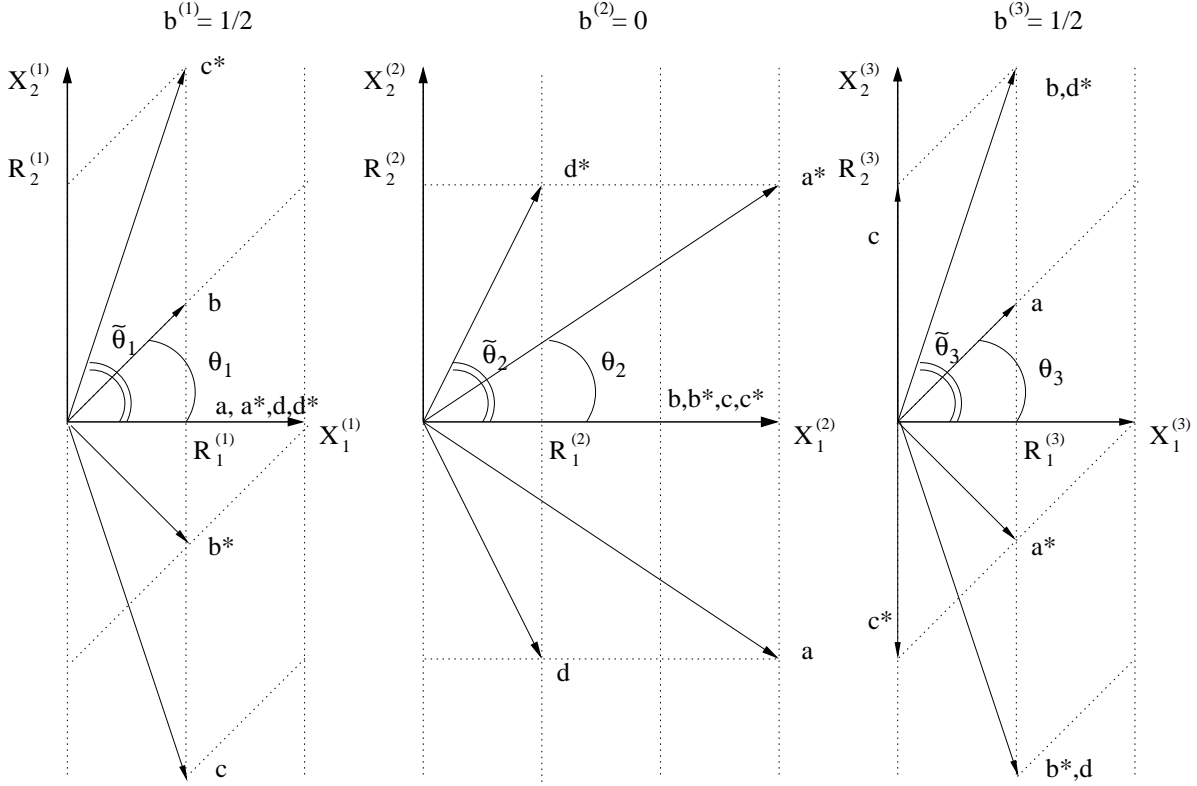}
\caption{Definition of the angles between the different branes on the three tori. We have selected a particular setting where $n_a^{(2)},n_b^{(1)},n_c^{(1)},n_d^{(2)} > 0$, $\tilde\eps = -\epsilon = 1$ and $\beta^1 = 1/2$, $\beta^2= 1$.}
\label{angles}
\end{figure}

As already noted in \cite{Aldazabal:2000dg}, in general it is possible to vary the compact radii in order to get rid of all tachyons of a given model. One can do a general analysis of sufficient conditions for absence of tachyons in the standard model examples of previous sections, which are parametrized in terms of $\beta^{1}, \b^2, \rho$ and the integers $n_a^{(2)}, n_b^{(1)}, n_c^{(1)}$ and $n_d^{(2)}$. Let us define the angles 
\bea
\theta_1 \ = \ \frac{1}{\pi} 
{\rm cot}^{-1}\frac{n_b^{(1)} R_1^{(1)}}{\beta^1 R_2^{(1)}} \ ,\quad
\theta_2 \ = \ \frac{1}{\pi} 
{\rm cot}^{-1}\frac{n_a^{(2)} R_1^{(2)}}{\beta^2 R_2^{(2)}} \ ,\quad
\theta_3 \ = \ \frac{1}{\pi} 
{\rm cot}^{-1}\frac{2 R_1^{(3)}}{\rho R_2^{(3)}}\ , \nonumber \\
{\tilde {\theta}_1} \ = \ \frac{1}{\pi} 
{\rm cot}^{-1}\frac{|n_c^{(1)}| R_1^{(1)}}{3\rho \beta^1 R_2^{(1)}}\ ,\quad
{\tilde {\theta}_2} \ =\ \frac{1}{\pi} 
{\rm cot}^{-1}\frac{\rho n_d^{(2)} R_1^{(2)}}{\beta^2 R_2^{(2)}} \ ,\quad
{\tilde {\theta}_3} \ = \ \frac{1}{\pi} 
{\rm cot}^{-1}\frac{2R_1^{(3)}}{3\rho R_2^{(3)}}
\label{angulos}
\eea
where $R^{(i)}_{1,2}$ are the compactification radii for the three $i=1,2,3$ tori. \footnote{As can be seen in figure \ref{angles}, $R^{(i)}_{1}$ are not compactification radii in a strict sense if $b^{(i)} \neq 0$ but their projection onto the $X^{(i)}_{1}$ direction.} The geometrical meaning of the angles is depicted in figure \ref{angles}.

%
Angles at all the intersections may be written in terms of those six angles which depend on the parameters of the particular model and the relative radii. We have four (possibly light) scalars $t_i, i = 1, 2, 3, 4$ at each of the 7 independent types of intersections, thus altogether 28 different scalar masses. Since all these 28 masses  can be written in terms of the above 6 angles, it is obvious that the masses are not all independent. Thus for example one finds:

\beq
\begin{array}{c} \vspace{0.1cm}
m^2_{ab}(t_2)+m^2_{ac}(t_3) \ =\ m^2_{ab*}(t_2)+m^2_{ac*}(t_3)
\ =\ m^2_{bd}(t_2)+m^2_{cd}(t_3) \nonumber \\ \vspace{0.1cm}
m^2_{ab}(t_1)+m^2_{ac}(t_4) \ =\ m^2_{ab*}(t_1)+m^2_{ac*}(t_4)
\ =\ m^2_{bd}(t_1)+m^2_{cd}(t_4) \nonumber \\ \vspace{0.1cm}
m^2_{ab}(t_1)+m^2_{ac*}(t_2) \ =\ m^2_{ab*}(t_1)+m^2_{ac}(t_2)
\ =\ m^2_{bd}(t_2)+m^2_{cd*}(t_1) \nonumber \\ \vspace{0.2cm}
m^2_{ab}(t_2)+m^2_{ac*}(t_1) \ =\ m^2_{ab*}(t_2)+m^2_{ac}(t_1)
\ =\ m^2_{bd}(t_1)+m^2_{cd*}(t_2) 
\end{array}
\label{sumrules}
\eeq
for the choice $\tilde\eps = 1$, and where the substitution $m^2_{cd} \leftrightarrow m^2_{cd*}$ must be performed if $\tilde\eps = -1$. These identities give interesting relationships among the squark and slepton partners of usual fermions. Due to these kind of constraints the 28 conditions for absence of tachyons may be reduced to only 14 general conditions (see Appendix II of \cite{Ibanez:2001nd}).

In order to get an idea of how easy is to get a tachyon-free configuration in one of the standard model examples of the previous section let us consider a particular case. Consider a model with $\rho = \beta^1 = \beta^2 = 1$, $\tilde\eps = -\epsilon = 1$ and with $n_a^{(2)} = 2$, $n_b^{(1)} = n_d^{(2)} = 0$ and $n_c^{(1)} = 1$. The wrapping numbers of the four stacks of branes are thus:
\bea
N_a=3 \ &\ (1,0)  (2,-1) (1,1/2)  \nonumber \\
N_b=2 \ &\ (0,1)  (1,0) (1,3/2)  \nonumber \\
N_c=1 \ &\ (1,-3)  (1,0) (0,1)  \nonumber \\
N_d=1 \ &\ (1,0)  (0,1) (1,3/2)    
\label{solu2} 
\eea
This verifies all the conditions to get just the SM gauge group with three quark/lepton generations. The first tadpole condition may be fulfilled by adding, e.g., 5 parallel branes with wrapping numbers $(1,0)(1,0)(2,0)$. Now, in this case one has $\theta_1 = 1/2 > {\tilde {\theta_1}}$, $\theta_2 = 1/2$ and many of the equations in the Appendix II of \cite{Ibanez:2001nd} are trivially satisfied. Then one can check that there are no tachyons at {\it any } of the intersections as long as:
\bea
\theta _2 \ +\ {\tilde {\theta_3}} \ -\ \theta_3 \ & \geq & \frac{1}{2}
\nonumber \\
\tilde{\theta_1} \ \geq \ {\tilde {\theta_3}}
\label{condi}
\eea
which may be easily satisfied for wide ranges of the radii. Similar simple expressions are obtained in other examples. For instance, a model within the first family in table \ref{minimal} with $n_a^{(2)} = 0, n_b^{(1)} = -1, n_c^{(1)} = \tilde\eps = 1, n_d^{(2)} = 1$ and $\rho=1/3$, $\beta^1 = 1/2,\beta^2 = 1$ has no tachyons as long as the two conditions ${\tilde \theta}_1+{\tilde \theta}_3-\theta_3\geq  1/2$ and ${\tilde \theta}_1+{\tilde \theta}_2-{\tilde \theta_3}\geq 1/2$ are fulfilled. Again, this happens for wide ranges of radii.

\subsection{The Higgs sector and electroweak symmetry breaking}

Up to now we have ignored the existence or not of the Higgs system required for the breaking of the electroweak symmetry as well as for giving masses to quarks and leptons. In general, we expect Higgs particles to arise from string states between Left and Right stacks, as this sector connects fermions of opposite chirality. Indeed, looking at the $U(1)$ charges of quarks and leptons in table \ref{SMcontent}, we see that possible Higgs fields coupling to quarks come in four varieties with charges under $Q_b, Q_c$, and hypercharge given in table \ref{higgsses}.

\begin{table}[htb] 
\renewcommand{\arraystretch}{1.3}
\begin{center}
\begin{tabular}{|c|c|c|c|}
\hline  
 Higgs   &  $Q_b$  &  $Q_c$   & Y \\
\hline\hline $h_1$ & 1  &  -1 & 1/2  \\
\hline $h_2$ &   -1    &  1  &  -1/2   \\
\hline\hline $H_1$ & -1  &  -1 & 1/2  \\        
\hline $H_2$ &   1    &  1  &  -1/2   \\    
\hline \end{tabular}
\end{center} \caption{ Electroweak Higgs fields \label{higgsses}}
\end{table}

Now, the question is whether for some configuration of the branes such  Higgs fields  appear in the light spectrum. Indeed that is the case. The Left stacks ($b,b^*$) are parallel to the Right stacks ($c,c^*$) along the second torus and hence their intersection number vanishes. However, there are open strings which stretch in between both sets of branes and which lead to light scalars when the distance $Z$ in the second torus is small. In particular there are the scalar states 
\bea
{m_{H^{\pm}}}^2 \ &=&\ {{Z^2_{bc*}}\over {4\pi^2\alpha^{\prime  2}}}
\ \pm  \ \frac{1}{2\alpha^{\prime}}\left||\vartheta^1_{bc*}|-|\vartheta^3_{bc*}|\right| \quad = \quad m_H^2 \pm m_B^2,
\nonumber \\
 {m_{h^{\pm}}}^2 \ &=&\ { {Z^2_{bc}}\over {4\pi^2\alpha^{\prime  2}}}\  
\ \pm \ \frac{1}{2\alpha^{\prime}}\left||\vartheta^1_{bc}|-|\vartheta^3_{bc}|\right|  \quad = \quad m_h^2 \pm m_b^2,
\label{Higgsmasses}
\eea
where $Z^2_{bc}$ ($Z^2_{bc^*}$)  is the distance$^2$ (in $\alpha '$ units) in transverse space along the second torus, and $\vt^1$, $\vt^3$ are the relative angles between stacks $b$ and $c$ (or $b$ and $c^*$) in the first and third complex planes. These four scalars have precisely the quantum numbers of the Higgs fields $H_i$ and $h_i$ in the table. The $H_i$'s come from the $bc$* sector whereas the $h_i$ come from the $bc$ sector. In addition to these scalars there are two fermionic partners at each of $bc$ and $bc$* sectors
{\bea
\begin{array}{cc}
{\rm \bf State} \quad & \quad {\bf Mass^2} \\
(-1/2+\vartheta^1, \mp 1/2 , -1/2+\vartheta^3, \pm 1/2 ) &  {\rm (Mass)}^2 =
  { {Z^2}\over {4\pi^2\alpha^{\prime 2}}}\  \\
\label{Higgsinomasses} 
\end{array}
\eea}
where we are supposing $\vt^1, \vt^3 > 0$. This Higgs system may be understood as massive $\N=2$ Hypermultiplets containing respectively the $h_i$ and $H_i$ scalars along with the above fermions. The above scalar spectrum can be interpreted as arising from the following mass terms in the effective potential:
\bea
& & {m_{H^+}}^2(H^+)^*H^+\ +\ \left(
\begin{array}{c}
+\leftrightarrow - \\ H\leftrightarrow h
\end{array}
\right)\ +\ h.c. \nonumber \\
& & = 
\left(H_1^* \ H_2 \right)\cdot
{\bf M} \cdot
\left(
\begin{array}{c}
H_1 \\ H_2^*
\end{array}
\right)\ 
+\ \left(h_1^* \ h_2 \right) \cdot
{\bf m} \cdot
\left(
\begin{array}{c}
h_1 \\ h_2^*
\end{array}
\right)\ + \ h.c.,
\eea
where
\beq
{\bf M} =
\left(
\begin{array}{cc}
m_H^2 & m_B^2 \\
m_B^2 & m_H^2
\end{array}
\right), 
\quad 
{\bf m} = 
\left(
\begin{array}{cc}
m_h^2 & m_b^2 \\
m_b^2 & m_h^2
\end{array}
\right), 
\eeq
and the fields $H_i$, $h_i$ $i= 1,2$ of table \ref{higgsses} are thus defined as
\beq
H^{\pm}={1\over2}(H_1^* \pm H_2), \quad \quad h^{\pm}={1\over2}(h_1^* \pm h_2),
\eeq

Notice that each of the Higgs systems have a quadratic potential similar to that of the MSSM. In our case the mass parameters of the potential have an interesting  geometrical  interpretation in terms of the brane distances and intersection angles. 

What are the sizes of the Higgs mass terms? The values of $m_H$ and $m_h$ are controlled by the distance between the  $b,c,c^*$ branes in the second torus. These values are in principle free parameters and hence one can make these parameters arbitrarily small compared to the string scale $M_s$. That is not the case of the $m_B^2,m_b^2$ parameters. We already mentioned that all scalar mass terms  depend on only 6 angles in this class of models. This is also the case here, these angles are not independent but one finds (using also eq.(\ref{sumrules})):
\bea
m^2_{B1}\ =&m^2_{Q_L}(t_2)+m^2_{U_R}(t_3) \ =\ m^2_{q_L}(t_2)+m^2_{D_R}(t_3)
\ =\ m^2_{L}(t_2)+m^2_{N_R}(t_3) \nonumber \\
m^2_{B2}\ =&m^2_{Q_L}(t_1)+m^2_{U_R}(t_4) \ =\ m^2_{q_L}(t_1)+m^2_{D_R}(t_4)
\ =\ m^2_{L}(t_1)+m^2_{N_R}(t_4) \nonumber \\
m^2_{b1}\ =&m^2_{Q_L}(t_1)+m^2_{D_R}(t_2) \ =\ m^2_{q_L}(t_1)+m^2_{U_R}(t_2)
\ =\ m^2_{L}(t_1)+m^2_{E_R}(t_2) \nonumber \\
m^2_{b2}\ =&m^2_{Q_L}(t_2)+m^2_{D_R}(t_1) \ =\ m^2_{q_L}(t_2)+m^2_{U_R}(t_1)
\ =\ m^2_{L}(t_2)+m^2_{E_R}(t_1) \nonumber \\
& m^2_B\ = min\{m^2_{B1},m^2_{B2}\}\ ;\ \ m^2_b = min\{m^2_{b1},m^2_{b2}\}
\label{sumrules2}
\eea
when $\tilde \eps =1$, and the same expressions but changing $m^2_B \lraw m^2_b$, when $\tilde \eps = -1$. \footnote{Roughly speaking, the change sign $\tilde\eps \lraw -\tilde\eps$ amounts to interchange $c \lraw c^*$ everywhere and the relabeling of scalar states $t_1 \lraw t_3$, $t_2 \lraw t_4$ arising at intersections of any stack with the stack $c$. It is also equivalent to choosing the alternative hypercharge of footnote \ref{alter} and relabeling the massless fields accordingly.} Thus if one lowers the $m_{B,b}^2$ parameters, some other scalar partners of quarks and leptons have also to be relatively light, and one cannot lower $m^2_{B,b}$ below present limits of these kind of scalars at accelerators.

Notice, however, that if the geometry is such that one approximately has $m^2_H\ =\ m^2_B$ (and/or $m^2_h\ = \ m^2_b$ ) there appear scalar flat directions along $\langle H_1 \rangle\ =\ \langle H_2\rangle$ ($\langle h_1\rangle\ =\ \langle h_2\rangle$) which may give rise to electroweak symmetry breaking at a scale well below the string scale. Obviously, this requires the string scale to be not far above the weak scale, i.e., $M_s = 1-few$ TeV since otherwise substantial fine-tuning would be needed.  Let us also point out that the particular Higgs coupling to the top-quark (either $h_1$ or $H_1$) will in general get an additional one-loop  negative contribution to its mass$^2$ in the usual way \cite{Ibanez:fr}.

Let us have a look now at the number of Higgs multiplets which may appear in the class of toroidal models discussed in previous sections. Notice first of all that the number $n_H$ ($n_h$) of Higgs sets of type $H_i$($h_i$) are given by the number of times the branes $b$ intersect with the branes $c$ ($c^*$) in the first and third tori:
\beq
n_h\ =\  {\beta ^1} |n_c^{(1)}+ 3\tilde\eps\rho n_b^{(1)}|, \quad  \quad
n_H\ =\ {\beta ^1} |n_c^{(1)}-3\tilde\eps\rho n_b^{(1)}|
\label{numhiggs}
\eeq
The simplest Higgs structure is obtained in the following cases:

\begin{itemize}

\item {\it  Higgs system of the MSSM}

From (\ref{numhiggs}) one sees that the minimal set of Higgs fields is obtained when either $n_H = 1$, $n_h = 0$ or $n_H = 0$, $n_H = 1$. For both of those cases it is easy to check that, after imposing condition (\ref{condhiper}), one is left with two families of models with $\rho = 1/3,\beta^1 = 1/2$ depending on a single integer $n_a^{(2)}$, on $\beta^2$ and on the phases of $|n_b^{(1)}| = |n_c^{(1)}| = 1$. These solutions are shown in the first two rows of table \ref{minimal}. The last column in the table shows the number $N_h$ of branes parallel to the orientifold plane one has to add in order to cancel global RR tadpoles (a negative sign means antibranes).
\begin{table}[htb]
\renewcommand{\arraystretch}{1.4}
\begin{center}
\begin{tabular}{|c|c|c|c|c|c|c|c|c|}
\hline
 Higgs & $\rho $ & $\beta^1$ & $\beta^2$ & $n_a^{(2)}$ 
& $n_b^{(1)}$ & $n_c^{(1)}$ & $n_d^{(2)}$  & $N_h$ \\
\hline\hline  
$n_H = 1, n_h = 0$ & 1/3 & 1/2 & $\beta^2$ & $n_a^{(2)}$ 
& $\pm 1$ & $ \pm\tilde\eps$ & $\pm \frac{1}{\beta^2}-n_a^{(2)}$ 
& $4\beta^2\left(1-n_a^{(2)}\right)$ \\
\hline\hline  
$n_H = 0, n_h = 1$ & 1/3 & 1/2 & $\beta^2$ & $n_a^{(2)}$ 
& $\pm 1$ & $\pm \tilde\eps$ & $\pm \frac{1}{\beta^2}-n_a^{(2)}$ 
& $4\beta^2\left(1-n_a^{(2)}\right) \mp 1$  \\
\hline\hline  
$n_H = 1, n_h = 1$  & 1 & 1 & $\beta^2$ & $n_a^{(2)}$ 
& 0 & $\pm \tilde\eps$ 
& $\frac{1}{3}\left(\pm \frac{2}{\beta^2}-n_a^{(2)}\right)$
& $\beta^2\left(8-\frac{4n_a^{(2)}}{3}\right) \mp \frac{1}{3}$ \\
\hline  
$n_H = 1, n_h = 1$ & 1/3 & 1 & $\beta^2$ & $n_a^{(2)}$ 
& 0  & $\pm \tilde\eps$
& $\pm \frac{2}{\beta^2}-n_a^{(2)}$ 
& $\beta^2 \left(8-{4n_a^{(2)}}\right) \mp 1$ \\
\hline \end{tabular}
\end{center} \caption{Families of models with the minimal Higgs content.
\label{minimal}}
\end{table}

As we will discuss in the following subsection, the minimal choice with $n_H=1,n_h=0$ is particularly interesting 
\footnote{It is amusing  that in this class of solutions with $n_a^{(2)}=1$ the SM sector is already tadpole free and one does not need to add extra non-intersecting branes, i.e., $N_h=0$. Thus the SM is the only gauge group of the whole model.} 
from the point of view of Yukawa couplings since the absence of the Higgs $h_i$ could be at the root of the smallness of neutrino masses. The opposite situation with $n_H=0$ and $n_h=1$ is less interesting since charged leptons would not get sufficiently large masses. 

\item
{\it Double MSSM Higgs system}

The next to minimal set is having $n_H = n_h = 1$. After imposing condition (\ref{condhiper}) one finds two families of such models depending on the integer $n_a^{(2)}$ and on $\beta^2$. They are shown in the last two rows of table \ref{minimal}. 

\end{itemize}

Let us finally comment that having a minimal set of Higgs fields would automatically lead to absence of flavour-changing neutral currents (FCNC) from Higgs exchange. In the case of a double Higgs system one would have to study in detail the structure of Yukawa couplings in order to check whether FCNC are sufficiently suppressed.

\subsection{Yukawas and gauge coupling constants}

As we discussed in the previous section, there are four possible varieties of Higgs fields $h_i,H_i$ in this class of models. The Yukawa couplings among the SM fields in table \ref{SMcontent} and the different Higgs fields which are allowed by the symmetries have the general form:
\bea
y^U_jQ_LU_R^j h_1 \ +\ y^D_jQ_LD_R^jH_2 \ +  \nonumber \\
y^u_{ij}q_L^iU_R^j H_1 \ +\ y^d_{ij}q_L^iD_R^jh_2 \ + \nonumber \\ 
y^L_{ij}L^iE_R^jH_2  \ +\  y^N_{ij}L^iN_R^jh_1 \ +\  h.c. 
\label{yuki}
\eea
where $i=1,2$ and $j=1,2,3$. Which of the observed quarks (i.e., whether a given left-handed quark inside $Q_L$ or $q_L^i$) fits into the multiplets will depend on which are the mass eigenstates of the quark and lepton mass matrices after diagonalization. These matrices depend on the Yukawa couplings in the above expression.   

The pattern of quark and lepton masses thus depends both on the vevs of the Higgs fields $h_i,H_i$ and on the Yukawa coupling constants and both dependences could be important in order to understand the observed hierarchical structure. 
In particular it could be that e.g., only one subset of the Higgs fields could get vevs. So let us consider two possibilities in turn.

\begin{itemize}

\item {\it Minimal set of Higgs fields}

This is for example the case in the  situation with $n_H = 1$, $n_h = 0$ described in the previous section in which only the $H_1,H_2$ fields appear. Looking at (\ref{yuki}) we see that only two $U$-quarks and one $D$-quark would get masses in this way. Thus one would identify them with the top, charm and $b$-quarks. In addition there are also masses for charged leptons. Thus, at this level, the $s,d,u$-quarks would remain massless, as well as the neutrinos. 

In fact this is not a bad starting point. The reason why the $H_1, H_2$ fields do not couple to these other fermions is because such couplings would violate the $U(1)_b$ symmetry (see the table). On the other hand strong interaction effects will break such a symmetry and one expects that they could allow for effective Yukawa couplings of type $Q_LU_R^j H_1$ and $q_L^iD_R^jH_2$ at some level. These effective terms could generate the current $u,d,s$-quark masses which are all estimated to have values of order or smaller than $\Lambda_{QCD}$.

Concerning neutrino masses, since Lepton number is a (perturbative) exact symmetry in these models, Majorana masses are forbidden and there can only be Dirac neutrino masses. The origin of neutrino (Dirac) masses could be quite interesting. One expects them to be  much more suppressed since neutrinos do not couple directly to strong interaction effects (which are the source of $U(1)_b$ symmetry breaking). In particular, there are in general dimension six operators of the form $\alpha '(L N_R)(Q_L U_R)^*$. These come from the exchange of massive string states and are consistent with all gauge symmetries. Plugging the $u$-quark chiral condensate, neutrino masses of order
\beq
m_{\nu } \ \propto   \frac{\langle u_Ru_L \rangle}{M_s^2}
\label{neutrinomass}
\eeq
are obtained.\footnote{The presence of Dirac neutrino masses of this order of magnitude from this mechanism looks like a general property of low string scale 
models.} For $\langle u_Ru_L \rangle$ $\propto $ $(200\ MeV)^3$ and $M_s$ $\propto$ $1-10$ TeV one gets neutrino masses of order $0.1-10$ eV's, which is consistent with oscillation experiments. The smallness of neutrino masses would be thus related to the existence of a PQ-like symmetry ($U(1)_b$), which is broken by chiral symmetry breaking. Notice that the dimension six operators $\alpha '(L N_R)(Q_L U_R)^*$ may have different coefficients for different neutrino generations so there will be, in general, a non-trivial generation structure.

\item
{\it Double Higgs system}

In the case in which both type of Higgs fields $H_i$ and $h_i$ coexist, all quarks and leptons have in general Yukawa couplings from the start. The observed hierarchy of fermion masses would be a consequence of the different values of the Higgs fields and hierarchical values for Yukawa couplings. In particular, if the vev of the Higgs $h_i$ turn out to be small, the fermion mass structure would be quite analogous to the previous case. This could be the case if the Higgs parameters are such that the $h_i$ Higgsess were very massive.

\end{itemize}
  
To reproduce the observed fermion spectrum it is not enough with the different mass scales given by the Higgs vevs. Thus, for example, in the charged lepton sector all masses are proportional to the vev $\langle H_2 \rangle$ and the hierarchy of lepton masses has to arise from a hierarchy of Yukawa couplings. Indeed, as was pointed out in \cite{Aldazabal:2000cn}, in models with intersecting branes it is quite natural the appearance of hierarchical Yukawa couplings. We postpone the general study of Yukawa couplings in intersecting D-brane models to Chapter \ref{yukint}.

Concerning the gauge coupling constants, a dimensional reduction of the SYM theory living on the worldvolume of a stacks of D-branes $\a$ shows that its gauge coupling $g_\a$ is controlled by the volume that this brane wraps on the internal space. More specifically, in the case of a stack of $N_a$ D6-branes wrapped on a 3-cycle $\Pi_\a$ of an internal manifold $\M_6$, the gauge coupling constant of $SU(N_\a)$ is given by
\beq
{1 \over g_\a^2} = {M_s^3 \over (2\pi)^4 \lambda_{II}} 
{\rm Vol }\ (\Pi_\a),
\label{coupling}
\eeq
where $M_s = \a^{\prime \ -\frac 12}$ is the string scale and $\lam_{II}$ is type IIA coupling constant. In case we are computing the gauge coupling constant of the $U(1)_\a$ group also arising from this stack, whose generator will be taken to be Id$_{N_\a}$, we must multiply the r.h.s. of the above expression by $N_\a$.

In the particular case of factorisable cycles on $T^6$ the volume of $\Pi_a$ takes the simple form
\beq
{\rm Vol\ } (\Pi_\a)^2\ =\ 
((n_\a^{(1)}R_1^{(1)})^2+(m_\a^{(1)}R_2^{(1)})^2)
((n_\a^{(2)}R_1^{(2)})^2+(m_\a^{(2)}R_2^{(2)})^2)
((n_\a^{(2)}R_1^{(3)})^2+(m_\a^{(2)}R_2^{(3)})^2).
\label{length}
\eeq

Thus, in the case of the SM configurations described in the previous sections we have
\bea
{\alpha_{QCD}}^{-1} & = &
{{M_s^3}\over {(2\pi)^2 \lambda_{II}}} {\rm Vol\ } (\Pi_a), \\
{\alpha_{2}}^{-1} & = & 
{{M_s^3}\over {(2\pi)^2 \lambda_{II}}} {\rm Vol\ } (\Pi_\b), \\
\alpha_Y^{-1} & = & 
{{M_s^3}\over {(2\pi)^2 \lambda_{II}}}
\left( \frac{1}{36} {\rm Vol\ } (\Pi_a)+\frac{1}{4}{\rm Vol\ } (\Pi_c)
+\frac{1}{4}{\rm Vol\ } (\Pi_d) \right).
\label{coupsm}
\eea
where lengths are measured in string units. These are the tree level\footnote{Threshold corrections to these values have been analysed in \cite{Lust:2003ky}.}  values at the string scale. In order to compare with the low-energy data one has to consider the effect of the running of couplings in between the string scale $M_s$ and the weak scale. Notice that even if the string scale is not far away (e.g., if $M_s\propto 1-{\rm few} $ TeV) those loop corrections may be important if some of the massive states (gonions, windings or KK states) have masses in between the weak scale and the string scale. Thus in order to make a full comparison with experimental data one has to compute the spectra of those massive states (which depend on radii and intersection angles as well  as the wrapping numbers of the model considered).  As in the case of Yukawa couplings, a detailed analysis of each model is required in order to see if one can reproduce the experimental values. It seems however that there is sufficient freedom to accommodate the observed results for some classes of models.

\section{D5-brane models}

In the present section we will be interested in finding intersecting D5-branes models whose gauge group and matter content correspond to either the Standard Model (SM) or some Left-Right symmetric (LR) extension of it \cite{Mohapatra:1974hk,Senjanovic:1975rk,Mohapatra:1997hi,Mohapatra:1999vv}. Such low energy spectra must contain the following gauge group and fermionic content:
\beq
\begin{array}{ccc}
{\rm Standard\ Model} &  & {\rm Left}$-${\rm Right\ Model}
\vspace{0.1cm} \\ \vspace{0.1cm}
SU(3)_c \ti SU(2)_L \ti U(1)_Y 
& & SU(3)_c \ti SU(2)_L \ti SU(2)_R \ti U(1)_{B-L} \\
Q_L^i=(3,2)_{\frac 16} & \raw & Q_L^i=(3,2,1)_{1/3}\\ 
\left.
\begin{array}{c} 
U_R^i =({\bar 3},1)_{-\frac 23} \\
D_R^i=({\bar 3},1)_{\frac 13} 
\end{array} \right\}
& \raw & Q_R^i=(\bar 3,2,1)_{-1/3} \\
L^i=(1,2)_{-\frac 12} & \raw & L_L^i=(1,2,1)_{-1}  \\ 
\left.
\begin{array}{c} 
E_R^i=(1,1)_1 \\
N_R^i=(1,1)_0
\end{array} \right\}
& \raw & L_R^i=(1,1,2)_{1}
\end{array}
\label{content}
\eeq
where the index $i = 1,2,3$ labels the three different generations that have to be considered in each model. 

Again, we will follow the general philosophy described at the beginning of this chapter, and we will be considering a class of configurations where chiral fermions arise only in the bifundamental representations (\ref{bifundamentals}). Notice that, from the point of view of Left-Right unification, right-handed neutrinos must exist, as they complete the $SU(2)_R$ leptonic doublet that contains the charged right-handed leptons $E_R^i$. From the point of view of SM building, though, there is a priori no reason why we should consider having such representations in our fermionic content. However, as has been emphasized, when obtaining the chiral content of our theory just from fields transforming in bifundamental representations, such right-handed neutrinos naturally appear from anomaly cancellation conditions. Since in the present section we will construct our models from such `bifundamental' fermions, we will include these particles right from the start \footnote{For some intersecting branes SM constructions without right-handed neutrinos see for instance \cite{Blumenhagen:2001te,Cremades:2002cs}.}. Again, it can be shown that in this case where chiral fields transform in bifundamentals the simplest embedding of the SM or the LR extension will consist in a configuration of four stack of branes, as presented in table \ref{SMbranes2}.

\begin{table}[htb]
\renewcommand{\arraystretch}{2}
\begin{center}
\begin{tabular}{|c|c|c|c|}
\hline
Label & Multiplicity & Gauge Group & Name \\
\hline
\hline
stack $a$ & $N_a = 3$ & $SU(3) \times U(1)_a$ & Baryonic brane\\
\hline
stack $b$ & $N_b = 2$ & $SU(2)_L \times U(1)_b$ & Left brane\\
\hline
stack $c$ & $N_c = 
\left\{
\begin{array}{c}
2\\ 1
\end{array}
\right.$ & 
$\begin{array}{c}
SU(2)_R \ti U(1)_c \\ U(1)_c
\end{array}$ & Right brane\\
\hline
stack $d$ & $N_d = 1$ & $U(1)_d$ & Leptonic brane \\
\hline
\end{tabular}
\caption{Brane content yielding the SM or LR spectrum.
\label{SMbranes2}}
\end{center}
\end{table}

From the intersection of this D-brane content we must then reproduce the chiral spectra of (\ref{content}). In particular, we will try to reproduce the SM spectrum of table \ref{SMcontent}, whereas the LR spectrum we will look for will follow the pattern of table \ref{LRcontent}.

\begin{table}[htb] 
\renewcommand{\arraystretch}{1.5}
\begin{center}
\begin{tabular}{|c|c|c|c|c|c|c|c|}
\hline Intersection &
 Matter fields  &   &  $Q_a$  & $Q_b $ & $Q_c $ & $Q_d$  & $B-L$ \\
\hline\hline (ab) & $Q_L$ &  $(3,2,1)$ & 1  & -1 & 0 & 0 & 1/3 \\
\hline (ab*) & $q_L$   &  $2(3,2,1)$ &  1  & 1  & 0  & 0  & 1/3 \\
\hline (ac) & $Q_R$   &  $({\bar 3},1,2)$ &  -1  & 0  & 1  & 0 & -1/3 \\
\hline (ac*) & $q_R$   &  $2({\bar 3},1,2)$ &  -1  & 0  & -1  & 0 & -1/3 \\
\hline (bd) & $L_L$    &  $3(1,2,1)$ &  0  & -1  & 0  & 1 & -1  \\
\hline (cd) & $L_R$   &  $3(1,1,2)$ &  0  & 0  & 1  & -1  & 1  \\
\hline \end{tabular}
\end{center} \caption{Left-Right symmetric chiral spectrum and $U(1)$ charges. The $U(1)_{B-L}$ generator is defined as $Q_{B-L} = \frac 13 Q_a - Q_d$.}
\label{LRcontent}
\end{table}

\begin{figure}[ht]
\centering
\epsfxsize=2.2in
\hspace*{0in}\vspace*{.2in}
\epsffile{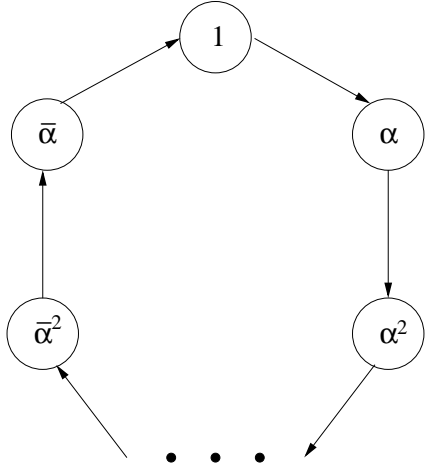}
\caption{Quiver diagram of a $\inte_N$ orbifold singularity. The nodes of such diagram represent the phases associated to each different gauge group in the theory, whereas each arrow represents a chiral fermion transforming in a bifundamental of the two groups it links.}
\label{quiverZN}
\end{figure}

In order to realise such spectra as the chiral content of a concrete configuration of D5-branes we must again impose some topological constraints on  our models. Unlike the case of D6-branes, where all the spectrum information is encoded on the intersection numbers, we must also consider the orbifold structure of our configuration\footnote{It is possible, however, to define a `generalized intersection number' {\bf I}$_{ab}$ which contains both informations, see Appendix \ref{qbasis}}. Such structure can be easily encoded in a quiver diagram as shown in figure \ref{quiverZN}. 

We have seen that a D-brane configuration living on an orbifold singularity can be locally described by quotienting the theory by a discrete group $\Gamma$, which is acting on both an ambient space $\cpx^n$ and on the CP degrees of freedom. To each $\Gamma$ action we can associate a quiver diagram \cite{Douglas:1996sw,Johnson:1996py,Douglas:1997de,Uranga:2000ck}. Each node of such diagram will represent an irreducible representation (irreps) of $\Gamma$, whereas the arrows connecting the nodes represent invariant fields under combined geometric and gauge actions. In general, the $\G$ action $\g_g$ on the Chan-Paton degrees of freedom can be written as a direct sum of such irreps, and the gauge groups that will arise from it will correspond to a product of unitary groups, each one
associated to a definite irreps. In our specific setup $n = 1$ and $\Gamma = \inte_N$, so each irreps of $\G$ is one-dimensional and can be associated to a $N^{th}$-root of unity. Indeed, any $\inte_N$ generator action on the Chan-Paton degrees of freedom can be written on the form (\ref{gamma}), where several such phases are involved. Without loss of generality, we will consider that each brane $a, b, c, d$ has a $\g_\om$ matrix proportional to the identity, that is $\g_{\om,i} = \a^n {\bf 1}_{N_i}$, so that it will give rise to just one unitary gauge group $U(N_i)$. We will represent this by locating that brane $i$ on the node corresponding to the irreps $\a^n$. Notice that, in an orientifold theory, the mirror brane $i^*$ will then be placed in the node $\bar \a^n$.

Chiral fields can also be easily identified in this diagram by arrows connecting the nodes. These arrows will always link two different nodes, so that if there is some brane content in both of them we will find a fermion transforming under the corresponding gauge groups. The sense of the arrow will denote the chirality that such representation has. In our conventions the positive sense represents left fermions. This arrow structure can be easily read from the chiral spectrum in (\ref{spectrum5ab}), giving rise to the cyclic quiver depicted in figure \ref{quiverZN}. Notice that this simple spectrum comes from a plain orbifold singularity. In this case every chiral field will transform in bifundamental representations of two gauge groups with contiguous phases. When considering orientifold singularities, however, we should also include the mirror branes on the picture, and more `exotic' representations may arise.

There are, in principle, many different ways of obtaining the desired chiral spectrum (\ref{content}) from the brane content of table \ref{SMbranes2}. Furthermore, the details of the construction will depend on the specific model (SM or LR) and on the $\inte_N$ quiver under consideration. There are, however, some general features of the construction that can be already addressed at this level.

\begin{itemize}

\item 
In both SM and LR models, chiral fermions must arise in a very definite pattern. Namely, we need left and right-handed quarks, so we must consider matter arising from the intersections of the {\it baryonic} brane with both the {\it left} and {\it right} branes. We must avoid, however, lepto-quarks which may arise from some intersection with the {\it leptonic} brane. The same considerations must be applied to the latter. This pattern can be easily achieved in D5-branes configurations by placing both $b$ and $c$ (or $c^*$) branes on the same node of the $\inte_N$ quiver, while $a$ and $d$ in some contiguous node. Since, in order to achieve the spectra of tables \ref{SMcontent} and \ref{LRcontent}, we must consider non-trivial $ab$, $ab^*$, $ac$ and $ac^*$ sectors, we must place the stack $a$ either in the phase $1$ or in the phase $\a$, while stacks $b$, $c$ must be in the other one. This restricts our search to essentially two different distributions of branes, which are shown in figure \ref{2quivers}.

\begin{figure}[ht]
\centering
\epsfxsize=6in
\hspace*{0in}\vspace*{.2in}
\epsffile{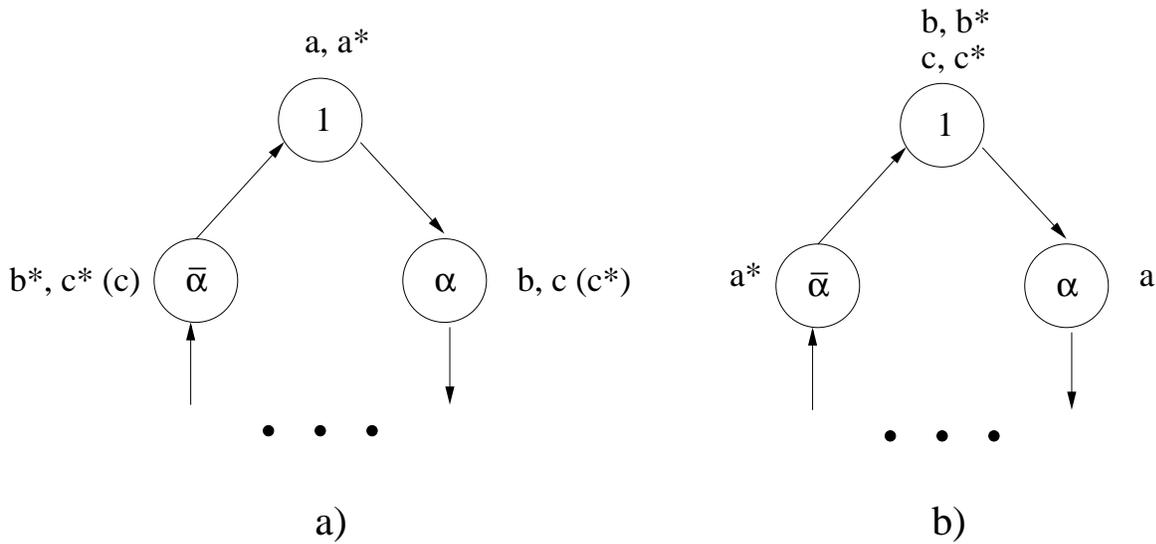}
\caption{Two possible embeddings of the brane content
of a SM or LR configuration.}
\label{2quivers}
\end{figure}

\item 
Given these two possibilities, it is now easy to guess which intersection numbers must we impose in order to achieve the desired spectra. Indeed, the modulus of an intersection number, say $I_{ab}$, will give us the multiplicity of this sector. This implies that, in order to have the desired number of left-handed quarks, we must impose $|I_{ab}| = 1$, $|I_{ab^*}| = 2$ as can be read directly from tables \ref{SMcontent} and \ref{LRcontent}. \footnote{We could have alternatively imposed $|I_{ab}| = 2$, $|I_{ab^*}| = 1$, giving an equivalent spectrum.}. On the other hand, we will have to choose the sign of these intersection numbers in order to properly fix the chirality of our fermions. These signs will be different for each distribution of branes considered in figure \ref{2quivers}, since chirality also depends on the arrow structure of the quiver diagram. For instance, we should impose $I_{ab} = 1$, $I_{ab*} = -2$ in the $a$)-type of quiver in this figure, while $I_{ab} = -1$, $I_{ab*} = -2$ in the $b$)-type. Similar reasoning  applies  to  other intersections involving branes $b$ and $c$.

\item 
Finally, we are interested in getting all of our chiral matter from bifundamental representations. Thus, we must avoid the appearance of fermions transforming as Symmetrics and Antisymmetrics that might appear from the general spectrum (\ref{spectrum5ab*}). This will specially arise in $\inte_3$ models, where we will have to impose $I_{ii^*} = 0$ for those branes in the $\a$ node. 

\end{itemize}

\subsection{D5 Standard Models \label{D5SM}}

\begin{figure}[ht]
\centering
\epsfxsize=4.5in
\hspace*{0in}\vspace*{.2in}
\epsffile{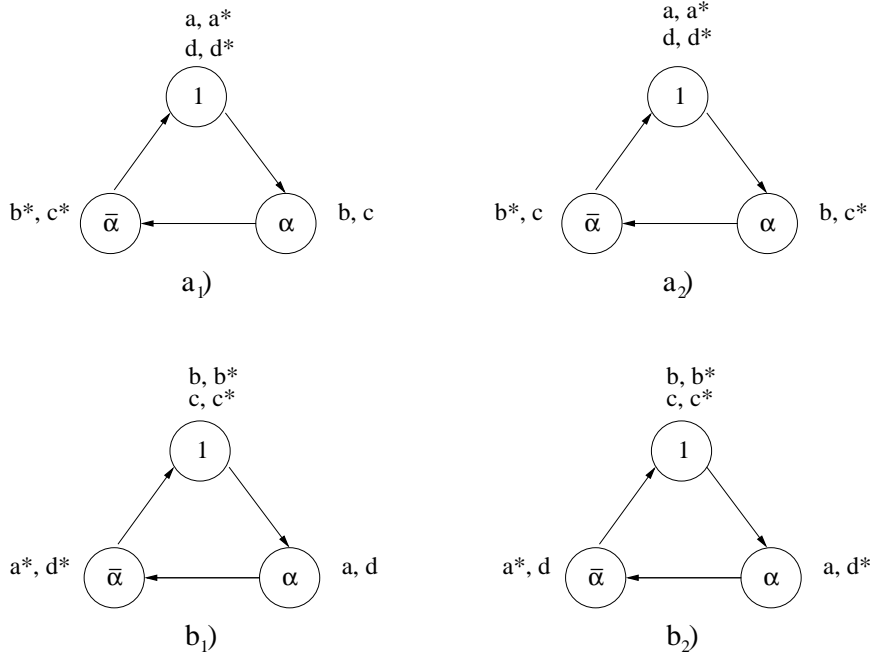}
\caption{Four possible embeddings of the brane content of a SM configuration in a $\inte_3$ quiver.}
\label{SMfig}
\end{figure}

Let us give an example that shows how the SM structure can be implemented on D5-branes configurations. The simplest choice for such example is the $\inte_3$ singularity, which is the smallest $\inte_N$ quiver that provides non vector-like spectra. Imposing the chirality pattern discussed above give us four different ways of embedding the SM spectrum, each of them depicted in figure \ref{SMfig}. In order to achieve a SM configuration, we must impose the intersection numbers that will give us the desired matter content. As discussed above, these will depend on the particular $\inte_3$ quiver considered. Let us first consider the quiver $a_1$). In table \ref{SMatab} we show the general class of solutions for the wrapping numbers that will provide us with such fermionic spectrum.

\begin{table}[htb] 
\renewcommand{\arraystretch}{2.5}
\begin{center}

\begin{tabular}{|c||c|c|c|}
\hline
 $N_i$  &  $(n_i^{(1)},m_i^{(1)})$  &  $(n_i^{(2)},m_i^{(2)})$ 
& $\gamma_{\om,i}$ \\
\hline\hline 
$N_a = 3$ & $(n_a^{(1)}, \eps \b^1)$  &  $(3,- \oh \eps \tilde \eps)$ & 
$1_3$  \\
\hline 
$N_b = 2$ & $(1/\b^1, 0)$ &
$(\tilde \eps, -\oh \eps)$ &  $\a 1_2$   \\
\hline 
$N_c = 1$ & $(1/\b^1, 0)$ & $(0, \eps)$ & $\a$ \\
\hline 
$N_d=1$ & $(n_d^{(1)}, 3\eps \b^1)$ & $(1, \oh\eps\tilde\eps) $ & $1$ \\
\hline
$N_h$ & $(\eps_h/\b^1, 0)$ & $(2,0)$ & $1_{N_h}$  \\
\hline
\end{tabular}

\caption{D5-branes wrapping numbers and CP phases giving rise to a SM spectrum the $\inte_3$ quiver of fig \ref{SMfig}.$a_1$). The solution is parametrized by $n_a^{(1)}, n_d^{(1)} \in \inte$, $\eps, \tilde\eps = \pm 1$ and $\b^1 = 1 - b^{(1)} = 1, 1/2$. Notice that the second torus has to be tilted, hence $\b^2 = 1/2$.
\label{SMatab}}
\end{center}
\end{table}

Notice that for the sake of generality we have added a new stack of $N_h$ branes to our initial configuration, yielding an extra $U(N_h)$ gauge group. However, the wrapping numbers and the CP phase of such brane have been chosen in such a way that no extra chiral matter arises from its presence. Since no chiral fermion is charged under the gauge group of this brane, the stack $h$ is a sort of hidden sector of the theory. This is strictly true, however, only from the fermion content point of view, and generically some scalars with both SM and $U(N_h)$ quantum numbers may  appear.

Having achieved the fermionic spectrum of table \ref{SMcontent}, our low energy field theory will be automatically free of cubic chiral anomalies. In order to have a consistent compactification, however, we must impose the stronger tadpole cancellation conditions. Interestingly enough, most of the conditions in (\ref{tadpoleO5b}) turn out to be trivially satisfied by this brane content, the only non-trivial one being the first condition, that now reads
\beq
9n_a^{(1)} + n_d^{(1)} - \frac{\tilde\eps}{\b^1} + 2N_h \frac{\eps_h}{\b^1}
= -8.
\label{tadSMZ3a}
\eeq

Let us now analyse the $U(1)$ structure of such model. As described in the previous chapter, couplings of gauge bosons to twisted RR fields will give rise to GS counterterms that will cancel the residual $U(1)$ anomalies. We are particularly interested in couplings (\ref{dualcouplingsO5}), that tell us which gauge bosons are becoming massive by this mechanism. In the $\inte_3$ orientifold case there is only one independent twisted sector, so only four couplings are relevant. By considering the brane content above we find that these couplings are
\beq
\begin{array}{rcl}
B_2^{(1)} & \wedge  & \ c_1\ (\a -\a^2) {{2\tilde\eps}\over {\beta^1}}
F_{b} \\
D_2^{(1)} & \wedge  & \ c_1 
\left( \eps \tilde\eps (-3n_a^{(1)} F_a\ +\ n_d^{(1)} F_d)\ +\
\frac{\eps}{\b^1}(F_b\ -\ F_c) \right) \\
E_2^{(1)} & \wedge  & \ c_1\ 6\eps\b^1 (3F_a\ +\  F_{d})
\end{array}
\label{bfsSMZ3a}
\eeq
the coupling to the $C_2^{(1)}$ field being trivially null. In general, such couplings will give mass to three linearly independent combinations of $U(1)$'s, leaving just one $U(1)$ as a true Abelian gauge symmetry of the spectrum. Among these massive $U(1)$'s, two are model-independent, and correspond to the `anomalous' combinations $U(1)_b$ and $3U(1)_a+U(1)_d$ characteristic of this fermionic spectrum. The third one, however, will depend on the specific model considered. Indeed, we find that the generator of the massless $U(1)$ is given by
\beq
Q_0 = Q_a - 3 Q_d - 3\tilde\eps \b^1 (n_a^{(1)} + n_d^{(1)}) Q_c,
\label{masslessSMZ3a}
\eeq
so if we further impose to our class of models the condition
\beq
\tilde\eps \b^1 \left(n_a^{(1)} + n_d^{(1)}\right) = 1,
\label{condSMZ3a}
\eeq
then we find that this massless Abelian gauge group precisely corresponds to the hypercharge, which in these models is given by $U(1)_Y = \frac 16 U(1)_a - \frac 12 U(1)_c - \frac 12 U(1)_d$.

Notice that the whole of this construction is quite analogous to the one described in the previous section. Indeed, we have imposed the same chiral spectrum, again arising from bifundamental representation of four stack of branes. After imposing some conditions regarding tadpoles and the Abelian gauge structure, we are finally led to a compactification yielding just the gauge and fermionic spectrum of the Standard Model (and possibly  some hidden sector of the theory given by stack $h$).

\subsubsection{Scalars and tachyons in the spectrum}

As explained in Section 2.1, at the intersection of pairs of D5-branes with  the {\it same} CP phase there may appear scalar tachyons with masses given in (\ref{sector5ab}). Since stacks $b,c$ and their mirrors are parallel along the first 2-torus, they generically do not intersect. On the other hand there may be tachyons at the intersections $(aa^*),(dd^*)$, $(ad),(ad^*)$ plus possibly others involving the hidden branes $h$. One can get rid of many of these tachyons by appropriately choosing some discrete parameters and the compactification radii. Consider for instance the following choice of parameters:
\beq
n_a^{(1)} = n_d^{(1)}= -1,\quad N_h=0,\quad  \tilde \epsilon = -1, \quad 
\b^1 = \oh.
\label{choice}
\eeq
With this choice it is easy to check that the tadpole cancellation conditions (\ref{tadSMZ3a}) are verified and the standard hypercharge is the only $U(1)$ remaining at the massless level. Furthermore, the $h$ brane is not needed in order to cancel tadpoles, this hidden sector thus being absent. Now, the angles formed by the branes $d,a$ with the orientifold plane on the two tori are given by
\bea
\theta_a^1 = \eps\ \left(\pi - tg^{-1}\left(\frac{U^1}{2}\right) \right) \ 
& ; & 
\theta_a^2 = \eps\  tg^{-1}\left(\frac{U^2}{6}\right) 
\nonumber \\
\theta_d^1 = \eps\ \left(\pi - tg^{-1}\left(\frac{3U^1}{2}\right) \right) \ 
& ; & 
\theta_d^2 = - \eps\  tg^{-1}\left(\frac{U^2}{2}\right)
\label{angulillos}
\eea
respectively. Here $U^i = R_2^{(i)}/R_1^{(i)}$, $i=1, 2$. Now, the angles formed by such branes with their mirrors is given by $\vt_{ad}^i \equiv - 2 \th_{ad}^i$ mod $2\pi$, so for $U^1 = U^2/3$ one gets $\arr \vt_{ad}^1 \arr = \arr \vt_{ad}^2 \arr$, and according to (\ref{sector5ab}) the scalars in $(aa^*)$ and $(dd^*)$ cease to be tachyonic and become massless \footnote{Actually, according to (\ref{spectrum5ab*}), scalars in the sector $(dd^*)$ transform in the antisymmetric representation of $U(N_d) =  U(1)$, thus being absent for any choice of angles.}. The only tachyonic scalars in the spectrum persist in the $ad$ and $ad^*$ intersections which have mass$^2$:
\beq
m^2_{ad}\ =\ m^2_{ad^*}\ = -\ \frac 1\pi 
tg^{-1}\left({{U^2}\over 6}\right)M_s^2 \ .
\label{tachyoncognazo}
\eeq
In table \ref{escalares} we present the lightest scalar spectrum arising from branes $a$, $d$ and their mirrors when the particular choice (\ref{choice}) is made.
\begin{table}[htb]
\renewcommand{\arraystretch}{1.3}
\begin{center}
\begin{tabular}{|c|c|c|}
\hline
 Sector   &  Representation  & $\a'$ mass$^2$ \\
\hline\hline ($aa^*$) & 
$\begin{array}{c}
4\b^1\ (3,1)_{1/3} \\
2\b^1\ (6,1)_{1/3} \\
\end{array}$
  & 0  \\
\hline ($ad$) & $4\b^1$ $(3,1)_{2/3}$ & 
$\pm \frac 1\pi tg^{-1}\left({{U^2}\over 6}\right)$   \\  
\hline ($ad^*$) & $4\b^1$ $(3,1)_{-1/3}$ & 
$\pm \frac 1\pi tg^{-1}\left({{U^2}\over 6}\right)$   \\  
\hline \end{tabular}
\end{center} 
\caption{Lighter scalar excitations arising from the brane content with phase $1$ in table \ref{SMatab}, under the choice of parameters (\ref{choice}).
\label{escalares}}
\end{table}

Note however that all the above scalar masses are tree level results and that, since the models are non-SUSY, there are in general important one-loop contributions to the scalar masses. Those will be particularly important for the coloured objects like the scalars in $(ad)$, $(ad^*)$ sectors which are color triplets. Those one-loop corrections may be estimated from the effective field theory (one gauge boson exchange) and yield \cite{Aldazabal:2000cn}
\beq
\Delta m^2(\mu ) \ =\ \sum_a { {4C_F^a \alpha_a(M_s) }\over {4\pi }} M_s^2
f_a \log(M_s/\mu) \ +\ \Delta M^2_{KK/W}
\label{massloop}
\eeq
where the sum on $a$ runs over the different gauge interactions under which the scalar transforms and $C_F^a$ is the eigenvalue of the quadratic Casimir in the fundamental representation. Here $\Delta M^2_{KK/W}$ denotes further contributions which may appear from the Kaluza-Klein, winding and string  excitations if they are substantially lighter than the string scale $M_s$. The function $f_a$ is given by
\beq
f_a \ =\  { {2+b_a{{\alpha_a(M_s)}\over {4\pi }} t}\over
{1+b_a{{\alpha_a(M_s)}\over {4\pi }}t } }
\eeq
where $t=2\log(M_s/\mu)$ and $b_a$ are the coefficients of the one-loop $\beta $-functions. These corrections are positive and may easily overcome  the tree level result if $U^2$ is not too large. This is analogous to the one-loop contribution to squark masses in the MSSM in which for large gaugino masses the one-loop contribution clearly dominates over the tree-level soft masses (see, e.g., ref.\cite{Martin:1997ns,Ibanez:1992rk} and references therein). Thus in this class of models, apart from the fermion spectrum of the SM, one expects the presence of some extra relatively light (of order the electroweak scale) coloured scalars.

\subsubsection{Electroweak symmetry breaking}

The Higgs sector in this class of theories is relatively similar to the one in the D6-brane standard models. Consider in particular the SM configuration described in the previous subsections. Here, the only light scalar with the quantum numbers of a Higgs boson lives in the $bc$ sector. Branes $b$ and $c$ are parallel in the first torus, but if the distance  $Z_{bc}$  between the branes in that torus is set to zero the branes intersect at an angle 
\beq
\pi \vt_{bc}^2 = \eps\tilde \eps \left({{\pi}\over 2}
+ tg^{-1}\left(\frac{U^2}{2}\right)\right)\ ,
\eeq
and at those intersections complex scalar doublets appear with masses
\beq
m^2_{H^{\pm}} \  =\ {{Z_{bc}^2}\over {4\pi } } M_s^2 \  \pm \
{{M_s^2}\over 2} |\vt_{bc}^2|.
\label{masshiggs}
\eeq

There are in fact two scalar doublets with quantum numbers as in table \ref{higgsses2},
\begin{table}[htb] 
\renewcommand{\arraystretch}{1.3}
\begin{center}
\begin{tabular}{|c|c|c|c|}
\hline
 Higgs   &  $Q_b$  &  $Q_c$   & Y \\
\hline\hline $H_1$ & 1  &  -1 & 1/2  \\
\hline $H_2$ & -1  & 1 &  -1/2   \\
\hline \end{tabular}
\end{center} \caption{Electroweak Higgs fields \label{higgsses2}}
\end{table}
and defined as
\beq
H^{\pm}={1\over2}(H_1^*\pm H_2).  
\eeq

The intersection number of these branes in the second torus is equal to $\pm 1$ so that only one copy of this Higgs system appears. Thus in the present model we have the same minimal Higgs sector as in the MSSM. As may be seen from eq.(\ref{masshiggs}) as the distance $Z_{bc}$ decreases the Higgs doublets become tachyonic, giving rise to EW symmetry breaking. This is quite similar to the process of EW symmetry breaking in the D6-brane models of last section (see also \cite{Cremades:2002cs}), in which it may be described as brane recombination of a $b$ brane and a $c$ brane into a single recombined brane $e$. Note that, although one-loop positive corrections as given in eq.(\ref{massloop}) will in general be present also for the Higgs fields, one also expects large negative contributions from the usual one-loop top-quark contribution which will again favour EW symmetry breaking \cite{Ibanez:fr}. 
  
To sum up, the brane content of table \ref{SMatab} give us an example of how an SM construction can be achieved by means of intersecting D5-branes. This particular class of models shares many features already present in the D6-branes models as the discussion at the beginning of this chapter already suggests, whereas some important novelties do also appear. Notice that in this section we have restricted ourselves to only one possible quiver configuration of figure \ref{SMfig}. Some other inequivalent constructions can also be performed from the rest of the quivers in that figure, their discussion being postponed to subsection \ref{extra}.

\subsection{D5 Left-Right Symmetric Models}

Quite analogously, the LR structure can also be implemented in a D5-brane construction. To show this, let us again consider a $\inte_3$ orbifold. Since the chirality pattern is the same for both SM and LR configurations, the possible brane distributions will again be those of figure \ref{SMfig}. Let us consider now the quiver $a_2$). The brane content with LR spectrum for such quiver is shown in table \ref{LRa2tab}.

\begin{table}[htb]
\renewcommand{\arraystretch}{2.5}
\begin{center}

\begin{tabular}{|c||c|c|c|}
\hline
$N_i$ & $(n_i^{(1)},m_i^{(1)})$ & $(n_i^{(2)},m_i^{(2)})$ & $\gamma_{w,i}$ \\
\hline\hline
$N_a=3$ & $(n_a^{(1)}, \eps\b^1)$ & $(1/\rho, -\oh\eps\tilde\eps)$ & $1_3$ \\
\hline
$N_b=2$ & $ (1/\b^1, 0)$ & $(\tilde\eps, -\frac{3\rho}{2}\eps)$ & $\a 1_2$ \\
\hline
$N_c=2$ & $ (1/\b^1, 0)$ & $(\tilde\eps, -\frac{3\rho}{2}\eps)$ & $\a^2 1_2$ \\
\hline
$N_d=1$ & $(n_d^{(1)}, \eps\b^1/\rho )$ & $(1, \frac{3\rho}{2}\eps\tilde\eps)$ 
& $1$ \\
\hline
$N_h$ & $(\eps_h/\b^1,0)$  &  $(2,0)$ & $1_{N_h}$  \\
\hline
\end{tabular}
\caption{D5-branes wrapping numbers and CP phases yielding a LR spectrum in the $\inte_3$ orbifold of figure \ref{SMfig}.$a_2$). Solutions are parametrized by $n_a^{(1)}, n_d^{(1)} \in \inte$, $\eps, \tilde\eps = \pm 1$, $\beta^1 = 1-b^{(1)} = 1, 1/2$ and $\rho=1,1/3$.
\label{LRa2tab}}
\end{center}
\end{table}

Notice that branes $b$ and $c$ belong in fact to the same stack of four branes, with a non-trivial CP action on it. From the point of view of gauge fields, however, each one is a separate sector. Tadpole cancellation conditions are, as usual, almost satisfied when imposing this wrapping numbers. The only non-trivial conditions that remains is
\beq
\frac{3n_a^{(1)}}{\rho} \ -\ \frac{2 \tilde\eps}{\b^1} \ +\ n_d^{(1)}\ 
+\ 2 N_h \frac{\eps_h}{\beta^1} \ = \ -8 .
\label{tadLRZ3}
\eeq

On the other hand, we must also compute the couplings to RR twisted fields, which in this case are
\beq
\begin{array}{rcl}
B_2^{(1)} & \wedge  & \ c_1 (\a -\a^2) {{2\tilde\eps}\over {\b^1} }
(F_{b} - F_{c}) \\
D_2^{(1)} & \wedge  & \ c_1 \left( {{3\rho}\eps \over {\b^1}}\ 
(F_{b} \ +\ F_c) 
-\ 3\eps\tilde\eps (n_a^{(1)} F_a\ - \ \rho  n_d^{(1)} F_d) \right) \
\nonumber \\
E_2^{(1)} & \wedge  &  \ c_1 \frac{2\eps\b^1}{\rho}(3F_a\ +\  F_{d})
\end{array}
\label{bfsLRZ3}
\eeq

This $B \wedge F$ couplings will again give mass to three of the four $U(1)$ gauge bosons initially present in our spectrum. If we impose the condition
\beq
n_a^{(1)} \ =\ - 3 \rho n_d^{(1)},
\label{condLRZ3}
\eeq
then the only generator with null coupling to these fields is $Q_0 = Q_a - 3Q_d$, which corresponds to $U(1)_{B-L}$. After imposing this condition, tadpoles (\ref{tadLRZ3}) become
\beq
4 n_d^{(1)}  \ + \ \frac{1}{\beta^1} \left(\tilde\eps - N_h \eps_h \right) \ = \ 4,
\label{tadlrZ3b}
\eeq
so the extra brane $h$ will be generically necessary in order to satisfy tadpoles. 

For completeness, let us give an explicit solution of (\ref{tadlrZ3b}). Consider the following choice of parameters:
\bea
n_d^{(1)} = N_h = 1 & \stackrel{(\ref{condLRZ3})}\Longrightarrow 
& n_a^{(1)} = -3\rho \nonumber \\
& \eps_h = \tilde \eps,
\label{choice2}
\eea
which now give us a non-trivial $h$ sector with gauge group $U(1)$. Following the same considerations as in the previous SM construction, we see that the angles the branes $a$, $d$ and $h$ form with the orientifold plane are
\beq
\begin{array}{cc}
\theta_a^1 = \eps\ \left(\pi - tg^{-1}
\left(\frac{\b^1}{3\rho}U^1\right) \right) & 
\theta_a^2 = - \eps\tilde\eps\  tg^{-1}\left(\frac{\rho}{2}U^2\right) \\
\theta_d^1 = \eps\ tg^{-1}\left(\frac{\b^1}{\rho}U^1\right) &
\theta_d^2 = \eps\tilde\eps\  tg^{-1}\left(\frac{3\rho}{2}U^2\right) \\ 
\th_h^1 = \frac{\pi}{2} (1 - \tilde\eps) & \th_h^2 = 0
\label{angulillos2}
\end{array}
\eeq
where again $U^i = R_2^{(i)}/R_1^{(i)}$, $i=1,2$. Under the choice $U^1 = \frac{3\rho^2}{2\b^1} U^2$, some of the potential tachyons in these sectors will become massless, as for instance those arising from ($aa^*$) intersections. However, just as in the previously discussed SM construction some tachyons will remain at ($ad$), ($ad^*$) intersections, and some other new tachyons involving the brane $h$. Again, as in the previous SM case, one-loop contributions to the scalar masses may easily overcome the tachyonic contribution.

One can also find an interesting family of left-right symmetric models with no open string tachyons already at the tree-level. Indeed, it is quite easy to generalize the Left-Right symmetric spectrum for a $\inte_N$ orbifold with odd $N > 3$. As an example, let us take the brane content of table \ref{wnumbersZN}, which corresponds to a particular case of figure \ref{2quivers}.$a$), and that will again give us the spectrum of table \ref{LRcontent}.
\begin{table}[h]
\renewcommand{\arraystretch}{2.5}
\begin{center}

\begin{tabular}{|c||c|c|c|}
\hline
 $N_i$  &  $(n_i^{(1)},m_i^{(1)})$  &  $(n_i^{(2)},m_i^{(2)})$  & $\gamma_{w,i}$ \\
\hline\hline
$N_a=3$ & $(1/\b^1,0)$ & $(\eps,\oh\tilde\eps)$ & $1_3$ \\
\hline
$N_b=2$ & $(n_b^{(1)},-\tilde\eps\b^1)$ & $ (3, \oh\eps\tilde\eps)$ & $\a 1_2$ \\
\hline
$N_c=2$ & $(n_c^{(1)},\tilde\eps\b^1)$ & $ (3, \oh\eps\tilde\eps)$ & $\a 1_2$ \\
\hline
$N_d=1$ & $(1/\b^1,0)$ & $(-3\eps, \oh\tilde\eps)$ & $1$ \\
\hline
\end{tabular}
\caption{D5-branes wrapping numbers and CP phases yielding a LR spectrum in a $\inte_N$. Solutions are parametrized by $n_b^{(1)}, n_c^{(1)} \in \inte$, $\eps, \tilde\eps = \pm 1$ and $\beta^1=1-b^{(1)} = 1, 1/2$.
\label{wnumbersZN}}
\end{center}
\end{table}
As in our previous LR example, tadpoles will be canceled by means of a hidden-brane sector, which in this $\inte_N$ case will consist of a brane system as shown in table \ref{hiddenZN}.
\begin{table}[htb]
\renewcommand{\arraystretch}{2}
\begin{center}

\begin{tabular}{|c||c|c|c|}
\hline
$N_i$ & $(n_i^{(1)},m_i^{(1)})$ & $(n_i^{(2)},m_i^{(2)})$ & $\gamma_{w,i}$ \\
\hline\hline
$N_{h1}$ & $(\eps_{h1}/\b^1,0)$ & $(n_h^{(2)},m_h^{(2)})$ & $\a$  \\
\hline
$N_{h2}$ & $(\eps_{h2}/\b^1, 0)$ & $(2,0)$ & $\a^3$  \\
\hline
$\vdots$ & $\vdots$ & $\vdots$  & $\vdots$  \\
\hline
$N_{hs}$ & $(\eps_{hs}/\b^1, 0)$ & $(2,0)$ & $\a^{2s-1}$  \\
\hline
\end{tabular}

\caption{Hidden brane system in a ${\bf Z_N}$ orbifold singularity.
\label{hiddenZN}}
\end{center}
\end{table}
There one has $\eps_{hi} = \pm 1$ and the value of $s$ is fixed by tadpole conditions. Consistency conditions in (\ref{tadpoleO5b}) are now easily satisfied. Indeed, second and fourth conditions are already satisfied with this brane content, while the third amounts to imposing
\beq
\eps \tilde\eps \left(n_b^{(1)} + n_c^{(1)}\right) + \eps_{h1} N_{h1} {m_h^{(2)} \over \beta^1} = 0.
\label{tadLRZNa}
\eeq
As shown in the previous chapter, the first of these conditions can be expressed as (\ref{generatortwist}), from where we can read that we must also impose
\bea
 6 \left(n_b^{(1)} + n_c^{(1)}\right) + \eps_{h1} N_{h1} {n_h^{(2)} \over \beta^1} = \eta 16 
\label{tadLRZNb} \\
\eps_{hi} N_{hi} {2 \over \beta^1} = \eta 16, \  \ {\rm (i=2,\cdots,r)},
\label{tadLRZNc}
\eea
where $r$ and $\eta$ have been defined in (\ref{eta}). Last condition actually implies $s = r$, $\eps_{hi} = \eta$ and $N_{hi} = 8\b^1$, for $i > 1$. Let us also compute the couplings to RR 2-form twisted fields which will render some of these $U(1)$ gauge bosons massive. Even if there are in principle 2($N-1$) such fields, most of their couplings are redundant, so we will still have some massless $U(1)$'s in our gauge group. Indeed, these couplings are
\beq
\begin{array}{rcl}
B_2^{(k)} & \wedge & c_k \left( (\alpha^k - \bar\alpha^k)  
\left[ 6(n_b^{(1)} F_b + n_c^{(1)} F_c) + 
\eps_{h1} N_{h1} {n_h^{(2)} \over \beta^1} F_{h1} \right]
+ \sum_{i=2}^r (\alpha^{ik} - \bar\alpha^{ik}) \eta 16 F_{hi} \right) \\
C_2^{(k)} &\wedge & c_k (\alpha^k - \bar\alpha^k) \eps\beta^1 (- F_b + F_c) 
\\
D_2^{(k)} &\wedge & c_k 
\left( {2\tilde\eps \over \beta^1} (3F_a + F_d) + (\alpha^k + \bar\alpha^k) 
\left[\eps\tilde\eps (n_b^{(1)} F_b + n_c^{(1)} F_c) 
+ \eps_{h1} N_{h1} {m_h^{(2)} \over \beta^1} F_{h1} \right] \right)
\\
E_2^{(k)} & \wedge & \ c_k (\alpha^k + \bar\alpha^k) 6\b^1 
\tilde\eps (- F_b + F_c) 
\end{array}
\label{bfslrZN}
\eeq

Imposing tadpole conditions (\ref{tadLRZNa}), (\ref{tadLRZNb}) and (\ref{tadLRZNc}) is easy to see that the only linear combination of abelian groups that does not couple to any RR field is just $U(1)_{B-L} = \frac 13 U(1)_a - U(1)_d$, providing us with another example of Left-Right symmetric model. This family of configurations yielding the same spectrum for arbitrary odd-ordered $\inte_N$ orientifold seems quite interesting, since it gives us a family of $\inte_N$ models with $N$ arbitrarily large. In addition they may have an open-string tachyonless spectrum. For instance, by the choice of discrete parameters
\beq
\begin{array}{ll}
n_b^{(1)} = n_c^{(1)} =\eta, & N_{h1} = 4 \b^1, \\ 
\eps_{h1} = \eta, & (n_h^{(2)}, m_h^{(2)}) = (1, -\oh\eps\tilde\eps),
\end{array}
\label{solZN}
\eeq
conditions (\ref{tadLRZNa}), (\ref{tadLRZNb}) and (\ref{tadLRZNc}) are satisfied, and the compactification radii can also be chosen to avoid any tachyonic excitation. Indeed, our potential tachyons will arise only from ($bh1$) and ($ch1$) intersections whose characteristic angles are
\beq
\pi |\vt_{bh1}^1| = \pi |\vt_{ch1}^1| =
tg^{-1}\left(\b^1 U^1 \right)\ ;\
\pi |\vt_{bh1}^2| = \pi |\vt_{ch1}^2| =
tg^{-1}\left(\frac{U^2}{6}\right) 
+ tg^{-1}\left(\frac{U^2}{2}\right),
\eeq
so by appropriately choosing the complex structure moduli we can achieve $|\vt_{bh1}^1| = |\vt_{bh1}^2|$ and $|\vt_{ch1}^1| = |\vt_{ch1}^2|$, finding a one-parameter family of tachyonless open-string spectra.

Let us end this subsection by recalling an apparent phenomenological shortcoming of the class of left-right symmetric models built here. Eventually we would like to break the gauge symmetry down to the Standard Model one and, in order to do that, we need to give a vev to a right-handed doublet of scalars with non-vanishing lepton number. No such scalars are present in the lightest spectrum of the particular models constructed here. It would be interesting to find other examples in which correct gauge symmetry breaking is feasible.

\subsection{Some extra D5-brane SM models \label{extra}}

Althought in subsection \ref{D5SM} we have focussed on a very particular class of D5-branes configurations in a $\inte_3$ orbifold, there are other possibilities when constructing models giving rise to just the SM fermionic spectrum. Indeed, the brane content of table \ref{SMatab} corresponds to the brane distribution of figure \ref{SMfig}.$a_1$), while in principle any of these four figures is valid. For completeness, in this subsection we consider the other three possibilities.

After imposing the analogous constraints to the rest of the $\inte_3$ quivers of figure \ref{SMfig}, we find that the distribution $a_2$) give us a totally equivalent class of models to the one already presented, whereas $b_1$) and $b_2$) give us two new different families of configurations. Let us first consider  the $\inte_3$ quiver in fig. \ref{SMfig}.$b_1$). The wrapping numbers giving the same SM spectrum of table \ref{SMcontent} are shown in table \ref{SMbtab}.

\begin{table}[htb] 
\renewcommand{\arraystretch}{2.5}
\begin{center}

\begin{tabular}{|c||c|c|c|}
\hline
 $N_i$    &  $(n_i^{(1)},m_i^{(1)})$  &  $(n_i^{(2)},m_i^{(2)})$   & $\gamma_{\om,i}$ \\
\hline\hline 
$N_a = 3$ & $(1/\b^1, 0)$  &  $(\eps, -\oh\tilde \eps)$ & 
$\a 1_3$  \\
\hline 
$N_b = 2$ & $(n_b^{(1)}, \tilde\eps\b^1)$ &
$(1, -\frac 32 \eps\tilde\eps)$ &  $1_2$   \\
\hline 
$N_c = 1$ & $(n_c^{(1)}, 3\eps\b^1)$ & $(0, 1)$ & $1$ \\
\hline 
$N_d=1$ & $(1/\b^1, 0)$ & $(\eps, \frac 32 \tilde\eps) $ & $\a$ \\
\hline
$N_h$ & $(\eps_h/\b^1, 0)$ & $(2,0)$ & $1_{N_h}$  \\
\hline
\end{tabular}

\caption{D5-branes wrapping numbers and CP phases giving rise to a SM spectrum in the $\inte_3$ quiver of fig. \ref{SMfig}.$b_1$). The solution is now parametrized by $n_b^{(1)}, n_c^{(1)} \in \inte$, $\eps, \tilde\eps = \pm 1$, and $\b^1 = 1 - b^{(1)} = 1, 1/2$.
\label{SMbtab}}
\end{center}
\end{table}

Just as before, tadpoles are almost automatically satisfied, the only condition to be imposed is
\beq
n_b^1 = -4 + \frac{1}{\b^1} (\eps - N_h\eps_h).
\label{tadSMZ3b}
\eeq

The $U(1)$ structure is quite similar as well, again with three non-trivial couplings to RR twisted fields, now given by
\beq
\begin{array}{rcl}
B_2^{(1)} & \wedge  & \ c_1\ (\a -\a^2)\ \frac{\eps}{\b^1}
\ (3F_a\ +\ F_{d})\\
D_2^{(1)} & \wedge  & \ c_1 
\left( \frac{3 \tilde\eps}{2\b^1} (F_a\ -\ F_d)\ -\ 6 n_b^{(1)} \eps\tilde\eps F_b\ 
+\ 2 n_c^{(1)} F_c \right) \\
E_2^{(1)} & \wedge  & \ c_1\ 4 \tilde\eps \b^1 F_b
\end{array}
\label{bfsSMZ3b}
\eeq
The massless $U(1)$ will again be model-dependent
\beq
Q_0 = Q_a - 3 Q_d - \frac{3\tilde\eps}{n_c^{(1)}\b^1} Q_c,
\label{masslessSMZ3b}
\eeq
and getting the hypercharge as the unique massless $U(1)$ amounts to requiring
\beq
n_c^{(1)} = \frac {\tilde\eps}{\b^1}\ \Rightarrow\ \b^1 = 1,  
\label{condSMZ3b}
\eeq
since $n_c^{(1)}$ has to be an integer. A simple solution is $N_h = 3$, $\eps = -\eps_h = 1$. This implies setting  $n_h^{(1)} = 0$, and then we have a single Higgs system as in table \ref{higgsses2}.
 
Considering now the quiver in figure \ref{SMfig}.$b_2$) give us another family of configurations. Looking for the same spectrum than in table \ref{SMcontent}, we find the wrapping numbers of table \ref{SMb2tab}.
\begin{table}[htb] 
\renewcommand{\arraystretch}{2.5}
\begin{center}

\begin{tabular}{|c||c|c|c|}
\hline
 $N_i$    &  $(n_i^{(1)},m_i^{(1)})$  &  $(n_i^{(2)},m_i^{(2)})$   & $\gamma_{\om,i}$ \\
\hline\hline 
$N_a = 3$ & $(1/\b^1, 0)$  &  $(\eps, -\oh\tilde \eps)$ & 
$\a 1_3$  \\
\hline 
$N_b = 2$ & $(n_b^{(1)}, \tilde\eps\b^1)$ &
$(1, -\frac 32 \eps\tilde\eps)$ &  $1_2$   \\
\hline 
$N_c = 1$ & $(n_c^{(1)}, 3\eps\b^1)$ & $(0, 1)$ & $1$ \\
\hline 
$N_d=1$ & $(1/\b^1, 0)$ & $(-\eps, -\frac 32 \tilde\eps) $ & $\a^2$ \\
\hline
$N_h$ & $(\eps_h/\b^1, 0)$ & $(2,0)$ & $1_{N_h}$ \\
\hline
\end{tabular}

\caption{D5-branes wrapping numbers and CP phases giving rise to a SM spectrum in the $\inte_3$ quiver of fig. \ref{SMfig}.$b_2$). The solution is now parametrized by $n_b^{(1)}, n_c^{(1)} \in \inte$, $\eps, \tilde\eps = \pm 1$, and $\b^1 = 1 - b^{(1)} = 1, 1/2$.
\label{SMb2tab}}
\end{center}
\end{table}
In this case tadpoles read:
\beq
2 n_b^{(1)} = -8 + \frac{1}{\b^1} (\eps - 2 N_h\eps_h)\ \Rightarrow\ \b^1 = \oh.
\label{tadSMZ3a2}
\eeq
The $U(1)$ couplings are:
\beq
\begin{array}{rcl}
B_2^{(1)} & \wedge  & \ c_1\ (\a -\a^2)\ \frac{\eps}{\b^1}
\ (3F_a\ +\ F_{d})\\
D_2^{(1)} & \wedge  & \ c_1 
\left( \frac{3 \tilde\eps}{2\b^1} (F_a\ {\bf +}\ F_d)\ 
-\ 6 n_b^{(1)}  \eps\tilde\eps F_b\ +\ 2 n_c^{(1)} F_c \right) \\
E_2^{(1)} & \wedge  & \ c_1\ 4 \tilde\eps \b^1 F_b
\end{array}
\label{bfsSMZ3a2}
\eeq
The massless $U(1)$ will now be
\beq
Q_0 = Q_a - 3 Q_d + \frac{3\tilde\eps}{2 n_c^{(1)}\b^1} Q_c,
\label{masslessSMZ3a2}
\eeq
and getting the hypercharge as the unique massless $U(1)$ amounts to requiring
\beq
n_c^{(1)} = - \frac {\tilde\eps}{2\b^1} = - \tilde\eps.
\label{condSMZ3a2}
\eeq
Unlike the previous SM D5-brane constructions, the lightest scalars and/or tachyons are now colour singlets.

\subsection{Low-energy spectrum beyond the SM} 

Let us summarize the lightest (open string) spectra in the class of SM D5-brane constructions:

\begin{itemize}

\item {\it Fermions}
 
The only massless fermions are the ones of the SM (plus right-handed neutrinos). In particular, unlike the case of D6-branes, there are no gauginos in the lightest spectrum. 

\item{\it Gauge bosons} 

There are only the ones of the SM (or its left-right extension). There are in addition three extra  massive (of order the string scale) $Z_0$'s, two of them anomalous and the other being the extra $Z_0$ of left-right symmetric models. As discussed in ref.\cite{Ghilencea:2002da} for a string scale of order a few TeV the presence of these extra $U(1)$'s may be amenable to experimental test. In fact already present constraints from electroweak precision data (i.e., $\rho $-parameter) put important bounds on the mass of these extra gauge bosons.  

\item{\it Scalars in the D5-branes  bulk}

There are two copies of scalars in the adjoint representation of $SU(3)\times SU(2)\times U(1)_a\times U(1)_b\times U(1)_c\times U(1)_d$, as given in eq.(\ref{spectrum5aa}). These will include a couple of colour octets and $SU(2)_L$ triplets plus eight singlets. The vevs of the latter parametrize the  locations of 
the four stacks of branes along the two tori ($4\times 2$ parameters) and hence are moduli at the classical  level. The colour octets and $SU(2)$ triplets get masses at one loop as given in eq.(\ref{massloop}).

\item {\it Scalars at the intersections} 

These are model dependent. In the SM example described in some detail in section (3.1) there are colour triplets and sextets (from $(aa*)$) and  colour triplets {\it `leptoquarks'} (from $(ad), (ad^*)$) (see table \ref{escalares}).  Again their leading contribution to their masses should come from eq.(\ref{massloop}). These scalars are not stable particles, they decay into quarks and leptons through Yukawa couplings. In the SM examples in subsection \ref{extra} the scalars at the intersections are colour singlets.

\item {\it SM Higgs doublets}   

There are sets of Higgs doublets as in table \ref{higgsses2} with a multiplicity which is model dependent. In the example of section (3.1) the multiplicity is one and hence we have the same minimal Higgs sector as in the MSSM.

\end{itemize}

The above states constitute the lightest states in the brane configuration. At the massive level there will appear Kaluza-Klein replicas for the gauge bosons as well as stringy winding and oscillator states (gonions). Compared to the spectra of intersecting D6-brane models the present spectrum is quite simpler, since the fermions and gauge bosons of the SM do not have any kind of SUSY partner.

Note that the structure of the $U(1)$ gauge bosons in D5-brane models is remarkably similar to that of the D6-brane models. This similarity is dictated by the massless chiral fermion spectrum in both classes of models which is identical, i.e., the fermions of the SM. In particular baryon number is a gauged symmetry ($U(1)_a$) which remains as a global symmetry in perturbation theory once the corresponding $U(1)$'s become massive. This naturally guarantees proton stability.

Concerning the closed string sector, the $\inte_N$ projection kills all fermionic partners of the untwisted sector. We will have the graviton plus a number of untwisted moduli field as well as untwisted RR-fields. The twisted closed string sector is relevant to anomaly cancellation.

\subsection{Lowering the string scale}

Any of the D6 and D5-brane models we have constructed in this chapter are non-supersymmetric, i.e., there is not a choice of wrapping numbers and compactification radii that allows for a common supersymmetry to be preserved by the whole configuration. This fact is manifest in D5-brane models, where already the orbifold twist is non-supersymmetric. Regarding D6-brane orientifold models, it can be shown that the only RR tadpole cancelling supersymmetric configuration that can be constructed is the trivial one, where all the D6-branes lie in the same homology cycle of the O6-plane\footnote{This is particular of the class of orientifold models discussed in the present work, where due to the quotient $T^{2n}/(1+ \OR)$ a single orientifold plane exist, wrapping a factorisable $n$-cycle of $T^{2n}$. See \cite{Blumenhagen:2000wh,Cremades:2002te}.}. A similar statement applies for D4-brane orientifold models. On the other hand, plain toroidal or orbifold constructions are always non-supersymmetric as well, since cancellation of RR tadpoles implies no net RR D-brane charge, so we must always include an `antibrane' somewhere in the configuration in order to cancel tadpoles.

From the phenomenological point of view, one of the main consequences of this lack of supersymmetry is the fact that the string scale $M_s$ has to be of the order of 1-10 TeV, so that the scalar Higgs fields triggering EW symmetry breaking do not get excesive one-loop correction to their mass. That is, since supersymmetry cannot be of any help, we must avoid the well-known hierarchy problem between the EW scale and the Planck scale by lowering the string scale down to the TeV region. In principle, if we are dealing with constructions where the gauge interactions of the SM are localized at some D-brane worldvolumes, this lowering of the string scale can be done while maintaining the experimentally measured four-dimensional Planck mass $M_p= 1.18 \times 10^{19}$ GeV. For this we need that gravitation (closed strings) propagates in a very large volume, much larger than gauge interactions (open strings) \cite{Arkani-Hamed:1998rs,Antoniadis:1998ig}. 

As noted in \cite{Blumenhagen:2000wh}, this mechanism cannot be applied directly to D6-brane toroidal models. The point is that there are no torus direction simultaneously transverse to all D6-branes. Hence, if some of the compact radii $R_{1,2}^{(i)}, i = 1,2,3$ of $T^6$ are made large, the worldvolume of some D6-branes will necessarily become large as well and, by (\ref{KK}), some charged KK modes living on the branes would become very light. In \cite{Aldazabal:2000dg} it was proposed a way in which one can have a low string scale compatible with the four-dimensional large Planck mass. The idea is that the 6-torus could be small while being connected to some very large volume manifold. For example, one can consider a region of the 6-torus away from the D6-branes, cut a ball and gluing a throat connecting it to a large volume manifold. In this way one would obtain a low string scale model without affecting directly the brane structure discussed in the previous sections. Actually, this problem seems very specific to toroidal models, whereas the general philosophy for getting the SM spectrum described at the beginning of this section can be applied to D6-brenes wraping general cycles on a compact manifold $M_6$. Indeed, it seems easy to conceive more general type IIA Calabi-Yau compactifications where the 3-cycles wrapped by the Standard Model branes are localized in a small region of the Calabi-Yau, and may be of small volume even for large Calabi-Yau volume. Some steps towards explicit realisation of this idea have been taken in \cite{Blumenhagen:2002wn,Uranga:2002pg}.

\begin{figure}
\centering
\epsfxsize=4in
\hspace*{0in}\vspace*{.2in}
\epsffile{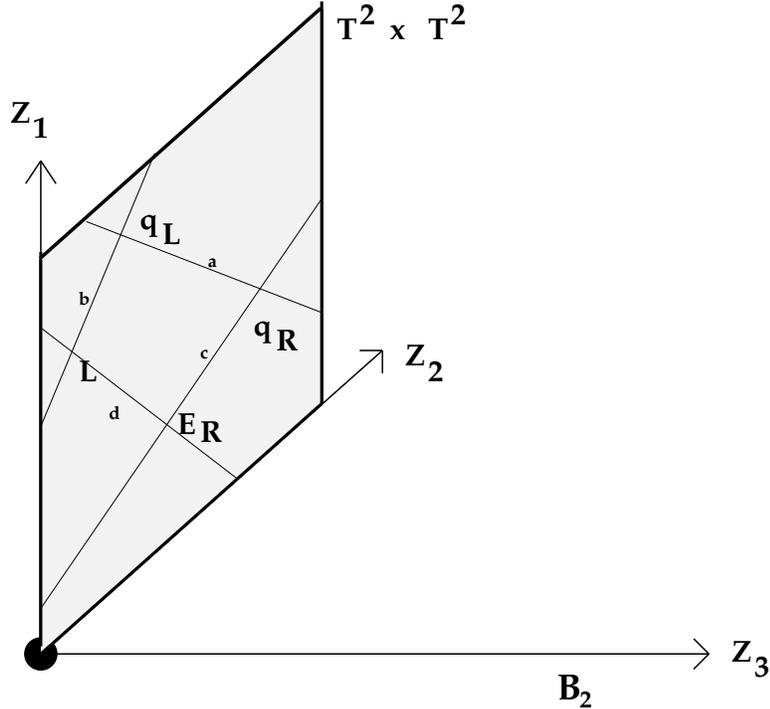}
\caption{Intersecting D5-world set up. The $Z_i$, $i=1,2,3$ represent complex compact dimensions. The D5-branes $a,b,c,d$ (corresponding to the gauge group $U(3)\times U(2)\times U(1)\times U(1)$) wrap cycles on ${\bf T^2\times T^2}$. At the intersections lie quarks and leptons. This system is transverse to a 2-dimensional compact space  ${\bf B_2}$ (e.g., ${\bf T^2/Z_N}$) whose volume may be quite large so as to explain $M_p>>M_s$. This would be a D-brane realisation of the scenario in \cite{Arkani-Hamed:1998rs,Antoniadis:1998ig}.}
\label{braneworld}  
\end{figure}

On the other hand, in D5 and D4-brane models the standard approach for lowering the string scale can be directly implemented, since there are two, respectively four, real dimensions which are transverse to the whole configuration of intersecting D-branes, and hence can be made as large as needed without affecting the gauge sector of the theory. Let us show explicitly how this mechanism works in the case of the D5-brane models constructed in this section. Indeed, in the present examples the compact space has the form ${\bf T^4\times B_2}$, and the D5-branes sit at a ${\bf C/Z_N}$ singularity in ${\bf B_2}$  and wrap two-cycles on ${\bf T^4}$. Let us denote by $V_4$ the volume of ${\bf T^4}$ and by $V_2$ that of the manifold ${\bf B_2}$. Then the Planck scale after dimensional reduction to four dimensions is given by 
\beq
M_p \ =\  {2\over \lambda }  M_s^4 \sqrt{V_4V_2} 
\eeq

In order to avoid too light KK/Winding modes in the worldvolume of the D5-branes let us assume $V_4\propto 1/M_s^4$. Then one has 
\beq
V_2 \ =\  {  { M_p^2 \lambda ^2 } \over {4 M_s^4} }
\eeq
and one can accommodate a low string scale $M_s\sim 1$ TeV by having the volume $V_2$  of the 2-dimensional manifold ${\bf B_2}$ large enough (i.e., of order $(mm.)^2$). For a pictorial view of this explicit D-brane realisation of the proposal in \cite{Arkani-Hamed:1998rs,Antoniadis:1998ig} see figure \ref{braneworld}.

\section{D4-brane models}

In principle, similar constructions to the ones obtained for D6 and D5-branes yielding realistic spectra could be envisaged for intersecting D4-brane orientifols models. These chiral models would also be non-supersymmetric, but the same mechanism that works for D5-branes would allow to consider a string scale $M_s \sim 1$ TeV. Moreover, on these models we can conceive constructing $C^2/\inte_N$ orbifolds wich preserve some bulk supersymmetry, and hence the closed string tachyons present in D5-brane models would be absent here. 

However, in orientifold D4-brane constructions it turns out to be difficult to follow {\em exactly} the approach described at the beginning of this chapter and to obtain models with just the SM fermion spectrum. Let us, however, present a example that illustrates some model-building features of this class of constructions. Since the $\inte_3$ orientifold case has already been explored in \cite{Honecker:2002hp}, let us consider a supersymmetric $\inte_5$ orientifold model whose brane content is shown in table \ref{D4model}.

\begin{table}[htb]
\renewcommand{\arraystretch}{2.5}
\begin{center}

\begin{tabular}{|c||c|c|}
\hline
$N_i$ & $(n_i,m_i)$ & $\gamma_{w,i}$ \\
\hline\hline
$N_a=3$ & $(2, 0) $ & $\a 1_3$ \\
\hline
$N_b=2$ & $ (1, -\frac 32)$ & $\a^2 1_2$ \\
\hline
$N_c=2$ & $ (-1, \frac 32)$ & $\a^2 1_2$ \\
\hline
$N_d=1$ & $(2, 0) $ & $\a 1_3$ \\
\hline
$N_h = 4$ & $(2, 0)$ & $1_{4}$  \\
\hline
\end{tabular}
\caption{Example of a D4-branes LR model in a $\inte_5$ orbifold.
\label{D4model}}
\end{center}
\end{table}

As usual, the brane content of this model consist of four D$4$-branes $a, b, c, d$, again identified with those of table \ref{SMbranes2}, plus some hidden brane $h$. The gauge group is $SU(3) \ti SU(2) \ti SU(2) \ti U(1)^4 \ti [U(4)_h]$, which is the LR gauge group extended by three abelian groups and one hidden $U(4)_h$. The chiral matter content of such model is given in table \ref{D4fermions}

\begin{table}[htb]
\renewcommand{\arraystretch}{1.5}
\begin{center}
\begin{tabular}{|c|c|c|c|c|c|c|c|}
\hline Intersection &
 Matter fields  &   &  $Q_a$  & $Q_b $ & $Q_c $ & $Q_d$  & $B-L$ \\
\hline
\hline ($ab$) & $Q_L$ & $3(3,2,1)$ & 1 & -1 & 0 & 0 & 1/3 \\
\hline ($ac$) & $Q_R$ & $3({\bar 3},1,2)$ & -1  & 0 & 1 & 0 & -1/3 \\
\hline ($bd$) & $L_L$ & $3(1,2,1)$ &  0  & -1 & 0 & 1 & -1 \\
\hline ($cd$) & $L_R$ & $3(1,1,2)$ &  0 & 0 & 1 & -1 & 1 \\
\hline ($bc^*$) & $H$ & $3(1,2,2)$ &  0  & 1 & 1 & 0 & 0 \\
\hline ($bb^*$) & $A_i$  & $3(1,1,1)$ & 0 & -2 & 0 & 0 & 0  \\
\hline ($cc^*$) & $S_i$  & $3(1,1,3)$ & 0 & 0 & -2 & 0 & 0  \\
\hline \end{tabular}
\end{center} \caption{Extended Left-Right symmetric chiral spectrum arising from the $\inte_5$ D4-branes model of table \ref{D4model}. The $U(1)_{B-L}$ generator is defined as $Q_{B-L} = \frac 13 Q_a - Q_d$.}
\label{D4fermions}
\end{table}

Notice that this particular example does not follow the general philosophy described above, where every chiral fermion arised from a bifundamental representation and the matter content was thus described by table \ref{LRcontent}. Instead, we now find some extra chiral fermions that can be identified with three Higgssino-like particles, whereas some exotic matter transforming as singlets ($A_i$) and symmetrics of $SU(2)_R$ ($S_i$) do also appear. The only light bosonic sector arises from branes $b$ and $c$ giving us a Higgs-like particle that can become tachyonic if we approach both branes close enough. No extra chiral matter nor scalars arise from the hidden sector of the theory.

It is easy to see that this brane content satisfies both twisted tadpole conditions (\ref{tadpoleO4n}) and (\ref{tadpoleO4m}). Interestingly enough, it also satisfies untwisted tadpoles conditions, so when embedding such model in a compact four-dimensional manifold $B$ no extra brane content would be needed.

Finally, by computing the couplings (\ref{dualcouplingsO4}) that mediate the GS mechanism, we can check that two of the abelian gauge groups are in fact massive, the only massless linear combinations being $U(1)_{B-L} = \frac 13 U(1)_a - U(1)_d$ and $U(1)_b + U(1)_c$, just as in our $\inte_N$ D$5$-branes constructions above.

\chapter{Supersymmetry and calibrations \label{SUSY}}

As we pointed out at the end of last chapter, the toroidal, orbifold and orientifold configuration we have been considering in this work are in general non-supersymmetric, unless we consider non-chiral trivial configurations of D-branes. Indeed, even if we consider a compactification where supersymmetry remains unbroken in the closed string sector (as happens in toroidal and orientifold D6-brane configurations) RR tadpole conditions imply that in a chiral D-brane configuration supersymmetry will be broken in the open string sector of the theory. However, it is still legitimate to ask oneself is some open string subsector will preserve some amount of supersymmetry and, in general, what are the conditions for a set of D-branes to preserve a common supersymmetry when wrapping a generic compact manifold $\M$. This question was answered in \cite{Berkooz:1996km} for the case of D$p$-branes at angles in a flat target space. In this chapter we will rederive such result from the perspective of the effective field theory, an later we will try to generalize it to arbitrary (supersymmetric) compactifications by means of calibrated geometry. Finally, we will briefly comment on some applications of these concepts for computing effective field theory quantities and present a model with the chiral spectrum of the MSSM.

Notice that, being non-supersymmetric, the class of theories presented will generically yield chiral $\N = 0$ four-dimensional effective theories. From the phenomenological point of view this is quite appealing, at least if one manages to lower the string scale to the TeV region (see previous chapter). However, one is then faced to deal with the difficult issue of supersymmetry breaking in string theory, which generically implies the existence of uncanceled NSNS tadpoles and leaves the full stability of the configuration as an open question. Hence, computation of NSNS tadpoles in the models under study might be an important issue of the whole construction. Calibrated geometry will turn out to be a useful tool for computing such NSNS tadpoles, either in flat geometries or in more general compactifications.

\section{Supersymmetries on $T^6$}

In Chapter \ref{spectrum}, when computing the $D=4$ low energy spectrum living at the intersection of two flat D6-branes $a$ and $b$, we saw that besides a $D=4$ massless fermion several light scalars also appeared at the intersection. Contrary to the fermion, the squared mass of these scalars depends on the intersection angles $\vt^i_{ab}$, $i = 1, 2, 3$, as (\ref{scalars}) shows, and it can either be positive, negative or null. If the last possibility holds for one of these scalars, then it will pair with the massless fermion in a $\N=1$ chiral multiplet. Moreover, the whole tower of massive states living at the $ab$ intersection will rearrange itself into $\N=1$ supermultiplets, as was shown at the end of Chapter \ref{spectrum}. This fact is not a mere coincidence, but signals that the pair of intersecting D6-branes $a$ and $b$ make up a supersymmetric system. 

In fact, we can characterize such preserved $\N=1$ supersymmetry by means of a four-dimensional vector $\tilde r \in (\inte + \med)^4$ with opposite GSO projection. This vector is defined by $\tilde r \equiv (r_{NS}+v_\vt) - (r_R+v_\vt) = r_{NS} - r_{R}$, where $r_R+v_\vt$ and $r_{NS}+v_\vt$ represent the massless fermionic and scalar states at the intersection, respectively, and is hence determined by (\ref{fermion2}) and the massless state in (\ref{scalars}). Indeed, we are doing nothing but describing in bosonic language a supersymmetry generator $Q_{\tilde r}$, which transforms a massless fermionic state into a massless bosonic state:
\beq
Q_{\tilde r} \ |r + v_\vt \rangle_R \ = \ |r + \tilde r + v_\vt \rangle_{NS}.
\label{SUSYgenerator}
\eeq
This relation between fermions and bosons will not only hold at the massless level of the spectrum, but on the whole $ab$ sector of the theory.

What is the geometry behind such $\N=1$ system? Recall that the angles $\vt_{ab}^i$ encode the precise $U(3)$ rotation relating the D6-brane $a$ to the D6-brane $b$ locus by
\bea
R_{ab}: \left(
\begin{array}{c}
Z^1 \\ Z^2 \\ Z^3
\end{array}
\right) \mapsto
\left(
\begin{array}{ccc}
e^{i\pi\vt^1} & 0 & 0 \\
0 & e^{i\pi\vt^2} & 0 \\
0 & 0 & e^{i\pi\vt^3}
\end{array}
\right)
\cdot
\left(
\begin{array}{c}
Z^1 \\ Z^2 \\ Z^3
\end{array}
\right)
\label{rotation3}
\eea
where $Z^i = X^{2i} + i X^{2i + 1}$ are the three complex coordinates on $T^2 \ti T^2 \ti T^2$. If $Q_{\tilde r}$ is a good supersymmetry of the system, then the associated spinor should be invariant under (\ref{rotation3}), that is, it must satisfy $R^2_{ab} Q_{\tilde r} = Q_{\tilde r}$ \cite{Blumenhagen:2000wh}. This condition amounts precisely to $\med \sum_i \vt_{ab}^i {\tilde r}_i \in \inte$, which in turn guarantees that one of the scalar in (\ref{scalars}) will have vanishing squared mass\footnote{Notice that we are taking $|\vt_{ab}^i| < 1$, so that $\med \sum_i \vt_{ab}^i {\tilde r}_i = -1, 0, 1$.}. In addition, this condition also guarantees that $R_{ab} \in SU(3)$ (or some isomorphic embedding of this group into the rotation group $SO(6)$). Actually, this was precisely the condition found in \cite{Berkooz:1996km} for two D6-branes to preserve a supersymmetry. In general, if two D$p$-branes intersect at $d$ angles, they will preserve a common supersymmetry if they are related by a $SU(d)$ rotation.

\begin{figure}[ht]
\centering
\epsfxsize=3.5in
\epsffile{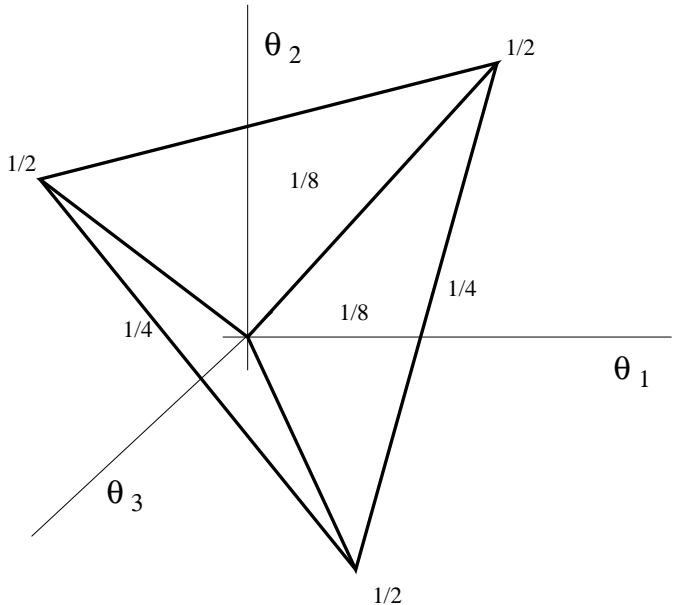}
\caption{Different supersymmetry regions in terms of the intersecting angles of two D6-branes. The fraction on each face, edge or vertex of the tetrahedron represents the amount of preserved supersymmetry.}
\label{tetrahedron}
\end{figure}

Notice that in the spectrum of two factorisable intersecting D6-branes there are several light scalars which may become massless, and this signals that two D6-branes may preserve {\em different} $\N=1$ supersymmetries depending on the intersection angles. Actually, even two of them may become massless at the same time, which implies that a higher amount of $\N = 2$ supersymmetry is preserved by the composite system of two D6-branes. Let us, for instance, consider the angle parameter space $0 < \vt^i_{ab} < 1$. The region where some supersymmetry is preserved is a codimension one locus with the shape of a tetrahedron, depicted in figure \ref{tetrahedron}. On the walls of the tetrahedron $\N=1$  $D=4$ supersymmetry is preserved, which is $\frac 18$ of the bulk supersymmetry, and each wall corresponds to a different vector $\tilde r$. On the edges of the tetrahedron two such walls will meet, hence there will be two massless scalars at the intersection and the system will preserve $\N = 2$. Finally, at the vertices of the tetrahedron the sector $ab$ of the theory will preserve a total amount of $\N = 4$ supersymmetry, which means that both D6-branes are parallel in the three complex dimensions.\footnote{Notice that, in fact, each of these four vertices can be identified with the same point, since rotations $(\vt^1, \vt^2, \vt^3, 0)$ and $(\vt^1 \pm 1, \vt^2 \pm 1, \vt^3, 0)$, etc., correspond to the same action (\ref{rotation3}) on both scalars and spinors. Correspondingly, the wrapping numbers $[(n_a^{(1)}, m_a^{(1)}) \otimes (n_a^{(2)}, m_a^{(2)}) \otimes (n_a^{(3)}, m_a^{(3)})]$ and $[(-n_a^{(1)}, -m_a^{(1)}) \otimes (-n_a^{(2)}, -m_a^{(2)}) \otimes (n_a^{(3)}, m_a^{(3)})]$, etc., should also be identified, since they are equivalent descriptions of the same homology class $[\Pi_a] \in H_3(T^6,\inte)$.} In the region of the parameter space which is inside the tetrahedron all the scalars in (\ref{scalars}) have positive squared mass, so although the system of two D6-branes is not supersymmetric, it cannot decay to another one which lowers the energy, at least by considering a small deformation of the D6-branes worldvolumes. Outside the tetrahedron some of the four scalars will become tachyonic, namely those which were massless at the closest wall. As mentioned before, this open string tachyon signals that the system of the two D6-branes is not stable, at that it can lower its energy by decaying to another system with same RR charges by condensing the tachyon at the intersection. Geometrically, this amounts to a smoothing of the intersection and recombining the two intersecting branes to a third one in the homology class $[\Pi_a] + [\Pi_b]$. For D-branes intersecting in flat non-compact space $\cpx^n$ this problem is well understood, since it amounts to the angle criterion of Nance and Lawlor \cite{Nance,Lawlor,Harvey} (see \cite{Acharya:1998en,Figueroa-O'Farrill:1998su} for a physical view of this problem and \cite{Blumenhagen:2000eb} for this recombination process in the context of D-branes). In compact spaces as $T^6$, however, it still remains an open issue.

What is the physical meaning of all this? Notice that each factorisable D6-brane is a $\med$ BPS state of type IIA theory. This means that it preserves one half of the total $\N=8$ bulk supersymmetry, in terms of $D=4$ non-compact physics. Hence, we can associate to each D6-brane an $\N=4$ superalgebra inside $\N=8$. If we now take two of such D6-branes, the supersymmetry preserved by their intersection will be the intersection of the two $\N=4$ superalgebras inside $\N=8$, which may be the full $\N=4$, some $\N=2$ or $\N=1$ common subalgebra or none. Indeed, this simple idea can be made more precise, and we will try to do so in the next sections, by means of calibrated geometry.

\section{Calabi-Yau and Special Lagrangian geometry \label{CY&sL}}

Let us make a detour from $T^6$ geometry and consider a more general class of compactifications involving type IIA intersecting D6-branes. Namely, let us consider the target space $M_4 \ti \M_6$, with configurations of D6-brane stacks filling $M_4$ and wrapping 3-cycles on $\M_6$, which is taken to be a compact Riemannian manifold. As was shown in section \ref{general}, the main features of a D6-brane configuration, such as gauge group, chiral fermions, and tadpole conditions can be encoded in topological properties of $\M_6$. Hence in principle we could envisage constructing D6-brane models analogous to the ones in the previous chapter, yielding the SM spectrum, once the homology space $H_3(\M_6,\inte)$ was fully understood. 

Notice that, so far, we have not imposed any particular constraint on our manifold $\M_6$, except that it must be compact so that we recover four-dimensional gravity at low energies. We may now require the closed string sector to be supersymmetric. This amounts to impose that $\M_6$, seen as a Riemannian manifold with metric $g$, has a holonomy group contained in $SU(3)$. This is quite an analogous condition for the system of two D6-branes to preserve a supersymmetry and, in fact, both amount to impose that one $D=4$ supersymmetry generator remains invariant under all the possible rotations on the internal space. Now, such a manifold with $SU(3)$ holonomy can be equipped with a complex structure $J$ and a holomorphic volume 3-form $\O_c$ which are invariant under the holonomy group, i.e., covariantly constant. This promotes $\M_6$ to a Calabi-Yau three-fold, or ${\bf CY_3} = (\M_6,g,J,\O)$.\footnote{For reviews on Calabi-Yau geometry see, e.g., \cite{Huebsch:nu,Joyce:2001xt}.} Moreover, $g$ and $J$ define a K\"ahler 2-form $\om$ which satisfies the following relation with the volume form
\beq
\frac{\om^3}{3!} = \left( \frac i2\right)^3 \O_c \wedge \ov \O_c.
\label{omegas}
\eeq
Given a real 3-form $\O$ normalized as this, we can always take $\O_c = e^{i\th} \O$ for any phase $\th$ as a solution of (\ref{omegas}). The K\"ahler form $\om$ can also be complexified to $\om_c$, by addition of  a non-vanishing $B$-field. Both $\O$ and $\om$ will play a central role  when considering the open string sector.

Notice, however, that we have not imposed Hol$(\M_6)$ to be {\em exactly} $SU(3)$. \footnote{This is an important phenomenological restriction when considering, e.g., perturbative heterotic compactifications. This is not longer the case on type II or type I theories, where bulk supersymmetry can be further broken by the presence of D-branes.} We may consider, for instance, $\M_6 = T^2 \ti {\bf K3}$, whose holonomy group is contained in $SU(2)$. In this case, there is not a unique invariant 3-form but two linearly independent ones, both satisfying (\ref{omegas}). In general, the number of (real) covariantly constant 3-forms of $\M_6$ satisfying (\ref{omegas}) indicates the amount of $D = 4$ supersymmetries preserved under compactification. In a ${\bf CY_3}$ in the strict sense, with Hol$(\M_6) = SU(3)$, the gravity sector yields $D = 4$ $\N = 2$ under compactification, and this fact is represented by the existence of a unique complex volume form $e^{i\th} \O$. Indeed, $\th$ parametrizes the $U(1)$ of $\N = 1$ superalgebras inside $\N = 2$ \cite{Douglas:1999vm}. Correspondingly, compactification on $T^2 \ti {\bf K3}$ yields a $D = 4$ $\N = 4$ gravity sector. There are some other consequences when considering manifolds of lower holonomy. For instance, Hol$(\M_6) = SU(3)$ implies that $b_1(\M_6) = 0$ \cite{Huebsch:nu}, while this might not be the case for lower holonomy, as the example $T^2 \ti {\bf K3}$ shows.

Let us now turn to the open string sector, represented by type IIA D6-branes. Intuitively, a dynamical object as a D-brane will tend to minimize its tension while conserving its RR charges. In our geometrical setup, this translates into the minimization of Vol($\Pi_a$) inside the homology class $[\Pi_a]$. A particular class of volume-minimizing objects are calibrated submanifolds, first introduced in \cite{Lawson}. The area or volume of such submanifolds can be computed by integrating on its $p$-volume a (real) closed $p$-form, named calibration,\footnote{We say that a closed $k$-form $\vp$ is a calibration on a Riemann manifold $\M$ if, for every oriented $k$-plane $W$ on $\M$ we have $\vp\arrowvert_{W} \leq {\rm Vol}(W)$. Let us consider an oriented $k$-submanifold $\S$ of $\M$, $T_x\S$ being its tangent space at each point $x \in \S$. $\S$ is called a calibrated submanifold with respect to $\vp$ if $\vp\arrowvert_{T_x\S} = {\rm Vol}(T_x\S)$, for every  $x \in \S$. The interesting property of calibrated submanifolds is that they are volume-minimizing objects in their homology class, see \cite{Joyce:2001xt}.} defined on the ambient space $\M_6$. Both $\O$ and $\om$ are calibrations in a ${\bf CY_3}$. Submanifolds calibrated by $\O$ are named special Lagrangian \cite{Joyce:2001xt} while those calibrated by $\om$ are holomorphic curves. Being a 3-form, $\O$ will calibrate 3-cycles where D6-branes may wrap. A 3-cycle $\Pi_a$ calibrated  by Re($e^{i\th}\O$) will have a minimal volume on $[\Pi_a]$, given by Vol($\Pi_a$) $ = \int_{\Pi_a}$ Re($e^{i\th}\O$), and will be said to  have phase $\th$. We can also characterize such calibration condition by
\beq
\om|_{\Pi_a} \equiv 0 \quad {\rm and} \quad 
{\rm Im} (e^{i\th}\O)|_{\Pi_a} \equiv 0.
\label{cali2}
\eeq
The middle-homology objects that satisfy the first condition in (\ref{cali2}) are named Lagrangian submanifolds and, although they are not volume minimizing, play a central role in symplectic geometry. 

A D6-brane $a$ whose 3-cycle $\Pi_a$ wraps a special Lagrangian (sL) submanifold  does not only minimize its volume but, as shown in \cite{Becker:1995kb}, also preserves  some amount of supersymmetry. Hence, such D6-brane is a BPS state of type IIA theory compactified in a ${\bf CY_3}$. Notice that the standard properties of calibrated submanifolds give evidence of this fact. Indeed, given a calibration $\O_{\tilde r}$ we can associate the RR charge of this object to
\beq
Q_{\tilde r} (\Pi_a) = \left| \int_{\Pi_a} \O_{\tilde r} \right|.
\label{RRcharge}
\eeq
Notice that $Q_{\tilde r}$ only depends on the homology class $[\Pi_a] \in H_3(\M_6, \inte)$, as we would expect from RR charges from the point of view of $D=4$ physics. On the other hand, the NSNS charge can be associated with the tension of $\Pi_a$. If we take the assume that no electric nor magnetic fluxes are induced on the worldvolume of such D6-brane, which will usually be the case, then tension will be proportional to the volume of $\Pi_a$. Moreover, if we take the mild assumption that $\Pi_a$ is a Lagrangian submanifold of $\M_6$, then 
\beq
{\rm Vol} (\Pi_a) = \int_{\Pi_a} \left| \O_{\tilde r} \right|.
\label{NSNScharge}
\eeq
From these definitions, we readily see that a well-known property of calibrations
\beq
Q_{\tilde r} (\Pi_a) \leq {\rm Vol} (\Pi_a),
\label{BPSbound}
\eeq
can be interpreted as the BPS bound. The objects which saturate this bound are the special Lagrangians under $\O_{\tilde r}$, i.e., BPS states of the compactification. Moreover, if a set of objects $\Pi_a$, $a = 1, \dots, K$ are all calibrated under the volume form $\O_{\tilde r}$ and with the same phase $\th_{\tilde r}$, then we can compute the tension of the composite system to be
\beq
{\rm Vol} \left(\sum_a [\Pi_a] \right) = \int_{\sum_a [\Pi_a]} {\rm Re} \left(e^{i\th_{\tilde r}} \O_{\tilde r} \right) = \sum_a \int_{[\Pi_a]} {\rm Re} \left(e^{i\th_{\tilde r}} \O_{\tilde r} \right) = \sum_a {\rm Vol} (\Pi_a),
\label{nobinding}
\eeq
i.e., the sum of tensions of each component. Thus there is no binding energy, which is also a well-known fact in supersymmetric systems. We see from this last computation that the phase $\th$ of a special Lagrangian is also relevant. Actually, when defining the RR charge in (\ref{RRcharge}) we have simplified matters a little bit. Let us for instance consider a Calabi-Yau threefold in the strict sense, where only one complex volume form $\O$ exists. In this case a sL with phase $\th$ will be a $\med$ BPS state of the $D=4$ $\N = 2$ bulk superalgebra, and the $\N=1$ subalgebra preserved will be given by $Q = e^{i\th} \bar Q$. Indeed, the phase $\th$ is nothing but the phase of the central charge $Z$ of this BPS state, and should then be part of the definition of the RR charge of $\Pi_a$ (notice that $\th$ only depends on $[\Pi_a]$ and on the complex structure moduli of the ${\bf CY_3}$).

In a more general setup, where more than one complex volume form $\O$ may exist, it is a natural question which amount of $D=4$ supersymmetry will preserve a D6-brane wrapping an arbitrary sL. The precise amount is given by the number of independent real volume forms Re$(e^{i\th_{\tilde r}}\O_{\tilde r})$ that calibrate the 3-cycle which, if the holonomy of $\M_6$ is contained in $SU(2)$, may be more than one. Moreover, the pairs $(\O_{\tilde r}, \th_{\tilde r})$ label the subalgebra that is preserved by this supersymmetric cycle inside the full bulk superalgebra. This can be seen by relating the Killing spinor equations for the corresponding D6-brane with some tailor-made calibrations, which will turn out to be the previous volume forms \cite{Gibbons:1998hm}.

\subsection{The six-torus revisited}

In light of the previous general discussion, let us again consider compactifications of type IIA on $T^6$. The flat six-torus is a very special case of ${\bf CY_3}$, where the holonomy group is trivial. Hence, dimensional reduction of the closed string sector will yield a $D=4$ $\N=8$ theory. This is consistent with the fact that, given a complex structure $(z_1, z_2, z_3)$ on $T^6$ and the induced K\"ahler form $\om = \frac i2 \sum_{i=1}^3 d z^i \wedge \bar d z^i$, we can find four independent 3-forms satisfying (\ref{omegas})
\beq
\begin{array}{ccccc}
\Omega_{(+,+,+)} & = & dz^1\wedge dz^2\wedge dz^3 & = & 
(dx^1 + \tau^{(1)} dy^1)\wedge(dx^2 + \tau^{(2)} dy^2)\wedge 
(dx^3 + \tau^{(3)} dy^3), \vspace{0.1cm} \\
\Omega_{(+,-,-)} & = & dz^1\wedge d\bar z^2\wedge d\bar z^3 & = & 
(dx^1 + \tau^{(1)} dy^1)\wedge(dx^2 + \bar\tau^{(2)} dy^2)\wedge 
(dx^3 + \bar\tau^{(3)} dy^3),\vspace{0.1cm} \\
\Omega_{(-,+,-)} & = & d\bar z^1\wedge dz^2\wedge d\bar z^3  & = & 
(dx^1 + \bar\tau^{(1)} dy^1)\wedge(dx^2 + \tau^{(2)} dy^2)\wedge 
(dx^3 + \bar\tau^{(3)} dy^3),\vspace{0.1cm} \\
\Omega_{(-,-,+)} & = & d\bar z^1\wedge d\bar z^2\wedge dz^3 & = & 
(dx^1 + \bar\tau^{(1)} dy^1)\wedge(dx^2 + \bar\tau^{(2)} dy^2) \wedge
(dx^3 + \tau^{(3)} dy^3),
\end{array}
\label{omegasT6}
\eeq
where on the second equality we have assumed a factorisable six-torus $T^2 \ti T^2 \ti T^2 $ and imposed the natural complex structure $dz^{i} = dx^i + \tau^{(i)} dy^i$. 

Is easy to check that a D6-brane wrapped on a factorisable 3-cycle
\beq
[\Pi_a] = \left(n_a^{(1)} [a_1] + m_a^{(1)} [b_1]\right) \otimes
\left(n_a^{(2)} [a_2] + m_a^{(2)} [b_2]\right) \otimes
\left(n_a^{(3)} [a_3] + m_a^{(3)} [b_3]\right),
\label{factorisable3}
\eeq
and minimizing its volume is a flat Lagrangian 3-torus calibrated by the four volume forms above. Namely,
\beq
\int_{\Pi_a} \O_{\tilde r} = \left|\prod_{i=1}^3 \left(n_a^{(i)}+\tau^{(i)} m_a^{(i)} \right)\right| \cdot \prod_{i=1}^3 (2\pi R^{(i)}) \cdot e^{-i\th_{\tilde r}},
\label{zcharge}
\eeq
where $R^{(i)}$ is the torus radius on the $x^i$ axis, and the calibration phases $\th_{\tilde r}$ differ for each choice of $\tilde r$. Each flat factorisable 3-cycle is thus calibrated by four {\em real} volume forms, which indicates that it preserves a $D=4$ $\N=4$ supersymmetry under dimensional reduction. This is a well-known fact, since flat D6-branes are $\med$ BPS states of type IIA theory.

Moreover, we know from our previous discussion that a pair of factorisable D6-branes $a$ and $b$ will preserve a common supersymmetry if they are calibrated by the same real volume form Re$(e^{i\th_{\tilde r}} \O_{\tilde r})$, that is, if some of the phases $\th_{\tilde r}([\Pi_a])$ and $\th_{\tilde r}([\Pi_b])$ match. Since two factorisable D6-branes are hyperplanes related by a rotation $R$ on the three complex planes, this will always happen if a volume form $\O_{\tilde r}$ is invariant under $R$. Now, if in this factorised geometry we perform a rotation of the form (\ref{rotation3}) and describe it by a usual twist vector $v_\vt = (\vt^1, \vt^2, \vt^3, 0)$, then $\O_{(\eps_1,\eps_2,\eps_3)}$ will be left invariant by the action of $R$ if $\oh \sum_i  \vt^i \eps_i \in \inte$. We can actually associate to each of these forms two four dimensional vectors given by
\beq
\tilde r = \pm \oh (\eps_1,\eps_2,\eps_3,\prod_{i=1}^3\eps_i),
\label{r}
\eeq
so that
\beq
R \O_{\tilde r} = \O_{\tilde r} \iff {\tilde r} \cdot v_{\vt} \in \inte.
\label{invariance}
\eeq
These vectors $\tilde r$ are those in (\ref{SUSYgenerator}) which label $\N = 1$ supersymmetries preserved by intersecting D6-branes, i.e., the faces of the tetrahedron of figure \ref{tetrahedron}.

\subsection{Orientifolded geometry}

A general orientifold compactification of type IIA which naturally involves D6-branes will be of the form \cite{Blumenhagen:2002wn}
\beq
\frac {{\rm Type \ IIA \ on \ } \M_6}{\{1 + \O \bar\sig\}},
\label{genori}
\eeq
where $\bar\sig$ is a well-defined geometric involution with a fixed locus of real codimension three. If we impose the low energy theory to preserve some supersymmetry, then $\M_6$ must be a Calabi-Yau threefold, and then $\bar\sig$ turns into an antiholomorphic involution which can be expressed locally as
\beq
\bar\sig : z^i \mapsto \bar z^i, \ i = 1, 2, 3.
\label{antihol}
\eeq
At the fixed locus of this involution there will be located the O6-planes of the theory and, $\bar\sig$ being antiholomorphic, this fixed locus will be an special Lagrangian submanifold calibrated by the complex volume form(s) $\O_r$ with vanishing phase $\th_r$. Another consequence of (\ref{antihol}) is that for each D6-brane wrapping a 3-cycle $\Pi_a$ which is not invariant under the action of $\bar\sig$ we will have to include a mirror brane located on $\Pi_{a*} = \bar\sig(\Pi_a)$ in order to have a well-defined orientifold quotient. By the action of $\bar\sig$, is easy to see that if $\Pi_a$ wraps a sL then $\Pi_{a*}$ will also wrap one, and that the central charges of both branes will be conjugate to each other. In fact, in order to be a truly BPS state of the theory, $\Pi_a$ should have central charge with vanishing phase $\th_r$, being calibrated by the same volume form(s) $\O_r$ as the O6-plane. As a result, the number of calibrations and BPS states in (\ref{genori}) will be reduced with respect to the unorientifolded theory, which is a direct consequence that the number of bulk supersymmetries has been reduced.

Let us apply this general picture to the toroidal orientifold (\ref{dual}), choosing again a factorisable geometry. Each of the $8\b^1\b^2\b^3$ O6-planes of this compactification will wrap the homology class
\beq
[\Pi_{O6}] =
\left[\left({1}/{\b^1},0\right)\right] \otimes 
\left[\left({1}/{\b^2},0\right)\right] \otimes 
\left[\left({1}/{\b^3},0\right)\right].
\label{orienti}
\eeq
where $\beta^i = 1, \oh$ is related to the rectangular or tilted geometry in the $i^{th}$ torus by $\beta^i = 1 - b^{(i)}$ (see figure \ref{bflux}), and we use fractional wrapping numbers. The volume forms in this geometry are given by
\beq
\begin{array}{ccccc}
\Omega_{(+,+,+)} & = & dz^1\wedge dz^2\wedge dz^3 & = & 
(\tau^{(1)} dx^1 + i dy^1)\wedge(\tau^{(2)} dx^2 + i dy^2)\wedge 
(\tau^{(3)} dx^3 + i dy^3), \vspace{0.1cm} \\
\Omega_{(+,-,-)} & = & dz^1\wedge d\bar z^2\wedge d\bar z^3 & = & 
( \tau^{(1)} dx^1 + i dy^1)\wedge(\bar\tau^{(2)}dx^2 - i dy^2)\wedge 
( \bar\tau^{(3)} dx^3 - i dy^3),\vspace{0.1cm} \\
\Omega_{(-,+,-)} & = & d\bar z^1\wedge dz^2\wedge d\bar z^3  & = & 
(\bar\tau^{(1)} dx^1 - i dy^1)\wedge(\tau^{(2)} dx^2 + i dy^2)\wedge 
(\bar\tau^{(3)} dx^3 - i dy^3),\vspace{0.1cm} \\
\Omega_{(-,-,+)} & = & d\bar z^1\wedge d\bar z^2\wedge dz^3 & = & 
(\bar\tau^{(1)} dx^1 - i dy^1)\wedge(\bar\tau^{(2)} dx^2 - i dy^2) \wedge
(\tau^{(3)} dx^3 + i dy^3),
\end{array}
\label{omegasO6}
\eeq
where the complex structure on each torus is given by $\tau^{(j)} = {R_1^{(j)} \over R_2^{(j)}} + i b^{(j)}$. 

A factorisable D6-brane wrapped on (\ref{factorisable3}) will no longer be automatically $\med$ BPS.\footnote{Although the $aa$ sector of the theory will yield the massless content of $D=4$ $\N = 4$ SYM, we must include the sector $aa$* in the gauge theory on the D6-brane, which will break some supersymmetry unless $[\Pi_a] = [\Pi_{a*}] = [\Pi_{O6}]$.} In fact, in order to preserve some supersymmetry it will have to share some of its four real calibrations with those of $\Pi_{O6}$, which are the real part of (\ref{omegasO6}). That is, some of the phases in (\ref{zcharge}) will have to vanish. We conclude that, in fact, instead of four {\em complex} volume forms, only four {\em real} of them are `good' calibrations of the theory, in the sense that they count BPS states. This fits quite well with the fact that the T-dual type I theory compactified on $T^6$ preserves $D=4$ $N=4$ in the bulk. Notice as well that we can perform a further orbifold projection to our theory, as in \cite{Cvetic:2001nr,Blumenhagen:2002gw,Honecker:2003vq}. Then, some of the volume forms will not be invariant under the orbifold action, and will thus not be well-defined on the quotient space. Hence, they can no longer be considered as calibrations, and this reproduces the fact that the bulk preserves a lower amount of supersymmetry.

\section{Some applications}

Let us enumerate some of the relevant quantities that can be computed with the help of calibrated geometry in a specific intersecting D6-brane model:

\begin{itemize}

\item{\it Number of supersymmetries}

Being a BPS soliton of type IIA theory, a stack of $N_a$ D6-branes wrapping a special Lagrangian submanifold $\Pi_a$ will yield a $U(N_a)$ Supersymmetric Yang-Mills theory on its worldvolume. A natural question is which amount of $D=4$ SUSY the D6-brane effective field theory will have by dimensional reduction down to $D =  4$. As we have seen above, the precise amount is again given by the number of independent real volume forms Re$(e^{i\th_{\tilde r}}\O_{\tilde r})$ that calibrate the 3-cycle $\Pi_a$.

\item{\it Gauge kinetic function}

By simple dimensional reduction, the gauge coupling constant $g_a$ of this gauge theory can be seen to be controlled by Vol$ (\Pi_a)$ by the formula (\ref{coupling}). Now, if $\Pi_a$ is a special Lagrangian submanifold of the compact space $\M_6$, then this quantity can be computed by simply integrating the 3-form Re($e^{i\th}\O$) on the corresponding homology cycle. See \cite{Cremades:2002te} for some explicit computations on toroidal compactifications.

\item{\it NSNS tadpoles}

We have seen that RR tadpoles do cancel when the total RR charge of a configuration vanishes. The same statement holds for NSNS tadpoles, where now  NSNS charge means the tension, in our specific framework the volume, of the D6-branes and the O6-plane. Now, we know how to compute the tension of a D6-brane wrapping a Lagrangian submanifold by (\ref{NSNScharge}). If in addition we take D6-branes wrapping special Lagrangian submanifolds as building blocks of our configuration, then this expression gets simplified to Vol $(\Pi_a) = \left| \int_{\Pi_a} \O_{\tilde r} \right|$ for the appropriate choice of $\O_{\tilde r}$. Finally, an O6-plane will always be calibrated by Re$(\O_{\tilde r})$. Thus, calibrations allow us to compute NSNS tadpoles in a rather elegant manner.

\item{\it Scalar potential and Fayet-Iliopoulos parameters}

In \cite{Blumenhagen:2001te} it was pointed out that uncanceled NSNS tadpoles generate, at the disc level, a scalar potential for the complex structure moduli, which are fields arising from the NSNS sector of the theory. Such potential is proportional to the sum of tensions of the configurations, namely
\beq
V = {T_{D6} \over \lambda_{II}} 
\left( \sum_a N_a {\rm Vol} (\Pi_a) + Q_{O6} N_{O6} {\rm Vol} (\Pi_{O6})\right),
\label{potential}
\eeq
where $T_{D6} = (2\pi)^{-6} \a^{\prime -7/2}$ is the tension of a D6-brane, $\lam_{II}$ is the (ten dimensional) type II string coupling constant, $Q_{O6} = -4$  the relative NSNS charge with respect to the D6-branes and $N_{O6}$ the number of O6-planes in the theory. For small deviations from a supersymmetric configuration, this superpotential can be understood in terms of a D-term generated potential. Indeed, in a $\N=1$ supersymmetric field theory such a potential is given by
\beq
V_{D-{\rm term}} = \sum_a {1 \over 2 g_{U(1)_a}^2} \left(\sum_i q_a^i|\phi_i|^2 + \xi_a \right)^2,
\label{potentialFI}
\eeq
where $g_a$ is the gauge coupling of each $U(1)_a$ factor, $\xi_a$ is a FI parameter associated to it, and $\phi_i$ are scalars fields charged under $U(1)_a$ with charge $q_a^i$. In \cite{Cremades:2002te} it was shown how, in toroidal or orbifold compactifications, the term $\sum_a \xi_a^2/2g_a^2$ of this potential can be associated with the scalar potential (\ref{potential}). Let us now show it for a more general ${\bf CY_3}$ geometry. First consider a supersymmetric configuration, where (\ref{potential}) vanishes and every single object, D6-branes and O6-planes, are calibrated by a volume form Re $(\O_{\tilde r})$. Let us now slightly deform this configuration from the supersymmetric case, by varying the complex structure moduli while maintaining the fact that each D6-brane individually wraps a special Lagrangian $\Pi_a$, not calibrated by Re $(\O_{\tilde r})$ but by Re $(e^{i\th_a} \O_{\tilde r})$. The scalar potential can be computed to be
\bea
V & = & {T_{D6} \over \lambda_{II}} 
\left( \sum_a N_a \left| \int_{[\Pi_a]} \O_{\tilde r} \right| + Q_{O6} N_{O6} \int_{[\Pi_{O6}]} {\rm Re} (\O_{\tilde r}) \right) \nonumber \\
& = & {T_{D6} \over \lambda_{II}} 
\sum_a N_a \left( \int_{[\Pi_a]}\left|\O_{\tilde r}\right| - \int_{[\Pi_a]} {\rm Re} (\O_{\tilde r}) \right) \nonumber \\
& = & {T_{D6} \over \lambda_{II}} 
\sum_a N_a  \int_{[\Pi_a]}\left|\O_{\tilde r}\right| \cdot (1 - {\rm cos\ } \th_a) \nonumber \\
& = & \sum_a {1 - {\rm cos\ } \th_a \over g_{U(1)_a}^2 (2\pi \a^\prime)^2} \ \approx \ \sum_a {1 \over 2 g_{U(1)_a}^2} \left({\th_a\over 2 \pi \a^\prime}\right)^2,
\label{potential2}
\eea
where in the second equality we have used the RR tadpole condition $Q_{O6} N_{O6} + \sum_a N_a [\Pi_a] =0$, and in the last line we have used the definitions of $g_{U(1)_a}^2$, $T_{D6}$ and supposed $\th_a$ to be very small. The quantities $\th_a$ are nothing but the non-trivial phase of the central charge $Z_a$ of each D6-brane stack, after the small deformation of complex structure moduli. They can also be seen as the separation from a `tetrahedron wall'\footnote{Strictly speaking, in a proper Calabi-Yau threefold there is not several walls making up a tetrahedron, but only one of them. In the mathematical literature, this codimension one surface is known as Marginal Stability wall.} in the angle space, where the angles in figure \ref{tetrahedron} are to be measured from the relative position of the tangent planes of $\Pi_a$ and $\Pi_{O6}$ at their intersection. This latter vision connects with the FI-parameter interpretation in \cite{Kachru:1999vj}.

\item{\it Superpotential}

The volume form $\O$ is not the only calibration of a Calabi-Yau threefold, but also the K\"ahler form $\om$ is a calibration. Since it is a two-form, it will not calibrate D-branes in type IIA theory but worldsheet instantons. In a supersymmetric theory, these will generate nonperturbative corrections to the superpotential, as we will see in the next chapter.

\end{itemize}

\section{An MSSM-like example \label{example}}

Let us illustrate the above general discussion with one specific example. In order to connect with Standard Model physics as much as possible, we will choose an intersecting brane model with a semi-realistic chiral spectrum, namely, that of the MSSM. As has been pointed out in the previous chapter, it seems impossible to get an intersecting D6-brane model with minimal Standard Model-like chiral spectrum from plain toroidal or orbifold compactifications of type IIA string theory. One is thus led to perform an extra orientifold twist $\OR$ on the theory, $\Om$ being the usual worldsheet parity reversal and $\R$ a geometric antiholomorphic involution of the compact space $\M_6$. The set of fixed points of $\R$ will lead to the locus of an O6-plane \footnote{When dealing with orbifold constructions, several O-planes may appear. More precisely, each fixed point locus of $\R\iota$ with $\iota$ an element of the orbifold group satisfying $\R\iota^2 = 1$ will lead to an O-plane \cite{Blumenhagen:1999md,Blumenhagen:1999ev,Forste:2000hx}.}. 

In addition, $\R: \M_6 \raw \M_6$ will induce an action on the homology of $\M_6$, more precisely on $H_3(\M_6, \inte)$, where our D6-branes wrap.
\beq
\begin{array}{rc}
\R : & H_3(\M_6, \inte) \raw H_3(\M_6, \inte) \\
& [\Pi_\a] \mapsto [\Pi_{\a^*}]
\end{array}
\label{Rhomol} 
\eeq
Thus, as stated before, an orientifold configuration must consist of pairs ($\Pi_\a$, $\Pi_{\a^*}$). If $\R(\Pi_\a) \neq \Pi_\a$, then a stack of $N_\a$ D6-branes on $\Pi_\a$ will yield a $U(N_\a)$ gauge group, identified with that on $\Pi_{\a^*}$ by the action of $\OR$ (i.e., complex conjugation). If, on the contrary, $\R(\Pi_\a) = \Pi_\a$, the gauge group will be real ($SO(2N_\a)$) or pseudoreal ($USp(2N_\a)$).

We will use this simple fact when constructing our MSSM-like example. Indeed, notice that $USp(2) \cong SU(2)$, so in an orientifold setup $SU(2)_L$ weak interactions could arise from a stack of two branes fixed under $\R$. We will suppose that this is the case, which consists on a slight variation from the SM brane content of table \ref{SMbranes}. The new brane content is presented in table \ref{SMbranes3}.
\begin{table}[htb]
\renewcommand{\arraystretch}{2}
\begin{center}

\begin{tabular}{|c|c|c|c|} 
\hline 
Label & Multiplicity & Gauge Group & Name \\ 
\hline 
\hline 
stack $a$ & $N_a = 3$ & $SU(3) \times U(1)_a$ & Baryonic brane\\ 
\hline 
stack $b$ & $N_b = 1$ & $SU(2)$ & Left brane\\ 
\hline 
stack $c$ & $N_c = 1$ & $U(1)_c$ & Right brane\\ 
\hline 
stack $d$ & $N_d = 1$ & $U(1)_d$ & Leptonic brane \\ 
\hline 
\end{tabular} 
\end{center}
\caption{Brane content yielding an MSSM-like spectrum. Only one representative of each brane is presented, not including the mirror branes $a^*, b^*, c^*, d^*$. Although $N_b = 1$, the mirror brane of $b$ lies on top of it, so it actually be considered as a stack of two branes which, under $\Om$ projection, yield a $USp(2) = SU(2)$ gauge group.\label{SMbranes3}}

\end{table}

Given this brane content, we can construct an intersecting brane model with the chiral content of the Standard Model (plus right-handed neutrinos) just by considering the following intersection numbers
\beq 
\begin{array}{lcl} 
I_{ab}\ = \ 3, \\ 
I_{ac}\ = \ -3, & & I_{ac^*}\ =\ -3, \\ 
I_{db}\ = \ 3, \\ 
I_{dc}\ = \ -3, & & I_{dc^*}\ =\ -3,
\end{array} 
\label{intersec} 
\eeq 
all the other intersection numbers being zero (we have not included those involving $b$*$ = b$). This chiral spectrum and the relevant non-Abelian and $U(1)$ quantum numbers have been represented in table \ref{mssm}, together with their identification with SM matter fields.
\begin{table}[htb]
\renewcommand{\arraystretch}{1.5}
\begin{center}

\begin{tabular}{|c|c|c|c|c|c|c|}
\hline Intersection &
 SM Matter fields  & $SU(3) \ti SU(2)$  &  $Q_a$   & $Q_c $ & $Q_d$  & Y \\
\hline\hline (ab) & $Q_L$ &  $3(3,2)$ & 1   & 0 & 0 & 1/6 \\
\hline (ac) & $U_R$   &  $3( {\bar 3},1)$ & -1  &  1  & 0 & -2/3 \\
\hline (ac*) & $D_R$   &  $3( {\bar 3},1)$ &  -1  & -1    & 0 & 1/3
\\
\hline (db) & $L$    &  $3(1,2)$ &  0    & 0  & 1 & -1/2  \\
\hline (dc) & $N_R$   &  $3(1,1)$ &  0  & 1  &  -1  &  0  \\
\hline (dc*) & $E_R$   &  $3(1,1)$ &  0  & -1  & -1    & 1 \\
\hline 
\end{tabular}
\end{center}
\caption{Standard model spectrum and $U(1)$ charges. The hypercharge generator is defined as $Q_Y = \frac 16 Q_a - \frac 12 Q_c - \frac 12 Q_d$.\label{mssm}}
\end{table}

Is easy to see that this spectrum is free of chiral anomalies, whereas it has an anomalous $U(1)$ given by $3 U(1)_a + U(1)_d$. Such anomaly will be canceled by a Green-Schwarz mechanism, the corresponding $U(1)$ gauge boson getting a Stueckelberg mass. The two non-anomalous $U(1)$'s can be identified with $(B-L)$ and the 3$^{rd}$ component of right-handed weak isospin, just as in last chapter SM models. This implies that the low energy gauge group is in principle $SU(3) \ti SU(2) \ti U(1)_{B-L} \ti U(1)_R$, giving the SM gauge group plus an extra $U(1)$. However, as we have seen in previous SM models, it may well happen that non-anomalous $U(1)$'s get also a mass by this same mechanism, the details of this depending on the specific homology cycles $[\Pi_\a]$, $\a = a, b, c, d$. This implies that in some specific constructions we could have only the SM gauge group. The Higgs system, which should arise from the $bc$ and $bc^*$ sector, gives no net chiral contribution and thus it does not appear at this abstract level of the construction, its associated spectrum depending on the particular realisation of (\ref{intersec}) in terms of homology cycles (see below).

\begin{figure}[ht]
\centering
\epsfxsize=5.3in
\hspace*{0in}\vspace*{.2in}
\epsffile{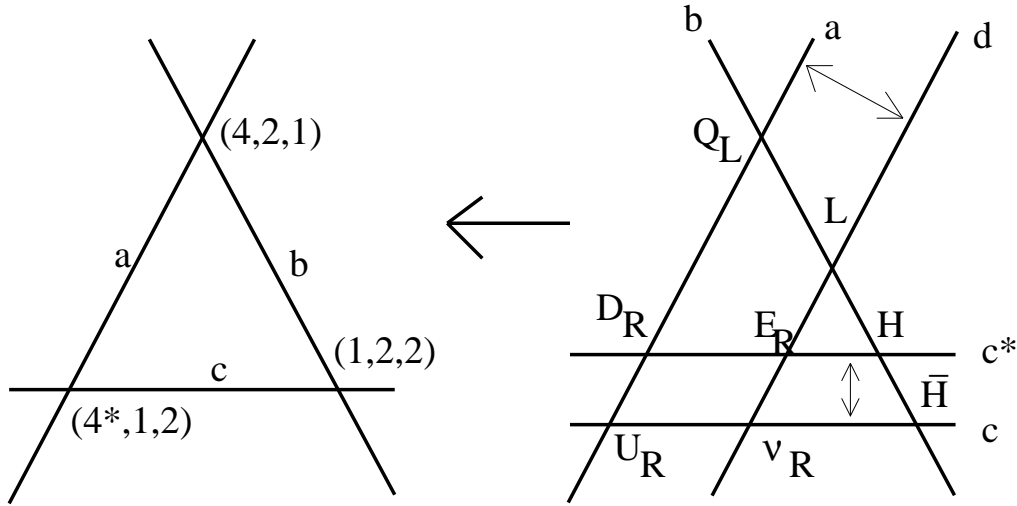}
\caption{Scheme of the model discussed in the text. Moving brane $c$ on top of its mirror $c$* one gets an enhanced $SU(2)_R$ symmetry. If in addition brane $d$ is located on top of brane $a$ one gets an enhanced $SU(4)$ Pati-Salam symmetry.}
\label{guayps}
\end{figure}

Notice that the intersection numbers (\ref{intersec}) allow for the possibility $[\Pi_{c}] =  [\Pi_{c^*}]$. This would mean that, at some points on the moduli space of configurations $\Pi_{c} =  \Pi_{c^*}$ and the  stack $c$ gauge group could be enhanced as $U(1)_c \raw SU(2)_R$, just as for stack $b$. We would then recover a left-right symmetric model, continuously connected to the previous Standard Model-like configuration. By the same token, we could have $[\Pi_{a}] =  [\Pi_{d}]$, so when both stacks lied on top of each other we would get an enhancement $SU(3) \ti U(1)_{B-L} \raw SU(4)$. Considering both possibilities, one is naturally led to a intersecting brane configuration yielding a Pati-Salam spectrum, as has been drawn schematically in figure \ref{guayps}.

Let us now give a specific realisation of such abstract construction. For simplicity, we will again consider a plain orientifold of type IIA compactified on a $T^6 = T^2 \ti T^2 \ti T^2$, with $\R: z^i \mapsto \bar z^i$ $i = 1,2,3$ being a simultaneous reflection on each complex plane. Our D6-branes will wrap factorisable cycles (\ref{factorisable3}) and the action of $\R$ on such 3-cycles will be given by $\R : [(n_\a^{(i)}, m_\a^{(i)})] \mapsto [(n_\a^{(i)}, -m_\a^{(i)})]$ $i=1,2,3$, where we are considering rectangular two-tori (i.e., $\b^i = 1$, $i=1,2,3$). Given this choice of geometry, the compactification involves eight different O6-planes, all of them wrapped on rigid 3-cycles in $[\Pi_{O6}] = \bigotimes_{i=1}^3 [(1,0)]^i$. 

A particular class of configurations satisfying (\ref{intersec}) in this specific setup is presented in table \ref{wnumbers}. A quick look at the wrapping numbers shows that this brane content by itself does not satisfy RR tadpole conditions $\sum_\a ([\Pi_\a] + [\Pi_{\a^*}]) = 32 [\Pi_{O6}]$, although it does cancel all kind of chiral anomalies arising from the gauge groups in table \ref{SMbranes3}. However, additional anomalies would appear in the worldvolume of D-brane probes as, e.g., D4-branes wrapping arbitrary supersymmetric 3-cycles \cite{Uranga:2000xp}. This construction should then be seen as a submodel embedded in a bigger one, where extra RR sources are included. These may either involve some hidden brane sector or NSNS field strength background fluxes \cite{Uranga:2002vk}, neither of these possibilities adding a {\it net} chiral matter content \cite{Cremades:2002cs}. As our main interest in this section is to give an example of D-brane sector where a $\N=1$ SUSY is preserved, we will not dwell on the details of such embedding.
\begin{table}[htb]
\renewcommand{\arraystretch}{2.5}
\begin{center}
\begin{tabular}{|c||c|c|c|}
\hline
 $N_\a$  &  $(n_\a^{(1)},m_\a^{(1)})$  &  $(n_\a^{(2)},m_\a^{(2)})$   
&  $(n_\a^{(3)},m_\a^{(3)})$ \\ 
\hline\hline $N_a=3$ & $(1,0)$  &  $(1/\rho , 3\rho )$ &
 $(1/\rho  ,  -3\rho )$  \\
\hline $N_b=1$ &   $(0, 1)$    &  $ (1,0)$  & 
$(0,-1)$   \\
\hline $N_c=1$ & $(0,1)$  & 
 $(0,-1)$  & $(1,0)$  \\
\hline $N_d=1$ &   $(1,0)$    &  $(1/\rho ,3\rho )$  &
$(1/\rho , -3\rho )$   \\
\hline \end{tabular}
\caption{D6-brane wrapping numbers giving rise to a the chiral spectrum of the MSSM. Here $\rho = 1,1/3$. The case $\rho = 1$ has been depicted in figure \ref{guay}.\label{wnumbers}}
\end{center}
\end{table}

Notice that this realisation satisfies the constraints $[\Pi_a] = [\Pi_d]$ and $[\Pi_c] = [\Pi_{c^*}]$. Moreover, both $[\Pi_b]$ and $[\Pi_c]$ have a $USp(2N_\a)$ gauge group when being invariant under the orientifold action. This can easily be seen, since in the T-dual picture they correspond to type I D5-branes, which by the arguments of \cite{Gimon:1996rq} have symplectic gauge groups. As a result, this configuration of D6-branes satisfies the necessary conditions for becoming a Pati-Salam model in a subset of points of its open string moduli space (i.e., brane positions and Wilson lines). 

In addition, by varying the complex structure on the second an third tori we can achieve the same $\N = 1$ SUSY to be preserved at each intersection. Indeed, the twist vectors relating the O6-plane and the D6-brane stack $\a$ by a rotation (\ref{rotation3}) are
\beq
\begin{array}{rcl} \vspace{0.1cm}
v_{O6\ a} & = & (0, \th^{ 2}, - \th^{ 3}, 0) \\\vspace{0.1cm}
v_{O6\ b} & = & (\med, 0 , -\med, 0) \\\vspace{0.1cm}
v_{O6\ c} & = & (\med, -\med, 0, 0) \\
\ v_{O6\ d} & = & (0, \th^{ 2}, - \th^{ 3}, 0)
\end{array}
\label{guaytwists}
\eeq
where $\th^i \equiv \frac 1\pi {\rm tg }^{-1} ( 3\rho^2 \frac{R_2^{(i)}}{R_1^{(i)}})$. So, if we set $R_2^{(2)}/R_1^{(2)} = R_2^{(3)}/R_1^{(3)} = \chi$, then $\th^2 = \th^3$ and the same $\N = 1$ SUSY will be preserved by each of the above rotations, since
\beq
\tilde r \cdot v_{O6\ \a} = 0 \quad {\rm for} \quad \tilde r = \med (+,+,+,+), \quad \a = a, b, c, d 
\label{N=1}
\eeq
and hence the first volume form in (\ref{omegasO6}) will be left invariant by the corresponding rotation $R_{O6\ \a}$. Since the twist vectors relating two D6-branes are additions and subtractions of those in (\ref{guaytwists}) (i.e., $v_{ab} = v_{O6\ b} - v_{O6\ a}$, etc.) the condition (\ref{N=1}) will hold for the whole configuration. The whole system in table \ref{mssm} will then be a sum of BPS states, all calibrated by Re $(\O_{(+,+,+)})$. As a result, each chiral fermion in table \ref{mssm} will be accompanied by a scalar superpartner, yielding an MSSM-like spectrum.

Let us finally discuss the Higgs sector of this model. As mentioned before, stacks $b$ and $c$ correspond, in a T-dual picture, to two (dynamical) D5-branes wrapped on the second and third tori, respectively. Both D5's yield a $SU(2)$ gauge group when no Wilson lines are turned on their worldvolumes and, if they are on top of each other in the first torus, the massless spectrum in their intersection amounts to a $\N = 2$ hypermultiplet in the representation $(2,2)$, invariant under CPT. This can also be seen as a $\N = 1$ chiral multiplet. Turning back to the branes at angles picture we see that the intersection number $[\Pi_b]\cdot [\Pi_c] = 0$ because stacks $b$ and $c$ are parallel in the first torus, while they intersect only once in the remaining two tori. This single intersection will give us just one copy of the $(2,2)$ $\N = 1$ chiral multiplet described above, whenever there exist the gauge enhancement to $SU(2) \ti SU(2)$. This will happen for stack $b$ when placed on top of any O6-plane on the second torus, and no Wilson line is turned on that direction. A similar story applies for stack $c$ in the third torus. Since we have no special interest in a gauge group $SU(2)_R$, we will consider arbitrary positions and Wilson lines for $c$ (see figure \ref{guay} for such a generic configuration). In that case, our $(2,2)$ chiral multiplet will split into two doublets charged as $2_{-1}$ and $2_{+1}$ under $SU(2) \ti U(1)_c$, which can be identified with the MSSM Higgs particles $H_u$ and $H_d$, respectively. In addition, there exists a Coulomb branch between stacks $b$ and $c$ ($c$*), which corresponds to either geometrical separation in the first torus, either different 'Wilson line' phases along the 1-cycle wrapped on this $T^2$.\footnote{The complex phases associated to the 1-cycles of stacks $b$ and $c$ cannot be called Wilson lines in the strict sense, as they do not transform in the adjoint of $SU(2)$ but in the antisymmetric.} From the point of view of MSSM physics, these quantities can be interpreted as the real and imaginary part of a $\mu$-parameter, which is the only mass term for both Higgs doublets allowed by the symmetries of the model\footnote{Indeed, the associated term in the superpotential has been computed in the T-dual picture of type I D5-branes in \cite{Berkooz:1996dw}, and shows the appropriate behaviour of a $\mu$-term.}.

After all these considerations, we see that the massless spectrum of table \ref{mssm} is that of the MSSM with a minimal Higgs set. Such a model was first presented in \cite{Cremades:2002qm}, where some of its phenomenology as FI-terms were briefly studied. In the next chapter, we will compute  the Yukawa couplings associated to such model which, as we will see, are particularly simple.

Notice that if we want to have a consistent compactification where RR tadpoles cancel, we must add extra RR sources to the brane content of table \ref{SMbranes3}. These extra sources will break the previous supersymmetry: some NSNS tadpoles will remain uncanceled and a scalar potential will be generated. In fact, is easy to see that NSNS tadpoles will generically not vanish in the kind of toroidal orientifolds discussed above, where due to the simple action $(1 + \OR)$ the O6-plane wraps a factorisable 3-cycle. Indeed, for NSNS tadpoles to vanish the scalar potential (\ref{potential}) should vanish as well. If all the D6-planes wrap the same factorisable 3-cycle, say (\ref{orienti}), this will be impossible unless the D6-branes all wrap $[\Pi_{O6}]$. However, this leads to a non-chiral theory, namely type I compactified on $T^6$. A solution is enlarging the orientifold group by an orbifold action, which implies the existence of several O6-planes wrapping different factorisable 3-cycles. This idea has been successfully applied in \cite{Cvetic:2001nr}, and later in \cite{Blumenhagen:2002gw,Honecker:2003vq} in order to construct $\N=1$ chiral compactifications from intersecting D6-branes.

\chapter{Yukawa couplings \label{yukint}}

In the quest for obtaining a realistic string-based model, generic properties of the low-energy effective Lagrangian such as $D=4$ chirality and unitary gauge groups are of fundamental importance. Once these have been found in a particular setup of string theory, there are still many other issues to face in order to reproduce some realistic physics at low energies. In particular, even if one manages to obtain a massless spectrum quite close to the Standard Model (or some extension of it), one is eventually faced with the problem of computing some finer data defining a Quantum Field Theory. These data may tell us how close are we of reproducing the SM which, as we know, is not a bunch of chiral fermions with appropriate quantum numbers, but an intricate theory with lots of well-measured parameters. The aim of this chapter is to address the computation of some of these finer data, namely Yukawa couplings, in the context of Intersecting Brane Worlds. 

\begin{figure}[ht]
\centering
\epsfxsize=3.4in
\hspace*{0in}\vspace*{.2in}
\epsffile{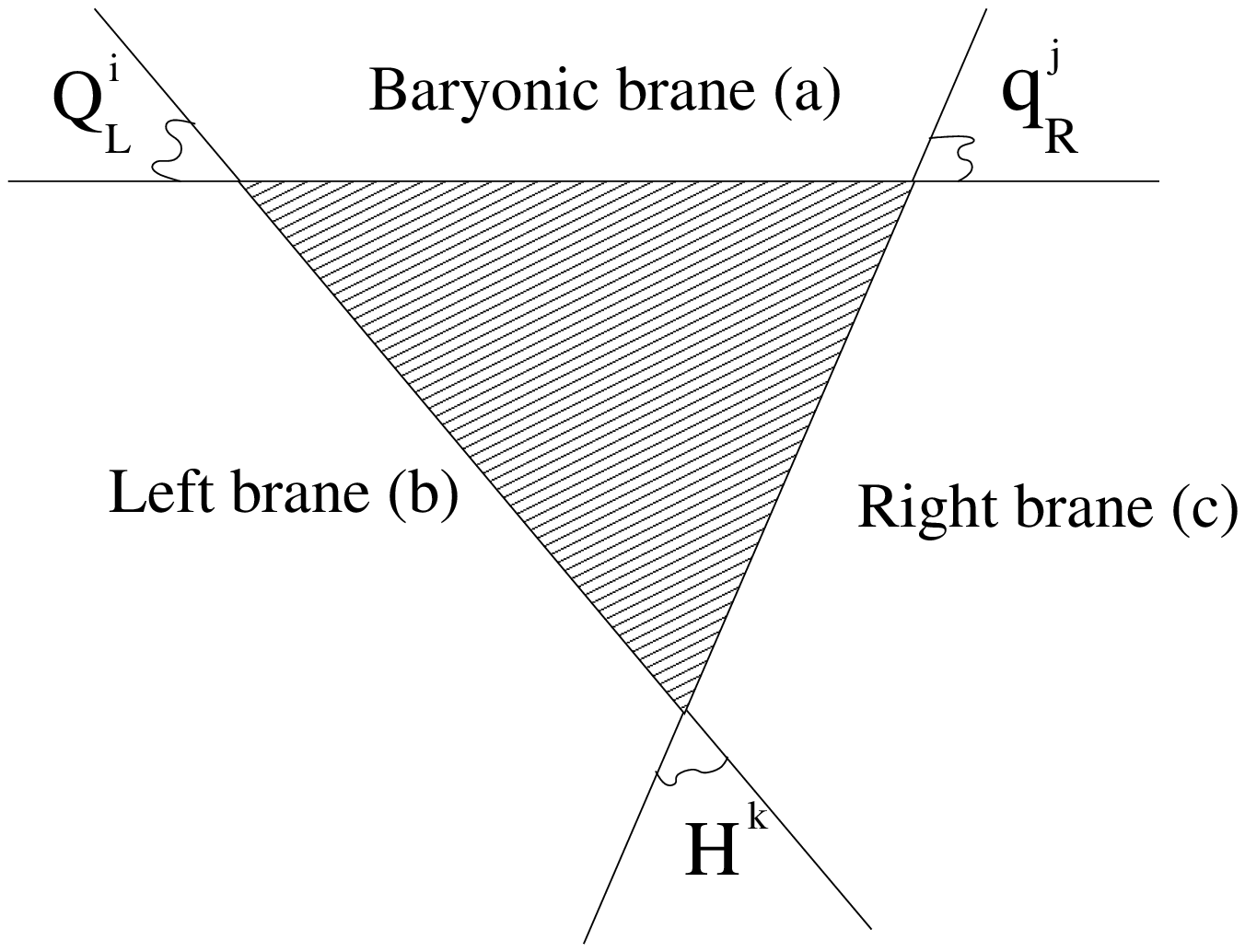}
\caption{Yukawa coupling between two quarks of opposite chirality and a Higgs boson.}
\label{yukis2}
\end{figure}

As advanced in \cite{Aldazabal:2000cn}, Yukawa couplings in intersecting D-brane models arise from open string worldsheet instantons connecting three D-brane intersections, in such a way that the open string states there located have suitable Lorentz and gauge quantum numbers to build up an invariant in the effective Lagrangian. This will usually involve the presence of three different D-branes, which determine the boundary conditions of the worldsheet instanton contributing to this Yukawa coupling, see figure \ref{yukis2}. 

Roughly speaking, the instanton contribution to the Yukawa coupling will be given by evaluating the classical action $e^{-S_{\rm cl}}$ on the surface of minimal area connecting the three intersections. As a result, Yukawa couplings will depend on several moduli of the theory, such as D-brane positions (open string moduli) and the compact manifold metric (closed string moduli). We will try to extract this dependence explicitly in the case where we have centered our attention throughout this thesis: D-branes wrapping factorisable $n$-cycles on $T^{2n}$. In doing so, the use of calibrated geometry will be crucial, since these worldsheet instantons turn out to be two-dimensional surfaces calibrated under the K\"ahler form $\om$. The general expression for a Yukawa coupling in this toroidal geometry involves products of theta functions, the moduli dependence being encoded in their parameters. This simple form will allow us to solve exactly the mass spectrum of particles after spontaneous symmetry breaking in some simple models, as the MSSM-like example of last chapter. Finally, we will comment on the deep connection between Yukawa couplings in intersecting brane worlds and Mirror Symmetry. 

\section{Intersecting brane models and Yukawa couplings \label{ibm&yuk}}

In this section we review intersecting D-brane models from a general viewpoint, collecting the necessary information for addressing the problem of Yukawa couplings. Most part of the effort on constructing phenomenologically appealing intersecting brane configurations has centered on simple toroidal and on orbifold/orientifold compactifications. Nevertheless, we have seen that important issues as massless chiral spectrum and tadpole cancellation conditions are of topological nature, thus easily tractable in more general compactifications where the metric may not be known explicitly. Following this general philosophy, we will introduce Yukawa couplings as arising from worldsheet instantons in a generic compactification. Although the specific computation of these worldsheet instantons needs the knowledge of the target space metric, many important features can be discussed at this more general level. In the next section we will perform such explicit computation in the simple case of toroidal compactifications, giving a hint of how these quantities may behave in a more general setup.

\subsection{The role of worldsheet instantons}

In the context of a general Calabi-Yau compactification of type IIA theory, a D6-brane wrapping a special Lagrangian (sL) submanifold does not only minimize its volume but also preserves some amount of supersymmetry, being a BPS stable object. Thus, it seems natural to consider such objects as building blocks of a generic intersecting D-brane configuration on ${\bf CY_3}$. In particular, we know that a stack of $N_\a$ D6-branes wrapping a sL $\Pi_\a$ will yield a $U(N_\a)$ Supersymmetric Yang-Mills theory on its worldvolume. A natural question is which amount of $D=4$ SUSY the D6-brane effective field theory will have. As we saw in the last chapter, the precise amount is given by the number of independent real volume forms Re$(e^{i\th}\Om)$ that calibrate the 3-cycle. We may, however, seek for a more topological alternative method.

McLean's theorem \cite{McLean} states that the moduli space of deformations  of a sL $\Pi_\a$ is a smooth manifold of (unobstructed) real dimension $b_1(\Pi_\a)$. String theory complexifies this space by adding $b_1(\Pi_\a)$ Wilson lines, obtained from the gauge field $U(N_\a)$ living on the worldvolume of the stack $\a$. In the low energy $D=4$ theory, this will translate into $b_1(\Pi_\a)$ massless complex scalar fields in the adjoint of $U(N_\a)$. Being in a supersymmetric theory, these fields will yield the scalar components of $D = 4$ supermultiplets. 

Let us consider the case $\N = 1$ (for a clear discussion on this see, e.g., \cite{Kachru:2000tg}). Here we find $b_1(\Pi_\a)$ chiral multiplets $\phi_j$ in the adjoint of $U(N_\a)$. Now, we may also seek to compute the superpotential $W$ of such $\N = 1$ theory which involves these chiral fields. By standard $\N = 1$ considerations, this superpotential cannot be generated at any order in $\a'$ perturbation theory, in accordance with the geometrical result of \cite{McLean}. Indeed, such superpotential will be generated non-perturbatively by worldsheet instantons with their boundary in $\Pi_\a$.

Superpotentials generated non-perturbatively by worldsheet instantons were first considered in closed string theory \cite{Dine:1986zy,Dine:bq}, while the analogous problem in type IIA open string has been recently studied in \cite{Aganagic:2000gs,Aganagic:2001nx,Acharya:2002ag,Kachru:2000ih,Kachru:2000an},  in the context of open string mirror symmetry. The basic setup considered is a {\it single} D6-brane  wrapping a sL $\Sig $  in a ${\bf CY_3}$, with $b_1(\Sig) > 0$. Worldsheet instantons are constructed by considering all the possible embeddings of a Riemann surface ${\cal S}_g$ with arbitrary genus $g$ on the target space $\M_6$ and with boundary on $\Sig$. In order to be topologically non-trivial, this boundary must be wrapped on the 1-cycles $\g_j$ that generate $b_1(\Sig)$, thus coupling naturally to the corresponding chiral multiplets. Moreover, in order to deserve the name instanton, this euclidean worldsheet embedding must satisfy the classical equations of motion. This is guaranteed by considering embeddings which are holomorphic (or antiholomorphic) with respect to the target space complex structure, plus some extra constraints on the boundary (Dirichlet conditions). In geometrical terms, this means that {\em worldsheet instantons must be surfaces calibrated by the K\"ahler form $\om$}. Calibration theory then assures the area minimality given such boundary conditions, which is what we would expect from naive Nambu-Goto considerations. As a general result, it is found that the superpotential of D6-brane theories is entirely generated by instantons with the topology of a disc, while higher-genus instantons correspond to open string analogues of Gromov-Witten invariants.

So we find that, in case of $\N = 1$ D6-branes on a ${\bf CY_3}$, great deal can be extracted from calibrated geometry of the target space $\M_6$. Whereas the gauge kinetic function $f_{ab}$ can be computed by evaluating the volume form $\Om$ on the worldvolume $\Sig$ of the brane (see last chapter), the superpotential can be computed by integrating the K\"ahler form $\om$ on the holomorphic discs with boundary on $\Sig$. The former only depends on the homology class $[\Sig] \in H_3(\M_6, \inte)$. The latter, on the contrary, is given by a sum over the relative homology class $H_2^D(\M_6,\Sig)$, that is, the classes of 2-cycles on $\M_6$ with boundary on $\Sig$ (the superscript $D$ means that we only consider those 2-cycles with the topology of a disc). Notice that, $\M_6$ being compact, the disc instantons may wrap multiple times. Although in principle one may need the knowledge of the metric on $\M_6$ in order to compute both, much can be known about the form of the superpotential by considerations on Topological String Theory. For our purposes, we will contempt to stress two salient features:
\begin{itemize}

\item The superpotential depends on the target space metric only by means of K\"ahler moduli, and is independent of the complex structure \cite{Brunner:1999jq}.

\item If we see those K\"ahler moduli as closed string parameters, the dependence of the superpotential is roughly of the form
\beq
W = \pm \sum_{n=1}^\infty {e^{-n\Phi} \over n^2},
\label{super}
\eeq
where $n$ indexes the multiple covers of a disc with same boundary conditions, and $\Phi$ stands for the open string chiral superfield \cite{Ooguri:1999bv}. The $\pm$ sign corresponds to holomorphic and antiholomorphic maps, respectively.

\end{itemize}

Given these considerations, is easy to see that no superpotential will be generated for a chiral superfield associated to a 1-cycle $\g$ of $\Sig$ which is also non-contractible in the ambient space $\M_6$, since no disc instanton exist that couples to such field. Notice that this could never happen in a ${\bf CY_3}$ on the strict sense, since in this case $b_1(\M_6) = 0$. In manifolds with lower holonomy, however, it may well be the case that $b_1(\M_6) > 0$, and so a D6-brane could have in its worldvolume a complex scalar not involved in the superpotential (\ref{super}). We expect such scalars to give us the scalar content of the vector supermultiplet, thus indicating the degree of supersymmetry on the worldvolume effective theory. This seems an alternative method for computing the amount of supersymmetry that such a D6-brane preserves. A clear example of the above argument is constituted by $\M_6 = T^2 \ti T^2 \ti T^2$ and $\Sig$ a Lagrangian $T^3$ (the so-called factorisable branes). Here, each of the three independent 1-cycles on $\Sig$ is non-contractible in $\M_6$, so our SYM theory will yield three complex scalars not involved in the superpotential. But these scalar fields fill in the precise content of a $D = 4$ $\N = 4$ vector multiplet, which is the amount of SUSY those branes preserve.

\subsection{Yukawa couplings in intersecting D-brane models}

\begin{figure}[ht]
\begin{center}
\begin{tabular}{ll}
\\
\hskip -0.5cm
\epsfig{file=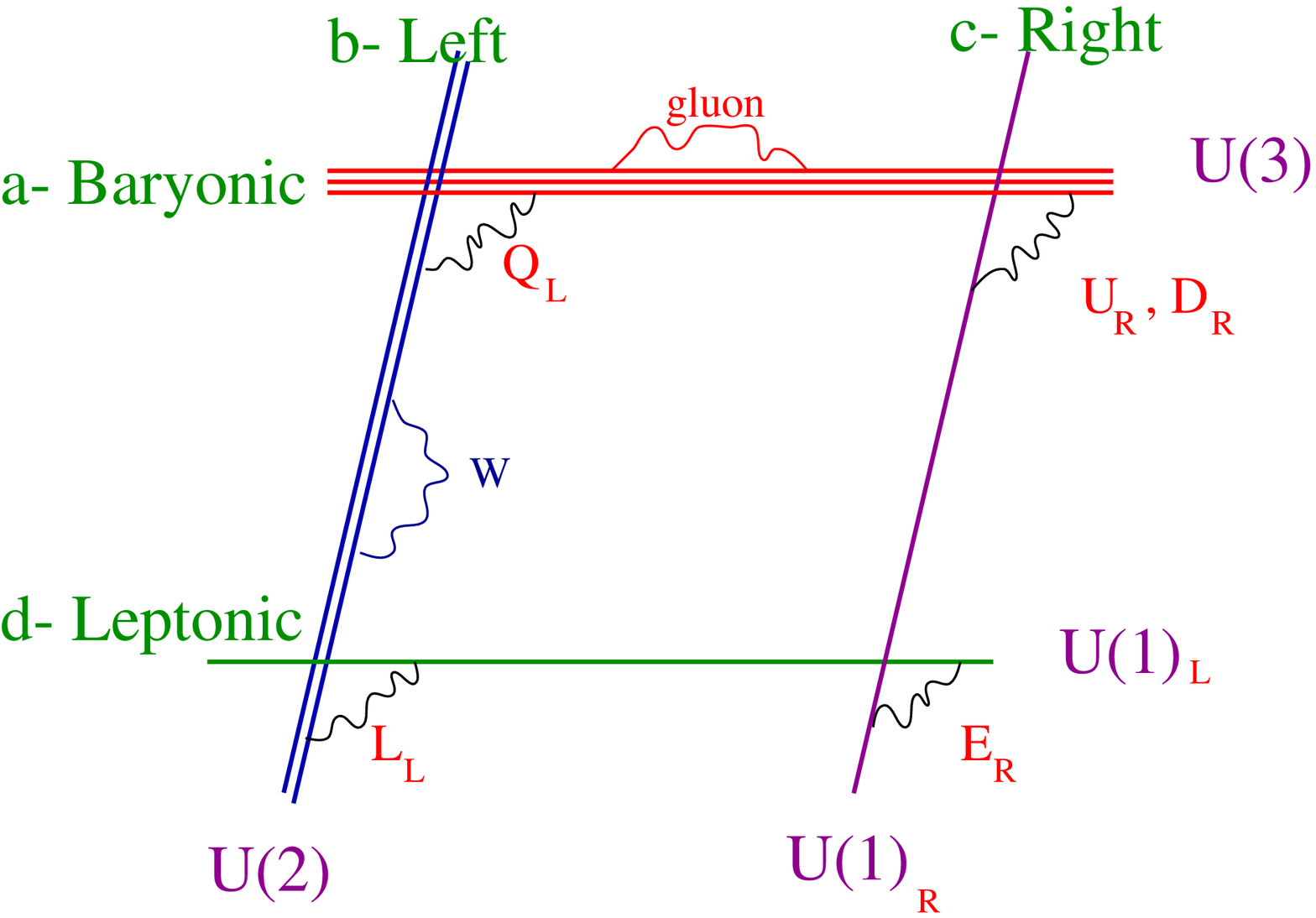, height=5.5cm} &
\epsfig{file=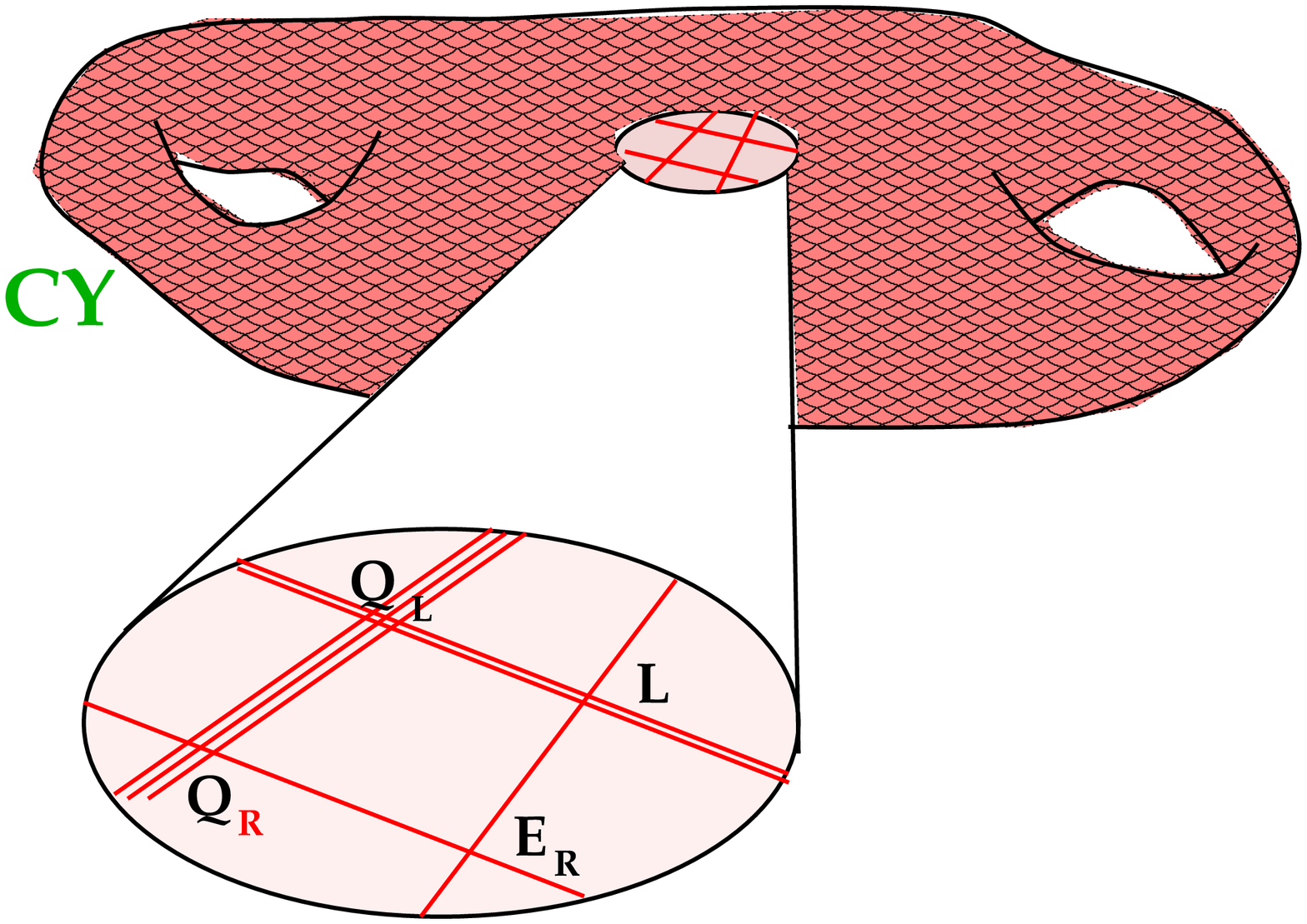, height=5.5cm}\\
\hskip 3.1truecm
{\small (a)}            &
\hskip 3.65truecm
{\small (b)}\\
\end{tabular}
\end{center}
\caption{The Standard Model at intersecting D-branes. a) Four stacks of branes, {\it baryonic, left, right } and {\it leptonic} are needed to get all quark and leptons at the intersections. b) The SM branes may be wrapping cycles on a, e.g., ${\bf CY_3}$ manifold, with appropriate intersection numbers so as to yield the SM chiral spectrum of table \ref{SMcontent} or \ref{mssm}.}
\label{sm}
\end{figure}

Up to now, we have only considered superpotentials arising from one single stack of D6-branes. In the intersecting brane world picture we have given above, however, chiral matter in the bifundamental arises from the intersection of two stacks of branes, each with a different gauge group. It thus seems that, in order to furnish a realistic scenario, several stacks of branes are needed. In fact, given the semi-realistic model-building philosophy considered in Chapter \ref{models}, it seems that a minimal number of four stacks of branes are necessary in order to accommodate the chiral content of the Standard Model in bifundamentals. 
Notice that the general discussion there applies equally well to a Calabi-Yau compactification, so we would again have the stacks named as {\it Baryonic (a)}, {\it Left (b)}, {\it Right (c)} and {\it Leptonic (d)}, which multiplicities $N_a = 3$, $N_b = 2$, $N_c = 1$ and $N_d = 1$ and wrapping special Lagrangian submanifolds of a ${\bf CY_3}$. The SM gauge group and $U(1)$ structure will appear just as in the toroidal case, and SM chiral fermions will naturally arise from pairs of intersecting stacks. This scenario has been depicted schematically in figure \ref{sm}.

Notice that considering a full D6-brane configuration instead of one single brane makes the supersymmetry discussion more involved. Although each of the components of the configuration (i.e., each stack of D6-branes) is wrapping a special Lagrangian cycle and thus yields a supersymmetric theory on its worldvolume, it may well happen that two cycles do not preserve a common supersymmetry. In a ${\bf CY_3}$ of $SU(3)$ holonomy this picture is conceptually quite simple. There only exist one family of real volume forms $\Om$ parametrized by a phase $e^{i\th}$. Two sL's $\Pi_\a$, $\Pi_\b$ will preserve the same supersymmetry if they are calibrated by the same real 3-form, that is, if $\th_\a = \th_\b$ in (\ref{cali2}). In this case, a chiral fermion living at the intersection $\Pi_\a \cap \Pi_\b$ will be accompanied by a complex scalar with the same quantum numbers, filling up a $\N = 1$ chiral multiplet \footnote{Departure from the equality of angles will be seen as  Fayet-Iliopoulos terms in the effective $D=4$ field theory. Contrary to the superpotential, these FI-terms are predicted to depend only on the complex structure moduli of the ${\bf CY_3}$. These aspects have been explored in \cite{Douglas:1999vm,Kachru:1999vj} in the general case, and computed from the field theory perspective in the toroidal case in \cite{Cremades:2002te}.}. In manifolds of lower holonomy, however, there are far more possibilities, since many more  SUSY's are involved. Consideration of such possibilities lead to the idea of  Quasi-Supersymmetry in \cite{Cremades:2002te,Cremades:2002cs} (see \cite{Klein:2002vu,Klein:2002jr} for related work). In order to simplify our discussion, we will suppose that all the branes preserve the same $\N = 1$ superalgebra, although our results in the next section seem totally independent of this assumption.

\begin{figure}
\centering
\epsfxsize=6in
\hspace*{0in}\vspace*{.2in}
\epsffile{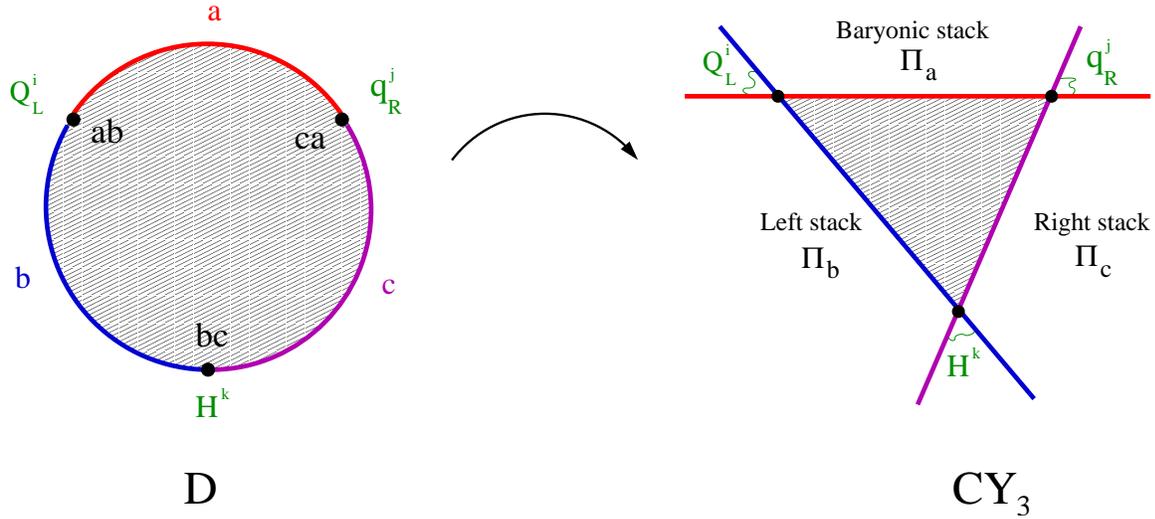}
\caption{Yukawa couplings as euclidean maps from the worldsheet.}
\label{yukis3}
\end{figure}

It was noticed in \cite{Aldazabal:2000cn} that, in the context of intersecting brane worlds, Yukawa couplings between fields living at brane intersections will arise from worldsheet instantons involving three different boundary conditions (see figure \ref{yukis3}). Let us, for instance, consider a triplet of D6-brane stacks and suppose them to be the Baryonic, Left and Right stacks, wrapping the sL's $\Pi_a$, $\Pi_b$ and $\Pi_c$ respectively. By computing the quantum numbers of the fields at the intersections, we find that fields $Q_L^i \in \Pi_a \cap \Pi_b$ can be identified with Left-handed quarks, $q_R^j \in \Pi_c \cap \Pi_a$ with Right-handed quarks and finally $H^k \in \Pi_b \cap \Pi_c$ with Higgs particles. A Yukawa coupling in SM physics will arise from a coupling between these three fields. In our context, such trilinear coupling will arise from the contribution of open worldsheet instantons with the topology of a disc and with three insertions on its boundary. Each of these insertions corresponds to an open string twisted vertex operators that changes boundary conditions, so that finally three different boundaries are involved in the amplitude. 

From the target space perspective, such amplitude will arise from an (euclidean) embedding of the disc in the compact manifold $\M_6$, with each vertex operator mapped to the appropriate intersection of two branes (generically a fixed point in $\M_6$) and the disc boundary between, say, $ca$ and $ab$ to the worldvolume of the D6-brane stack $a$, etc. Such mapping $D \rightarrow \M_6$ has been schematically drawn in figure \ref{yukis3}. 

Notice that an infinite family of such maps exist. However, only a subfamily satisfies the classical equations of motion. Also only a subfamily will contribute to the superpotential. By standard arguments, these correspond to holomorphic embeddings of $D$ on $\M_6$ with the boundary conditions described above, that is, worldsheet disc instantons. Just as in the previous case of one single D6-brane, these instantons correspond to surfaces calibrated by the K\"ahler form $\om$, hence of minimal area. In the specific setup discussed in \cite{Aldazabal:2000cn}, the geometry of intersecting brane worlds is reduced to each stack wrapping (linear) 1-cycles on a $T^2$. It is thus easy to see that in this case the target space worldsheet instanton has a planar triangular shape. This, however, will not be the general shape for a holomorphic curve, even in the familiar case of higher dimensional tori with a flat metric (see Appendix \ref{hdhd}).

More concretely, we expect the Yukawa couplings between the fields  $Q_L^i$, $q_R^j$ and $H^k$ to be roughly of the form:
\beq
Y_{ijk} = h_{qu} \sum_{\vec{n} \pm \in H_2^D(\M_6,\cup_\a \Pi_\a,ijk)} 
{d_{\vec{n}} \ e^{-{A_{ijk}(\vec{n}) \over 2\pi \a^\prime}} 
e^{-2\pi i \phi_{ijk}(\vec{n})}}.
\label{yukabs}
\eeq
Here $\vec{n}$ is an element of the relative homology group $H_2(\M_6, \cup_\a \Pi_\a; \inte)$, that is, a 2-cycle in the Calabi-Yau $\M_6$ ending on $\cup_\a \Pi_\a = \Pi_a \cup \Pi_b \cup \Pi_c$. We further impose this 2-cycle to have the topology of a disc, and to connect the intersections $i \in \Pi_a \cap \Pi_b$, $j \in \Pi_c \cap \Pi_a$, $k \in \Pi_b \cap \Pi_c$ following the boundary conditions described above. Given such a topological sector indexed by $\vec{n}$, we expect a discrete number of holomorphic discs to exist, and we have indicated such multiplicity by $d_{\vec{n}}$. The main contribution comes from the exponentiation of $A_{ijk}(\vec{n}) = \int_{\vec{n}} \om$, which is the target-area of such 'triangular' surface, whereas $\phi_{ijk}(\vec{n})$ is the phase the string endpoints pick up when going around the disc boundary $\partial D$ (see next section). As in (\ref{super}), the sign depends on the discs wrapping holomorphic or antiholomorphic maps. Finally, $h_{qu}$ stands for the contribution coming from quantum corrections, i.e., fluctuations around the minimal area semiclassical solution. Just as in the closed string case \cite{Hamidi:1986vh,Dixon:1986qv}, we expect such contributions to factorise from the infinite semiclassical sum.

At this point one may wonder what is the detailed mechanism by which the chiral fermions get their mass. That is, one may want to understand what is the D-brane analogue of the Higgs mechanism in this intersecting brane picture. The right answer seems to be {\it brane recombination}, studied from a geometrical viewpoint by Joyce \cite{Joyce:1999tz}, and later in terms of D-brane physics in \cite{Kachru:1999vj,Blumenhagen:2000eb,Uranga:2002ag}. The connection of such phenomenon to the SM Higgs mechanism was addressed in \cite{Cremades:2002cs}. Here we will briefly sketch this line of thought from a general viewpoint. Consider two D6-branes wrapping two sL's $\Pi_\a$ and $\Pi_\b$ on a ${\bf CY_3}$ $\M_6$, and further assume that they have the same phase, i.e., both are calibrated by the same real volume form $\Om_\th$ and thus preserve (at least) a common $\N = 1$ in $D = 4$. From the geometrical viewpoint, they lie in a marginal stability wall of $\M_6$. This implies that we can marginally deform our configuration by `smoothing out' the intersections $\Pi_a \cap \Pi_\b$, combining the previous two sL's into a third one $\Pi_\g$. This family of deformations will all be calibrated by the same real volume form $\Om_\th$, so that the total volume or tension of the system will be invariant. From the field theory point of view, this deformation translates into giving non-vanishing v.e.v.'s to the massless scalar fields at the intersections.

\begin{figure}
\centering
\epsfxsize=6.2in
\hspace*{0in}\vspace*{.2in}
\epsffile{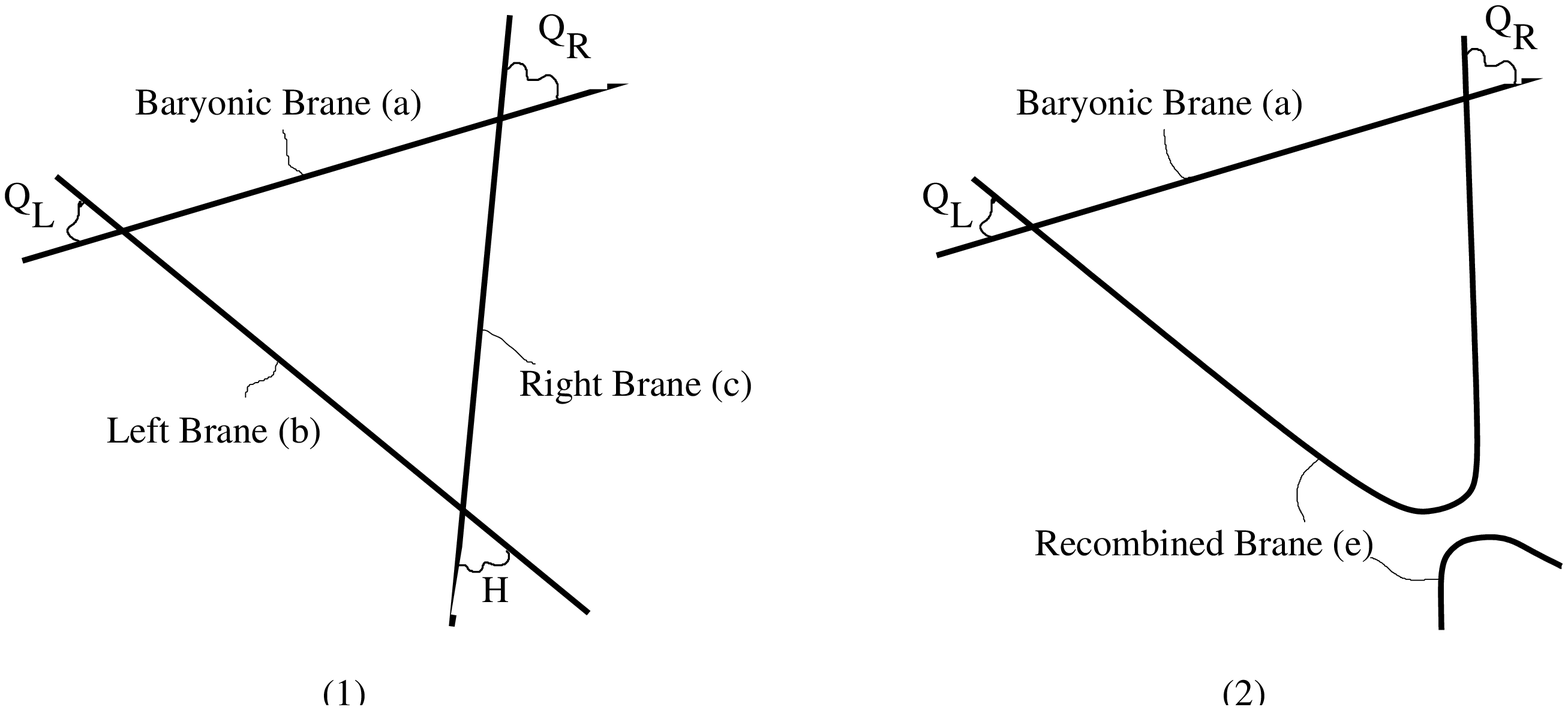}
\caption{Picture of the recombination. (1) Before the recombination, the worldsheet instantons connecting the $Q_L$, $Q_R$ and Higgs multiplets corresponds to holomorphic discs with their boundaries embedded in three different branes. (2) After giving a v.e.v. to $H$, stacks $b$ and $c$ have recombined into a third one $e$.If the recombination is soft enough, the number of chiral fermions at the intersections will not vary. However, they will get mass terms by holomorphic discs that connect fermions of opposite chirality, having its boundary on stacks $a$ and $e$.}
\label{recombination}
\end{figure}

Let us then consider the recombination of our SM stacks $b$ and $c$ into a third one $e$. By the above discussion, this correspond to giving a v.e.v. to Higgs multiplets living on $\Pi_b \cap \Pi_c$, so we expect that this implies a mass term for our chiral fermions. Indeed, in intersecting brane worlds, the chirality condition that prevents fermions from getting a mass is encoded in the non-vanishing topological intersection number of two branes, such as $[\Pi_a] \cdot [\Pi_b]$, $[\Pi_c] \cdot [\Pi_a] \neq 0$ that give us the number of net chiral quarks. Notice, however, that upon brane recombination, we will have 
\beq
\left|[\Pi_a] \cdot [\Pi_e]\right| = \left|[\Pi_a] 
\cdot ([\Pi_b] + [\Pi_c])\right| \leq \left|[\Pi_a] 
\cdot [\Pi_b]\right| + \left|[\Pi_a] \cdot [\Pi_c]\right|,
\label{recom}
\eeq
which implies that the number of `protected' chiral fermions decreases if $[\Pi_a] \cdot [\Pi_b]$ and $[\Pi_a] \cdot [\Pi_c]$ have opposite sign, that is, yield fermions of opposite chirality. In semi-realistic models we usually have $[\Pi_a] \cdot [\Pi_e] = 0$, so we expect every quark to get a mass. 

Since we are in a supersymmetric situation, we are allowed to perform an arbitrary small deformation from the initial configuration where the branes were not recombined. Upon such 'soft recombination', the actual number of intersections will not change, i.e., $\#(\Pi_a \cap \Pi_e) = \#(\Pi_a \cap \Pi_b) + \#(\Pi_a \cap \Pi_c)$. This implies that left and right-handed quarks will still be localized at intersections of $a$ and $e$. They will get, however, a mass term from a worldsheet instanton connecting each pair of them, now involving only two different boundaries. This situation has been illustrated in figure \ref{recombination}.

Before closing this section, let us mention that the discussion of Yukawa couplings, involving three or more stacks of branes, is intimately related to the previous discussion involving one single D-brane. Indeed, given a supersymmetric configuration of three stacks of D6-branes, we could think of slightly smoothing out each single intersection between each pair of them, thus recovering one single D6-brane wrapping a special Lagrangian. Now, by our general considerations of the superpotential of one single brane, we know that such superpotential will only depend on closed string K\"ahler moduli, and that will have the general form (\ref{super}). We expect the same results to hold in the case of the superpotential involving Yukawa couplings before recombination. In the next section we will compute such trilinear couplings for the simple case of Lagrangian $T^n$ wrapping $n$-cycles on $T^{2n}$, and see that they indeed satisfy such conditions.

\section{The general form of Yukawa couplings in toroidal models \label{toroidal}}

In this section we derive the general expression for Yukawa couplings in toroidal and factorisable intersecting brane configurations. By this we mean that the compact manifold will be a factorisable flat torus $\M_6 = T^{2n} = \otimes_{r=1}^n T_r^2$, whereas D-branes will be wrapping Lagrangian factorisable $n$-cycles, that is, those that can be expressed as a product of $n$ 1-cycles $\Pi_\a = \otimes_{r=1}^n (n_\a^{(r)},m_\a^{(r)})$, one on each $T^2$. Such $n$-cycles have the topology of $T^n$ and, if we minimize their volume in its homology class, they are described by hyperplanes quotiented by a torus lattice. This implies, in particular, that the intersection number between two cycles is nothing but the (signed) number of intersections, that is, $\#(\Pi_\a \cap \Pi_\b) = |[\Pi_\a] \cdot [\Pi_\b]|$.  Such class of configurations are known in the literature as {\it branes at angles} \cite{Berkooz:1996km}. Although the discussion in the previous sections is centered on the case $n = 3$, we have seen that semi-realistic models can also be constructed involving $n = 1, 2$. For completeness, we derive our results for arbitrary $n$.

\subsection{Computing Yukawas on a $T^2$ \label{twotorus}}

The simplest case when computing a sum of worldsheet instantons comes, as usual, from D-branes wrapping 1-cycles in a $T^2$, that is, branes intersecting at {\it one} angle. Let us then consider three of such branes, given by
\beq
\begin{array}{ccc}
\quad [\Pi_a] = [(n_a, m_a)] & \raw & z_a = R \cdot (n_a + \tau m_a) 
\cdot x_a  \\
\quad [\Pi_b] = [(n_b, m_b)] & \raw & z_b = R \cdot (n_b + \tau m_b) 
\cdot x_b \\
\quad [\Pi_c] = [(n_c, m_c)] & \raw & z_c = R \cdot (n_c + \tau m_c) 
\cdot x_c
\end{array}
\label{cycles}
\eeq
where $(n_\a,m_\a) \in \inte^2$ denote the 1-cycle the brane $\a$ wraps on $T^2$. Since the manifold of minimal volume in this homology class is given by a straight line with the proper slope, we can associate a complex number $z_\a$ to each brane, which stands for a segment of such 1-cycle in the covering space $\cpx$. Here $\tau$ is the complex structure of the torus and $x_\a \in \real$ an arbitrary number. We fix the area of $T^2$ (the K\"ahler structure, if we ignore the possibility of a B-field) to be $A = R^2 \ \pim \tau$. The triangles that will contribute to a Yukawa coupling involving branes $a$, $b$ and $c$ will consist of those triangles whose sides lie on such branes, hence of the form $(z_a, z_b, z_c)$. To be an actual triangle, however, we must impose that it {\it closes}, that is
\beq
z_a + z_b + z_c = 0.
\label{closure}
\eeq
Since $n_\a$, $m_\a$ can only take integer values, (\ref{closure}) can be 
translated into a Diophantine equation, whose solution is
\bea
\begin{array}{c}
x_a = \left(I_{bc}/d\right) \cdot x \\ 
x_b = \left(I_{ca}/d\right) \cdot x \\ 
x_c = \left(I_{ab}/d\right) \cdot x
\end{array}
& \ {\rm with} &
\begin{array}{c}
x = \left(x_0 + l\right) \\
x_0 \in \real, \ l \in \inte \\
d = g.c.d. \left( I_{ab}, I_{bc}, I_{ca} \right) 
\end{array}
\label{diophan}
\eea
where $I_{\a\b} = [\Pi_\a] \cdot [\Pi_\b] = n_\a m_\b - n_\b m_\a$ stands for the intersection number of branes $\a$ and $\b$, and $x_0$ is a continuous parameter which is fixed for a particular choice of intersection points and brane positions, being a particular solution of (\ref{closure}). If, for instance, we choose branes $a$, $b$ and $c$ to intersect all at the same point, then we must take $x_0 = 0$. The discrete parameter $l$ then arises from triangles connecting different points in the covering space $\cpx$ but the same points under the lattice identification that defines our $T^2$. In the language of section \ref{ibm&yuk}, $l$ indexes the elements of the relative homology class $H_2^D (T^2, \Pi_a \cup \Pi_b \cup \Pi_c, ijk)$. We thus see that a given Yukawa coupling gets contributions from an infinite (discrete) number of triangles indexed by $l$. 

\begin{figure}
\centering
\epsfxsize=6in
\hspace*{0in}\vspace*{.2in}
\epsffile{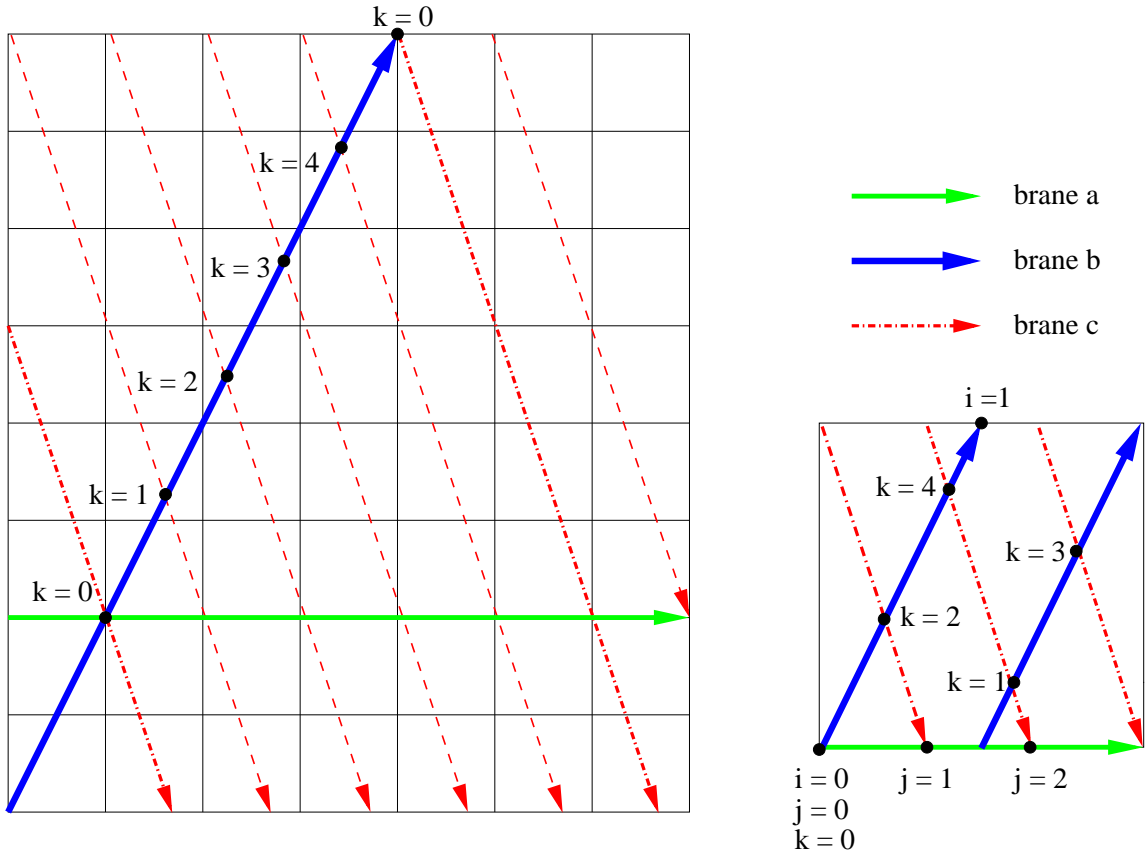}
\caption{Relevant intersections and triangles for the three 1-cycles $(n_a,m_a) = (1,0)$, $(n_b,m_b) = (1,2)$ and  $(n_c,m_c) = (1,-3)$. In the left figure we have depicted these 1-cycles intersecting on the covering space $\cpx$. The dashed lines represent various images of the brane $c$ under torus translations. The indexing of the $bc$ intersections by the integer $k$ coincides with expressions (\ref{diophan}) and (\ref{ijkd=1}). In the right figure we have depicted a single fundamental region of the torus, and have indexed every intersection. Notice that we have chosen that all branes intersect at the origin, and the special choice of a square lattice for the complex structure. The results, however, are general.}
\label{tri}
\end{figure}

Let us describe the specific values that $x_0$ can take. First notice that each pair of branes will intersect several times, each of them in a different point of $T^2$. Namely, we can index such intersection points by
\beq
\begin{array}{cc}
i = 0,1, \dots, |I_{ab}| -1, & i \in \Pi_a \cap \Pi_b \\
j = 0,1, \dots, |I_{ca}| -1, & j \in \Pi_c \cap \Pi_a \\
k = 0,1, \dots, |I_{bc}| -1, & k \in \Pi_b \cap \Pi_c
\label{indices}
\end{array}
\eeq
In general, $x_0$ must depend on the particular triplet $(i,j,k)$ of intersection points and on the relative positions of the branes. For simplicity, let us take the triplet of intersections $(0,0,0)$ to correspond to a triangle of zero area. That is, we are supposing that the three branes intersect at a single point, which we will choose as the origin of the covering space (see figure \ref{tri}). Then it can be shown that, given the  appropriate indexing of the intersection points, there is a simple expression for $x_0$ given by
\beq
x_0 (i,j,k) = \frac{i}{I_{ab}} + \frac{j}{I_{ca}} + \frac{k}{I_{bc}},
\label{ijkd=1}
\eeq 
where $i$, $j$ and $k$ are defined as in (\ref{indices}) \footnote{Notice that, since for a given triplet $(i,j,k)$ we must consider all the solutions $x_0 (i,j,k) + l$, $l \in \inte$, the index $i$ is actually defined mod $|I_{ab}|$, same for the others indices.}. In this latter expression we are supposing that $d = 1$, that is, that $I_{ab}$, $I_{bc}$ and $I_{ca}$ are coprime integers. This guarantees that there exist a triangle connecting every triplet $(i,j,k)$, and also a simple expression for $x_0$. The case $d \neq 1$ will be treated below. An illustrative example of the above formula is shown in figure \ref{tri}, where a triplet of 1-cycles intersecting at the origin have been depicted, both in a square torus and in its covering space, and the intersections have been indexed in the appropriate manner so that (\ref{ijkd=1}) holds. 

What if these branes do not intersect all at the origin? Let us consider shifting the positions of the three branes by the translations $\eps_a$, $\eps_b$ and $\eps_c$, where $\eps_\a$ is the transversal distance of the brane $\a$ from the origin measured in units of $A/ ||\Pi_\a||$, in clockwise sense from the direction defined by $\Pi_\a$. Then is easy to see that (\ref{ijkd=1}) is transformed to
\beq
x_0 (i,j,k) = \frac{i}{I_{ab}} + \frac{j}{I_{ca}} + \frac{k}{I_{bc}}
+ \frac{I_{ab} \eps_c + I_{ca} \eps_{b} + I_{bc} \eps_{a}}
{I_{ab} I_{bc} I_{ca}}.
\label{ijkd=1eps}
\eeq 
Notice, however, that we can absorb these three parameters into only one, to be defined as $\tilde\eps = (I_{ab} \eps_c + I_{ca} \eps_{b} + I_{bc} \eps_{a})/I_{ab} I_{bc} I_{ca}$. This was to be expected since, given the reparametrization invariance present in $T^2$, we can always choose branes $b$ and $c$ to intersect at the origin, and then the only freedom comes from shifting the brane $a$ away from this point.

Given this solution, now we can compute the areas of the triangles whose vertices lie on the triplet of intersections $(i,j,k)$ (we will say that this triangle 'connects' these three intersections), by using the well-known formula
\beq
A(z_a,z_b) = \oh \sqrt{|z_a|^2 \cdot |z_b|^2 - (\preal z_a \bar{z_b})^2 }.
\label{area}
\eeq 

Then we find that
\bea
A_{ijk}(l) & = & \oh (2\pi)^2 A |I_{ab} I_{bc} I_{ca}| 
\ (x_0 (i,j,k) + l)^2 \nonumber \\
& = & \oh (2\pi)^2 A |I_{ab} I_{bc} I_{ca}| 
\left(\frac{i}{I_{ab}} + \frac{j}{I_{ca}} + 
\frac{k}{I_{bc}} + \tilde\eps + l \right)^2,
\label{area2}
\eea
where $A$ represents the K\"ahler structure of the torus, and we have absorbed all the shift parameters into $\tilde\eps$. The area of such triangle may correspond to either an holomorphic or an antiholomorphic map from the disc. From (\ref{diophan}), we see this depends on the sign of $I_{ab}I_{bc}I_{ca}$, so we must add a real phase $\sig_{abc} = {\rm sign } (I_{ab}I_{bc}I_{ca})$ to the full instanton contribution.

We can finally compute the corresponding Yukawa coupling for the three particles living at the intersections $(i,j,k)$:
\beq
Y_{ijk}  \sim  \sig_{abc} \sum_{l \in \inte}{\rm exp} 
\left( - {A_{ijk}(l) \over 2\pi \a^\prime} \right) 
\label{yukiT2}
\eeq

This last quantity can be naturally expressed in terms of a modular theta function, which in their real version are defined as
\beq
\vt \left[
\begin{array}{c}
\d \\ \phi
\end{array}
\right] (t) = \sum_{l \in \inte} q^{\oh (\d + l)^2} 
\ e^{2\pi i (\d + l)\phi}, \ \ q = e^{-2\pi t}.
\label{theta}
\eeq

Indeed, we find that (\ref{yukiT2}) can be expressed as such theta function with parameters
\bea
\d & = & \frac{i}{I_{ab}} + \frac{j}{I_{ca}} + \frac{k}{I_{bc}}
+ \frac{I_{ab} \eps_c + I_{ca} \eps_{b} + I_{bc} \eps_{a}}
{I_{ab} I_{bc} I_{ca}}, \\
\phi & = & 0, \\
t & = & \frac{A}{\a^\prime} |I_{ab} I_{bc} I_{ca}|.
\label{paramT2}
\eea

\subsubsection{Adding a B-field and Wilson lines}

It is quite remarkable that we can express our Yukawa couplings in terms of a simple theta function. However, reached this point we could ask ourselves why it is such a specific theta function. That is, we are only considering the variable $t$ as a real number, instead of a more general parameter $\k \in \cpx$, and we are always setting $\phi = 0$. These two constraints imply that our theta functions are strictly real. From both the theoretical an phenomenological point of view, however, it would be interesting to have a Yukawa defined by a complex number. 

These two constraints come from the fact that we have considered very particular configurations of branes at angles. First of all we have not considered but tori where the B-field was turned off. This translates into a very special K\"ahler structure, where only the area plays an important role. In general, if we turn on a B-field, the string sweeping a two-dimensional surface will not only couple to the metric but also to this B-field. In a $T^2$, since the K\"ahler structure is the complex field
\beq
J = B + iA,
\label{kahler2}
\eeq
we expect that, by including a B-field, our results (\ref{paramT2}) will 
remain almost unchanged, with the only change given by the substitution 
$A \raw (-i) J$. this amounts to changing our parameter $t$ to a complex one 
defined as 
\beq
\k = \frac{J}{\a^\prime} |I_{ab} I_{bc} I_{ca}|.
\label{cpxt}
\eeq

Our second generalization is including Wilson lines around the compact directions that the D-branes wrap. Indeed, when considering D-branes wrapping 1-cycles on a $T^2$, we can consider the possibility of adding a Wilson line around this particular one-cycle. Since we do not want any gauge symmetry breaking, we will generally choose these Wilson lines to correspond to group elements on the centre of our gauge group, i.e., a phase \footnote{Notice that, although Wilson 
lines may produce a shift on the KK momenta living on the worldvolume of the brane, they never affect the mass of the particles living at the intersections, in the same manner that shifting the position of the branes does not affect them.}. 

Let us then consider a triangle formed by D-branes $a$, $b$ and $c$ each wrapped on one different 1-cycle of $T^2$ and with Wilson lines given by the phases $exp(2\pi i \th_a)$,  $exp(2\pi i \th_b)$ and $exp(2\pi i \th_c)$, respectively. The total phase that an open string sweeping such triangle picks up depends on the relative longitude of each segment, and is given by
\beq
e^{2\pi i x_a \th_a} \cdot e^{2\pi i x_b \th_b} \cdot e^{2\pi i x_c \th_c} 
= e^{2\pi i \left(I_{bc} \th_a + I_{ca} \th_b + I_{ab} \th_c \right) x}.
\label{wilson}
\eeq

Finally, we will consider both possibilities: having a B-field and some Wilson lines. In order to express our results we need to consider the complex theta function with characteristics, defined as 
\beq
\vt \left[
\begin{array}{c}
\d \\ \phi
\end{array}
\right] (\k) = \sum_{l \in \inte} 
e^{\pi i (\d + l)^2 \k} \ e^{2\pi i (\d + l) \phi }.
\label{thetacpx}
\eeq

Our results for the Yukawa couplings can then be expressed as such a function with parameters\footnote{Notice that this implies that Yukawa couplings will be generically given by complex numbers, which is an important issue in order to achieve a non-trivial CKM mixing phase in semirealistic models. }
\bea
\d & = & \frac{i}{I_{ab}} + \frac{j}{I_{ca}} + \frac{k}{I_{bc}}
+ \frac{I_{ab} \eps_c + I_{ca} \eps_{b} + I_{bc} \eps_{a}}
{I_{ab} I_{bc} I_{ca}}, \\
\phi & = & I_{ab} \th_c + I_{ca} \th_b + I_{bc} \th_a,  \\
\k & = & \frac{J}{\a^\prime} |I_{ab} I_{bc} I_{ca}|.
\label{paramT2cpx}
\eea

\subsubsection{Orientifolding the torus}

Let us now deal with the slight modification of the above toroidal model which consist on performing an orientifold projection on the torus. Namely, we quotient the theory by $\OR$, where $\Om$ is the usual worldsheet orientation reversal and $\R: z \mapsto \bar z$ is a $\inte_2$ action on the torus. This introduces several new features, the most relevant for our discussion being

\begin{itemize}

\item There appears a new object: the O-plane, which lies on the horizontal axis described by $\{\pim z = 0 \}$ in the covering space $\cpx$.

\item In order to consider well-defined constructions, for each D-brane $a$ in our configuration we must include its image under $\OR$, denoted by $\Om\R a$ or $a^*$. These mirror branes will, generically, wrap a cycle $[\Pi_{a^*}]$ different from $[\Pi_a]$, of course related by the action of $\R$ on the homology of the torus.

\end{itemize}

This last feature has a straightforward consequence, which is the proliferation of sectors as $ab$, $ab^*$, etc. Indeed, if we think of a configuration involving D-branes $a$, $b$ and $c$, we can no longer bother only about the triangle $abc$, but we must also consider $abc^*$, $ab^*c$ and $ab^*c^*$ triangles (the other possible combinations are mirror pairs of these four\footnote{We will not bother about triangles involving a brane and its mirror, as $abb^*$, for purely practical reasons. The results of this section, however, are easily extensible to these cases.}). Once specified the wrapping numbers of the triangle $abc$ all the others are also fixed. Since  our formulae for the Yukawas are not very sensitive to the actual wrapping number of the 1-cycles but only to the intersection numbers, we do not expect these to appear in the final expression. Notice, however, that if we specify the position of the brane $a$ the position of its mirror $a^*$ is also specified. Hence, shifts of branes should be related in the four triangles. This can be also deduced from the first item above. Since we have a rigid object lying in one definite 1-cycle, which is the O-plane, translation invariance is broken in the directions transverse to it, so we have to specify more parameters in a certain configuration. In this case of $T^2$ this means that if we consider that the three branes intersect at one point, we must specify the 'height' ($\pim z$) of such intersection. 

So our problem consists of, given the theta function parameters of the triangle $abc$, find those of the other three triangles. First notice that, if we actually consider the three branes $a$, $b$ and $c$ intersecting at one point, then the same will happen for the triangles $abc^*$, $ab^*c$ and $ab^*c^*$. Then by our previous results on triangles on a plain $T^2$ we see that the theta parameters will be given by
\bea
\d_{abc} & = & \frac{i}{I_{ab}} + \frac{j}{I_{ca}} + \frac{k}{I_{bc}},\\
\k_{abc} & = & \frac{J}{\a^\prime} |I_{ab} I_{bc} I_{ca}|,
\label{paramoriabc}
\eea
for the $abc$ triangle and
\bea
\d_{ab^*c} & = & \frac{i^*}{I_{ab^*}} + 
\frac{j}{I_{ca}} + \frac{k^*}{I_{b^*c}},\\
\k_{ab^*c} & = & \frac{J}{\a^\prime} |I_{ab^*} I_{b^*c} I_{ca}|,
\label{paramoriab*c}
\eea
for the $ab^*c$ triangle, etc. Notice that $i$ and $i^*$ are different indices which label, respectively, $ab$ and $ab^*$ intersections. 

A general configuration will not, however, contain every triplet of branes intersecting at one point, and will also contain non-zero Wilson lines. As mentioned, once specified the relative positions and Wilson lines of the triangle $abc$ all the other triangles are also specified. By simple inspection we can see that a general solution is given by the parameters  
\bea
\d_{abc} & = & \frac{i}{I_{ab}} + \frac{j}{I_{ca}} + \frac{k}{I_{bc}}
+ \frac{I_{ab} \eps_c + I_{ca} \eps_{b} + I_{bc} \eps_{a}}
{I_{ab} I_{bc} I_{ca}}, \\
\phi_{abc} & = & I_{ab} \th_c + I_{ca} \th_b + I_{bc} \th_a,
\label{paramoriabceps}
\eea
for the triangle $abc$, and the parameters
\bea
\d_{ab^*c} & = & \frac{i^*}{I_{ab^*}} + \frac{j}{I_{ca}} 
+ \frac{k^*}{I_{b^*c}}
+ \frac{I_{ab^*} \eps_c + I_{ca} \eps_{b^*} 
+ I_{b^*c} \eps_{a}}{I_{ab^*} I_{b^*c} I_{ca}}, 
\\
\phi_{ab^*c} & = & I_{ab^*} \th_c + I_{ca} \th_{b^*} + I_{b^*c} \th_a,
\label{paramoriab*ceps}
\eea
for the triangle $ab^*c$, and similarly for the other two triangles. Here we have defined
\bea
\begin{array}{c}
\eps_{\a^*} = - \eps_\a \\
\th_{\a^*} = - \th_\a
\end{array}
 & & \a = a, b, c.
\label{mirrorparam}
\eea

\subsubsection{The non-coprime case}

Up to now, we have only consider a very particular class of Yukawa couplings: those that arise from intersecting D-branes wrapping 1-cycles on a $T^2$. Furthermore, we have also assumed the constraint $d = g.c.d. (I_{ab}, I_{bc}, I_{ca}) = 1$, that is, that the three intersection numbers are coprime. The non-coprime case is, however, the most interesting from the phenomenological point of view\footnote{This is no longer true when dealing with higher-dimensional cycles as, e.g., $n$-cycles wrapped on $T^{2n}$ for $n = 2,3$. In those cases, requiring that the brane configurations have only one Higgs particle imposes the coprime condition $d =1$ on each separate torus.}. In this section, we will try to address the non-coprime case. Although no explicit formula is given, we propose an ansatz that has been checked in plenty of models.

\begin{figure}
\centering
\epsfxsize=5in
\hspace*{0in}\vspace*{.2in}
\epsffile{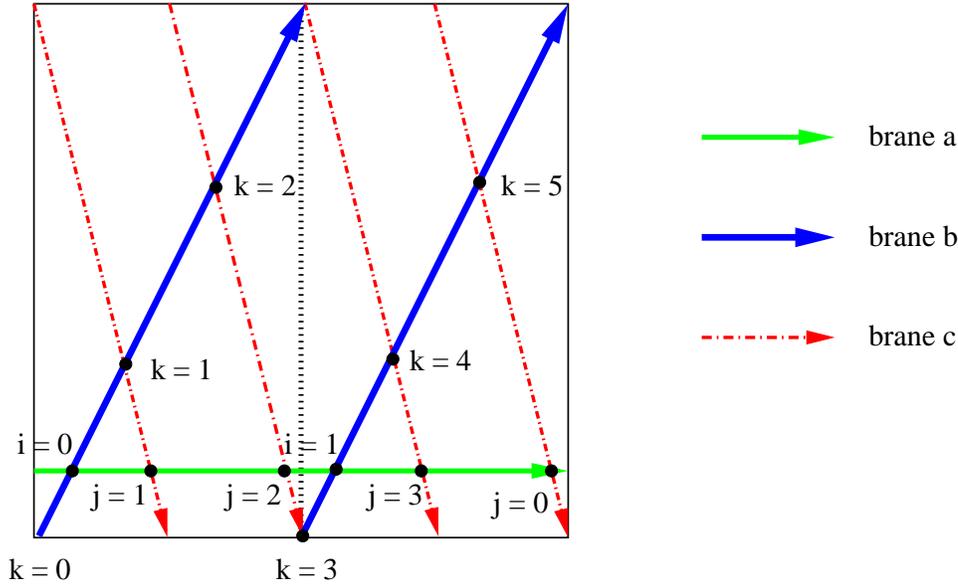}
\caption{Relevant intersections 
and triangles for the three 1-cycles $(n_a,m_a) = 
(1,0)$, $(n_b,m_b) = (1,2)$ and  $(n_c,m_c) = (1,-4)$. 
Notice that the fundamental region of the torus has two 
identical regions, exactly matching with $d = g.c.d. (I
_{ab}, I_{bc}, I_{ca}) = 2$. Also notice that a triangle 
exists connecting the vertices $(i,j,k)$ if and only if 
$i + j + k =$ even.}
\label{tri2}
\end{figure}

A particular feature of the configurations where $d > 1$ is that not every triplet of intersections $(i,j,k)$ is connected by a triangle. Indeed, from solution (\ref{diophan}) we see that a pair of intersections $(i,j)$ from $(ab,ca)$ will only couple to $|I_{bc}|/d$ different $bc$ intersections, same for the other pairs. Similarly, one definite intersection from $bc$ will couple to $|I_{ab}I_{ca}|/d^2$ $(i,j)$ pairs of $(ab,ca)$ intersections. This can be seen in figure \ref{tri2}, where a particular example of non-coprime configuration is shown.

In this same figure we can appreciate another feature of these configurations, which is that the fundamental region of the torus divides in $d$ identical copies. That is, the intersection pattern of any of these regions exactly matches the others. This is a direct consequence from the Diophantine solution (\ref{diophan}).

Let us now formulate the ansatz for this more general class of configurations. It consist of two points:
\begin{itemize}

\item A Yukawa coupling can be expressed as a complex theta function, whose parameters are
\bea
\d & = & \frac{i}{I_{ab}} + \frac{j}{I_{ca}} + \frac{k}{I_{bc}}
+ \frac{d \cdot \left(I_{ab} \eps_c + I_{ca} \eps_{b} + 
I_{bc} \eps_{a}\right)}{I_{ab} I_{bc} I_{ca}} + \frac{s}{d},
\label{paramT2nc1}\\
\phi & = & \left( I_{ab} \th_c + I_{ca} \th_b + I_{bc} \th_a \right) / d,  
\label{paramT2nc2}\\
\kappa & = & \frac{J}{\a^\prime} {|I_{ab} I_{bc} I_{ca}| \over d^2} 
\label{paramT2nc3}
\eea
where $s \equiv s(i,j,k) \in \inte$ is a linear function on the integers $i$, $j$ and $k$. 

\item A triplet of intersections $(i,j,k)$ is connected by a family of triangles, that is, has a non-zero Yukawa, if and only if
\beq
i + j + k \equiv 0 \ {\rm mod\ } d.
\label{condition}
\eeq

\end{itemize}

Notice that this ansatz correctly reduces to the previous solution in the coprime case, i.e., when $d = 1$. Indeed, in that case the actual value of $s$ becomes unimportant for the evaluation of the theta function and the condition (\ref{condition}) is trivially satisfied by any triplet $(i,j,k)$.

\subsection{Higher dimensional tori}

Having computed the sum of holomorphic instantons in the simple case of 1-cycles in a $T^2$, let us now turn to the case of $T^{2n} = T^2 \times \dots \times T^2$. Since we are dealing with higher-dimensional geometry, surfaces are more difficult to visualize and computations less intuitive. We will see, however, that the final result is a straightforward generalization of the previous case. A $T^{2n}$ is a very particular case of ${\bf CY_n}$ manifold. Such manifolds are equipped with both a K\"ahler 2-form $\om$ and a volume $n$-form $\Om$ that satisfy
\beq
\frac{\om^n}{n!} = (-1)^{n(n-1)/2} \left( \frac i2\right)^n \Om \wedge \ov \Om.
\label{omegasn}
\eeq
In the particular case of a flat factorisable $T^{2n}$ we may take them to be
\beq
\om = \frac i2 \sum_{r=1}^n d z_r \wedge d\bar z_r \quad {\rm and} \quad
\Om = \preal \left( e^{i\th} d z_1 \wedge \dots \wedge d z_n \right).
\label{calit2n}
\eeq
As can easily be deduced from the discussion of Chapter \ref{SUSY}, these are not the only possible choices for $\Om$, but there are actually 2$^{n-1}$ independent complex $n$-forms satisfying (\ref{omegasn}), all suitable as calibrations. In particular, for suitable phases $\th_j$ ($j = 1, \dots, 2^{n-1}$) they all calibrate the so-called factorisable $n$-cycles, that is, the $n$-cycles that are Lagrangian $T^n$ and can be expressed as a product of $n$ 1-cycles $\Pi_\a = \otimes_{r=1}^n (n_\a^{(r)},m_\a^{(r)})$, one on each $T^2$. We will focus on configurations of branes on such factorisable cycles. Notice that these factorisable constructions, which yield branes intersecting at $n$ angles as in \cite{Berkooz:1996km}, are not the more general possibility. They are, however, particularly well-suited for extending our previous analysis of computation of Yukawas on a $T^2$. Indeed, the closure condition analogous to (\ref{closure}) can be decomposed into $n$ independent closure conditions, such as 
\beq
z_a^{(r)} + z_b^{(r)} + z_c^{(r)} = 0, \ \ r = 1, \dots, n,
\label{closure2n2}
\eeq
where $\a$ labels the corresponding $T^2$. Then we can apply our results from plain toroidal configurations to solve each of these $n$ Diophantine equations. The solution can then be expressed as three vectors $z_a, z_b, z_c \in \cpx^n$:
\beq
\begin{array}{ccc}
\quad [\Pi_a] = \bigotimes_{r = 1}^n [(n_a^{(r)}, m_a^{(r)})] 
& \quad \raw \quad & 
z_a =  \left( z_a^{(1)}, z_a^{(2)}, \dots, z_a^{(n)} \right), \\
\quad [\Pi_b] =  \bigotimes_{r = 1}^n [(n_b^{(r)}, m_b^{(r)})] 
& \quad \raw  \quad & 
z_b = \left( z_b^{(1)}, z_b^{(2)}, \dots, z_b^{(n)} \right), \\
\quad [\Pi_c] = \bigotimes_{r = 1}^n [(n_c^{(r)}, m_c^{(r)})] 
& \quad \raw \quad & 
z_c = \left( z_c^{(1)}, z_c^{(2)}, \dots, z_c^{(n)} \right).
\end{array}
\label{vectorscpx}
\eeq
Just as in (\ref{cycles}) and (\ref{diophan}), each entry is given by
\bea
\begin{array}{c}
z_a^{(r)} = R^{(r)} \cdot \left(n_a^{(r)} + \tau^{(r)} m_a^{(r)}\right) 
I_{bc}^{(r)} x^{(r)} / d^{(r)} \\ 
z_b^{(r)} = R^{(r)} \cdot \left(n_b^{(r)} + \tau^{(r)} m_b^{(r)}\right)  
I_{ca}^{(r)} x^{(r)} / d^{(r)} \\ 
z_c^{(r)} = R^{(r)} \cdot \left(n_c^{(r)} + \tau^{(r)} m_c^{(r)}\right)  
I_{ab}^{(r)} x^{(r)} / d^{(r)}
\end{array}
& \quad {\rm with} &
\begin{array}{c}
x^{(r)} = \left(x_0^{(r)} + l^{(r)} \right) \\
x_0^{(r)} \in \real, \ l^{(r)} \in \inte \\
d^{(r)} = g.c.d. \left( I_{ab}^{(r)}, I_{bc}^{(r)}, I_{ca}^{(r)} \right) 
\end{array}
\label{diophan2n}
\eea
where $\tau^{(r)}$ denotes the complex structure of the corresponding two-torus, and the area of such is given by $A^{(r)} = (R^{(r)})^2 \cdot \pim \tau^{(r)}$. The intersection number of two cycles is simply computed as $I_{ab} = [\Pi_a] \cdot [\Pi_b] = \prod_{r = 1}^n I_{ab}^{(r)}$, where $I_{ab}^{(r)} = ( n_a^{(r)} m_b^{(r)} - n_b^{(r)} m_a^{(r)} )$ denotes the intersection number of cycles $a$ and $b$ on the $r^{th}$ $T^2$. Notice that now, each triplet of intersections $(i,j,k)$ is described by the multi-indices
\beq
\begin{array}{cc}
i = (i^{(1)}, i^{(2)}, \dots, i^{(n)}) \in \Pi_a \cap \Pi_b , 
& \quad i^{(r)} = 0, \dots, |I_{ab}^{(r)}| -1, \\
j = (j^{(1)}, j^{(2)}, \dots, j^{(n)}) \in \Pi_c \cap \Pi_a, 
& \quad j^{(r)} = 0, \dots, |I_{ca}^{(r)}| -1, \\
k = (k^{(1)}, k^{(2)}, \dots, k^{(n)}) \in \Pi_b \cap \Pi_c , 
& \quad k^{(r)} = 0, \dots, |I_{bc}^{(r)}| -1,
\end{array}
\label{multiindices}
\eeq
Correspondingly, each particular solution $x_0^{(r)}$ will depend on the triplet $(i^{(r)}, j^{(r)}, k^{(r)})$, and also on the corresponding shifting parameters. Namely,
\beq
x_0^{(r)} = \frac{i^{(r)}}{I_{ab}^{(r)}} 
+ \frac{j^{(r)}}{I_{ca}^{(r)}} 
+ \frac{k^{(r)}}{I_{bc}^{(r)}} 
+ \frac{d^{(r)} \cdot \left(I_{ab}^{(r)} \eps_c^{(r)} 
+ I_{ca}^{(r)} \eps_{b}^{(r)} + I_{bc}^{(r)} \eps_{a}^{(r)}\right)}
{I_{ab}^{(r)} I_{bc}^{(r)} I_{ca}^{(r)}}
+ \frac{s^{(r)}}{d^{(r)}}.
\label{ijkgeneral}
\eeq

Having parametrized the points of intersection in terms of the positions of the branes, it is now an easy matter to compute what is the area of the holomorphic surface that connects them. Recall that such a surface must have the topology of a disc embedded in $T^{2n}$, with its boundary embedded on the worldvolumes of $\Pi_a$, $\Pi_b$ and $\Pi_c$ (see figure \ref{yukis3}). Furthermore, in order to solve the equations of motion and contribute to the superpotential, it must be calibrated by $\om$ or, equivalently, parametrized by an (anti)holomorphic embedding into $T^{2n}$. We will discuss the existence and uniqueness of such surface in Appendix \ref{hdhd}. For the time being, we only need to assume that it exist, since by properties of calibrations we know that its area is given by the direct evaluation of $\om$ on the relative homology class $H_2 (T^{2n}, \Pi_a \cup \Pi_b \cup \Pi_c, ijk) = \otimes_{r = 1}^n H_2 (T^2_r, \Pi_a^{(r)} \cup \Pi_b^{(r)} \cup \Pi_c^{(r)}, i^{(r)}j^{(r)}k^{(r)})$, indexed by the $n$ integer parameters $\{ l^{(r)} \}_{r = 1}^n$. Since $\om$ is essentially a sum of K\"ahler forms for each individual $T^2$, i.e., $\om = \sum_r \om_{T^2}^{(r)}$, this area is nothing but the sum of the areas of the triangles $(i^{(r)}j^{(r)}k^{(r)})$ defined on each $T^2$:
\beq
A(z_a,z_b) = \sum_r A(z_a^{(r)},z_b^{(r)}) = 
\oh (2\pi)^2 \sum_r  A^{(r)} 
\left|I_{ab}^{(r)} I_{bc}^{(r)} I_{ca}^{(r)}\right| 
\ \left(x_0^{(r)} + l^{(r)} \right)^2,
\label{areat2n}
\eeq
where we have used our previous computations (\ref{area}) and (\ref{area2}) relative to the case of $T^2$.

In order to compute the full instanton contribution, we must exponentiate such area as in (\ref{yukiT2}) and then sum over all the family of triangles. Notice that we must now sum over the whole of $n$ integer parameters $\{ l^{(r)} \}_{r = 1}^n$, one for each $T^2$. We thus find
\bea
Y_{ijk}  & \sim &  \sig_{abc}
\sum_{\{l^{(r)}\} \in \inte^n}{\rm exp} 
\left( - {A_{ijk}(\{l^{(r)}\}) \over 2\pi \a^\prime} \right) =
\sig_{abc} \sum_{\{l^{(r)}\} \in \inte^n}{\rm exp} 
\left( - {\sum_\a A_{i^{(r)}j^{(r)}k^{(r)}}(l^{(r)}) 
\over 2\pi \a^\prime} \right) \nonumber \\
& = & \sig_{abc} \prod_r \sum_{l^{(r)} \in \inte} {\rm exp} 
\left( - {A_{i^{(r)}j^{(r)}k^{(r)}}(l^{(r)}) 
\over 2\pi \a^\prime} \right) = \sig_{abc}
\prod_r
\vt \left[
\begin{array}{c}
\d^{(r)} \\ 0
\end{array}
\right] (t^{(r)}),
\label{yukiT2n}
\eea
with $\d^{(r)} = x_0^{(r)}$ and $t^{(r)} =  A^{(r)}/\a^\prime |I_{ab}^{(r)} I_{bc}^{(r)} I_{ca}^{(r)}|$ as these {\it real} theta functions parameters. Here, $\sig_{abc} = \prod_r \sig^{(r)}_{abc} = \prod_r {\rm sign } (I_{ab}^{(r)} I_{bc}^{(r)} I_{ca}^{(r)}) = {\rm sign } (I_{ab}I_{bc}I_{ca})$. We thus see that for the case of higher dimensional tori, we obtain a straightforward generalization in terms of the $T^2$ case. Namely, the sum over worldsheet instantons is given by a product of theta functions. 

Given this result, is now an easy matter to generalize it to the case of non-zero $B$-field and Wilson lines. In order not to spoil the supersymmetric condition on D-branes wrapping sL's, we will add a non-vanishing $B$-field only in the dimensions transverse to them, that is, on the planes corresponding to each $T^2$. This complexifies the K\"ahler form to
\beq
 J^{(r)} = B_{(2r,2r+1)} + i A^{(r)}.
\label{cpxt2n}
\eeq
In the same manner, adding Wilson lines will contribute with a complex phase to the instanton amplitude. It can be easily seen that this phase will have the form
\beq
\prod_{r = 1}^n {\rm exp}\left( 2\pi i \left(I_{bc}^{(r)}
 \th_a^{(r)} + I_{ca}^{(r)} \th_b^{(r)} + I_{ab}^{(r)}
 \th_c^{(r)} \right) \cdot \left(x_0^{(r)} + l^{(r)} \right) \right),
\label{Wilson2n}
\eeq
where $\th_a^{(r)}$ correspond to a Wilson line of stack $a$ on the 1-cycle wrapped on the $r^{th}$ $T^2$. These two sources of complex phases nicely fit into the definition of complex theta functions.

To sum up, we see that the Yukawa coupling for a triplet of intersections $(i,j,k)$ decomposed as in (\ref{multiindices}) will be given by
\beq
Y_{ijk} \sim \sig_{abc}
\prod_{r = 1}^n
\vt \left[
\begin{array}{c}
\d^{(r)} \\ \phi^{(r)}
\end{array}
\right] (\k^{(r)}),
\label{totalyuki}
\eeq
with parameters
\bea
\d^{(r)} & = & \frac{i^{(r)}}{I_{ab}^{(r)}} 
+ \frac{j^{(r)}}{I_{ca}^{(r)}} 
+ \frac{k^{(r)}}{I_{bc}^{(r)}} 
+ \frac{d^{(r)} \cdot \left(I_{ab}^{(r)} \eps_c^{(r)} 
+ I_{ca}^{(r)} \eps_{b}^{(r)} + I_{bc}^{(r)} \eps_{a}^{(r)}\right)}
{I_{ab}^{(r)} I_{bc}^{(r)} I_{ca}^{(r)}}
+ \frac{s^{(r)}}{d^{(r)}},
\label{paramT2ncpx1}\\
\phi^{(r)} & = & 
\left(I_{ab}^{(r)} \th_c^{(r)} + 
I_{ca}^{(r)} \th_b^{(r)} + 
I_{bc}^{(r)} \th_a^{(r)}\right) / d^{(r)}, 
\label{paramT2ncpx2}\\
\k^{(r)} & = & 
\frac{J^{(r)}}{\a^\prime} 
\frac{|I_{ab}^{(r)} I_{bc}^{(r)} I_{ca}^{(r)}|}{(d^{(r)})^2}
\label{paramT2ncpx3}
\eea

\subsection{Physical interpretation}

Let us summarize our results. A Yukawa coupling between fields on the intersection of factorisable $n$-cycles $\Pi_a$, $\Pi_b$ and $\Pi_c$ on a factorisable $T^{2n}$ is given by 
\beq
Y_{ijk} = h_{qu} \sig_{abc}
\prod_{r = 1}^n
\vt \left[
\begin{array}{c}
\d^{(r)} \\ \phi^{(r)}
\end{array}
\right] (\k^{(r)}),
\label{totalyuki2}
\eeq
where $h_{qu}$ stands for the quantum contribution to the instanton amplitude. Such contributions arise from fluctuations of the worldsheet around the volume minimizing holomorphic surface. Given a triplet of factorisable cycles $abc$, the geometry of the several instantons are related by rescalings on the target space, so we expect these contributions to be the same for each instanton connecting a triplet $(i,j,k) \in (\Pi_a\cap \Pi_b,\Pi_c\cap \Pi_a, \Pi_b\cap \Pi_c)$, in close analogy with its closed string analogue \cite{Hamidi:1986vh,Dixon:1986qv}. Moreover, such quantum contributions are expected to cancel the divergences that arise for small volumes of the compact manifold. Indeed, notice that by using the well-known property of the theta-functions
\beq
\vt \left[
\begin{array}{c}
\d \\ \phi
\end{array}
\right] (\k)
=
(-i\k)^{-1/2}
e^{2\pi i \d \phi}
\vt \left[
\begin{array}{c}
\phi \\ -\d
\end{array}
\right] (-1/\k),
\label{small}
\eeq
and taking $\k^{(r)} = \frac{i A^{(r)}}{\a^\prime} |I_{ab}^{(r)} I_{bc}^{(r)} I_{ca}^{(r)}|$, we see that $Y_{ijk}$ diverges as $( {\rm Vol} (T^{2n})/\a^\prime)^{-1/2}$.

Another salient feature of (\ref{totalyuki2}) involves its dependence in closed and open string moduli of the D-brane configuration. Notice that the only dependence of the Yukawa couplings on the closed string moduli enters through $J^{(r)}$, the K\"ahler structure of our compactification. Yukawa couplings are thus independent of the complex structure, which was to be expected from the general considerations of the previous section. On the other hand, the open string moduli are contained in the theta-function parameters $(\d^{(r)}, \phi^{(r)})$. If we define our complex moduli field as in \cite{Kachru:2000an}, we find
\beq
\Phi_a^{(r)} = J \eps_a^{(r)} + \th_a^{(r)}
\label{omoduli}
\eeq
for the modulus field of D-brane wrapping $\Pi_a$, on the $r^{th}$ two-torus. By considering the K\"ahler moduli as external parameters, we recover Yukawa couplings which resemble those derived from a superpotential of the form (\ref{super}). Notice that not all the moduli are relevant for the value of the Yukawa couplings,  $2n$ of them decouple from the superpotential, as they can be absorbed by translation invariance in $T^{2n}$. The instanton generated superpotential will thus depend on $(K-2)n$ open moduli, where $K$ is the number of stacks of our configuration. In the orientifold case, only $n$ of such moduli decouple, so we have $(K-1)n$ such moduli.

As a final remark, notice that our formula (\ref{totalyuki2}) has been obtained for the special case of a diagonal K\"ahler form $\om$. In the general case we would have
\beq
\om = \frac{i}{2} \sum_{r, s} a_{rs} \ dz_r \wedge d\bar z_s,
\label{gomega}
\eeq
so by evaluating $\om$ in the relative homology class we would expect an instanton contribution
\bea
Y_{ijk} & \sim & \sig_{abc} \
\vt \left[
\begin{array}{c}
\vec{\d} \\ \vec{\phi}
\end{array}
\right] ( A)  \nonumber \\
& = & \sig_{abc} \ \sum_{\vec{m} \in \inte^n} 
e^{i\pi (\vec{m} + \vec{\d}) \cdot A \cdot (\vec{m} + \vec{\d})}
e^{2\pi i (\vec{m} + \vec{\d}) \cdot  \vec{\phi}}
\label{multith}
\eea
where $A$ is an $n \times n$ matrix related to (\ref{gomega}) and the intersection numbers of the $n$-cycles, and $\vec{\d},\ \vec{\phi}\  \in \real^n$ have entries defined by (\ref{paramT2ncpx1}, \ref{paramT2ncpx2}, \ref{paramT2ncpx3}). We thus see that the most general form of Yukawa couplings in intersecting brane worlds involves multi-theta functions, again paralleling the closed string case.

\section{Yukawa couplings in the MSSM-like example}

Let us analyse the Yukawa couplings of the MSSM-like example described in section \ref{example}. Notice that, although we have given a explicit realisation of the intersection numbers(\ref{intersec}) by specifying the wrapping numbers of each stack of branes, the mere knowledge of the intersection numbers $I_{\a\b}^{(r)}$ on each $T^2_r$ would have been enough for computing the Yukawa couplings in this model. Indeed, the general formula (\ref{totalyuki}) only depends on these topological invariant numbers, plus some open string and closed string moduli. 

Let us first concentrate on the quark sector of the model, which involves the triplets of branes $abc$ and $abc$*. These correspond to the Up-like and Down-like quark Yukawas, respectively, as can be checked in table \ref{mssm}. Since stacks $b$ and $c$ are parallel in the first torus, the relative position and Wilson lines here do not affect the Yukawas (only the $\mu$-parameter). The Yukawa couplings will be given by the product of two theta functions, whose parameters depend on the second and third tori moduli. Let us take the option $\rho = 1$ in table \ref{wnumbers}. By applying formulae (\ref{paramT2ncpx1}) and (\ref{paramT2ncpx2}) we easily find these parameters for the triplet $abc$ to be
\beq
\begin{array}{rcl}
\d^{(2)} & = & 
- \frac {i^{(2)}}{3} - \left( \frac{\eps_a^{(2)}}{3} + \eps_c^{(2)} \right), \\
\d^{(3)} & = &
- \frac{j^{(3)}}{3} + \left(\frac{\eps_a^{(3)}}{3} - \eps_b^{(3)} \right) 
- \frac{\eps_c^{(3)}}{3},
\end{array}
\label{param1}
\eeq
\beq
\begin{array}{rcl}
\phi^{(2)} & = & - \left( \th_a^{(2)} + 3 \th_c^{(2)} \right), \\
\phi^{(3)} & = & \left( \th_a^{(3)} - 3 \th_b^{(3)} \right) - \th_c^{(3)},
\end{array}
\label{param2}
\eeq
where we have set $\eps_b^{(3)}$, $\th_b^{(3)}$ both to zero, in order to have the enhancement $U(1)_b \raw SU(2)_L$. In fact, their value must be frozen to either $0$ or $1/2$, so there are several possibilities, but all of them can be absorbed into redefinitions of the other continuous parameters.Since the $abc$* triplet is related to $abc$ by orientifold reflection of one of its stacks, we can simply deduce its parameters by replacement $(\eps_c^{(3)},\th_c^{(3)}) \mapsto (-\eps_c^{(3)},-\th_c^{(3)})$, and $j^{(3)} \mapsto j^{(3)*}$ as the rules of section 3.1.2 teach us.

\begin{figure}
\centering
\epsfxsize=6.5in
\hspace*{0in}\vspace*{.2in}
\epsffile{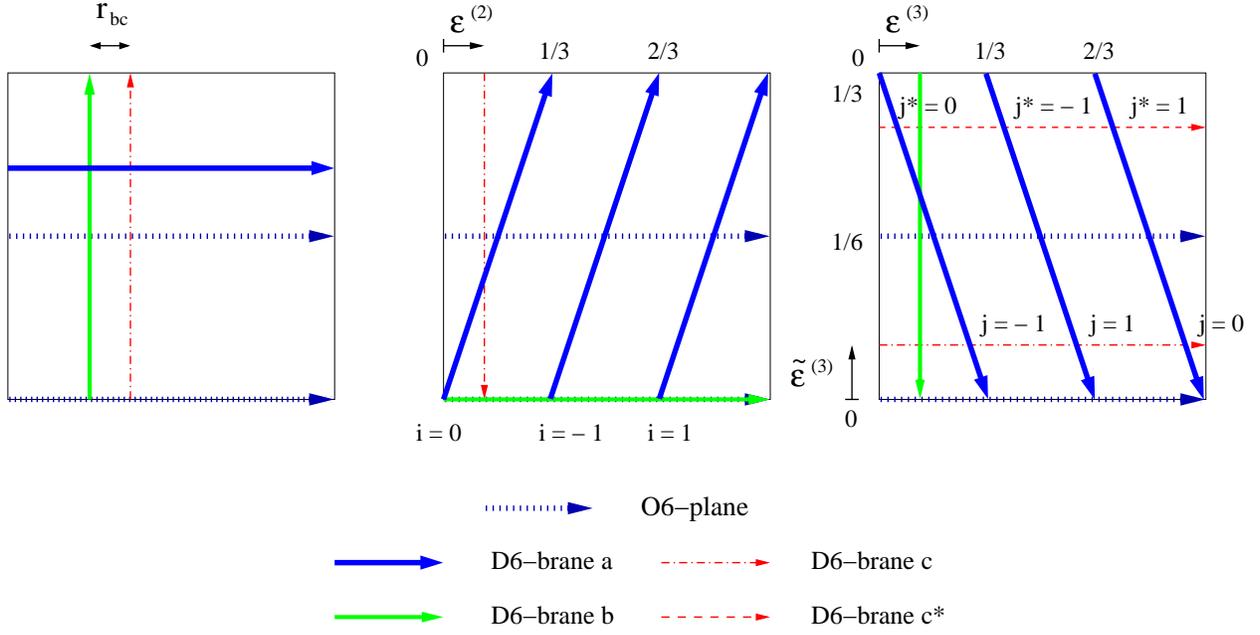}
\caption{Brane configuration corresponding to the MSSM-like model described in the text, for the choice $\rho = 1$. For simplicity, we have not depicted the leptonic stack nor the mirror $a$* stack.}
\label{guay}
\end{figure}

Since these open string moduli $\eps$ and $\th$ appear in very definite combinations, we can express everything in terms of new variables. These can be interpreted as the linear combination of chiral fields living on the branes worldvolumes that couple to matter in the intersections through Yukawa couplings. The discrete indices $i$, $j$, $j$*, which label such matter at the intersections, have also been redefined for convenience. The final result is presented in table \ref{moduli}, and the geometrical meaning of these new variables is shown in figure \ref{guay}. Notice that, by field redefinitions, we can always take our open string moduli to range in $\eps \in [0,1/3)$ and $\th \in [0,1)$.

\begin{table}[htb] 
\renewcommand{\arraystretch}{1.75}
\begin{center}
\begin{tabular}{|c||c||c|}
\hline
 & $abc$ & $abc$* \\
\hline
\hline
$\d^{(2)}$ 
& $\frac i3 + \eps^{(2)}$ 
& $\frac i3 + \eps^{(2)}$ \\
\hline
$\d^{(3)}$ & 
$\frac j3 + \eps^{(3)} + \tilde\eps^{(3)} $ 
& $\frac {j^*}{3} + \eps^{(3)} - \tilde\eps^{(3)} $  
\\
\hline
\hline
$\phi^{(2)}$ & 
$\th^{(2)}$ & $\th^{(2)}$ \\
\hline
$\phi^3$ & $\th^{(3)} + \tilde\th^{(3)}$ & $\th^{(3)} - \tilde\th^{(3)}$
\\
\hline
\end{tabular}
\end{center} 
\caption{\small Parameters in the MSSM-like model of table \ref{wnumbers}, for the case $\rho = 1$. $i$ labels left-handed quarks, whereas $j$, $j$* label up and down-like right-handed quarks respectively.
\label{moduli}}
\end{table}

Considering the leptonic sector involves triplets $dbc$ and $dbc$*. Now, since the stack $d$ is similar to the $a$, the above discussion also apply to this case, and the only change that we have to make is considering new variables $(\eps^{(2)}_l, \eps^{(3)}_l; \th^{(2)}_l, \th^{(3)}_l)$ instead of $(\eps^{(2)}, \eps^{(3)}; \th^{(2)}, \th^{(3)})$. Notice that the difference of this two sets of variables parametrizes the breaking $SU(4) \raw SU(3) \ti U(1)_{B-L}$, whereas $(\eps_c^{(3)},\th_c^{(3)})$ parametrize $SU(2)_R \raw U(1)_c$ breaking. 

On the other hand, Yukawa couplings depend only on two closed string parameters, namely the complex K\"ahler structures on the second and third tori, through $\k^{(r)} = 3 J^{(r)}/\a' = 3 \chi (R^{(r)})^2/\a'$, $r = 2, 3$. Since the index $i$ is an index labeling left-handed quarks, whereas $j$, $j$* label up-like and down-like right-handed quarks, our Yukawa couplings will be of the form
\beq
Y_{ij}^U Q_L^i H_u U_R^j, \quad \quad Y_{ij*}^D Q_L^i H_d D_R^{j*},
\label{yukawas}
\eeq
with Yukawa matrices
\beq
\begin{array}{c}
Y_{ij}^U \sim 
\vt \left[
\begin{array}{c}
\frac i3 + \eps^{(2)} \\ \th^{(2)}
\end{array}
\right] \left({3 J^{(2)}  \over \a^\prime}\right) 
\times 
\vt \left[
\begin{array}{c}
\frac j3 + \eps^{(3)} + \tilde\eps^{(3)}\\ 
\th^{(3)} + \tilde\th^{(3)}
\end{array}
\right] \left({3 J^{(3)}  \over \a^\prime}\right), \\
Y_{ij*}^D \sim 
\vt \left[
\begin{array}{c}
\frac i3 + \eps^{(2)} \\ \th^{(2)}
\end{array}
\right] \left({3 J^{(2)}  \over \a^\prime}\right) 
\times 
\vt \left[
\begin{array}{c}
\frac {j*}{3} + \eps^{(3)} - \tilde\eps^{(3)}\\ 
\th^{(3)} - \tilde\th^{(3)}
\end{array}
\right] \left({3 J^{(3)}  \over \a^\prime}\right).
\end{array} 
\label{yukmatrices}
\eeq
We can apply analogous arguments for the case $\rho = 1/3$ in table \ref{wnumbers}. The final result is
\beq
Y_{ij}^U \sim 
\vt \left[
\begin{array}{c}
\frac j3 + \eps^{(2)} \\ \th^{(2)}
\end{array}
\right] \left({3 J^{(2)}  \over \a^\prime}\right)
\times 
\vt \left[
\begin{array}{c}
\frac i3 + \eps^{(3)} + \tilde\eps^{(3)}\\ 
\th^{(3)} + \tilde\th^{(3)}
\end{array}
\right] \left({3 J^{(3)}  \over \a^\prime}\right)
\label{yukmatrices2}
\eeq
for the up-like couplings, whereas the down-like ones are obtained form (\ref{yukmatrices2}) by the replacement $(\eps_c^{(3)},\th_c^{(3)}) \mapsto (-\eps_c^{(3)},-\th_c^{(3)})$.

The quark and lepton mass spectrum can be easily computed from these data. Indeed, let us consider the quark mass matrices proportional to (\ref{yukmatrices}), and define
\beq
\begin{array}{rcl}
a_i & \equiv & 
\vt \left[
\begin{array}{c}
\frac i3 + \eps^{(2)} \\ \th^{(2)}
\end{array}
\right] \left({3 J^{(2)}  \over \a^\prime}\right),
\\
b_j & \equiv &
\vt \left[
\begin{array}{c}
\frac j3 + \eps^{(3)} + \tilde\eps^{(3)}\\ 
\th^{(3)} + \tilde\th^{(3)}
\end{array}
\right] \left({3 J^{(3)}  \over \a^\prime}\right),
\\
\tilde b_{j*} & \equiv &
\vt \left[
\begin{array}{c}
\frac {j*}{3} + \eps^{(3)} - \tilde\eps^{(3)}\\ 
\th^{(3)} - \tilde\th^{(3)}
\end{array}
\right] \left({3 J^{(3)}  \over \a^\prime}\right).
\end{array}
\label{defin}
\eeq
Then, the Yukawa matrices can be expressed as
\beq
Y^U \sim A \cdot 
\left(
\begin{array}{ccc}
1 & 1 & 1 \\
1 & 1 & 1 \\
1 & 1 & 1
\end{array}
\right)
\cdot B,
\quad \quad
Y^D \sim A \cdot 
\left(
\begin{array}{ccc}
1 & 1 & 1 \\
1 & 1 & 1 \\
1 & 1 & 1
\end{array}
\right)
\cdot \tilde B.
\label{redef}
\eeq

\beq
\begin{array}{ccc}
A = \left(
\begin{array}{ccc}
a_1 &     &  \\
    & a_0 &  \\
    &     &  a_{-1}
\end{array}
\right) \quad
&
B = \left(
\begin{array}{ccc}
b_1 &     &  \\
    & b_0 &  \\
    &     &  b_{-1}
\end{array}
\right) \quad
&
\tilde B = \left(
\begin{array}{ccc}
\tilde b_1 &     &  \\
    & \tilde b_0 &  \\
    &     &  \tilde b_{-1}
\end{array}
\right).
\end{array}
\label{defin2}
\eeq

In order to compute the mass eigenstates, we can consider the hermitian, definite positive matrix $Y\cdot Y^\dag$ and diagonalize it. Let us take, for instance, $Y^U$. We find
\beq
Y^U \cdot (Y^U)^\dag \sim \Tr (B\cdot \bar B) \quad
A \cdot
\left(
\begin{array}{ccc}
1 & 1 & 1 \\
1 & 1 & 1 \\
1 & 1 & 1
\end{array}
\right)
\cdot \bar A,
\label{square}
\eeq
where bar denotes complex conjugation. This matrix has one nonzero eigenvalue given by 
\bea
\lam^U = \Tr (A\cdot \bar A) \ \Tr (B\cdot \bar B),
& \quad & 
| \lam^U \rangle = {A \over \sqrt{\Tr (A\cdot \bar A)}} \cdot 
\left(
\begin{array}{c}
1 \\ 1 \\ 1
\end{array}
\right),
\label{masseigenval}
\eea
and two zero eigenvalues whose eigenvectors span the subspace
\beq
\bar A^{-1} \cdot 
\left[{\rm Ker}
\left(
\begin{array}{ccc}
1 & 1 & 1 \\
1 & 1 & 1 \\
1 & 1 & 1
\end{array}
\right)
\right]
\bigcup
{\rm Ker} \bar A.
\label{zeroeigenval}
\eeq
Similar considerations can be applied to $Y^D$, and the results only differ by the replacement $B \raw \tilde{B}$. This provides a natural mass scale between up-like and down-like quarks:
\beq
{m_U \over m_D} \sim 
\sqrt{\frac{\Tr (B\cdot \bar B)}{\Tr (\tilde{B}\cdot \bar{\tilde{B}})}}
\eeq
(we should also include $\langle H_u \rangle / \langle H_d \rangle$ in order to connect with actual quark masses). This ratio is equal to one whenever $\tilde \eps^{(3)} = - \tilde \eps^{(3)} \ {\rm mod} \ 1/3$ and $\tilde \th^{(3)} = - \tilde \th^{(3)} \ {\rm mod} \ 1$. These points in moduli space correspond precisely to the enhancement $U(1)_c \raw SU(2)_R$, where we would expect equal masses for up-like and down-like quarks. On the other hand, we find that the CKM matrix is the identity at every point in the moduli space. 

Thus we find that in this simple model only the third generation of quarks and leptons are massive. This could  be considered as a promising starting point for a phenomenological description of the SM fermion mass spectrum. One can conceive that small perturbations of this simple brane setup can give rise to smaller but non-vanishing masses for the rest of quarks and leptons as well as non-vanishing CKM mixing \footnote{Note, for example, that the symmetry properties of
the Yukawa couplings leading to the presence of two massless modes disappear in the presence of a small non-diagonal  component of the K\"ahler form $\omega$ as discussed at the end of section \ref{toroidal}.}. We postpone a detailed study of Yukawa couplings in semirealistic intersecting D-brane models to future work.

The previous discussion parallels for the case $\rho = 1/3$. Indeed, we find the same mass spectrum of two massless and one massive eigenvalue for each Yukawa matrix. The only difference arises from the CKM matrix, which is not always the identity but only for the special values of $(\tilde \eps^{(3)}, \tilde \th^{(3)})$ where the symmetry enhancement to $SU(2)_R$ occurs.

\section{Extension to elliptic fibrations}

Although so far we have concentrated on computing Yukawa couplings in toroidal compactifications, it turns out that the same machinery can be applied to certain D-brane models involving non-trivial Calabi-Yau geometries. Indeed, in \cite{Uranga:2002pg} a whole family of intersecting D6-brane models wrapping 3-cycles of non-compact ${\bf CY_3}$'s was constructed. The simplest of such local Calabi-Yau geometries was based on elliptic and $\cpx$* fibrations over a complex plane, parametrized by $z$. In this setup, gauge theories arise from D6-branes wrapping compact special Lagrangian 3-cycles which, roughly speaking, consist of real segments in the complex $z$-plane over which two $S^1$ are fibered. One of such $S^1$ is a non-contractible cycle in $\cpx$*, while the other wraps a $(p,q)$ 1-cycle on the elliptic fiber. The intersection of any such compact 3-cycles is localized on the base point $z=0$, where the $\cpx$* fibre degenerates to $\cpx \ti \cpx$. We refer the reader to \cite{Uranga:2002pg,Hori:2000ck,Hanany:2001py} for details on this construction. 

The important point for our discussion is that the geometry of any number of intersecting D6-branes can be locally reduced to that of intersecting 1-cycles on the elliptic fiber in $z=0$, that is, to that of $(p_\a, q_\a)$ cycles on a $T^2$. Moreover, due to this local geometry, any worldsheet instanton connecting a triplet of D6-branes will also be localized in the elliptic fiber at $z=0$. The computation of Yukawas in this ${\bf CY_3}$ setup then mimics the one studied in subsection \ref{twotorus}, where we considered worldsheet instantons on a $T^2$. 

As a result, we find that the structure of Yukawa couplings computed in section \ref{toroidal}, which could be expressed in terms of (multi) theta functions, is in fact more general than the simple case of factorisable cycles in a $T^{2n}$. In fact, it turns out to be even more general than intersecting brane worlds setup. Indeed, as noticed in \cite{Uranga:2002pg}, this family of non-compact ${\bf CY_3}$ geometries is related, via mirror symmetry, to Calabi-Yau threefold singularities given by complex cones over del Pezzo surfaces. In turn, the intersecting D6-brane content corresponds to D3-branes sitting on such singularities.

\begin{figure}
\centering
\epsfxsize=6in
\hspace*{0in}\vspace*{.2in}
\epsffile{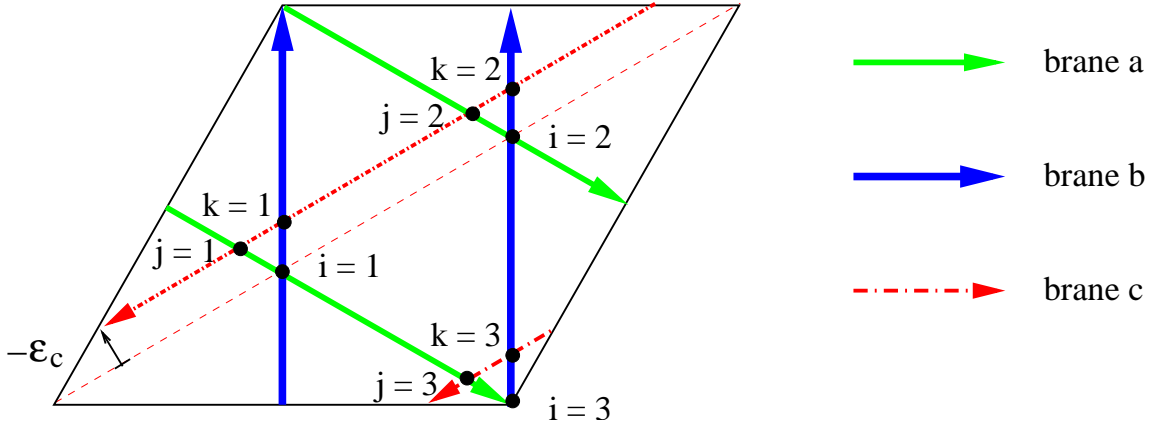}
\caption{D-brane configuration in (\ref{1cycles}), restricted to the elliptic fibre at $z=0$ in the base space. Due to the quantum $\inte_3$ symmetry of this configuration, each stack of branes must be related to the others by a $e^{\pm 2\pi i/3}$ rotation. Hence, the complex structure of the elliptic fibre is frozen. Notice that a triangle between vertices $i$, $j$, $k$ closes only if the condition $i + j + k \equiv 0$ mod $3$ is satisfied.}
\label{local1}
\end{figure}

Let us illustrate these facts with a simple example already discussed in \cite{Uranga:2002pg}, section 2.5.1. The brane content consist of three stacks of $N$ branes each, wrapping the 1-cycles
\beq
\Pi_a = (2,-1), \quad \Pi_b = (-1,2), \quad  \Pi_c = (-1,-1),
\label{1cycles}
\eeq
with intersection numbers $I_{ab} = I_{bc} = I_{ca} = 3$. This yields a simple $\N=1$ spectrum with gauge group $U(N)^3$ and matter triplication in each bifundamental. We have depicted the D-brane content of (\ref{1cycles}) in figure \ref{local1}, restricting ourselves to the elliptic fiber on the base point $z=0$. Notice that the complex structure of such $T^2$ is fixed by the $\inte_3$ symmetry that the whole configuration must preserve \cite{Uranga:2002pg}.

Notice also that the intersection numbers are not coprime, so the Yukawa couplings between intersections $i$, $j$, $k$ will be given by a theta function with characteristics (\ref{paramT2nc1}), (\ref{paramT2nc2}) and (\ref{paramT2nc3}), with $d = 3$. Given the specific choice of numbering of figure \ref{local1}, we can take the linear function $s$ to be $s = -k - 2j$. Moreover, not all the triplet of intersections are connected by an instanton, but they have to satisfy the selection rule
\beq
i + j + k \equiv 0 \ {\rm mod} \ 3.
\label{selection}
\eeq

This give us the following form for the Yukawa couplings in the present model
\beq
Y_{ij1} \sim 
\left(
\begin{array}{ccc}
A & 0 & 0 \\
0 & 0 & B \\
0 & C & 0
\end{array}
\right), \quad
Y_{ij2} \sim 
\left(
\begin{array}{ccc}
0 & 0 & C \\
0 & A & 0 \\
B & 0 & 0
\end{array}
\right), \quad
Y_{ij3} \sim 
\left(
\begin{array}{ccc}
0 & B & 0 \\
C & 0 & 0 \\
0 & 0 & A
\end{array}
\right),
\label{yukilocal}
\eeq
with
\beq
A = 
\vt \left[
\begin{array}{c}
\eps/3 \\ 3\th
\end{array}
\right] (3 J/ \a'), \quad
B = 
\vt \left[
\begin{array}{c}
(\eps - 1)/3 \\ 3\th
\end{array}
\right] (3 J/ \a'), \quad
C = 
\vt \left[
\begin{array}{c}
(\eps + 1)/3 \\ 3\th
\end{array}
\right] (3 J/ \a'),
\label{thetas}
\eeq
and where we have defined the parameters $\eps = \eps_a + \eps_b + \eps_c,\ \th = \th_a + \th_b + \th_c \in [0,1)$.

A particularity of these elliptically fibered 3-cycles which the D6-branes wrap is that, topologically, they are 3-spheres. This means they are simply connected and, by \cite{McLean}, their moduli space is zero-dimensional. This means that the D-brane position parameter $\eps_\a$ is fixed, and the same story holds for $\th_\a$. Although frozen, we do not know the precise value of these quantities and, presumably, different values will correspond to different physics. 

\begin{figure}
\centering
\epsfxsize=6in
\hspace*{0in}\vspace*{.2in}
\epsffile{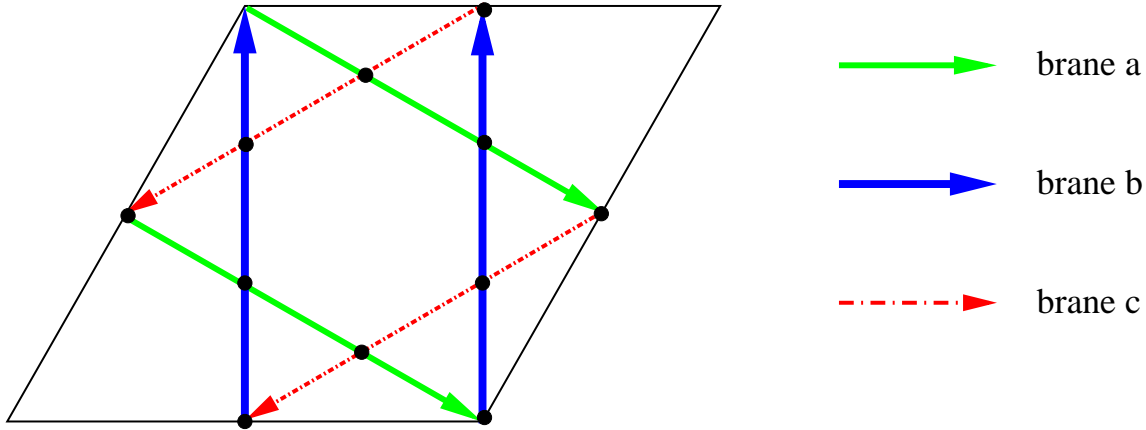}
\caption{Final D-brane configuration, with the brane positions fixed by (\ref{C=0}). Again we restrict to the elliptic fibre at $z=0$ in the base space.}
\label{local2}
\end{figure}

This simple model with matter triplication is in fact mirror to the $\cpx^3/\inte_3$ orbifold singularity and, indeed, the chiral matter content exactly reproduces the one obtained from D3-branes at that singularity, in $N$ copies of the fundamental representation \cite{Uranga:2002pg}. The superpotential of such mirror configuration is given by
\beq
W = \sum_{\{abc\}} \eps^{ijk} [\Phi_{ab}^i \Phi_{bc}^j \Phi_{ca}^k],
\label{superp}
\eeq
where $\{abc\}$ means that we have to consider all the cyclic orderings. This superpotential implies Yukawa couplings of the form $Y_{ijk} \sim \eps^{ijk}$. We see that we can reproduce such result in terms of the general solution (\ref{yukilocal}), only if one of the entries $A$, $B$ or $C$ vanishes. Let us take $C \equiv 0$, which can be obtained by fixing the theta-function parameters to be
\beq
\eps = \oh, \quad \quad \th = \frac{2m+1}{6}, \ m \in \inte.
\label{C=0}
\eeq
Is easy to see that this condition also implies that $|A| = |B|$. More precisely,
\beq
A = Z \cdot e^{2\pi i (m + \oh)\frac{1}{6}},
\quad \quad
B = Z \cdot e^{-2\pi i (m + \oh)\frac{1}{6}}, 
\quad \quad
Z \in \cpx.
\label{conq}
\eeq
Now, if we perform the relabeling 
\beq
\begin{array}{cc}
i:  & 1 \lraw 3 \\
j:  & 1 \lraw 2 \\
k:  & 2 \lraw 3 
\end{array}
\label{relab}
\eeq
(which preserves the condition (\ref{selection})) we are led to Yukawa couplings of the form
\beq
Y_{ij1} \sim 
Z \cdot \left(
\begin{array}{ccc}
 0 & 0 & 0 \\
 0 & 0 & \bar \om \\
 0 & \om & 0
\end{array}
\right), \quad
Y_{ij2} \sim 
Z \cdot \left(
\begin{array}{ccc}
0 & 0 & \om \\
0 & 0 & 0 \\
\bar \om & 0 & 0
\end{array}
\right), \quad
Y_{ij3} \sim 
Z \cdot \left(
\begin{array}{ccc}
0 & \bar \om & 0 \\
\om & 0 & 0 \\
0 & 0 & 0
\end{array}
\right),
\label{yukilocal2}
\eeq
where $\om = {\rm exp} (2\pi i (m + 1/2)\cdot 1/6)$. By taking the choice $m = 1$, we obtain $\om = i$, so that $Y_{ijk} \sim \eps^{ijk}$, as was to be expected from (\ref{superp}). There are, however, two other inequivalent choices of $\th$, given by $m = 0, 2$. Is easy to check that these two values yield the superpotentials corresponding to the two choices of $\cpx^3/(\inte_3 \times \inte_3 \times \inte_3)$ orbifold singularity with discrete torsion, which have the same chiral spectrum as a plain $\cpx^3/\inte_3$ orbifold. \footnote{It is, however, far from clear that these configurations are actually mirror to orbifolds $\cpx^3/(\inte_3 \times \inte_3 \times \inte_3)$ with discrete torsion. Further checks involving, e.g., matching of moduli spaces should be performed to test this possibility.}. We present such final configuration in figure \ref{local2}.

\section{Yukawa versus Fukaya}

In the previous section we have shown how, combined with mirror symmetry, the computation of worldsheet instantons between chiral fields in intersecting D-brane models can yield a powerful tool to compute Yukawa couplings in more general setups as, e.g., D-branes at singularities. The purpose of the present section is to note that computation of Yukawas and other disc worldsheet instantons is not only a tool, but lies at the very heart of the definition of mirror symmetry. The precise context to look at is Kontsevich's homological mirror symmetry conjecture \cite{Konty}, performed before the importance of D-branes was appreciated by the physics community. This proposal relates two a priori very different constructions in two different $n$-fold Calabi-Yau manifolds $\M_6$ and $\cw$, which are dual (or mirror) to each other. $\M_6$ is to be seen as a $2n$-dimensional symplectic manifold with vanishing first Chern class, while $\cw$ shall be regarded as an $n$-dimensional complex algebraic manifold. From the physics viewpoint these are the so-called A-side and B-side of the mirror map, respectively. The structure of complex manifold allows us to construct the derived category of coherent sheaves from $\cw$, which classifies the boundary conditions in the B-twisted topological open string theory model on such manifold, hence the spectrum of BPS branes on the B-side \cite{Douglas:2000gi}. On the other hand, the symplectic structure on $\M_6$ naturally leads to the construction of Fukaya's category. Kontsevich's proposal amounts to the equivalence of both categories. \footnote{Actually, it identifies a properly enlarged (derived) version of Fukaya's category to a full subcategory of coherent sheaves. Since we are only interested on the A-side of the story, we will not deal on these subtle points regarding the mirror map.}

Intersecting D6-brane configurations as described in \ref{CY&sL} lie on the A-side of the story. Hence, they should be described by Fukaya's category. This seems indeed to be the case and, in the following, we will try to describe the physical meaning of Fukaya's mathematical construction from the point of view of intersecting D-brane models, paying special attention to the role of worldsheet instantons.

Roughly speaking, a category is given by a set of objects and morphisms between them. The objects of the category we want to construct are pairs of the form $\cl_\a = (\Pi_\a, \ce_\a)$, of special Lagrangian submanifolds $\Pi_\a$ of $\M_6$, endowed with unitary local systems $\ce_\a$, or flat $U(N_\a)$ gauge bundles. This is precisely the set of objects used as building blocks in our ${\bf CY_n}$ intersecting brane configurations, each object associated to a stack of $N_\a$ D-branes \footnote{In the general construction, Fukaya considers a different (larger) set of objects in $\M_6$, which are equivalence classes of Lagrangian submanifolds identified by Hamiltonian diffeomorphisms. The precise relation between the moduli space of these two set of objects is still an open problem, but in some simple cases as Lagrangian tori in $\M_6 = T^{2n}$ they can be shown to be equal \cite{theta}.}. The morphisms between a pair of objects are generated by the points of intersection of the two sL submanifolds, more precisely
\beq
Hom (\cl_\a, \cl_\b) = {\# (\Pi_\a \cap \Pi_\b)} \otimes Hom (\ce_\a, \ce_\b),
\label{morph}
\eeq
where $Hom (\ce_\a, \ce_\b)$ represents a morphism of gauge bundles. The morphisms between two objects, then, are naturally associated to the set of chiral fields living at the intersections of two D-branes, which transform under the bifundamental representation of the corresponding gauge groups.

Actually, (\ref{morph}) can be converted to a {\it graded} set of morphisms. That is, each intersection $p$ can be attached with an index $\eta(p) \in \inte$, named Maslov index. The physical meaning of such index is related to how one must describe the spectrum of particles at the intersections from the underlying CFT viewpoint, i.e., two sets of fields arising from two elements of $\Pi_\a \cap \Pi_\b$ may look locally similar, but related by some spectral flow that the global theory must take into account. In the case of Lagrangian tori, i.e., factorisable D-branes, in $T^{2n}$ all the intersections have the same Maslov index, which can be defined as \cite{theta}
\beq
\eta (p)_{\a\b} = \sum_{r=1}^n [\th_{\a\b}^r],
\label{maslov}
\eeq
where $[\th_{\a\b}^r]$ is (the integer part of) the angle between $\Pi_\a$ and $\Pi_\b$ on the $r^{th}$ two-torus, measured in the counterclockwise sense and in units of $\pi$. A quick look at the case $n = 3$ (D6-branes) reveals that $\eta (p)$ mod $2$ can be associated to the chirality of the corresponding field at the intersection $p$. In fact, this grading was already present from the very beginning since, actually, (\ref{morph}) is only valid whenever $\Pi_\a$ and $\Pi_\b$ intersect transversally in $\M_6$. In case their intersection is not a point, we should replace (\ref{morph}) by the cohomology complex $H^*(\Pi_\a \cap \Pi_\b, Hom (\ce_\a, \ce_\b))$, with its associate grading. We finally find that the morphisms of our construction are given by the graded set
\beq
Hom^i (\cl_\a, \cl_\b) \simeq \bigotimes_{p \in \Pi_\a \cap \Pi_\b} 
H^{i-\eta(p)} (\Pi_\a \cap \Pi_\b, Hom (\ce_\a, \ce_\b)).
\label{graded}
\eeq

In general, categories not only consist of objects and morphisms between them, but also of compositions of morphisms. In \cite{Fuka} Fukaya proved that, actually, an involved structure can be defined in the previous described category. This structure is based on the properties of an $A^\infty$ algebra, which is a generalization of a differential graded algebra, and makes Fukaya's category into an $A^\infty$ category. The definition of such category involves
\begin{itemize}

\item A class of objects

\item For any two objects $X$, $Y$ a \inte-graded Abelian group of morphisms $Hom(X,Y)$

\item Composition maps
\beq
m_k : Hom(X_1,X_2) \otimes  Hom(X_2,X_3) \dots \otimes  Hom(X_k,X_{k+1}) \raw  Hom(X_1,X_{k+1}),
\nonumber
\eeq
of degree $2-k$ for all $k \geq 1$, satisfying the $n^{th}$ order associativity conditions

$\sum_{r,s} (-1)^\eps m_{n-r+1} (a_1 \otimes \dots \otimes a_{s-1} \otimes m_r(a_s \otimes \dots \otimes a_{s+r-1}) \otimes a_{s+r} \otimes \dots \otimes a_n) = 0$

for all $n \geq 1$. Here $\eps = (s+1)k + s \left(n + \sum_{j=1}^{s-1} {\rm deg} (a_j)\right)$.

\end{itemize}

Notice that the associative condition for $n=1$ reduces to $m_1$ being a degree one operator $m_1: Hom(X,Y) \raw Hom(X,Y)$ such that $m_1 \circ m_1 = 0$, hence it defines a coboundary operator which makes $Hom(X,Y)$ a cochain complex (e.g., just as the exterior derivative in the deRham complex). Furthermore, the second relation ($n=2$) means that $m_2: Hom(X,Y) \times Hom(Y,Z) \raw Hom(X,Z)$ is a cochain homomorphism, which induces a product on cohomology groups (e.g., analogous to a wedge product in deRham cohomology). 

Up to now we have described the two first items of the previous definition, and is in the third one where worldsheet instantons will play a role. Indeed, the composition maps $m_k$ are constructed by considering holomorphic maps $\phi: D \raw \M_6$, taking the boundary of $D$ to $k+1$ Lagrangian submanifolds $\cl_\a$ in $\M_6$ and $k+1$ marked points $z_\a$ in the cyclic order of $\partial D$ to the intersection points $p_\a \in \Pi_\a \cap \Pi_{\a+1}$ (see figure \ref{yukis3} for the case $k = 2$). Two of these maps are regarded  as equivalent if related by a disc conformal automorphism $Aut(D) \simeq PSL(2,\real)$. This is precisely the definition of a euclidean worldsheet connecting $k+1$ fields at the intersection of $k+1$ D-branes and satisfying the classical equations of motion. Let us fix $k+1$ intersections $p_\a \in \Pi_\a \cap \Pi_{\a+1}$, and the corresponding $k+1$ matrices $t_\a \in Hom(\ce_\a|_{p_\a}, \ce_{\a+1}|_{p_{\a+1}})$. If we choose them to have the appropriate degree, the moduli space of holomorphic inequivalent maps connecting them will be discrete, so we can sum over all of such $\phi$ to obtain the quantity
\beq
C((p_1,t_1), \dots,(p_k,t_k), p_{k+1}) = \sum_{\phi} \pm 
\ e^{2\pi i \int_{\phi(D)} \om}
\ P e^{\oint_{\phi(\partial D)} \b},
\label{instanton}
\eeq
where the $\pm$ sign stands for holomorphic and antiholomorphic maps, respectively and $\om$ is the complexified K\"ahler form in $\M_6$. Finally, $P$ stands for a path-ordered integration on the boundary of the disc, with $\b$ is being the connection of the flat bundles along the boundaries in $\cl_\a$'s, and the matrices $t_i$ inserted at the boundary marked points \cite{Polishchuk:1998db}. The r.h.s. of (\ref{instanton}) is thus a homomorphism of $\ce_1$ to $\ce_{k+1}$, and we can define the composition maps $m_k$ as
\beq
m_k ((p_1,t_1), \dots,(p_k,t_k)) = 
\sum_{p_{k+1} \in \Pi_1 \cap \Pi_{k+1}}
C((p_1,t_1), \dots,(p_k,t_k), p_{k+1}) \cdot p_{k+1}.
\label{comp}
\eeq

From these definitions, is easy to see that the computation of the composition maps $m_k$ is equivalent to computing the worldsheet instanton correction to the superpotential involving $k+1$ chiral fields at the D-branes intersections. Indeed, (\ref{instanton}) is nothing but a generalization of (\ref{yukabs}) for $k+1$ disc insertions. In the case $k=1$, we are evaluating the instantons connecting two chiral fields at the intersections of the same pair of D-branes $\cl_1$, $\cl_2$, just as in figure \ref{recombination} (2). Notice that $m_1$ has degree $1$, so it can be regarded as a chirality-changing operator, i.e., a mass term. If we consider D-brane configurations consisting of Lagrangian $n$-tori on $T^{2n}$ its action is trivial, $m_1 \equiv 0$, but this will not be the case in general. Computing the cohomology
\beq
{\{ {\rm Ker} 
\left(m_1: Hom^{i} (\cl_\a, \cl_\b)
\raw Hom^{i+1} (\cl_\a, \cl_\b) \right) \}
\over
\{ {\rm Im} 
\left(m_1: Hom^{i-1} (\cl_\a, \cl_\b)
\raw Hom^{i} (\cl_\a, \cl_\b) \right) \} },
\label{cohom}
\eeq
is equivalent to computing the massless fields in our D-brane configuration. On the other hand, the case $k=2$ coincides with the general computation of Yukawa couplings between chiral fields at sL's intersections described in section \ref{ibm&yuk}. Indeed, the action of $m_2$ involves three objects of our configurations, hence three D-branes $(a, b, c)$, and three different morphisms between them, hence a triplet of intersections $(i,j,k)$. For each such choice we can define an element of the form (\ref{instanton}), which represents a Yukawa coupling. Indeed, in the phenomenological setup of figure \ref{yukis3}, if we let $i,j$ index left- and right-handed quarks, respectively, the matrix of maps $(M_2)_{ij} = m_2((p_i,t_i),(p_j,t_j))$ is equivalent to the Yukawa matrix of such chiral fields. In the same manner, the composition maps $m_k$, $k \geq 3$ encode the corrections to the superpotential involving non-renormalizable higher dimensional couplings.

\begin{table}[htb] 
\renewcommand{\arraystretch}{1.7}
\begin{center}
\begin{tabular}{ccc} 
\hline 
\hline
$A^\infty$ category & Fukaya Category & Intersecting D-branes  \\ 
\hline 
\hline
\vspace{-0.25cm}
Objects & special Lagrangian submanifolds $\Pi_\a$ & D-brane stacks \\
& endowed with unitary systems $\ce_\a$ & with flat gauge bundles \\
\hline
\vspace{-0.25cm}
Morphisms &  $Hom(\ce_\a, \ce_\b)$ generated by & 
Chiral matter in bifundamentals \\
& intersection points $p_{\a\b} \in \Pi_\a \cap \Pi_\b$ & 
localized at the intersections\\
\hline
\vspace{-0.25cm}
Grading & Maslov index $\eta(p_{\a\b})$ & Spectral flow index \\
& & (mod $2$ $=$ chirality) \\
\hline
Composition maps & Holomorphic discs & 
Worldsheet instanton corrections \\
& $m_1$ & mass operator \\
& $m_2$ & Yukawa coupling \\
\vspace{-0.25cm}
& $m_k$, $k \geq 3$ & non-renormalizable couplings \\
& & involving $k+1$ chiral fields \\
\hline
\end{tabular}
\end{center}
\caption{\small Fukaya-Yukawa dictionary.\label{corresp}}
\end{table}

We have summarized in table \ref{corresp} the physical interpretation of the basic quantities conforming Fukaya's category. This list is not meant to be complete. In fact, it turns out that Fukaya's categories are not always well defined, and not until very recently has a complete definition of such construction been provided \cite{Fu1,Fu2}. The computation of such category in the simplest case of a Calabi-Yau one-fold, i.e., the elliptic curve, was performed in \cite{Polishchuk:1998db}, where the match with the derived category of coherent sheaves in a T-dual torus was also performed, proving Kontsevich's conjecture in this case. The extension to higher-dimensional symplectic tori followed in \cite{theta}. The similarities of the results in those papers with the computations of section \ref{toroidal} are not an accident, but a sample of the deep mathematical meaning of the computation of Yukawa couplings in intersecting D-brane models.

\chapter{Conclusions}

In this work we have presented a detailed study of a class of constructions based on Superstring Theory, baptized in the literature as Intersecting Brane Worlds. The main motivation of this work is of phenomenological nature, in particular obtaining string-based constructions providing an effective field theory as close as possible to the Standard Model of Particle Physics. Throughout our research, however, we have faced many issues of theoretical interest, mainly regarding D-brane physics. We have, as well, paved the way towards constructing new classes of string theory models. 

In this research we have followed a `Bottom-up philosophy'. That is, we have mainly considered the features of the construction affecting the gauge sector of the theory, not bothering about other sectors such as gravitation. This philosophy assumes that, when performing a string compactification, the gauge sector of the effective theory is not sensitive to the whole construction, but only to the details regarding some part of it. This assumption is indeed true in many D-brane models, in particular in Intersecting Brane Worlds, where gauge theories are localized in the worldvolumes of some D$p$-branes involved in the construction. Hence, in order to achieve a D-brane string-based model with a realistic low-energy gauge sector, we only have to care about the details of the construction affecting the local physics of the open string sector.

One of the main issues of our constructions is chirality, which naturally arises from D-brane intersections. We have introduced the simplest system of D-branes where chiral fermions arise, which is flat D-branes intersecting at angles, and then considered several kinds of compactifications involving them. These include D6-branes wrapping 3-cycles on a flat $T^6$, D5-branes wrapping 2-cycles on $T^4$ and finally D4-branes wrapping 1-cycles on $T^2$. In the last two cases $T^{2n}$ has to seat in an orbifold singularity of the kind $\cpx^{3-n}/\inte_N$ in order to get a $D=4$ chiral spectrum. We have also considered orientifolded versions of these constructions. The effective field theory spectrum for a generic configuration of intersecting D-branes has been deduced in each case, paying special attention to the open string sector, where gauge theories and chiral fermions arise. We have been mainly concerned with massless and tachyonic string excitations, the latter indicating an instability of the configuration against D-brane recombination. In addition, we have discussed the appearance of light particles that may put phenomenological constraints to a model.

Given these six families of intersecting D-brane models, we have deduced the conditions they should satisfy in order to yield consistent string constructions. Specifically, we have studied cancellation of Ramond-Ramond tadpoles. We have seen that they restrict the topological data of a D-brane configuration, imposing conditions on the homology classes where D-branes wrap and on their orbifold charges. We have also seen how RR tadpole cancellation implies cancellation of chiral anomalies in the effective theory. In particular, we have computed non-Abelian, mixed and cubic $U(1)$ anomalies on each different construction. While the cancellation of non-Abelian anomalies is automatic once RR tadpoles conditions have been imposed, mixed and cubic $U(1)$ anomalies need of the presence of a generalized GS mechanism, mediated by closed string fields. As a direct consequence of the GS mechanism some Abelian factors of the gauge group will become massive, although they will remain as global symmetries of the effective Lagrangian.

After these general considerations, we have presented specific models involving D6 and D5-branes at angles. These are orientifold constructions which have the appealing feature of yielding just the chiral content and gauge group of a three-generation Standard Model already from the start, without any effective field theory assumption. There are a number of remarkable properties which seem more general than these specific toroidal and orbifold examples. {\it i)} Baryon and Lepton numbers are exact (gauged) symmetries in perturbation theory.  {\it ii)} There are three generations of right-handed neutrinos but no Majorana neutrino masses are allowed. {\it iii)} There is a gauged (generation dependent) Peccei-Quinn-like symmetry. All these facts depend on the $U(1)$ structure of the D-brane configuration, which ultimately depends on the topological data of the configuration. This implies that if we find a general model, say D6-branes on a ${\bf CY_3}$, with the same D-brane content and intersection numbers they will be valid as well.

The specific examples we have provided are non-supersymmetric configurations for all values of the closed string moduli. It thus seems that the only way to avoid the Hierarchy problem is to lower the string scale down to $1-10$ TeV. Four-dimensional Planck mass can be maintained at its experimental value by taking some dimensions transverse to the D-branes very large. Although this mechanism is hard to implement in D6-brane configurations wrapping a $T^6$, it is natural in D5-brane models, showing that it is feasible in a general construction. Notice that Baryon number conservation is particularly welcome in these low-string scenarios, in which up to now there was no convincing explanation for the absence of fast proton decay.

Supersymmetry breaking, however, is a difficult issue in superstring theory. In D6-brane models it will lead to the existence of NSNS tadpoles, whereas in D5-brane models it already appears as a twisted closed string tachyon. These features leave the full stability of the configuration as an open question. Motivated by this, we perform a study of the preserved supersymmetries for general D6-brane models in ${\bf CY_3}$ compactifications. We pay particular attention to the case of $T^6$, where the space of supersymmetric configurations between two D6-branes has the structure of a tetrahedron. We explain this structure in terms of bulk supersymmetries. The extension to general ${\bf CY_3}$ compactifications can be done by means of calibrated and special Lagrangian geometry. In general, we find that supersymmetry does not only depend on topological data of the construction, but also involves geometric data as the holonomy and complex structure of the ${\bf CY_3}$. We also find that many physical quantities of a supersymmetric effective field theory can be computed with the help of calibrated geometry. For instance, the number of $D=4$ supersymmetries in a D-brane gauge theory, the (tree level) value of the gauge kinetic function, and the Fayet-Iliopoulos parameters. Finally, we can also compute the superpotential of the theory, and in particular Yukawa couplings.

The computation of Yukawa couplings in Intersecting Brane Worlds has been addressed in the last part of this thesis. These couplings arise from worldsheet instantons with the topology of a disc and involving three different boundary conditions. We have computed them explicitly in the case of D-branes wrapping factorisable $n$-cycles on $T^{2n}$. We find that Yukawa couplings can be expressed as a product of theta functions, which depend on the K\"ahler class of the torus and on the open string moduli (D-brane positions and Wilson lines) but not on the complex structure. We have computed such Yukawa couplings in a specific example, based on D6-branes wrapping 3-cycles of $T^6$, and yielding the chiral spectrum of the MSSM. We have found that, in this specific example, only one generation of quarks and leptons do get a mass after EW symmetry breaking, which may be considered as a good starting point for a description of the experimental data.

We have also seen how the formulas that we have derived for D-branes intersecting at angles in toroidal compactifications may in fact be valid for more general constructions. Indeed, by means of Mirror Symmetry, we have related our results to the Yukawas of D3-branes sitting on complex cones over del Pezzo surfaces. In particular, we have rederived the Yukawa couplings of D3-branes on a $\cpx^3/\inte_3$ singularity. Actually, the computation of Yukawa couplings is not only a tool for matching mirror models, but lies at the very heart of the definition of Mirror Symmetry. The precise context where to look at is Kontsevich's Homological Mirror Symmetry conjecture, through the computation of Fukaya's category in a Calabi-Yau manifold. We have provided a dictionary between quantities on this mathematical construction and the associated physical objects.

To summarize, we have explored several issues regarding the proposal of Intersecting Brane Worlds as string-based models describing semi-realistic low-energy physics. We have found that they provide an interesting framework where just the Standard Model chiral content and gauge group can be obtained. They allow, as well, to translate physical quantities of the effective field theory into geometrical objects on the string construction, shedding new light onto their behaviour and significance. Finally, they relate phenomenological questions as the structure of quark and lepton masses and textures to very abstract mathematical notions as the construction of Fukaya's category in the context of Homological Mirror Symmetry.

\appendix

\chapter{Q-basis formalism \label{qbasis}}

Along the thesis, and specially in Chapter \ref{t&a}, we have made use of a general language for describing D-brane charges under Ramond-Ramond fields. This language has been baptized as $q$-basis formalism and, roughly speaking, consist of encoding the quantized RR charges of D-branes and O-planes in elements of a discrete vector space. In particular, to each D$(3+n)$-brane or O$(3+n)$-plane wrapping an $n$-cycle on $T^{2n}$ and sitting in a $\cpx^{3-n}/\inte_N$ singularity we can associate a vector $\q$ on $\inte^{2^n} \ti \inte_N$.The latter vector space classifies the RR charges of these objects. 

This formalisms allows to unify all the different constructions considered in this thesis (i.e., $n = 1,2,3$ in the setup above and orientifold twist of them) into a single picture. It allows, as well, to naturally extend some theoretical results, based on configurations of D-branes at angles, to D-branes wrapping general cycles. In addition, we have seen that RR tadpole conditions finally amount to the cancellation of the total RR charge in compact spaces. Hence the language of $q$-vectors is particularly well-suited for expressing them, giving us simple linear formulas for RR tadpoles. This language is also useful when checking the effect of RR tadpoles on cancellation of chiral anomalies in the effective theory, since all the computations that have to be done can be translated into linear algebra operations into the vector space of RR charges. We have made extensive use of this fact in Chapter \ref{t&a}.

As mentioned in the text, the $q$-basis formalism is inspired in the geometric interpretation of Chan-Paton charges in supersymmetric orbifold singularities. Indeed, a fractional D-brane with a definite Chan-Paton orbifold phase $\a^s$, can also be seen as a D-brane of higher dimension wrapped on a collapsed cycle $\Sig_s$ on the singularity. Is easy to generalize this same concept of collapsed cycles to non-supersymmetric singularities, see \cite{Uranga:2002pg}. In this picture, $q$-vectors represent nothing but elements on the vector space of homology classes on some resolved singularity, which classify RR charges of the D-branes in the large volume limit.

\section{General properties}

Let us first describe, in terms of the $q$-basis formalism, some linear operators which can be defined in a general D-brane system, and derive some algebraic properties of them valid in any particular representation. Let $\q_a$, $\q_b$ describe (the RR charges of) two D-branes $a$ and $b$. The $U(N_\a)$ gauge theory on each D-brane will be given by the number of D-branes, respectively $N_a$ and $N_b$, on each of the two stacks. The net chiral spectrum on the $ab$ sector will consist of massless fermions transforming in the bifundamental $(N_a,\bar N_b)$. The multiplicity of those will be given by the intersection number 
of both D-branes, defined as
\beq
\I_{ab} \equiv \q_a^{\ \dagger\ }\I \q_b \quad \in \quad \inte.
\label{numinter}
\eeq
Thus, in the $q$-basis formalism, the intersection form $\I$ will be given by a bilinear form. The sign of $\I_{ab}$ indicates the chirality of such fermions. Since the $ba$ sector of the theory contains the antiparticles of the $ab$ sector, we deduce that $\I_{ba} = - I_{ab}$ and $\I$ will be expressed as antihermitic matrix.

In orientifold models, each D-brane $a$ will have a mirror D-brane $a$*. Both are related by the action of a linear operator $\OR$, wich satisfies
\beq
\OR \q_a = \q_{a^*}, \quad (\OR)^2 = {\rm Id}.
\label{accionom}
\eeq
This operator has a geometrical action and also reverses orientation, thus relating $ab$ and $b$*$a$* sectors. These two sectors have to be identified under the orientifold quotient, in particular the chiral spectrum of both have to be the same, hence their intersection number:
\beq
\I_{ab} = \I_{b^*a^*} \quad \Raw \quad
\q_a^{\ \dagger\ }\I \q_b 
= \left( \OR\q_b \right)^{\dagger}\I\left( \OR\q_a \right)  \quad
\Raw \quad  \I \OR = - \OR^{\dagger}\I,
\label{demo}
\eeq
where we have made use of $\I$ being antihermitic and of (\ref{accionom}). From $\OR$ we can also define the linear operators $P_\pm \equiv \med \left({\rm Id} \pm \O \right)$, which satisfy
\bea
P_\pm^2 = P_\pm, & & P_\pm \cdot P_\mp = 0,
\label{proyectores}
\eea
being projectors on the space of $q$-vectors. From (\ref{demo}) we also see that 
\beq
P_\mp^{\dagger\ }\I = \I P_\pm,  
\label{prop}
\eeq
which will be a fundamental property when deducing the cancelation of anomalies from RR tadpole conditions. 

\section{D6-branes on $T^6$}

The case of D6-branes wrapping 3-cycles of $T^6$ is the simplest one, since no orbifold phases appear in the construction and hence RR charges are fully classified by homology classes in $H_3(T^6,\inte)$. Notice that, in principle, this vector space defines a lattice of dimension $b_3 = 20$. We will restric, however, to the subspace $\left( H_1(T^2,\inte)\right)^3$ spanned by {\it Lagrangian} 3-cycles on $T^2 \ti T^2 \ti T^2$ and which is a eight-dimensional sublattice of $H_3(T^6,\inte)$. A basis in such space is given by 
\begin{center}

\begin{tabular}{ccc}
$q$ comp. & 3-cycles & fact. comp. \\ \hline
$q_1$ & $[a_1] \times [a_2] \times [a_3]$ & $n^{(1)} n^{(2)} n^{(3)}$ \\
$q_2$ & $[b_1] \times [b_2] \times [b_3]$ & $m^{(1)} m^{(2)} m^{(3)}$ \\
$q_3$ & $[a_1] \times [b_2] \times [b_3]$ & $n^{(1)} m^{(2)} m^{(3)}$ \\
$q_4$ & $[b_1] \times [a_2] \times [a_3]$ & $m^{(1)} n^{(2)} n^{(3)}$ \\
$q_5$ & $[b_1] \times [a_2] \times [b_3]$ & $m^{(1)} n^{(2)} m^{(3)}$ \\
$q_6$ & $[a_1] \times [b_2] \times [a_3]$ & $n^{(1)} m^{(2)} n^{(3)}$ \\
$q_7$ & $[b_1] \times [b_2] \times [a_3]$ & $m^{(1)} m^{(2)} n^{(3)}$ \\
$q_8$ & $[a_1] \times [a_2] \times [b_3]$ & $n^{(1)} n^{(2)} m^{(3)}$ \\
\end{tabular}

\end{center}
where in the first column we have written down the components of the vector $\q$, in the second the corresponding 3-cycle in $T^{2} \ti T^{2} \ti T^{2}$ in such basis and finally, in the third column we have written the value that this vector $\q$ entry would have if coming from a D6-brane wrapping a factorizable 3-cycle $[\Pi]$, i.e., given by 
\beq
[\Pi_a] = \left(n_a^{(1)} [a_1] + m_a^{(1)} [b_1]\right) \otimes
\left(n_a^{(2)} [a_2] + m_a^{(2)} [b_2]\right) \otimes
\left(n_a^{(3)} [a_3] + m_a^{(3)} [b_3]\right).
\label{factorisable4}
\eeq
Notice that this is indeed the case for the explicit models that we have presented in the main text. The language of $q$-basis, however, allows to consider general cycles on the same footing.

On this subspace of the middle homology of $T^6$, the intersection matrix can be expressed as
\beq
\I =
\left(
\begin{array}{cccccccc}
0 & 1 & & & & & & \\
-1 & 0 & & & & & & \\
 & & 0 & 1 & & & & \\ 
 & & -1 & 0 & & & & \\
 & & & & 0 & 1 & & \\
 & & & & -1 & 0 & & \\ 
 & & & & & & 0 & 1 \\
 & & & & & & -1 & 0 \\
\end{array}
\right) 
\label{matrizD6}
\eeq
while the operator $\OR$ implementing the orientifold twist in the $q$-basis space will be given by
\beq
\OR = 
\left(
\begin{array}{cccccccc}
1 & 0 & 0 & 0 & 0 & 0 & 0 & 0\\
-b^{(1)}b^{(2)}b^{(3)} & -1 & -b^{(1)} & -b^{(2)}b^{(3)} & -b^{(2)} 
& -b^{(1)}b^{(3)} & -b^{(3)} & -b^{(2)}b^{(1)} \\
b^{(2)}b^{(3)} & 0 & 1 & 0 & 0 & b^{(3)} & 0 & b^{(2)} \\ 
-b^{(1)} &0 & 0 & -1 &0 &0 &0 &0 \\
b^{(1)}b^{(3)} &0 &0 &b^{(3)} & 1 & 0 &0 &b^{(1)} \\
-b^{(2)} &0 &0 &0 & 0 & -1 &0 &0 \\ 
b^{(1)}b^{(2)} &0 &0 &b^{(2)} &0 &b^{(1)} & 1 & 0 \\
-b^{(3)} &0 &0 &0 &0 &0 & 0 & -1 \\
\end{array}
\right) 
\eeq
where $b^{(i)} = 0,\med$ stands for the choice of either rectangular or tilted geometry in the $i^{th}$ two-torus (see figure \ref{bflux}). From these two matrices is possible to compute the proyectors $P_\pm$ and check that they satisfy the algebraic properties described above. It is also straightfordward to check that $\OR$ leaves invariant the class of the 3-cycle where the O6-plane wraps
\beq
[\Pi_{O6}] = \bigotimes_{i=1}^3 \left({1 \over 1 - b^{(i)}}[a_i] - 2 b^{(i)} [b_i]\right).
\label{ciclooriap}
\eeq
In general there will be $N_{O6} = 8\b^1\b^2\b^3$ such orientifold planes in the compactification, and their relative charge under untwisted RR fields compared to those of D6-branes is $- Q_{Op}  = - 4$.

Up to now we have considered the entries of the vector $\q$ to be given by integer numbers. This should be indeed the case if we take our $q$-basis to correspond to homology cycles. During the text, however, we have seen that it turns useful sometimes to express the wrapping numbers of a factorizable $n$-cycle in terms of fractional 1-cycles, defined in (\ref{frac}). We could then try to take such fractional 1-cycles as defining our $q$-basis, instead of the more natural choice above. This has the clear advantage of yielding a simpler orientiold action, which now reads
\beq
\OR' = 
\left(
\begin{array}{cccccccc}
1 & 0 & 0 & 0 & 0 & 0 & 0 & 0\\
0 & -1 & 0 & 0 & 0 
& 0 & 0 & 0 \\ 
0 & 0 & 1 & 0 & 0 & 0 & 0 & 0 \\ 
0 &0 & 0 & -1 &0 &0 &0 &0 \\
0 &0 &0 & 0 & 1 & 0 &0 & 0 \\
0 &0 &0 &0 & 0 & -1 &0 &0 \\ 
0 &0 &0 & 0 &0 & 0 & 1 & 0 \\
0 &0 &0 &0 &0 &0 & 0 & -1 \\
\end{array}
\right) 
\eeq
whereas the action of the projectors on a vector $\q$ is given by
\bea
P_+^{\ \prime} \q = \left(
\begin{array}{c}
n^{(1)} n^{(2)} n^{(3)} \\
0 \\
n^{(1)} m^{(2)} m^{(3)} \\
0 \\
m^{(1)} n^{(2)} m^{(3)} \\
0 \\
m^{(1)} m^{(2)} n^{(3)} \\
0
\end{array}
\right), & &
P_-^{\ \prime} \q = \left(
\begin{array}{c}
0 \\
m^{(1)} m^{(2)} m^{(3)} \\
0 \\
m^{(1)} n^{(2)} n^{(3)} \\
0 \\
n^{(1)} m^{(2)} n^{(3)} \\
0 \\
n^{(1)} n^{(2)} m^{(3)}
\end{array}
\right).
\label{accionproyD6}
\eea
Notice that this action reduces to the usual one when $b^{(i)}=0$ $\forall i$, since in this case fractional and natural 1-cycles are actually the same thing. The vector $\q$ describing the O6-plane does also have a simpler expression in this fractional choice of $q$-basis
\beq
N_{O6}^{\ \prime} \cdot Q_{O6}^{\ \prime} \cdot \q_{O6}^{\ \prime} = \left(
\begin{array}{c}
32 \\
0 \\
0 \\
0 \\
0 \\
0 \\
0 \\
0
\end{array}
\right).
\label{qoriD6}
\eeq

\section{D5-branes on $T^4 \ti \cpx/\inte_N$}

Let us now consider D5-branes wrapping 2-cycles of $T^4$ and sitting at the orbifold singularity $\cpx/\inte_N$. Now, the classification of RR charges will be a tensor product of two spaces. First let us consider the space homology of homology classes $H_2(T^4,\inte)$. We will again restrict to a sublattice of Lagrangian 2-cycles on $T^2 \ti T^2$, which is $\left(H_1(T^2,\inte)\right)^2$. The basis for such vector space can be taken to be
\begin{center}

\begin{tabular}{ccc}
comp. $q$ & 2-ciclo & comp. factorizables \\ \hline
$q_1$ & $[a_1] \times [a_2]$ & $n^{(1)} n^{(2)}$ \\
$q_2$ & $[b_1] \times [b_2]$ & $m^{(1)} m^{(2)}$ \\
$q_3$ & $[a_1] \times [b_2]$ & $n^{(1)} m^{(2)}$ \\
$q_4$ & $[b_1] \times [a_2]$ & $m^{(1)} n^{(2)}$ \\
\end{tabular}

\end{center}

Second, a fractional D5-brane sitting on an orbifold singularity will have associated a phase which is a power of $\a \equiv e^{2\pi i/N}$, encoding the action of the $\inte_N$ orbifold group on the Chan-Paton degrees of freedom. This phase will also enter the RR charge of such D5-brane under twisted RR fields. If a D5-brane $a$ wraps a 2-cycle $[\Pi_a]$ and its orbifold phase $\a^i$, then the corresponding $q$-vector will be denoted by
\beq
\q_{a,i} = \left(
\begin{array}{c}
q_1 \\ q_2 \\ q_3 \\ q_4
\end{array}\right) \a^i.
\label{qD5}
\eeq

Since now we are describing our $q$-vectors by (a discrete choice of) complex numbers, the intersection matrix will not longer be real but hermitian, and we will define it such that it satisfies $\q_{a,i}^{\ \dagger\ } \I \ \q_{b,j} \equiv I_{ab} \d(i,j)$, where $\d(i,j) = \d_{i,j+1} - \d_{i,j-1}$ encodes the chirality structure in a $\cpx/\inte_N$ quiver (see chapter \ref{spectrum}). Actually, if we want to express $\I$ as an actual matrix, then we must consider $q$-vectors of $4N$ components, al of them vanising except four given by (\ref{qD5}). We can now express the entries of $\I$ as $\I_{ri,sj} = I_{rs} \cdot \d(i,j) \a^i \bar \a^j$, where $r, s = 1, 2, 3, 4$ index the basis elements in $\left(H_1(T^2,\inte)\right)^2$ and $I_{rs}$ is the intersection matrix in such homology space
\beq
I = \left(
\begin{array}{cccc}
0 & 1 & 0 & 0 \\
1 & 0 & 0 & 0 \\
0 & 0 & 0 & -1 \\
0 & 0 & -1 & 0
\end{array}
\right).
\label{matrizD5}
\eeq
Let us, however, express our $q$-basis vectors with only four entries, as in (\ref{qD5}). The orientifold action will now act both on the homology and orbifold charges, the latter by complex conjugation:
\beq
\OR = \left(
\begin{array}{cccc}
1 & 0 & 0 & 0\\
b^{(1)}b^{(2)} & 1 & b^{(1)} & b^{(2)} \\
-b^{(2)} & 0 & -1 & 0 \\
- b^{(1)} & 0 & 0 & -1 
\end{array} 
\right) \otimes (\a \mapsto \bar\a),
\eeq
and so we will have the action $\OR \q_{a,i} = \q_{a^*,-i}$.

Finally, the language of fractional 1-cycles allows to express everything on a simpler way. In particular, the action of the projectors will be given by
\bea
P_+^{\ \prime} \q_{a,i} = \left(
\begin{array}{c}
n_a^{(1)} n_a^{(2)} (\a^i + \a^{-i}) \\
m_a^{(1)} m_a^{(2)} (\a^i + \a^{-i}) \\
n_a^{(1)} m_a^{(2)} (\a^i - \a^{-i}) \\
m_a^{(1)} n_a^{(2)} (\a^i - \a^{-i}) 
\end{array}
\right), & &
P_-^{\ \prime}  \q_{a,i} = \left(
\begin{array}{c}
n_a^{(1)} n_a^{(2)} (\a^i - \a^{-i}) \\
m_a^{(1)} m_a^{(2)} (\a^i - \a^{-i}) \\
n_a^{(1)} m_a^{(2)} (\a^i + \a^{-i}) \\
m_a^{(1)} n_a^{(2)} (\a^i + \a^{-i}) 
\end{array}
\right).
\label{accionproyD5}
\eea

There will be in general several O5-planes, all of them on the same 2-cycle of $T^2 \ti T^2$ fixed by the geometrical action of $\OR$ and with different orbifold phases. In terms of a fractional $q$-basis we can express its charges as
\beq
N_{O5}^{\ \prime} \cdot Q_{O5}^{\ \prime} \cdot \q_{O5,i} = \left(
\begin{array}{c}
16\eta \\
0 \\
0 \\
0 
\end{array}
\right) \a^i, \quad
N_{O5}^{\ \prime} \cdot Q_{O5}^{\ \prime} \cdot \q_{O5,-i} = \left(
\begin{array}{c}
16\eta \\
0 \\
0 \\
0 
\end{array}
\right) \bar\a^{i},
\eeq
\beq
\begin{array}{c} \vspace{0.2cm}
i = 1,3,\dots,{N-1 \over 2} \\
\eta = \left\{\begin{array}{l}
+1 \ {\rm if} \ N = -1  \ {\rm mod} \ 4\\
-1 \ {\rm if} \ N = 1 \ {\rm mod} \ 4
\end{array}\right.
\end{array}
\label{qoriD5}
\eeq
Notice that if we add all these RR charges we get a real number, as we would expect since it must be invariant under the action of $\OR$.

\section{D4-branes on $T^{2} \ti \cpx^{2}/\inte_N$}

The structure of the $q$-basis in configurations of D4-branes on $T^2 \ti \cpx^{2}/\inte_N$ is quite similar. The vector $\q_{a,i}$ will now have the simple expression
\beq
\q_{a,i} = \left(
\begin{array}{c}
n_a \\ m_a 
\end{array}\right) \a^i.
\label{qD4}
\eeq
where $(n_a,  m_a)$ are the usual wrapping numbers in a $T^2$, and $\a^i$ the $\inte_N$ orbifold phase. The intersection matrix is again defined by $\q_{a,i}^{\ \dagger\ } \I \q_{b,j} \equiv I_{ab} \d(i,j)$, where now $\d(i,j)  \equiv \d_{i,j+\frac{b_1+b_2}{2}} + \d_{i,j-\frac{b_1+b_2}{2}} - \d_{i,j+\frac{b_1-b_2}{2}} -  \d_{i,j-\frac{b_1-b_2}{2}}$, for a general orbifold twist $v_\om = \frac 1N (0,0,b_1, b_2)$. The orientifold action has also a simple expression:
\beq
\O = \left(
\begin{array}{cc}
1 & 0 \\
- b & 1 \\
\end{array} 
\right) \otimes (\a \mapsto \bar\a).
\eeq
Again, the action of the projectors has a simpler form in the basis of fractional 1-cycles, where is given by
\bea
P_+ \q_{a,i} = \left(
\begin{array}{c}
n_a (\a^i + \a^{-i}) \\
m_a (\a^i - \a^{-i}) 
\end{array}
\right), & &
P_- \q_{a,i} = \left(
\begin{array}{c}
n_a (\a^i - \a^{-i}) \\
m_a (\a^i + \a^{-i}) 
\end{array}
\right).
\label{accionproyD4}
\eea
Finally, one can deduce the vector of charges for the O4-planes in this fractional language. In case of supersymmetric twists, the phases of such O4-planes can be deduced from (\ref{decomp2}) and (\ref{iota}).

\chapter{K-theory constraints \label{Ktheory}}

In Chapter \ref{t&a} we have deduced the necessary conditions that any intersecting D-brane configuration must satisfy in order to yield a RR tadpole-free model. In the case of D6-branes at angles, these conditions reduce to (\ref{tadpoleD6}) in the toroidal case and to (\ref{tadpoleO6}) in the orientifolded case. These are equations in the homology space $H_3(T^6,\inte)$, which are obtained by factorisation of one-loop amplitudes or using the cohomological form of the equations of motion for the RR forms, as done in the text. These expressions suggest that RR charges are classified by vectors in $H_3(T^6,\inte)$, which is the essential idea when developing the $q$-basis formalism of Appendix \ref{qbasis}. We know, however, that D-brane RR charges should be classified by K-theory groups, rather than homology groups \cite{Witten:1998cd}. This fact may induce new constraints for RR tadpole cancellation, which are associated to K-theory charges invisible to homology. Indeed, it turns out that such constraints appear in the D6-branes orientifold models we have discussed in the text, giving us some extra (milder) constraints to the RR tadpole conditions already derived in Chapter \ref{t&a}. 

Although widely unnoticed in the literature, these constraints are satisfied by the intersecting D6-brane models built so far, in particular in those presented in the main text of this thesis. We have thus decided to discuss them in the present appendix, since they are not relevant for the main results of this work. The content of this appendix is mainly based on \cite{private}.

Let us then consider type IIA D6-branes wrapping factorisable 3-cycles of $T^2 \ti T^2 \ti T^2$, and impose the usual orientifold projection (\ref{dual}). We know that this theory is T-dual to type I string theory compactified in $T^2 \ti T^2 \ti T^2$, where magnetic fluxes are present on the worldvolumes of the D9-branes and where the two-tori are related by the T-dual maps of section \ref{Tdual}. In this type I picture, cancellation of RR charges amount to cancellation of induced D-brane charges. Indeed, in the presence of background fluxes, D9-branes carry charges under RR forms of all even degrees, due to the WZ world-volume couplings \cite{Li:1995pq,Douglas:1995bn}. The tadpole conditions deduced in Chapter \ref{t&a}
{\bea
\sum_a N_a \ n_a^{(1)}n_a^{(2)}n_a^{(3)} & = & 16 \label{tadpoleO6ap1} \\ 
\sum_a N_a \ n_a^{(1)}m_a^{(2)}m_a^{(3)} & = & 0 \label{tadpoleO6ap2}\\ 
\sum_a N_a \ m_a^{(1)}n_a^{(2)}m_a^{(3)} & = & 0 \label{tadpoleO6ap3}\\ 
\sum_a N_a \ m_a^{(1)}m_a^{(2)}n_a^{(3)} & = & 0 \label{tadpoleO6ap4}
\eea}
are now interpreted as the cancellation of D9-brane charge (eq.(\ref{tadpoleO6ap1})) and cancellation of D5$_i$-brane charge, with $i=1,2,3$ labeling the $T^2$ where the D5-brane is wrapped (eqs.(\ref{tadpoleO6ap2}),(\ref{tadpoleO6ap3}),(\ref{tadpoleO6ap4})). Notice that these are the objects charged under the type I RR fields $A_6$ and $A_{10}$.

There are however additional constraints, related to the existence of non-BPS D-branes of type I theory carrying non-trivial K-theory charges. Let us consider a type I non-BPS $\wh{{\rm D7}}$-brane, which is constructed as a pair of one D7 and one $\ov{{\rm D7}}$-brane in type IIB theory, exchanged by the action of world-sheet parity $\O$ \cite{Witten:1998cd}. Such non-BPS object carries a non-trivial $\inte_2$ charge, classified by K-theory and invisible in cohomology. Just as usual type I RR charges, in order to have a consistent RR tadpole-free compactification, these K-theory torsion charges must cancel globally in compact transverse spaces. In our setup we can define three different kinds of non-BPS $\wh{{\rm D7}}$-branes, denoted as $\wh{{\rm D7}}_i$, where $i=1,2,3$ labels the two-torus where the $\wh{{\rm D7}}$-brane does not wrap. K-theory conditions amount to requiring that the total $\wh{{\rm D7}}_i$-brane induced charge vanishes, hence imposing three extra conditions to the previous tadpoles. Since these charges take values in $\inte_2$, we actually only need to require an even number of induced $\wh{{\rm D7}}_i$-brane charge.

Let us first consider type I theory with no B-field on any of the internal dimensions, as well as its T-dual theory. A type IIB D7-brane wrapping the first and the second two-tori, denoted as D7$_3$, is given in the T-dual picture of type IIA D-branes at angles by a D6-brane wrapping the homology 3-cycle
\beq
[\Pi_{D7_3}] = [(1,0)] \otimes [(1,0)] \otimes [(0,1)],
\label{D73}
\eeq
whereas a type IIB $\ov{{\rm D7}}_3$-brane is given by
\beq
[\Pi_{\ov{D7}_3}] = [(1,0)] \otimes [(1,0)] \otimes [(0,-1)].
\label{antiD73}
\eeq
Notice that both objects are related by the orientifold action (\ref{conjugation}), and hence when modding out type IIA by $\OR$ they will be part of the single non-BPS object T-dual to $\wh{{\rm D7}}_3$. Although such pair does not contribute to the tadpole conditions (\ref{tadpoleO6ap1}-\ref{tadpoleO6ap4}) it does contribute to some RR fields carrying the $\inte_2$ K-theory charge present in the T-dual type I picture. In general, a stack of $N_a$ D6-branes wrapping an arbitrary factorisable 3-cycle (\ref{factorisable}) will also couple to such a RR field as $N_a \ n_a^{(1)}n_a^{(2)}m_a^{(3)}$, in units of $[\Pi_{D7_3}]$ charge, whereas its orientifold image will carry the corresponding $[\Pi_{\ov{D7}_3}]$ charge. We can apply the same considerations to the non-BPS charges of $\wh{{\rm D7}}_1$ and $\wh{{\rm D7}}_2$-branes. We finally obtain that the three extra K-theory conditions are given by
{\bea
\sum_a N_a \ m_a^{(1)}n_a^{(2)}n_a^{(3)} & = & {\rm even} \label{Kth1} \\ 
\sum_a N_a \ n_a^{(1)}m_a^{(2)}n_a^{(3)} & = & {\rm even} \label{Kth2}\\ 
\sum_a N_a \ n_a^{(1)}n_a^{(2)}m_a^{(3)} & = & {\rm even} \label{Kth3}
\eea}
where we are not summing over the orientifold images.

One may now wonder why should such conditions be important in order to guarantee the consistency of the configuration. Indeed, it is not difficult to construct a chiral model where conditions (\ref{tadpoleO6ap1}-\ref{tadpoleO6ap4}) are satisfied whereas (\ref{Kth1}-\ref{Kth3}) are not. The low-energy spectrum of this model will not have any kind of chiral anomalies, since, as we have seen in Chapter \ref{t&a}, they only depend on cancellation of usual tadpoles. As pointed out in \cite{Uranga:2000xp} a simple way to see such inconsistency is by introducing D-brane probes and looking at the effective theory on their worldvolumes. In the particular case of type I non-BPS $\wh{{\rm D7}}_i$-branes the appropriate probe to introduce is a type I (dynamical) D5$_i$-brane 
\footnote{If one may be concerned with the anomalies introduced by this probe in the configuration, i.e., a D5-brane will modify any of the tadpole conditions in (\ref{tadpoleO6ap2}-\ref{tadpoleO6ap4}). It then suffices to introduce an `antiprobe' (in this case an $\ov{{\rm D5}}_i$-brane) as well.},
yielding a $SU(2)$ gauge group on its worldvolume. Now, the corresponding $SU(2)$ $D=4$ effective theory will suffer from a global gauge anomaly if the number of $D=4$ Weyl fermions charged under such $SU(2)$ gauge group is odd, which will render this field theory inconsistent at the quantum level \cite{Witten:fp}. As noticed in \cite{Uranga:2000xp}, the absence of such anomaly precisely matches with the global cancellation of the $\inte_2$ $\wh{{\rm D7}}_i$-brane charge.

In principle, we could use the probe idea to look for extra RR tadpole conditions, associated to $\inte_2$ K-theory charges, in arbitrary orientifold compactifications. For this we only need to consider D-branes fixed under the orientifold action and giving rise to a $USp(2) \simeq SU(2)$ gauge group. Then we must compute the constraints arising from the potential global gauge anomaly in their worldvolumes, and finally we should interpret such constraints as tadpole conditions on a general D-brane configuration.

Let us apply this general procedure to type IIA D6-branes at angles on $T^2 \ti T^2 \ti T^2$, which is nothing but the T-dual framework to the the type I picture described above. Let us consider rectangular two-tori, i.e., $\b^1 = \b^2 = \b^3 = 1$. A $USp(2)$ gauge group can be obtained by the T-dual analogue of a type I D5$_3$-brane which, in terms of D6-branes at angles, corresponds to the homology class of 3-cycles
\beq
[\Pi_{D5_3}] = [(0,1)] \otimes [(0,1)] \otimes [(1,0)].
\label{D53}
\eeq
The net number of $D=4$ Weyl fermions charged under such $SU(2)$ gauge group will be given, as usual, by the intersection number of such D6-brane with each of the D6-branes of the configuration, namely
\beq
\sum_a N_a [\Pi_{D5_3}]\cdot [\Pi_a] = \sum_a N_a \ n_a^{(1)}n_a^{(2)}m_a^{(3)}.
\label{Ksum}
\eeq
By imposing this number to be even, we recover the constraint (\ref{Kth3}). Analogous considerations involving permutation of two-tori lead us to constraints (\ref{Kth1}) and (\ref{Kth2}), so we find that this simple procedure allow us to rederive the K-theory constraints in intersecting D6-brane configurations.

Let us now consider the case when one of the two-tori, say the third, is tilted ($\b^3= \med$). In this case the D6-branes yielding $USp(2)$ gauge groups are given by
{\beq
\begin{array}{rcl} \vspace{0.1cm}
\quad [\Pi_{D5_1}] & = & [(1,0)] \otimes [(0,1)] \otimes [(0,1)], \\ \vspace{0.1cm}
\quad [\Pi_{D5_2}] & = & [(0,1)] \otimes [(1,0)] \otimes [(0,1)], \\ 
\quad [\Pi_{D5_3}] & = & [(0,1)] \otimes [(0,1)] \otimes [(2,0)],
\end{array}
\label{D5bflux}
\eeq}
as well as their antibranes. Again we are using the language of fractional 1-cycles (\ref{frac}). There is actually one further way of getting a $USp(2)$ gauge group, which is by means of a D6-brane parallel to the O6-plane in all the six two-tori:
{\beq
\begin{array}{rcl}
\quad [\Pi_{D9}] & = & [(1,0)] \otimes [(1,0)] \otimes [(2,0)]. 
\end{array}
\label{D9bflux}
\eeq}
Indeed, such D6-brane (plus its orientifold image) is T-dual to a (dynamical) type I D9-brane with no magnetic fluxes induced on its worldvolume. Now, in this type I picture the previous tilted geometry translates into a discrete B-flux in the third two-torus (see section \ref{oricomp}). In such setup, it is possible to obtain either an $SO(2)$ or a $USp(2)$ gauge group in the D9-brane worldvolume, as has been shown in \cite{Witten:1997bs}.

The constraints arising from the probes (\ref{D5bflux}) and (\ref{D9bflux}) are
\beq
\begin{array}{ccccc} \vspace{0.1cm}
\sum_a N_a [\Pi_{D5_1}]\cdot [\Pi_a] & = & \sum_a N_a \ m_a^{(1)}n_a^{(2)}n_a^{(3)} & = & {\rm even}, \\ \vspace{0.1cm}
\sum_a N_a [\Pi_{D5_2}]\cdot [\Pi_a] & = & \sum_a N_a \ n_a^{(1)}m_a^{(2)}n_a^{(3)} & = & {\rm even}, \\ \vspace{0.1cm}
\sum_a N_a [\Pi_{D5_3}]\cdot [\Pi_a] & = & 2 \cdot \sum_a N_a \ n_a^{(1)}n_a^{(2)}m_a^{(3)} & = & {\rm even}, \\ 
\sum_a N_a [\Pi_{D9}]\cdot [\Pi_a] & = & 2 \cdot \sum_a N_a \ m_a^{(1)}m_a^{(2)}m_a^{(3)} & = & {\rm even}.
\end{array}
\label{Ksumbflux}
\eeq

Let us see how stringent these constraints are. Consider the first of these conditions, and notice that, given a particular stack $a$ of the configuration
\bea
N_a [\Pi_{D5_1}]\cdot [\Pi_a] = N_a \ m_a^{(1)}n_a^{(2)}n_a^{(3)}  =  {\rm odd} & \Longleftrightarrow  & 2 \cdot N_a \ m_a^{(1)}n_a^{(2)}m_a^{(3)} =  {\rm odd},
\label{iff}
\eea
by the very definition of (\ref{frac}) of the fractional 1-cycle $(n_a^{(3)},m_a^{(3)})$. As a consequence of this,
\bea
\sum_a N_a \ m_a^{(1)}n_a^{(2)}n_a^{(3)}  =  {\rm even} & \Longleftrightarrow  & 2 \cdot \sum_a N_a \ m_a^{(1)}n_a^{(2)}m_a^{(3)} =  {\rm even}.
\label{iff2}
\eea
However, by the usual tadpole constraints, more precisely by (\ref{tadpoleO6ap3}), this will always happen. As a result, the first potential K-theory constraint of (\ref{Ksumbflux}) is always satisfied if usual tadpole conditions are. Similar considerations can be taken for the rest of the constraints in (\ref{tadpoleO6ap3}) and so it seem that, when considering one tilted torus, the extra K-theory constrains are a consequence of usual tadpoles (\ref{tadpoleO6ap1}-\ref{tadpoleO6ap4}.

The last statement is not strictly correct, as a more detailed analysis shows. Let us consider the third constraint in (\ref{tadpoleO6ap3}). Following the same computations above, we arrive at the equality
\bea
2 \cdot \sum_a N_a \ n_a^{(1)}n_a^{(2)}m_a^{(3)}  =  {\rm even} 
& \Longleftrightarrow  &  
\sum_a N_a \ n_a^{(1)}n_a^{(2)}n_a^{(3)} =  {\rm even},
\label{iff3}
\eea
which seems to be guaranteed by tadpole condition (\ref{tadpoleO6ap1}). This is not actually true. Indeed, notice that in the sum (\ref{tadpoleO6ap1}) we are not considering the mirror image $a$* of each stack $a$. What shall we do when computing the contribution of a stack $b$ which is its own mirror image? This may happen when dealing with a stack of $N_b$ D6-branes wrapping the 3-cycle (\ref{D9bflux}) and being mapped to themselves under the $\OR$ action. In this case either an $SO(N_b)$ or a $USp(N_b)$ gauge group will arise. In case $N_b$ is even we just consider that we have $N_b'= N_b/2$ dynamical D6-branes in our configuration, and constraint (\ref{tadpoleO6ap1}) applies the same by just counting dynamical D6-branes. On the other hand, if $N_b$ is odd (which is a possibility that only orthogonal groups admit) not all the D6-branes can be paired into dynamical D6-branes, and there must be at least one which is a fractional D6-brane stuck at the orientifold fixed position. The fractional stacks giving rise to such $SO(N_b)$ gauge groups will contribute with $\med$ the usual contribution of a stack of dynamical D6-branes in (\ref{D9bflux}) and then (\ref{tadpoleO6ap1}) will read
\beq
\sum_a N_a \ n_a^{(1)}n_a^{(2)}n_a^{(3)} + \sum_b N_b \ = \ 16,
\label{tadpoleO6ap1b}
\eeq
where the second summation is taken over the branes fixed under $\OR$. Conditions (\ref{iff3}) and (\ref{tadpoleO6ap1b}) are incompatible only if $\sum_b N_b =$ odd. An this is exactly what the K-theory constraints in (\ref{Ksumbflux}) forbid to our configuration.

This result can be restated in a simpler form, namely {\it cancellation of RR charges forbid the existence of an odd number of gauge groups of the form $SO(2N+1)$, $N \in \inte$}. It can easily be checked that, when having more than one tilted torus, extra K-theory constraints also reduce to this fact. Notice that this statement is also true when no tilted geometry is present i.e., when $\b^1 = \b^2 = \b^3 = 1$, although in that case K-theory constraints impose stronger conditions.

Although we have discussed the particular case of D6-branes at angles, the computations above can be easily extended to the orientifold models involving D4 and D5-branes trapped at orientifold singularities, obtaining similar results.

\chapter{Higher dimensional holomorphic discs \label{hdhd}}

The aim of this section is to show the existence and uniqueness of (anti)holomorphic worldsheet instantons that connect factorisable $n$-cycles in $T^{2n}$. Contrary to Euclidean intuition, these volume-minimizing surfaces are not, in general, given by flat triangles of $\real^{2n}$ but by more complicated calibrated manifolds.

Let us first describe the problem more precisely. For simplicity, instead of dealing with the complex manifold $T^2 \ti \dots \ti T^2$ equipped with a complex structure, let us consider its covering space, given by $\cpx^n$. The factorisable $n$-cycles of (\ref{vectorscpx}) are mapped to affine Lagrangian $n$-planes given by 
\beq
L_\a  = \bigotimes_{r = 1}^n
\{ t^{(r)} \cdot z_\a^{(r)} + v_\a^{(r)} | 
t^{(r)} \in \real \},
\label{affine}
\eeq
where the fixed quantities $z_\a^{(r)}, v_\a^{(r)} \in \cpx$ define the affine equation of a line in the $r^{th}$ complex plane. Of course, we should also consider all the copies of such affine subspace under $T^{2n}$ lattice translations, which implies the modification $v_\a^{(r)} \mapsto v_\a^{(r)} + l_1^{(r)} + \tau^{(r)} l_2^{(r)}$, $l_1^{(r)}, l_2^{(r)} \in \inte$ in (\ref{affine}) (see figure \ref{tri} for an example of this). For our purposes, however, it will suffice to consider just a single copy given by (\ref{affine}).

\begin{figure}[ht]
\centering
\epsfxsize=6in
\epsffile{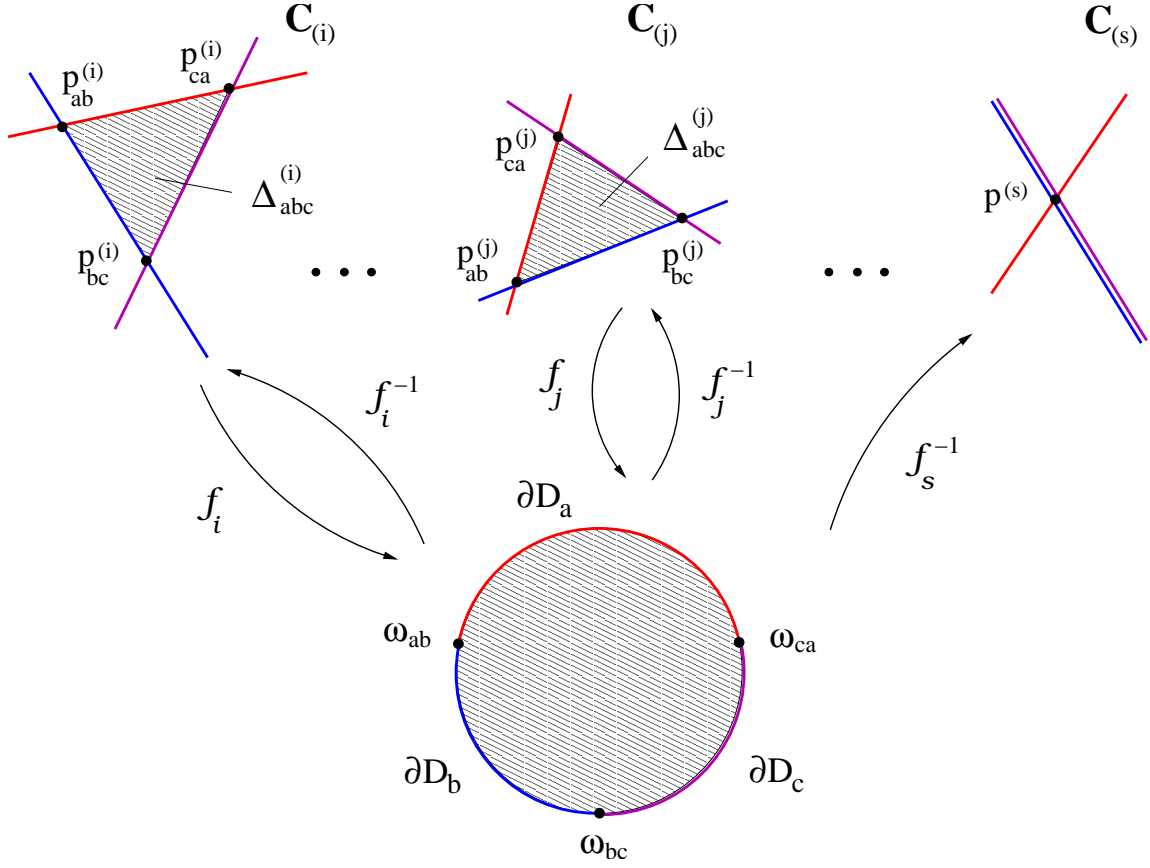}
\caption{Construction of higher dimensional holomorphic discs.}
\label{multimap}
\end{figure}

Let us now consider a triplet of such $n$-hyperplanes $(L_a, L_b, L_c)$ in $\cpx^n$. Two by two, these hyperplanes will either intersect at a single point in $\cpx^n$ either be parallel in $s$ complex dimensions. If that is the case, we will consider that they are on top on each other on such complex dimensions, intersecting in a Lagrangian hyperplane of real dimension $s < n$. We can visualize such geometry by projecting it into each complex dimension $\cpx_{(r)}$. This picture will look as three real lines $l_a^{(r)}, l_b^{(r)}, l_c^{(r)} \subset \cpx_{(r)}$ forming a triangle whose vertices are the projection of the hyperplanes' intersections. If two hyperplanes are parallel in such complex dimension, the corresponding triangle will be degenerate (see figure \ref{multimap}).

A worldsheet instanton will be described by a map $\vp : D \raw \cpx^n$ with the following properties:
\begin{itemize}

\item $\vp$ is either holomorphic or antiholomorphic.

\item 
$\vp(\om_{ab}) = L_a \cap L_b$,\quad 
$\vp(\om_{bc}) = L_b \cap L_c$,\quad
$\vp(\om_{ca}) = L_c \cap L_a$,
where $\om_{ab}, \om_{bc}, \om_{ca} \in \p D$ are counterclockwise ordered in $\p D$.

\item 
$\vp(\p D_{a}) \subset L_a$, \quad
$\vp(\p D_{b}) \subset L_b$, \quad
$\vp(\p D_{c}) \subset L_c$,
where $\phi(\p D_{a})$ is the part of $\p D$ between $\om_{ca}$ and $\om_{ab}$, etc.

\end{itemize}
Furthermore, since we are only interested in the embedded surface $\vp(D)$, we must quotient our space of solutions by $Aut(D) = PSL(2,\real)$.

%
%

Let us first focus on the simple case $n=1$, where the worldsheet instanton will have the form of a triangle $\D_{abc} \subset \cpx$. The Riemmann Mapping Theorem \cite{conway} asserts that there exist a one-to-one analytic function $f : \cpx \raw D$ such that the image of $Int(\D_{abc})$ is $Int(D)$. Moreover, this function is unique up to the group of conformal maps of the open unit disc into itself, which is given by 
\beq
{\rm Conf}(D) = \{\phi_\om^\tau(z) = \tau \cdot {z - \om \over 1 - \bar  \om z}
| \om \in Int(D), \tau \in \p D \}.
\label{confdisc}
\eeq
A several statement holding for antianalytic functions. If we also require the boundary of such regions to match under $f$, that is, if we require $f(\p \D_{abc}) = \p D$ following the second point above, then we will select one unique map from the whole family $\phi_\om^\tau \circ f$, since Con$(D) \cong PSL(2,\inte)$, and this latter group can be fixed by specifying the location of $\om_{ab}$, $\om_{bc}$ and $\om_{ca}$ in $\p D$. Moreover, the map $f$ will be either analytic or antianalytic, but not both. Finally, the third point is satisfied by continuity of $f$.

We have thus shown that, for the case $n = 1$ there exist a unique function, given by $f^{-1}$, that describes our worldsheet instanton. From this simple result we can derive the statement for general $n$. Indeed, let us consider the $n$ functions $f_{(r)}$ which are obtained from the above construction, now with the triangles $\D_{abc}^{(r)}$ which are defined from the lines $l_a^{(r)}, l_b^{(r)}, l_c^{(r)} \subset \cpx_{(r)}$. That is, we consider the same problem for each projection into $\cpx_{(r)}$. In case the triangle is degenerate (e.g., $l_b^{(s)} = l_c^{(s)}$ for some $s$) then such function does not exist, and we take $f^{-1}$ to be the constant map from $D$ to the common intersection point of the three lines. After this, is easy to see that the desired (anti)holomorphic surface is given by $\phi(D) = \vec f^{-1}  (D)$, where
\beq
\vec f^{-1}(\om) = \left(f_1^{-1}(\om), f_2^{-1}(\om), 
\dots, f_n^{-1}(\om) \right) \in \cpx^n,  \quad \quad \om \in D,
\label{surface}
\eeq
and that this surface satisfies all the three requirements above.

A consequence of the holomorphic properties of $\vec f$ is that  $S = \phi(D)$ will be a surface calibrated by the K\"ahler 2-form. It is natural to wonder which is the 'shape' of $S$ in terms of $\cpx^n$ geometry. A na\"{\i}ve guess would lead us to think that, being surface minimizing, it has a triangular shape. This will not, however, be the generic case. Indeed, let us consider the relatively simple case of two-complex dimensions. In this case, $\vec f^{-1} = \left(f_1^{-1}, f_2^{-1} \right)$ maps $D$ into $\D_{abc}^{1} \times \D_{abc}^{2} \subset \cpx^2$. We can see the surface $\vec f^{-1}  (D)$ embedded on $\cpx^2$ by considering the graph
\beq
(z_1, z_2) = (z_1 , f_2^{-1} \circ f_1 (z_1)), \quad z_1 \in  \D_{abc}^{1}.
\label{graph}
\eeq
Clearly, $f_2^{-1} \circ f_1 (\D_{abc}^{1}) = \D_{abc}^{2}$. This surface will be a triangle in $\cpx^2$ only if $f_2^{-1} \circ f_1$ is a constant complex number, that is, only if $\D_{abc}^{1}$ and $\D_{abc}^{2}$ are congruent triangles. A similar argument can be carried out for higher dimensions.


%

\end{document}